\documentclass[12pt,a4paper]{article}
\usepackage{geometry}
\usepackage[utf8]{inputenc}
\usepackage{amssymb}
\usepackage{amsfonts}
\usepackage{amsthm}
\usepackage{amsmath}
\usepackage{bbm}
\usepackage{hyperref}
\usepackage{xcolor}
\usepackage{natbib}
\usepackage{algorithm}
\usepackage{algpseudocode}
\usepackage{graphicx}
\usepackage{caption}
\usepackage[position=top]{subfig} 
\usepackage{booktabs}
\usepackage{tabularx}
\theoremstyle{remark}
\usepackage{setspace}
\usepackage{rotating}
\usepackage{adjustbox}

\hypersetup{
    colorlinks=true,
    linkcolor=blue,
    filecolor=magenta,      
    urlcolor=blue,
    citecolor = blue
    }
\usepackage[flushleft]{threeparttable}
\usepackage{titling}
\usepackage{tikz}
\usepackage{chngcntr} 

\usepackage{xcolor}
\newcommand{\E}[1]{\mathbb{E}}

\newcommand{\tabnotes}[1]{\caption*{\footnotesize{\textit{Notes.} #1}}}

\title{Copyright and Competition:\\Estimating Supply and Demand with Unstructured Data\thanks{We appreciate feedback from Steve Berry, Pierre Bod\'er\'e, Sophie Calder-Wang, Kevin Chen, Tim Christensen, Liran Einav, Phil Haile, Charles Hodgson, Alessandro Iaria, Michi Igami, Lorenzo Magnolfi, Julien Monardo, Lars Nesheim, Katja Seim, Jesse Shapiro, Wujin Sim, Jann Spiess, Pietro Tebaldi, Yanos Zylberberg, and participants at the World Congress 2025, Chicago Booth MLESC 2025, ACM EC 2025, AI at Yale Symposium 2025,  APIOC 2024, SweRIE 2024, 
and seminars at Yale and Stanford. We gratefully acknowledge Monotype Inc. for generously providing the dataset. We thank Oska Sheldon-Fentem for excellent research assistance.}}
\author{Sukjin Han\thanks{School of Economics, University of Bristol. \url{vincent.han@bristol.ac.uk}}  \and Kyungho Lee\thanks{Department of Economics, Yale University. \url{kyungho.lee@yale.edu}}}
\date{\today}

\begin{document}

\maketitle

\begin{abstract}
We study the competitive and welfare effects of copyright in creative industries in the face of cost-reducing technologies such as generative artificial intelligence. Creative products often feature \emph{unstructured} attributes (e.g., images and text) that are complex and high-dimensional. To address this challenge, we study a stylized design product---fonts---using data from the world’s largest font marketplace. We construct neural network embeddings to quantify unstructured attributes and measure visual similarity in a manner consistent with human perception. Spatial regression and event-study analyses demonstrate that competition is local in the visual characteristics space. Building on this evidence, we develop a structural model of supply and demand that incorporates embeddings and captures product positioning under copyright-based similarity constraints. Our estimates reveal consumers’ heterogeneous design preferences and producers’ cost-effective mimicry advantages. Counterfactual analyses show that copyright protection can raise consumer welfare by encouraging product relocation, and that the optimal policy depends on the interaction between copyright and cost-reducing technologies.
\vspace{0.1in}


\noindent \textit{Keywords:} Copyright, creative industries, unstructured data, embeddings, visual similarity, consumer demand, product positioning.
\end{abstract}

\section{Introduction}

Copyright policies play a pivotal role in protecting the intellectual property of creators and companies in the modern knowledge-based economy. These policies grant monopoly rights to creators and serve as gatekeepers in various sectors of the economy. This ranges from sectors as obvious as cultural industries (e.g., books, movies, music, illustrations) to less obvious ones like design industries (e.g., garments, automobiles, furniture, mobile applications). In recent years, these creative industries have witnessed the advent of a disruptive technology, namely, generative artificial intelligence (AI). Generative AI has begun to engage in a human-like creative process with significantly low cost, generating high-quality images, texts, sounds, and videos with scale and efficiency never seen before. This recent transformative trend thus calls for renewed attention to the role of copyright policies \citep{samuelson2023generative,de2024intellectual}.

The primary goal of this paper is to study competition in a market of creatively differentiated products and the role of copyright laws, especially in the context of cost-reducing technologies. A common feature of products with creative elements is that their key attributes are \emph{unstructured} and thus high-dimensional. Examples of such unstructured attributes are images and text, which are often the focus of copyright protection. Traditional economic models typically include these attributes as unobservables, because ``characteristics
such as style are inherently difficult to quantify but are frequent determinants of demand'' \citep[p.~243]{berry1994estimating}. However, this approach limits our ability to investigate copyright policies along the dimensions defined by these attributes. Explicitly quantifying these attributes and developing an economic model based on them are crucial steps towards achieving our research objectives. This is not a trivial undertaking. Products frequently possess complex unstructured attributes that are challenging to standardize, compare and analyze. As a result, mathematically characterizing copyright policies becomes a daunting task. Another challenge is that, for certain products, consumers may not value unstructured attributes as much as structured ones (e.g., product specifications), which could render the consideration of copyright of creative features less relevant. Finally, for many artistic products, markets are intrinsically thin \citep{baumol1986unnatural, ashenfelter2003auctions}.

To make progress on these challenges, we focus on a specific type of creative product—fonts—which possess several attractive features for our study. First, fonts are differentiated products where visual attributes mostly describe the characteristics of the product, highly predictive of its value and functionality. This is a feature unique to this particular product. Second, copyright issues have been important policy questions in this industry (e.g., \citealp{carroll1994protection,lipton2009c,manfredi2010sans,evans2013fonts}), as has the introduction of AI-assisted design of fonts (e.g., \citealp{zeng2019artificial,wang2020attribute2font}). Third, the product’s visual information is among the simplest of all design products (as the images are monochrome and standardized), which facilitates our dimension-reduction procedure. Owing to this visual simplicity, it is also relatively straightforward to interpret the unstructured attributes and the associated copyright policies within our economic model, a feature that is particularly useful for our counterfactual analyses. Fourth, fonts are ubiquitous and serve as intermediate goods for many final products (e.g., websites, mobile applications, printed materials), and therefore policies in this market have implications beyond the font market. Fifth, as reflected in the scale of our dataset, the fonts market is large with frequent productions and transactions. Finally, we view fonts as stylized products that capture essential aspects that many products in the market have in common, namely, design attributes and associated copyrights.

In this stylized market, we aim to understand: (i) the anatomy of competition among design products in terms of visual attributes, (ii) the role of copyright policy in protecting originality and ensuring the welfare of market participants, and (iii) the optimal level of permissible similarity (i.e., optimal variety), particularly in the presence of cost-reducing technologies such as generative AI. We use data from the world’s largest online font marketplace, which include information on nearly 33,000 fonts (created by font design firms known as foundries) and approximately 3,000,000 transactions spanning from 2014 to 2017. To achieve our goals, we initially represent font images as embeddings---low-dimensional normalized vectors---by training a modern convolutional neural network from scratch on our data with triplet loss \citep{schroff2015facenet, han2021shapes} and show the resulting representation aligns well with human perception. Given the embeddings, we characterize the competition of firms in the visual dimension as a spatial competition in the embedding space, namely in the \emph{visual characteristics space}.

Visual similarity that we compute as the Euclidean distance between embeddings serves as a crucial metric in our policy analyses. As detailed below, it forms the basis for modeling copyright policy, enabling us to conduct counterfactual analyses by varying policy stringency. Visual similarity is also a practically relevant concept, as real-world copyright infringement judgments are typically based on this criterion \citep{lemley2009our,balganesh2014judging}. Furthermore, many policymakers are particularly interested in regulating output similarity in the context of generative AI, because regulating AI's inputs (i.e., training data) may hamper innovation and competition.\footnote{For example, in July 2023, the Japanese government introduced copyright policy guidelines pertaining to generative AI. These guidelines permit the use of copyrighted materials as training inputs for AI \textit{without} permission. Instead, the guidelines enforce copyright policies through the application of existing similarity-based criteria for determining copyright infringement (\hyperlink{https://www.natlawreview.com/article/japanese-government-identified-issues-related-ai-and-copyrights}{https://www.natlawreview.com/article/japanese-government-identified-issues-related-ai-and-copyrights}).}

We begin with exploratory analyses to understand the nature of competition between products. Using the panel data and two distinct strategies to measure changes among competitors defined by proximity in the embedding space, we empirically demonstrate that firms in this market engage in \emph{local competition} in the visual characteristics space. We find that business stealing has significant and lasting impacts on sales and revenue, especially when entry occurs near the focal product. This suggests that a copyright policy providing local protection in the characteristics space would directly influence market competition and have significant welfare implications.

To study competition and welfare effects of copyright policy, we then develop an equilibrium model of supply and demand that integrates unstructured data. On the supply side, our model describes firms' location choices within the visual characteristics space as well as pricing and entry decisions. A copyright policy is modeled as imposing restrictions on the area of possible choices in the characteristics space, providing local protection to right holders. This modeling is made feasible though the embeddings we construct. On the demand side, we characterize consumers' heterogeneous preferences over visual attributes captured in the embeddings and recover flexible substitution patterns across different designs. Our modeling approach is not specific to the font market or image data; it can generally be applied to other markets and industries where products have unstructured features that are protected by policies.

Overall, our demand-side estimation results show that consumers tend to prefer products with high quality and functionality, prices decrease utility, and visual attributes are important determinants of consumer heterogeneity and substitution. In particular, the estimated model reveals that the degree of competition---as captured in consumers' substitution patterns---is effectively explained by the visual similarity measured through the embeddings. We show this explanatory power disappears when the visual attributes are dropped from the model. The estimated supply-side model indicates that the firm's development costs are low when mimicking close competitors and increase as products become more visually differentiated. This suggests \emph{mimicking externality}---the presence of incumbents reduces the fixed costs of visually similar entrants, an effect not internalized by the incumbents themselves. Such mimicry advantages complicate policymaking, since stricter copyright would restrict not only classic business stealing but also cost-effective mimicking \citep{mankiw1986free}. Therefore, the optimal degree of protection is ultimately an empirical matter, reflecting a welfare trade-off.

Using the structural model, we first assess how the stringency of copyright policy affects welfare. We find that as copyright policy becomes stricter (i.e., the protection boundary around each font expands) and infringers are removed, consumer surplus decreases, primarily due to the elimination of products with attributes preferred by consumers. However, when infringers are relocated outside the protection boundary, consumer surplus increases as the area with desirable product attributes is optimally filled through relocation.

The interplay between copyright policy and cost-reducing technologies is crucial in determining the optimal level of policy strictness and, consequently, overall welfare. We demonstrate that, in a counterfactual scenario where generative AI complements human designers, copyright protection can encourage firms to choose consumer-favorable locations in the characteristics space. Conversely, protection can also prevent mimicry advantages and thereby harm entrants. In a scenario where generative AI is a complete substitute, there is no mimicry advantage because firms can enter at low cost regardless of location, which in turn leads them to enter areas not associated with high consumer surplus. In this case, copyright plays different roles depending on whether the policymaker is concerned with aggregate or average social welfare.

This paper is structured as follows. We conclude this section by discussing our contributions to the relevant literature. Section \ref{sec:institution} introduces the institutional background of the fonts marketplace and copyright policies. Section \ref{sec:data} describes the marketplace data and presents stylized facts about the market. We then construct the neural network embeddings and the resulting visual characteristics space, and provide interpretations. Section \ref{sec:spatial} analyzes firms' spatial competition in the visual characteristics space. In Section \ref{sec:structural_model}, we build structural models and discuss issues of identification and estimation, followed by estimation results in Section \ref{sec:est-result}. Section \ref{sec:counterfactual} conducts counterfactual welfare simulation, and Section \ref{sec:conclusion} concludes.

\subsection{Related Literature} \label{sec:literature}

This research contributes to the literature employing high-dimensional unstructured data in the economics literature. For instance, \cite{gentzkow2010drives} use text data in the U.S. daily newspapers to construct an index of ideological slant in news. Based on the index they estimate consumer demand for newspapers and find that ideological preferences significantly influence newspaper demand. \cite{gentzkow2019measuring} measure political polarization by congressional speech textual data. In addition, \citet{hoberg2016text} propose product classification by creating a product location space, similar to the visual characteristics space in this paper, via 10-K product descriptions of firms. \citet{gentzkow2019text} provide a survey about textual data and its application in economics. In the realm of image data, \citet{glaeser2018big} use Google Street View data and predict economic outcomes of neighborhoods, and \citet{gorin2025} use images of paintings to predict historical economic outcomes. \citet{bajari2023hedonic} and \citet{compiani2023demand} use images and text to analyze markets, focusing on estimating hedonic prices and demand, respectively.  \citet{han2021shapes} use product images and construct embeddings to revisit market definitions and examine mergers. In contrast, the present paper develops supply and demand models that incorporate unstructured product attributes in order to conduct counterfactual analyses related to copyright policies.

This paper extends the literature on product positioning by introducing new forms of data and considering entry decisions in a high-dimensional characteristics space. Studies like \cite{berry1992estimation}, \citet{mazzeo2002product} and \cite{seim2006empirical} introduce models of firms' entry choices, utilizing cross-sectional variations in the number of firms across markets. \cite{jia2008happens} studies the entry decisions of large retailers in each location and their welfare implications for nearby small retailers. \citet{holmes2011diffusion} study Wal-Mart's location choices as a single-agent dynamic problem and trade-offs between the benefits of economies of density and cannibalization. \cite{fan2013ownership} proposes a merger analysis including firms' static product differentiation using U.S. newspaper market data. \cite{eizenberg2014upstream} studies the impact of upstream innovation, i.e. increases in CPU performance, on downstream product configuration choices in the computer market. \citet{wollmann2018trucks} investigates model-level entry and exit in the U.S. truck market, showing the importance of changes in product offerings in terms of welfare analysis. 

This paper also relates to the literature on the welfare trade-off engendered by property rights, which is a classic economic problem \citep{romer2002should, stiglitz2007economic}. Studies have used historical quasi-experimental variations to identify the effects of the copyright system on outcomes such as price, creation and quality.\footnote{For instance, \cite{li2018dead} find that the UK Copyright Act of 1814, which resulted in a differential increase in copyright length, increased prices. \cite{biasi2021effects} exploit the weakening of copyrights during World War II, highlighting that such dilution led to the creation of follow-on science, manifested as increased citations. \cite{giorcelli2020copyrights} show that the adoption of copyright policy in Italy, induced by Napol\'eon’s victories, leads to the creation and longevity of new operas.} Copyright protection is essential for the functioning of digital markets and new technology has generated policy challenges. Existing literature has paid attentions on piracy of digital products \citep{oberholzer2007effect, rob2006piracy, waldfogel2012copyright}. \footnote{To be specific, \cite{oberholzer2007effect} study the effect of file sharing on revenues in the music industry and conclude that file sharing resulted in a significant decline in music sales. \cite{rob2006piracy} use a sample of college students and report reduced expenditures on albums but an increase in consumer welfare due to downloading. \cite{waldfogel2012copyright} shows that the quality of music was not degraded due to the introduction of Napster. } To the best of our knowledge, our paper is the first attempt to address the question of permissible similarity for copyright protection in economics, a key concept in the copyright policy. 

Our study is also related to the literature on optimal product variety. It is theoretically well-documented that free entry may lead to social inefficiency \citep{dixit1977monopolistic,spence1976product2,spence1976product1,mankiw1986free,anderson1995oligopolistic}. This conclusion has motivated empirical researchers to examine inefficiency in markets and policy tools for achieving socially optimal levels. \cite{berry1999mergers, berry2001mergers} empirically demonstrate market inefficiency in the radio broadcasting market. They conclude that market concentration reduces entry yet increases product variety, using the 1996 Telecommunications Act as a quasi-experimental variation for relaxation of ownership restrictions. \cite{berry2016optimal} also report the existence of such inefficiency by extending the empirical model of \cite{berry1999mergers} with vertical differentiation of radio stations. \cite{sweeting2013dynamic} studies dynamic product positioning of radio stations and shows that high fees for music performance rights quickly decrease the number of music stations. We contribute to this literature by introducing novel mimicking externality that is particularly relevant in creative industries and by examining allowable similarity under copyright protection as a policy tool to enhance social welfare. Furthermore, \cite{de2024intellectual} emphasizes the importance of economically analyzing the welfare implications of copyright policies in the context of ``creative machines.'' We join this discourse by providing an analytical framework suited to address relevant counterfactual questions.

\section{Backgrounds} \label{sec:institution}

\subsection{Product Similarity and Copyright Policy}

The concept of \emph{substantial similarity} is fundamental in copyright infringement cases involving many kinds of creative works---including artistic, visual, and design products---and serves as a key criterion for establishing evidence of copying \citep{lemley2009our,balganesh2014judging}. According to \cite{lemley2009our}, court procedures for determining copyright infringement involve gathering and aggregating information from ``ordinary observers'' (i.e., consumers of copyrighted products) and experts knowledgeable about the characteristics of such products. This information is then used by an intellectual property judge to assess whether the ``total concept and feel'' of one product is substantially similar to another.\footnote{Court cases are abundant. A well-known example is Star Athletica, LLC v. Varsity Brands, Inc. (\hyperlink{https://harvardlawreview.org/print/vol-131/star-athletica-l-l-c-v-varsity-brands-inc/?utm_source=chatgpt.com}{U.S. Supreme Court, 2017}), which established that design or aesthetic features (such as graphic surface ornamentation) on ``useful articles'' (like clothing) can be copyrightable if those features are identifiable separately from the utilitarian aspects and are capable of existing independently of those utilitarian aspects. Another example is Emeco Industries, Inc. v. Restoration Hardware, Inc. (\hyperlink{https://www.nytimes.com/2012/10/11/garden/copying-classic-designs-is-the-focus-of-a-lawsuit-against-restoration-hardware.html}{U.S. District Court for N.D. Cal, 2012}), a legal dispute in which one furniture maker sued Restoration Hardware alleging it sold knock-off chair designs infringing on protected design rights.} As these procedure rely on human perception, the subjectivity and ambiguity of similarity judgments have faced criticism in the legal literature \citep{lemley2009our, balganesh2014judging}. Recently, legal scholars have begun to reexamine copyright issues in the context of generative AI \citep{lemley2023generative,sag2023copyright,sobel2024elements}.

The font industry offers a particularly compelling setting to examine copyright and questions of substantial similarity. Unlike many other creative industries, fonts occupy a dual role: they are creative works in their own right but also serve as essential intermediate goods in a vast array of final products (see Figure \ref{fig:fonts-examples} below). This unique position has placed the industry at the center of legal and policy debates, but no clear consensus on protection levels or enforcement mechanisms has emerged yet.\footnote{For example, although typeface designs are generally not protected under U.S. copyright law (Eltra Corp. v. Ringer), the U.S. Copyright Office has recognized the registration of computer programs that generate fonts (Adobe Systems, Inc. v. Southern Software, Inc.). In that case, the court in 1998 held that Adobe's Utopia font was protectable under copyright and Southern Software's Veracity font was substantially similar and infringing, raising questions about copyrightability and competition \citep{mezrich1998extension}. Another example is a 2001 lawsuit in the UK, where GreenStreet Technologies lost a High Court case to Linotype Library for distributing 122 infringing fonts in its collections, including Neue Helvetica (\hyperlink{https://www.pinsentmasons.com/out-law/news/typeface-copyright-decision-in-uk-high-court}{https://www.pinsentmasons.com/out-law/news/typeface-copyright-decision-in-uk-high-court}).} These challenges are further amplified by the advent of off-the-shelf and customized algorithms capable of generating new fonts \citep{wang2020attribute2font, xie2021dg, peong2024typographic}.

\subsection{Fonts}

\begin{figure}[htbp!]
    \centering
    \caption{Commercial Applications of Fonts}
    \subfloat[Product Packages and Brand Logos]{\includegraphics[width = 0.45\textwidth]{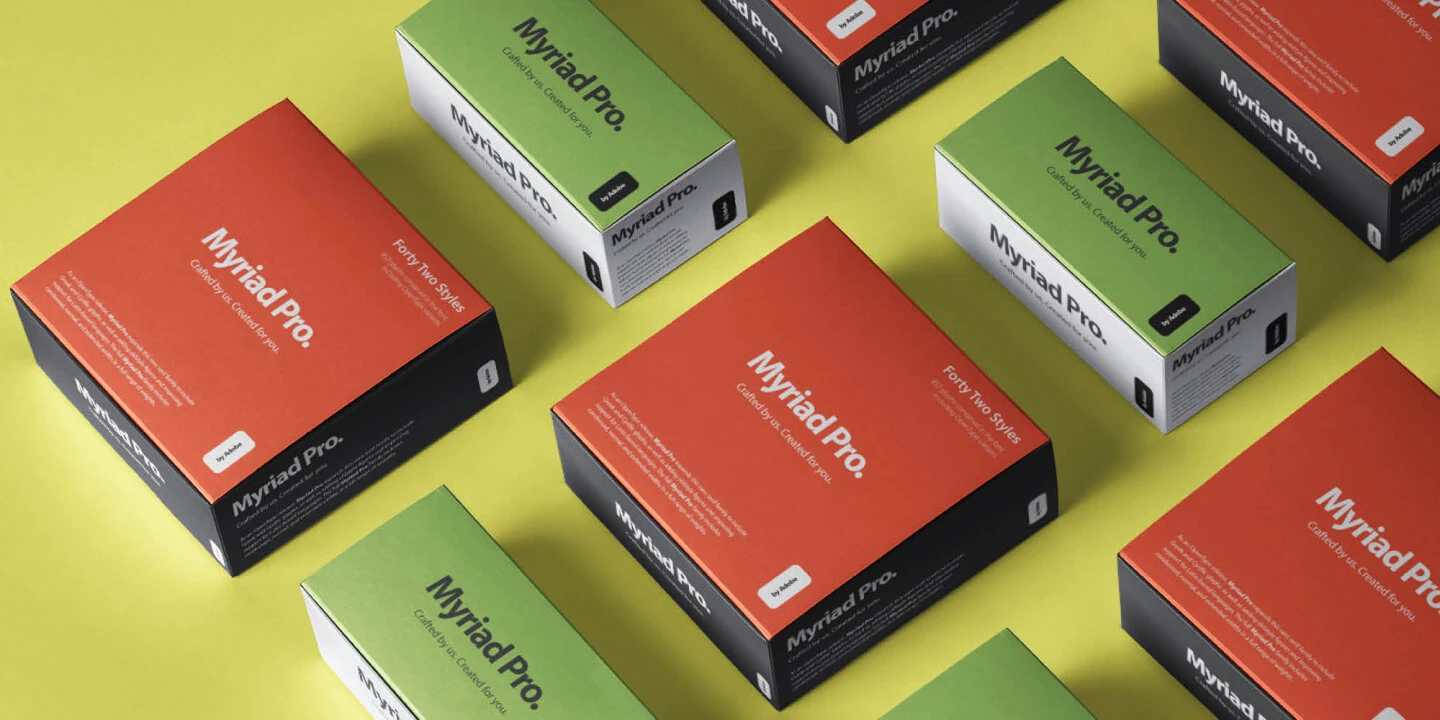}} 
    \subfloat[Books, Websites and Apps]{\includegraphics[width = 0.45\textwidth]{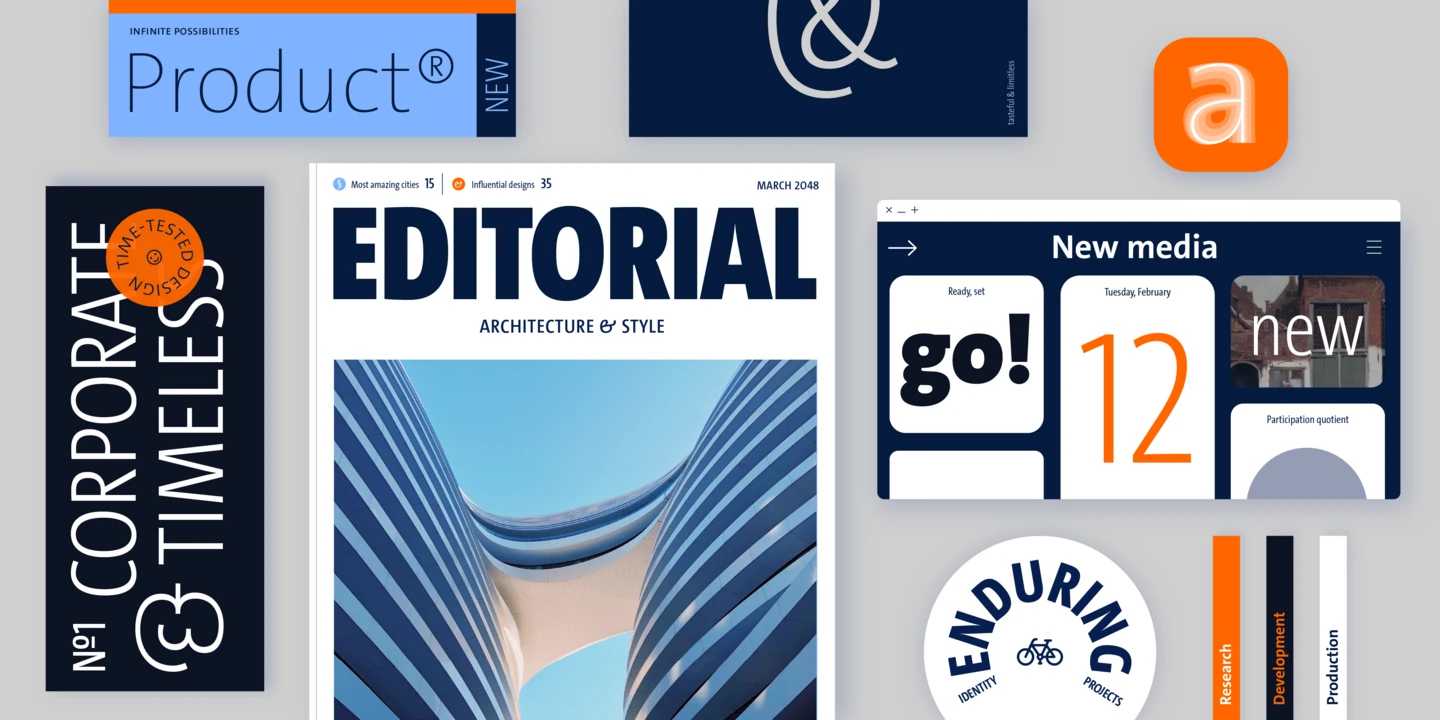}}\\
    \tabnotes{These figures show examples of commercial applications of fonts. Other examples include newspapers, posters, pamphlets, store signs, and other digital products.}
    \label{fig:fonts-examples}
\end{figure}

Fonts are recognized as software goods in the digital marketplace. The software delivers typefaces to the user and is purchased through downloads. The main consumers of fonts are designers, who use them in a wide array of commercial design projects. Examples of such design outputs include digital and printed materials (e.g., book covers and interiors, banners, advertising posters), packaging and store signs, as well as websites and mobile applications (Figure \ref{fig:fonts-examples}). In this sense, fonts serve as intermediate goods for final products and they are among the most ubiquitous objects encountered in daily life. Fonts are downloaded by consumers under specific types of licenses. For instance, a desktop license (used for printed materials) specifies the number of users who can install and use the font, while a web font license (used for websites) is based on the number of online views the font receives. In this market, the sellers are design firms known as foundries, which specialize in font production.

The market for fonts shares several characteristics with broader markets for creative goods. First, the key product attributes in this market are unstructured. This feature is related to the very reason copyright policies exist in this market. Among creative products, fonts possess arguably some of the simplest visual attributes, facilitating our analysis. Second, fonts have a relatively high fixed cost associated with font creation, which includes both the design of typefaces and the development of software. This aspect of high fixed cost associated creative production is common among markets with copyright protection \citep{waldfogel2012copyright}.

\begin{figure}[htbp!]
    \centering    
    \caption{Font Family, Style and Glyphs}
    \subfloat[Font Family Structure]{\includegraphics[width = 0.4\textwidth]{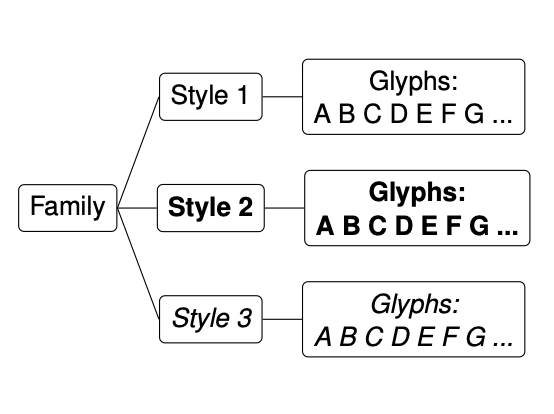}} 
    \subfloat[Example: Gilroy font family]{\includegraphics[width= 0.5\textwidth]{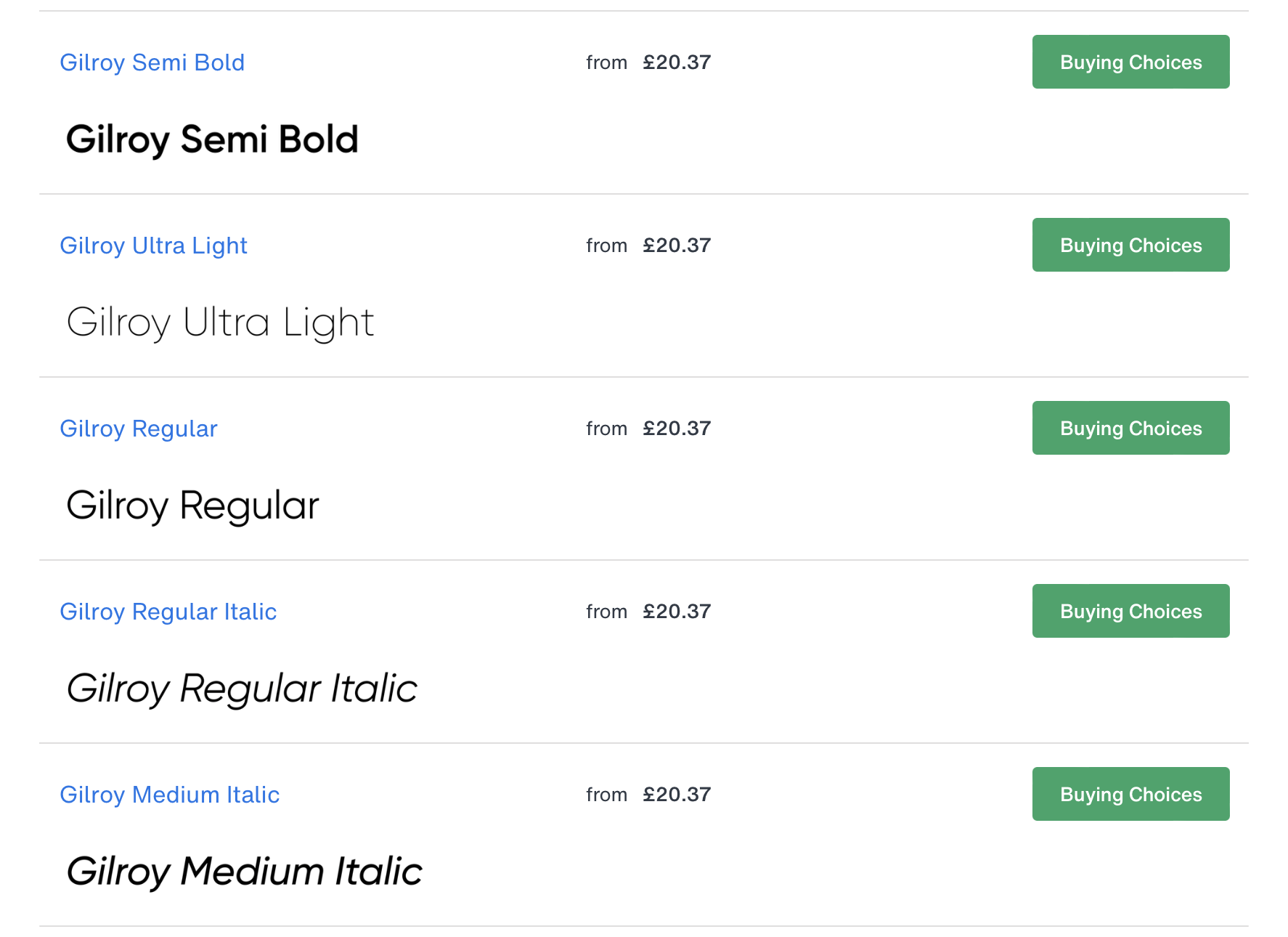}}  
    \label{fig:family-structure}
    \tabnotes{Panel (a) illustrates a nested structure of a font family, styles and glyphs. A family is a set of font styles, while individual styles within the family, such as italic or bold, are variation to the default style design. Glyphs are unique characters in each style. Panel (b) presents Gilroy font family as an example.}
\end{figure}

Fonts are organized into a hierarchical or nested structure that includes family, styles, and glyphs. A font family is a set of font styles that share common design traits, while individual styles within the family, such as italic or bold, introduce variation to the base design.\footnote{As detailed below, this structure becomes a key element in subsequent training data that enables us to use a triplet loss function in embedding construction.} The design process often begins with the creation of a default style, which serves as the foundation from which variations are developed. Glyphs are unique characters specific to each style, and the number of glyphs in a font family is often indicative of its functionality and quality. Figure \ref{fig:family-structure} illustrates the family structure and provides examples.

\subsection{MyFonts.com}

We consider MyFonts.com, the world’s largest online font marketplace, which offers approximately 33,000 different fonts. This market is a superset of all major global online font stores, all of which, including MyFonts.com, are owned by Monotype Inc.\footnote{A recent \emph{Freakonomics} podcast episode features MyFonts.com and explores the overall font industry: \href{https://freakonomics.com/podcast/fonts/}{https://freakonomics.com/podcast/fonts/}.} Panel (b) in Figure \ref{fig:family-structure}  presents an example of a webpage from MyFonts.com for a particular font family.

Related to our research questions, the platform owner, Monotype Inc., is highly concerned with preventing copyright infringement to maintain a well-functioning marketplace and foster competition. Their policy prohibits the acceptance and sale of fonts that are suspected of plagiarizing existing products.\footnote{See Monotype's policy on font plagiarism: \href{https://foundrysupport.monotype.com/hc/en-us/articles/360029957811-Font-Plagiarism}{https://foundrysupport.monotype.com/hc/en-us/articles/360029957811-Font-Plagiarism}.} Plagiarism is mainly determined by comparing the shapes of new and existing fonts; if a new product is ``nearly identical'' to an existing one, Monotype regards the new font as plagiarized and prevents its listing in the marketplace. 

\section{Data and Embeddings} \label{sec:data}

This section describes the datasets and embeddings we develop for the empirical analyses. The first dataset consists of market data, including transaction records, from which we construct panel data. The second dataset comprises product image data, from which we generate embeddings. Both the market and image data are sourced from MyFonts.com.

\subsection{Market Data}

Our transaction data spans from the second quarter of 2014 to the end of 2017 and can be likened to ``scanner data'' from retail shopping. Each order, identified by a unique ID, consists of one or more SKUs, each corresponding to either a font family or a single font style.\footnote{While the data begins in 2012, transactions involving desktop licenses are not recorded from 2012 to 2014. As desktop licenses account for the majority of transactions, we limit our data period from the second quarter of 2014 to the end of 2017, using earlier data for auxiliary purposes. Approximately 80\% of all transactions involve desktop license fonts.} Each order contains information such as consumer ID, transaction time, and revenue (i.e. subtotal). We also have information on registered consumers in the marketplace, which can be linked to transactions via consumer ID. This includes country, city, registration date, last purchase date, and total marketplace expenditure. 

In addition, we utilize firm and product data, which can be interconnected. This dataset provides important product-level information: list price, entry date, ownership, name, supported languages, number of glyphs per style, a family's default style, and product tags (i.e., short descriptive text assigned to each
font family by font designers and consumer, such as ``curly'' and ``geometric'').\footnote{List prices are recorded at the SKU level, not the family level. As list prices vary with the number of styles, we calculate a per-style list price for each family.} We also observe changes in list prices through new SKUs for the same family. Furthermore, we observe designers associated with each font creation. 

From the transaction data, we construct panel data structured by product (i.e., font family), license type, country and month. We only consider transactions involving desktop and web license types, which represent approximately 99\% of total transactions. Also, we focus on 12 countries that contribute the most to total sales and primarily use the Roman alphabet.\footnote{These countries are Australia, Austria, Canada, Finland, France, Germany, Italy, Netherlands, Sweden, Switzerland, United Kingdom, and United States of America.} We define a market to be a combination of country, month and license type.

\subsection{Descriptive Facts about the Market}

Table \ref{tab:desktop_stats_panel} presents descriptive statistics of the panel data. Overall, revenue, quantity, and prices are right-skewed, with large standard deviations.

\begin{table}[htbp!]
\centering
\caption{Descriptive Statistics of Panel Data}
{\small
\begin{tabular}{ll|rrr}
\toprule
License & Variables (Unit) & Observations & Mean & Std. Dev. \\ \midrule
Desktop & Revenue (\$) & 3,476,436 & 20.05 & 202.10 \\ 
& Quantity (Users) & 3,476,436 & 5.83 & 196.92 \\ 
& Sales Price (\$) & 3,476,436 & 9.97 & 10.25 \\ 
& List Price (\$) & 3,324,792 & 28.20 & 76.82 \\ 

\midrule
Web  & Revenue (\$) & 989,196 & 17.50 & 205.88 \\ 
& Quantity (1M Views) & 989,196 & 3,244 & 191,322 \\ 
& Sales Price (\$) & 989,196 & 12.44 & 12.31 \\ 
& List Price (\$) & 943,176 & 28.34 & 50.55 \\ \bottomrule
\end{tabular}
}
\tabnotes{This table contains descriptive statistics of panel data constructed from transaction records. The panel is four-way: product (font family), license type, country, and month. \$ stands for United States Dollar.}
\label{tab:desktop_stats_panel}
\end{table}
List prices typically remain constant over time. From the analysis of variances (Table \ref{tab:anova_results} in the Appendix), we find that product fixed effects account for approximately 99.2\% of the variation in list prices. This suggests that once a price is set at the time of introduction, it remains largely unchanged over time. In contrast, other variables, such as revenue, quantity, and sales prices, vary over time; product fixed effects explain only 13.4\%, 0.8\%, and 65.2\% of the variation in revenue, quantity, and sales prices, respectively. Differences between sales and list prices primarily arise from universal quantity discounts in the marketplace. For transactions without quantity discounts, the distributions of sales and list prices are similar, with sales prices slightly shifted to the left, possibly due to temporary discounts (Figure \ref{fig:dist-prices} in the Appendix).

The main consumers of the marketplace are small-sized businesses or individuals. The quantity units for desktop and web licenses correspond to the number of users who can install the software and the number of website views, respectively. Transactions involving one-user and five-user desktop licenses, as well as the 10,000-view web font license, account for approximately 87\% of all transactions (Table \ref{tab:qunatity-category} in the Appendix). This indicates that a significant portion of consumers are small-scale.

\begin{figure}[htbp!]
    \centering
    \caption{Descriptive Figures on Number of Entrants}
    \subfloat[Monthly Trend]{\includegraphics[width=0.5\linewidth]{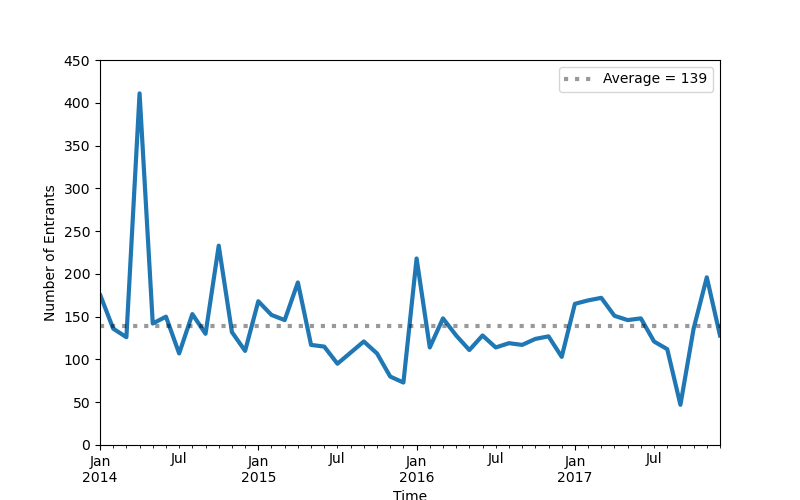}}
    \subfloat[Histogram]{\includegraphics[width=0.5\linewidth]{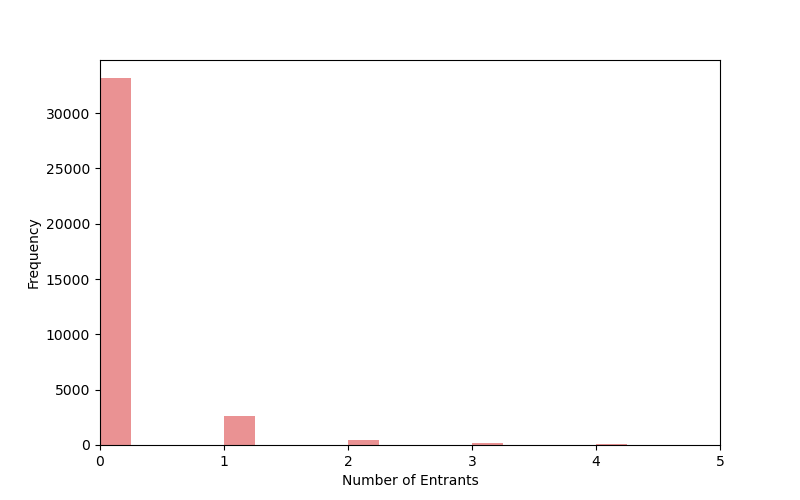}}
    \tabnotes{The panel (a) shows the monthly number of families that newly entered into the marketplace. The average number of entrants is 139, which is shown as the horizontal dot line. Panel (b) shows the histogram of the number of entrants across month-firm pairs.}
    \label{fig:num-entry}
\end{figure}

We also examine the entry behaviors of firms, as shown in Figure \ref{fig:num-entry}. Panel (a) illustrates the monthly trend of entrants (i.e., new font families) listed on the platform. There are constant entries in the marketplace, and the trend remains relatively stable over time, fluctuating around the average level, except for a few outliers. On average, there are 139 monthly entrants. Panel (b) shows the distribution of the number of entrants across month-firm pairs, revealing that it is uncommon for a firm to introduce more than a single product at a given point in time. We leverage these findings for modeling supply-side behaviors in Section \ref{sec:entry}.

\subsection{Embedding Construction with Product Images}\label{subsec:embedding}

In addition to the market data, we use font images as the main unstructured data. We transform product images into embeddings, a low-dimensional feature representation of images. More specifically, we construct embeddings for images of pangrams,\footnote{A pangram is a sentence that uses every alphabet character at least once. Each pangram image is a bitmap with $200\times 1000$ pixels.} because pangrams are what consumers typically see as product images in the marketplace. Embedding analysis is a machine learning method that transforms high-dimensional (or unstructured) data into low-dimensional vectors of numerical values.\footnote{For example, embedding analysis has been applied to text data in natural language processing to measure the similarity of words or sentences, among other uses.} To construct embeddings, we use a convolutional neural network (CNN) with a triplet loss function \citep{schroff2015facenet,han2021shapes}.\footnote{The CNN model is adept at capturing the visual properties of images because the method preserves the local features among pixels, as demonstrated in \cite{krizhevsky2017imagenet, simonyan2014very}.} Rather than using a general-purpose pre-trained model, we train a new neural network using our product images to ensure domain-specific performances \citep{yosinski2014transferable,huh2016makes}. The triplet loss function leverages the nested structure of font families (e.g., Figure \ref{fig:family-structure}) by minimizing the resulting Euclidean distance between fonts within the same family (i.e. positives) and maximizing the distance between fonts from different families (i.e. negatives). The learning with triplet loss is a special form of contrastive learning, which promotes both \emph{alignment} (i.e. closeness) of features from positive pairs and \emph{uniformity} of the induced distribution of the normalized features \citep{wang2020understanding}. These properties are crucial for the resulting embedding distance to be interpreted as a measure of visual similarity, with fonts of similar shapes appearing closer together in the embedding space.\footnote{See \cite{dell2025deep} for relevant discussions in a text analysis context.}

Each embedding is constructed to be a $128 \times 1$ vector normalized to have a unit length and thus lies in a 128-dimensional hyper-sphere. As the length of the embedding is normalized to one, the Euclidean distance and the cosine similarity distance have a one-to-one relationship. See Appendix \ref{subsec:embedding_construction} for the details of embedding construction. In all subsequent analyses, we use the embedding of the default style per font family. For reference, the minimum and maximum Euclidean distances of pairs of embeddings in our data are 0.0002 and 0.9563, respectively.

Finally, one can easily append an additional dimension-reduction layer in the network architecture to further reduce the dimension of the final embedding (e.g., \citealp{hinton2006reducing}). For the data visualization and structural economic analyses below, we use a PCA-type dimension reduction to ensure that the order of elements in the resulting 128-dimensional embeddings corresponds to the amount of explained variation by those elements. This improves embeddings' \emph{portability}, as we can simply select the first $K$ elements, where $K\le 128$ is the desired dimensionality for a given downstream analysis, instead of retraining the network each time a different embedding dimension is needed.

\subsection{Interpreting the Visual Characteristics Space}

Our measure of visual similarity is central to understanding competition and welfare. In this section, we examine how the constructed embeddings, and the resulting visual characteristics space, align with interpretable consumer perceptions. First, as mentioned above, this agreement is partly ensured by construction through the use of triplet loss. If consumers on average conceive fonts within the same family as more similar than fonts across different families, then embeddings constructed to minimize the triplet loss would capture visual similarity perceived by these consumers.\footnote{Consumers' perception over triplets are using in measuring product characteristics relevant to consumers. Relatedly, \cite{magnolfi2022triplet} propose to use embeddings constructed from reported perception of distances within a triplet, solicited from consumers in online surveys.} 

Second, to verify that the distance indeed corresponds to visual similarity, we examine changes in shapes along the embedding distances. Table \ref{tab:dist-examples} presents one instance of this examination. It presents examples of fonts and images according to their distance from the focal font, Minion. As shown in this example, we find that fonts tend to be more visually distinct as the pairwise distance increases.

\begin{table}[htbp!]
\centering
\caption{Examples of Fonts and Images by Distance from Focal Font (Minion)}
\begin{tabular}{lccc|c}
\toprule
Font Name &  Distance & Pangram Shape \\ \midrule

 Minion & 0.000  & \parbox{3in}{%
    \vspace{0.1em}
    \centering
    \includegraphics[width=3in]{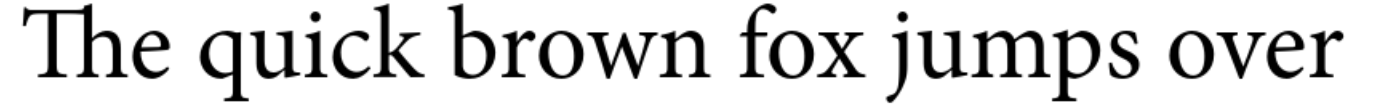}
    \vspace{-1.1em}
} \\  
Alia JY & 0.057 & \parbox{3in}{%
    \vspace{0.1em}
    \centering
    \includegraphics[width=3in]{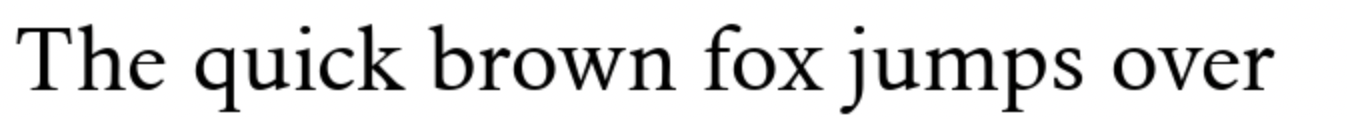}
    \vspace{-0.9em}
} \\

Garamond & 0.081 & \parbox{3in}{%
    \vspace{0.1em}
    \centering
    \includegraphics[width=3in]{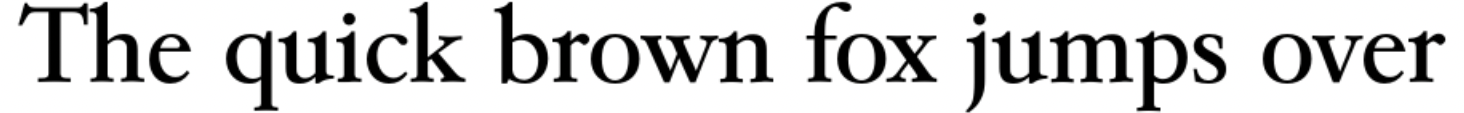}
    \vspace{-1.1em}
} \\

Bauhaus Bugler Soft & 0.090 & \parbox{3in}{%
    \vspace{0.1em}
    \centering
    \includegraphics[width=3in]{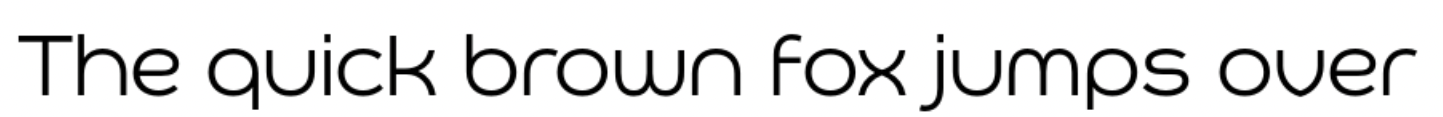}
        \vspace{-0.7em}
}\\  

Andrea Handwritting II & 0.149 & \parbox{3in}{%
    \vspace{0.1em}
    \centering
    \includegraphics[width=3in]{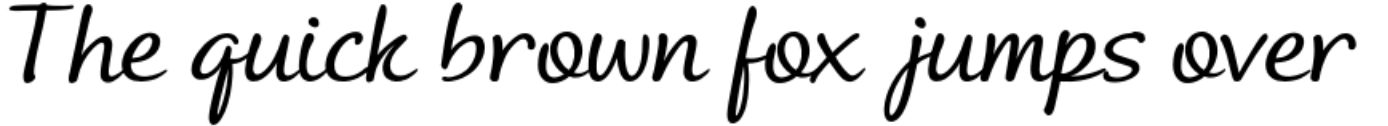} 
    \vspace{-1.3em}
}\\ 

Ruling Script & 0.375 & \parbox{3in}{%
    \vspace{0.1em}
    \centering
    \includegraphics[width=3in]{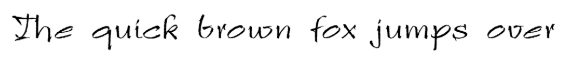}
    \vspace{-1.3em}
}\\ 

Scruff & 0.477 & \parbox{3in}{%
    \vspace{0.1em}
    \centering
    \includegraphics[width=3in]{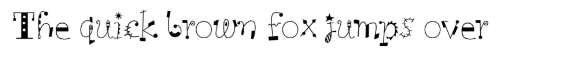}
    \vspace{-1.3em}
} \\
\bottomrule
\end{tabular}\tabnotes{This table displays examples of fonts alongside images of (the first part of) their corresponding pangrams. The pairwise Euclidean distances are calculated between the focal font (Minion) and each font listed in the table. The first column displays the names of the font families, the second column provides the calculated pairwise distances, and the third column exhibits the pangram images for each font. As a reference, in our data, the minimum and maximum Euclidean distances among products are 0.0002 and 0.9563, respectively.}
\label{tab:dist-examples}
\end{table}

\begin{figure}[htbp!]
    \centering
    \caption{Scatter Plot of Principal Components 1 and 2 with Sampled Shapes}
    \includegraphics[width=\textwidth]{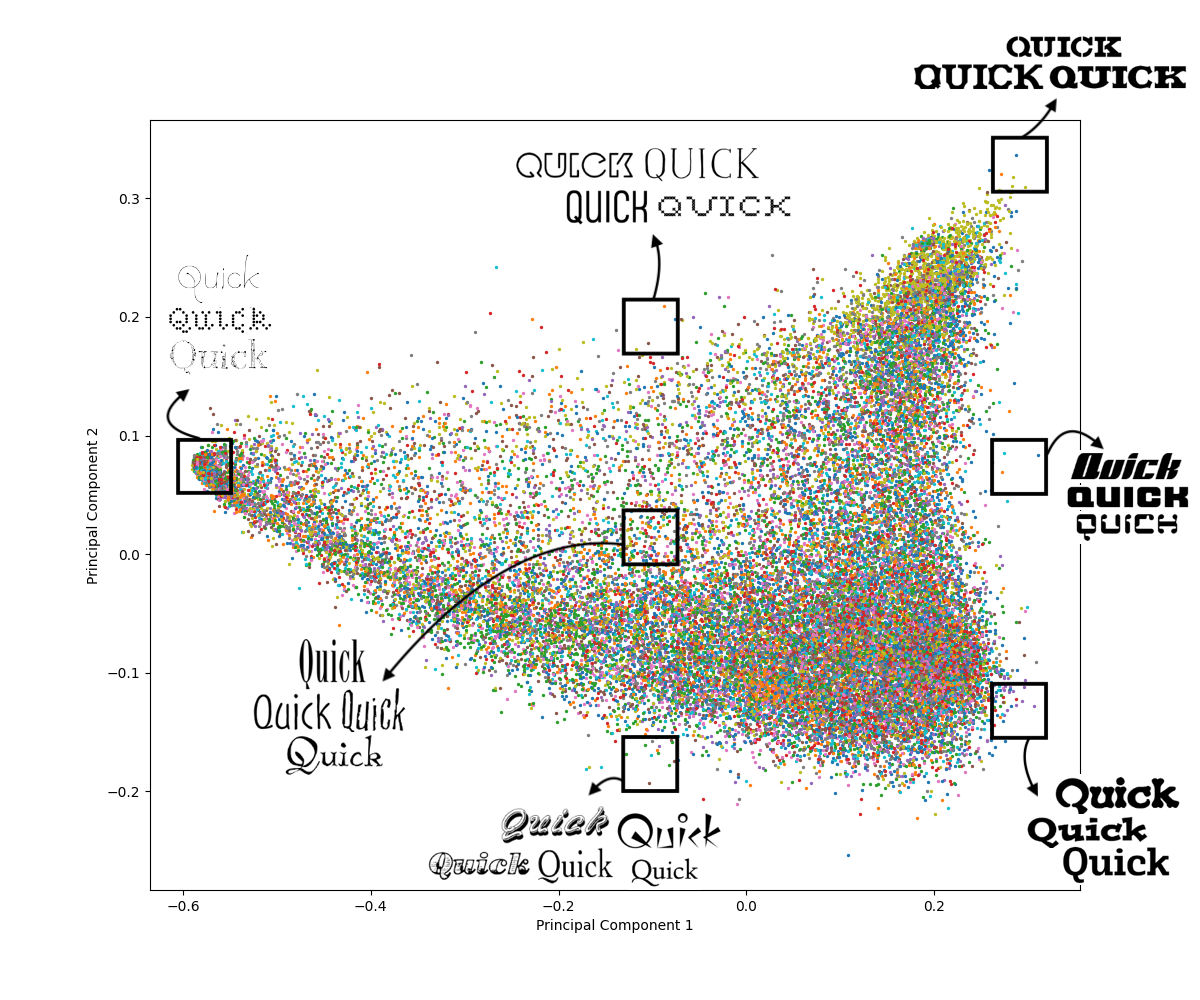}
    \tabnotes{This figure presents a scatter plot of principal components 1 and 2. Shapes sampled in seven different regions are also displayed. Each color represents a different firm owning a font product.}
    \label{fig:scatter-PC1-PC2}
\end{figure}

Third, we investigate how visual characteristics captured in the embeddings are aligned with human perception reflected in the product tags. To this end, we visualize the embeddings by using Principal Component Analysis (PCA).\footnote{Principal Components (PCs) have the well-known interpretation of capturing the largest \emph{orthogonal} variations of the embeddings. The scree plot in Figure \ref{fig:scree-plot} in the Appendix shows that most variations of the embeddings can be explained by just a few PCs, especially the first two.} Figure \ref{fig:scatter-PC1-PC2} displays the scatter plot of Principal Components (PCs) 1 and 2 along with the sampled font shapes in various locations.\footnote{We display sampled fonts as showing all would be hard to visualize.} Even in this low-dimensional space, similar designs are clustered together within the space, bolder shapes towards the right-hand side and more geometric shapes towards the top. The PCs, even in the two dimension, also have predictive power for the official product categories that classify shapes; see Figure \ref{fig:scatter-PCA-UMAP}(a) in the Appendix. 

We then show that these findings are consistent with the information contained in the tags. Product tags (e.g., ``bold,'' ``serif,'' and ``decorative'') are created by both sellers and consumers, namely, they are human-labeled text. We run the Lasso regression of each of the PCs on tag dummies constructed from product tags.\footnote{We use the Lasso regression due to the large number of consumer tags (about 29,000) compared to the number of all products (about 33,000) and the variable selection feature, yielding interpretability. The details of running the Lasso regression are discussed in Appendix \ref{sec:lasso-detail}.} Figure \ref{fig:wordcloud-lasso-1-2} presents the word clouds of tag dummies selected by Lasso (panels (a) and (b)), using the absolute value of the estimated coefficient as a weight, as well as top 5 tags (panels (c) and (d)). In panels (a) and (b), a tag dummy with a negative (positive) coefficient estimate is displayed on the left (right) side. The results suggest that PC 1 is associated with the ``boldness'' of the shape, consistent with Figure \ref{fig:scatter-PC1-PC2}.\footnote{In this exercise, we focus on the ``regular'' style, which are the representative style of a family. Therefore, ``boldness'' is a design feature inherent to the family, rather than a result of the ``bold'' style of the family.} PC 2 seems to capture ``display'' features, namely, design features that are seen in short-form and large-format applications (e.g., billboards or posters, headlines or headings in magazines or websites, and book covers). This is also consistent with Figure \ref{fig:scatter-PC1-PC2}, because geometric shapes are common in display fonts.

\begin{figure}[htbp!]
    \centering
    \caption{Pixel-Level Conditional Mean and Variance}
    \subfloat[Mean of Pixel Values]{\includegraphics[width=0.46\linewidth]{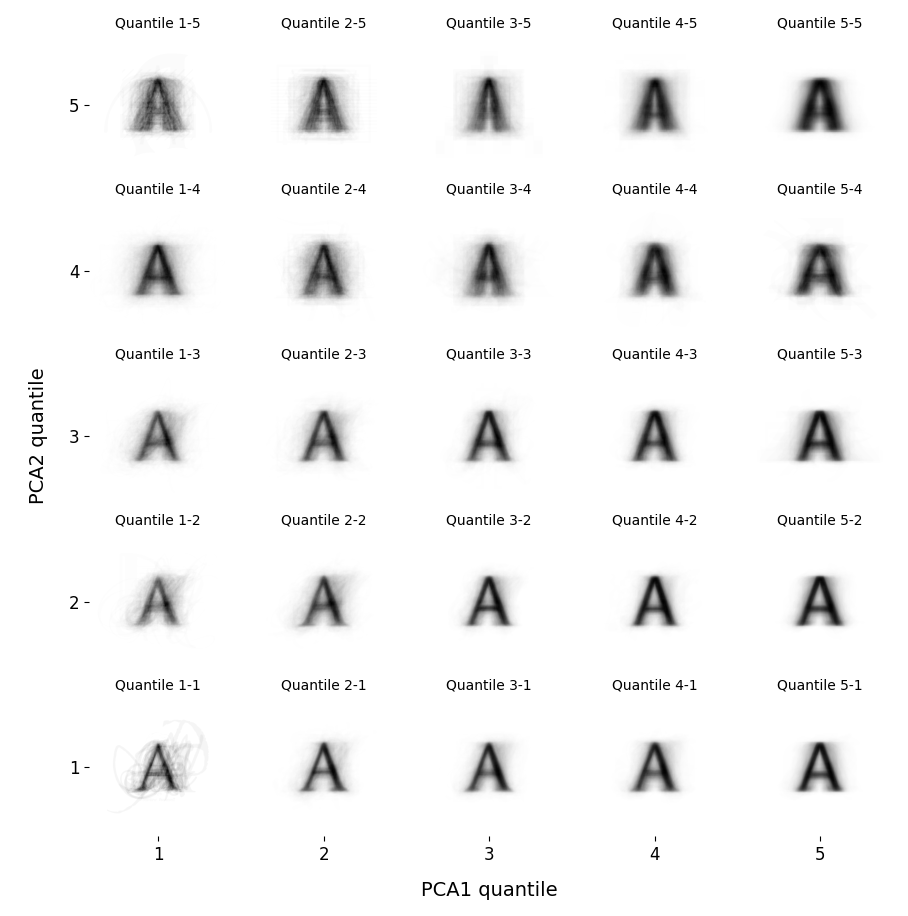}}
    \subfloat[Variance of Pixel Values]{$\quad\qquad$\includegraphics[width=0.5\linewidth]{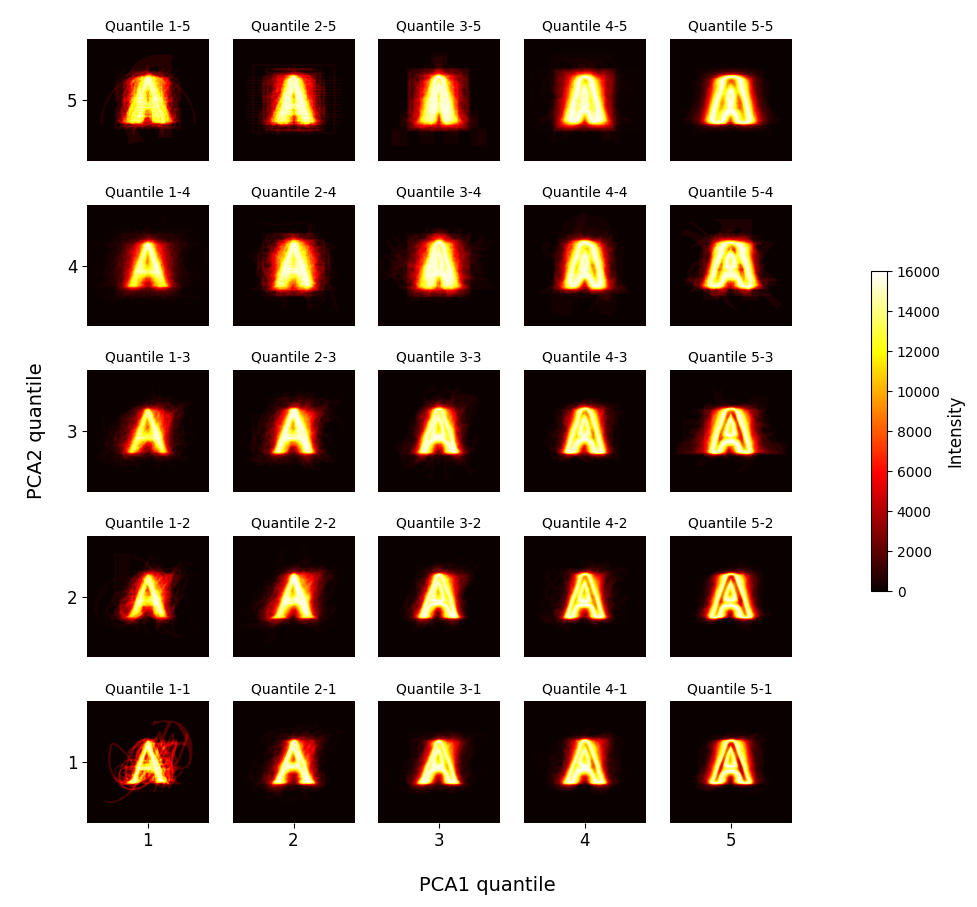}}
    \label{fig:pixel_mean_variance}
    \tabnotes{This figure presents the means and variances of pixel values of the letter `A', conditional on a range of values of principal components 1 and 2.}
\end{figure}

The interpretation of PCs 1 and 2 coincides with the patterns found in the pixel-level analysis. Figure \ref{fig:pixel_mean_variance} shows the pixel-level conditional mean and variance of product shapes (represented by the letter `A') for a given range of PC values. Figure \ref{fig:pixel_mean_variance}(a) suggests that as PC 1 increases, the font thickness also increases. This is confirmed by the low variance in the core of the letter in Figure \ref{fig:pixel_mean_variance}(b) due to increased overlap of thick fonts. Though it initially appears that PC 2 also controls thickness in Figure \ref{fig:pixel_mean_variance}(a), Figure \ref{fig:pixel_mean_variance}(b) indicates otherwise; the increased size of pixel clouds is due to the increased variation of shapes. This is consistent with the fact that display fonts can come in more variety of shapes.

Finally, on a side note, each color in Figure \ref{fig:scatter-PC1-PC2} represents a different firm; it appears that no particular area is overwhelmingly occupied by a small number of firms. This aspect is taken into account when building a supply-side structural model in Section \ref{sec:entry}.

\begin{figure}[htbp!]
    \centering
    \caption{Lasso Regression Results: Principal Components on Tags}
     \subfloat[Wordcloud (PC 1)]{\includegraphics[width=0.5\linewidth]{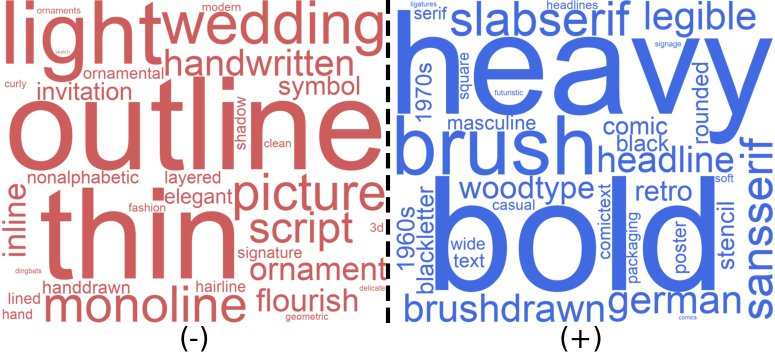}}     
    \subfloat[Wordcloud (PC 2)]{\includegraphics[width=0.5\linewidth]{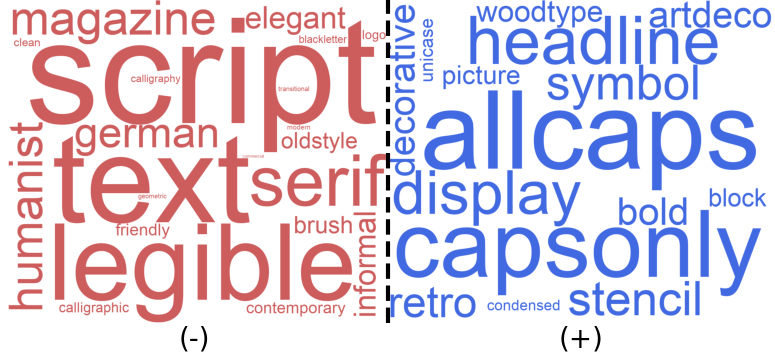}} \\
     \subfloat[Top 5 Tags (PC 1)]{\includegraphics[width=0.5\linewidth]{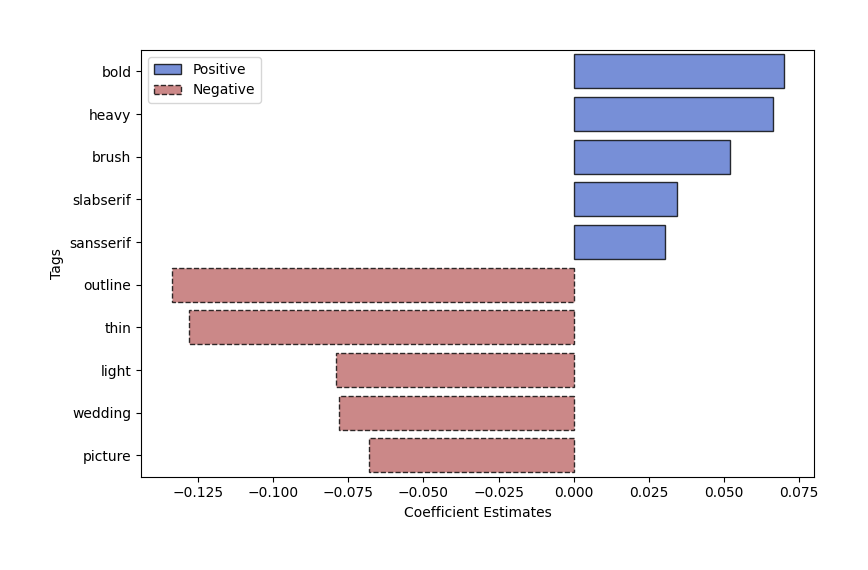}} 
    \subfloat[Top 5 Tags (PC 2)]{\includegraphics[width=0.5\linewidth]{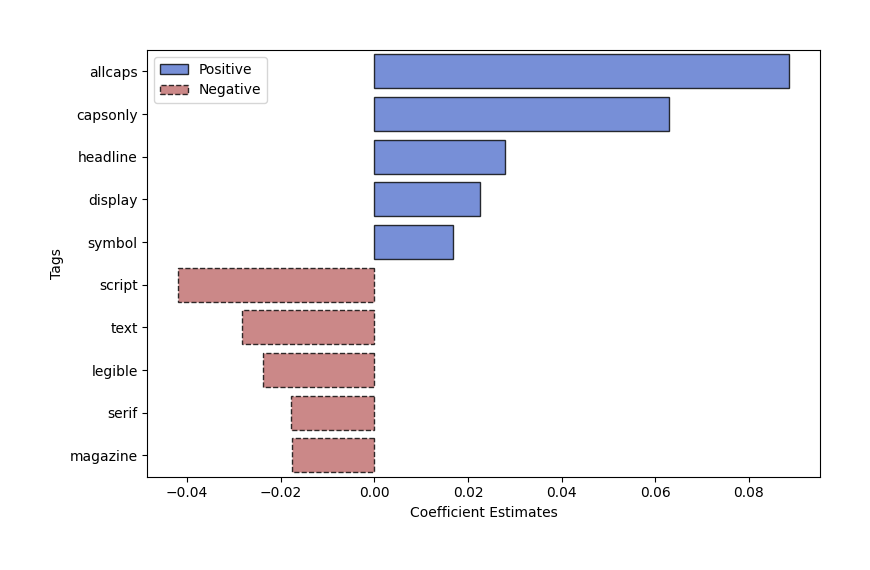}} 
    \label{fig:wordcloud-lasso-1-2}
    \tabnotes{This figure presents the results of the Lasso regression for each principal component. Panels (a) and (b) display word clouds for principal components 1 and 2, respectively, with word sizes weighted by the coefficient estimates. Panels (c) and (d) show the top 5 coefficient estimates for principal components 1 and 2, respectively.}
    \end{figure}

\section{Spatial Competition in Characteristics Space} \label{sec:spatial}
Given the neural network embeddings constructed from the image data, we can define the visual characteristics space of fonts as the subset of the embedding space. Then, each designer's design differentiation decision can be viewed as choosing a location in the characteristics space, potentially engaging in spatial competition with other designers. This conceptual framework is the basis for the paper's empirical analyses. 

As initial exploratory analyses, we investigate the nature of spatial competitions among designers. In particular we ask the following questions: Does visual similarity (calculated by the embeddings) matter for competition? Is competition local in the characteristics space? What are the effects of competition on firm outputs? We answer these questions through two analyses: (1) establishing descriptive relationship between the degree of spatial competition and market outcomes, such as revenues, sales quantity and prices, and (2) quantifying the causal business stealing effects of visually similar entrants on incumbents.

\subsection{Number of Spatial Competitors and Market Outcomes}\label{subsec:num_spatial_competitor}

To understand the relationship between the number of spatial competitors and market outcomes, we define the number of competitors within a radius range between $r$ and $r'$ ($r<r'$) around focal product $j$ at time $t$ as:
\begin{equation} \label{ball-def}
    R^{r,r'}_{jt} := \sum_{j' \in J_{t}}\mathbbm{1}\{r < \lVert x^{emb}_{j'} - x^{emb}_{j} \rVert_{2} < r' \} \text{ for } r,r' \in \mathbb{R},
\end{equation}where $x^{emb}_j\in\mathbb{S}^{128}$ is the embedding in the 128-dimensional hypershere $\mathbb{S}^{128}$ and $J_{t}$ is the set of products in period $t$. We use $R^{r,r'}_{jt}$ as a measure of the degree of spatial competition for a given distance range. The calculation of \eqref{ball-def} is illustrated in Figure \ref{fig:spatial_balls}. Table \ref{tab:B_stats} in the Appendix presents the descriptive statistics of the number of spatial competitors. The number of competitors varies due to both cross-sectional and time-series variations, and there is significant dispersion in the number of competitors.

\begin{figure}[htbp!]
    \centering
    \caption{Counting Competitors on Visual Characteristics Space}
    \includegraphics[width=4in]{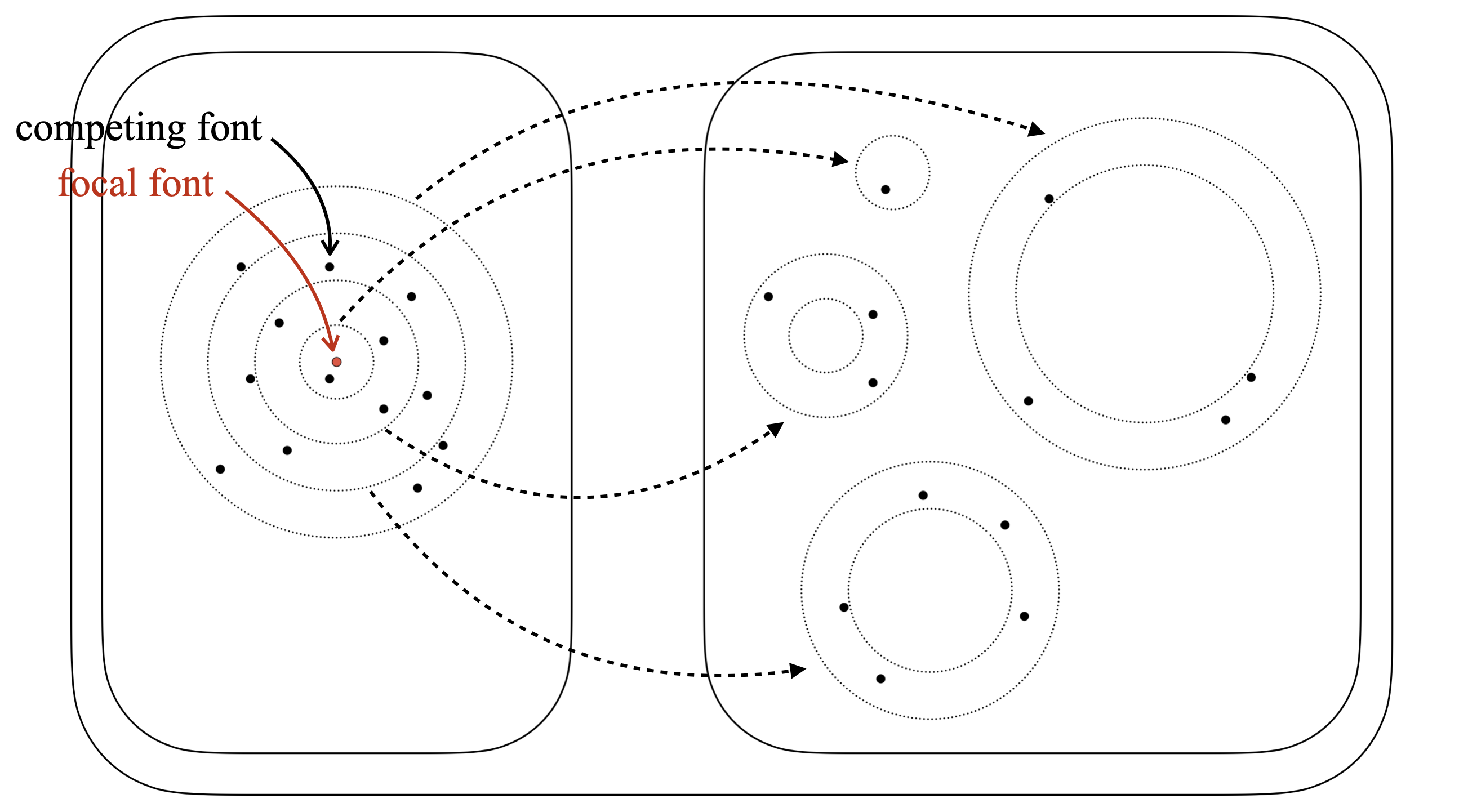}
    \tabnotes{For each focal product $j$ (the red dot on the left-hand side), $R^{r,r'}_{jt}$ counts the number of competitors (the black dots) located between two concentric circles with radii $r$ and $r'$ forming a radial area. We use the Euclidean pairwise distance.}
    \label{fig:spatial_balls}
\end{figure}

Using the number of spatial competitors, we write a regression equation for market outcomes as
\begin{equation} \label{eq:spatial-reg}
    y_{jlct} = \sum_{r \in \{0.1,0.2, 0.3, 0.4\}}\gamma_{r}R^{r - 0.1,r}_{jt} + \alpha_{j} + \alpha_{l} + \alpha_{c} + u_{jlct},
\end{equation}where $y_{jlct}$ is the $arsinh$ transformation of revenue and quantity, and the log of price.\footnote{We also consider an alternative specification by taking the log after adding 1 to revenue and quantity to accommodate zero-sale products. The results are qualitatively similar. Additionally,
to address potential bias due to universal quantity discounts in the marketplace, we use the list price per style of product $j$ as the price variable.} Here, $\alpha_{j}$, $\alpha_{l}$, and $\alpha_{c}$ are product, license type and country fixed effect terms, and $u_{jlct}$ is an error term. In this equation, $\gamma_{r}$ captures the relationship between the number of additional competitors within a given distance range and market outcomes, controlling for fixed effects. We normalize $R_{jt}^{r-0.1,r}$ by dividing them by 100, which defines $\gamma_{r}$ as the semi-elasticity for additional 100 products within a radius area.\footnote{Since the market comprises nearly 30,000 products, 100 products represent a relatively small portion of the total.} 

\begin{figure}[htbp!]
    \centering
    \caption{Spatial Regression Estimates ($\gamma_{r}$)}
     \includegraphics[width=4in]{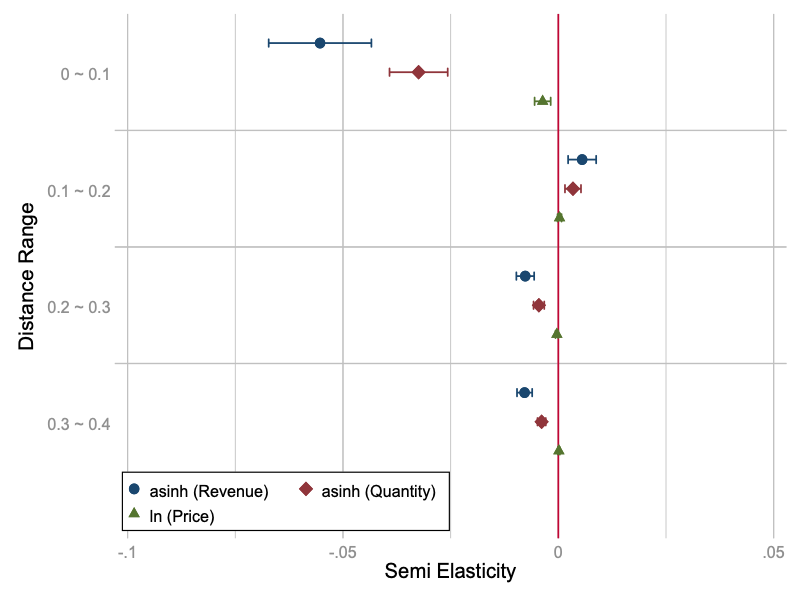}\label{fig:spatial-reg-result}
     \caption*{\footnotesize{\textit{Notes.} Coefficient estimates from regression \eqref{eq:spatial-reg} are presented. The radius of innermost ball is 0.1. Round-, diamond-, and triangle-shaped dots represent estimates for the revenue, quantity, and price variables, respectively. Solid lines indicate 95\% confidence intervals. Standard errors are clustered at the product level.}}
\end{figure}
Figure \ref{fig:spatial-reg-result} presents the regression results, which suggest that competition in the visual characteristics space significantly affects market outcomes and that such competition is \emph{local}. The average elasticity of revenue and quantity in response to additional 100 competitors within the innermost ball is around -0.40 and -0.24, respectively.\footnote{We approximate the elasticity by calculating $\frac{\partial y}{\partial R^{r-0.1,r}}\frac{R^{r-0.1,r}}{y} =  \gamma_{r} R^{r-0.1,r} \times \sqrt{1 + \frac{1}{y^{2}}}$ for the revenue and quantity variables under the $arsinh$-linear specification as shown in \cite{bellemare2020elasticities}. The elasticity approaches $\gamma_{r}R^{r-0.1,r}$ as $y$ increases.} Notably, these estimates are significantly larger than those for the outer rings. For prices, the coefficient estimates are near zero and not statistically significant, consistent with the fact that prices are not responsive in the market.\footnote{When price is the dependent variable, the model may be too saturated to control for product-level fixed effects. Therefore, we use firm-level dummies instead of product dummies, which yields qualitatively similar results; see the Appendix \ref{sec:more_tab_fig}.}

\subsection{Business Stealing of Visually Similar Entrants}

Motivated from the previous analysis which reveals that competition is local, we conduct an event study to estimate the causal business stealing effects of the entry of visually similar products. We are particularly interested in determining whether and to what extent the profits of an incumbent are reduced by such a local entry. This analysis complements the previous analysis, which does not explain how the post-entry of a new product affects market outcomes. Also, the analysis in this section is free from the choice of distance cutoffs used in the previous analysis.

First, we define the treatment as an indicator for a new entry occurring within the five visually closest products. That is, the treatment indicates that there is a change in the membership of five closest competitors due to entry. Let $T_{j}$ represent the first month when this treatment occurs for focal font $j$. If the treatment never happens, we set $T_{j} = \infty$. We then define the event dummies as $E^{s}_{jt} := \mathbbm{1}\{ t - T_{j} = s\}$ for $s \in \mathbb{Z}$ and specify an event study design:
\begin{equation} \label{eq:event-4w}
    y_{jlct} = \sum_{s = -5}^{9} \beta_{s} E_{jt}^{s} + \alpha_{f} + \alpha_{l}+ \alpha_{c} + \alpha_{t} + e_{jlct}, 
\end{equation}where $\alpha_{f}$, $\alpha_{l}$, $\alpha_{c} $, and $\alpha_{t}$ are fixed effects of firm, license type, country, and time, respectively, and $e_{jlct}$ is an error term. The dependent variables are the $arsinh$ transformation of revenue and quantity, and the log of price. We normalize regression results by setting $\beta_{-1} = 0$, following the standard event study exercises. In addition, we define the first and last event dummies to include the periods before and after, by setting $E_{jt}^{-5} = \mathbbm{1}\{ t - T_{j} \leq -5\}$ and $E_{jt}^{9} = \mathbbm{1}\{ t - T_{j} \geq 9\}$.

\begin{figure}[htbp!]
    \centering
    \caption{Event Study Estimates ($\beta_{s}$)}
    \subfloat[Revenue]{\includegraphics[width=0.5\textwidth]{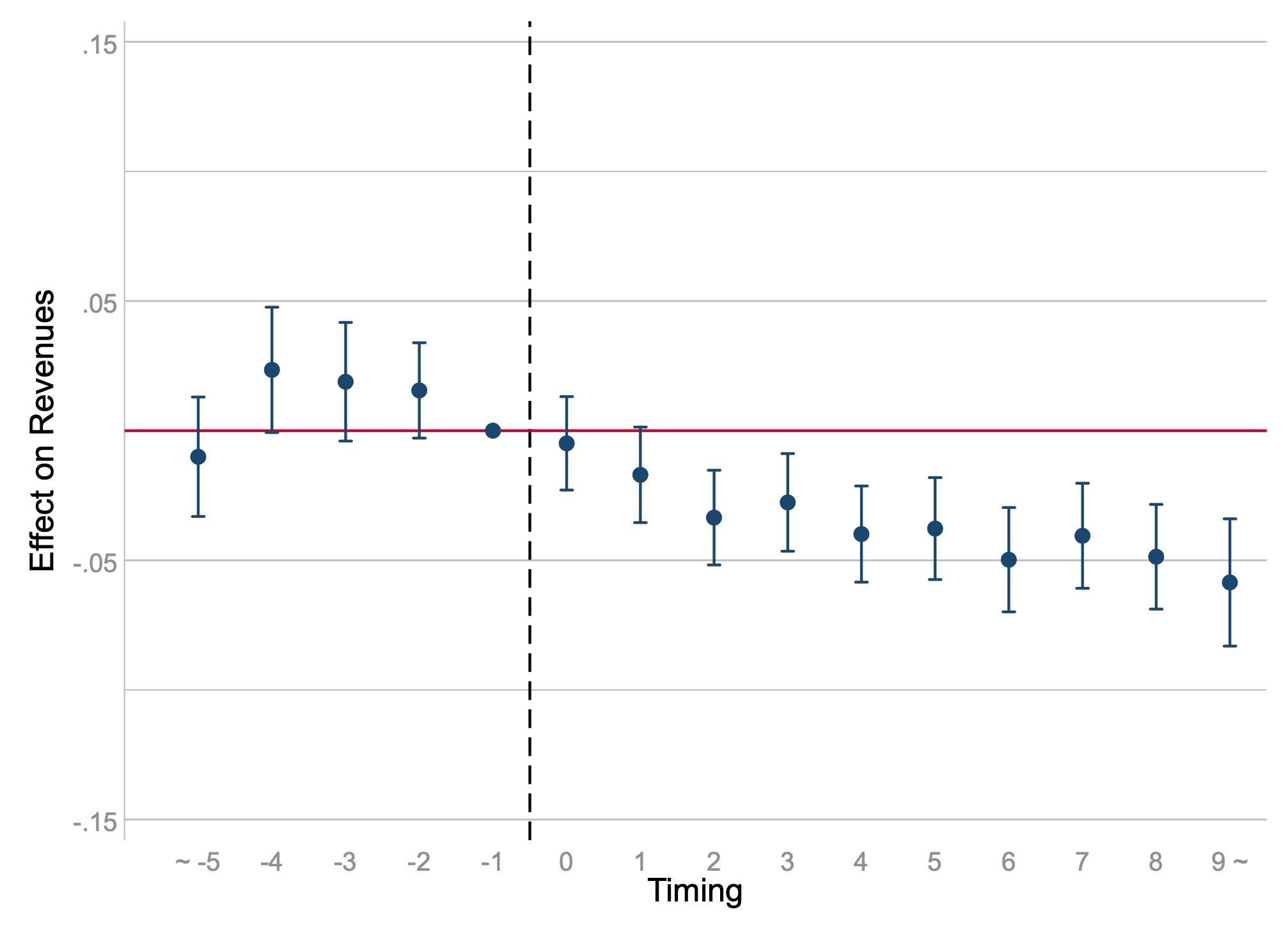}} \\
    \subfloat[Quantity]{\includegraphics[width=0.5\textwidth]{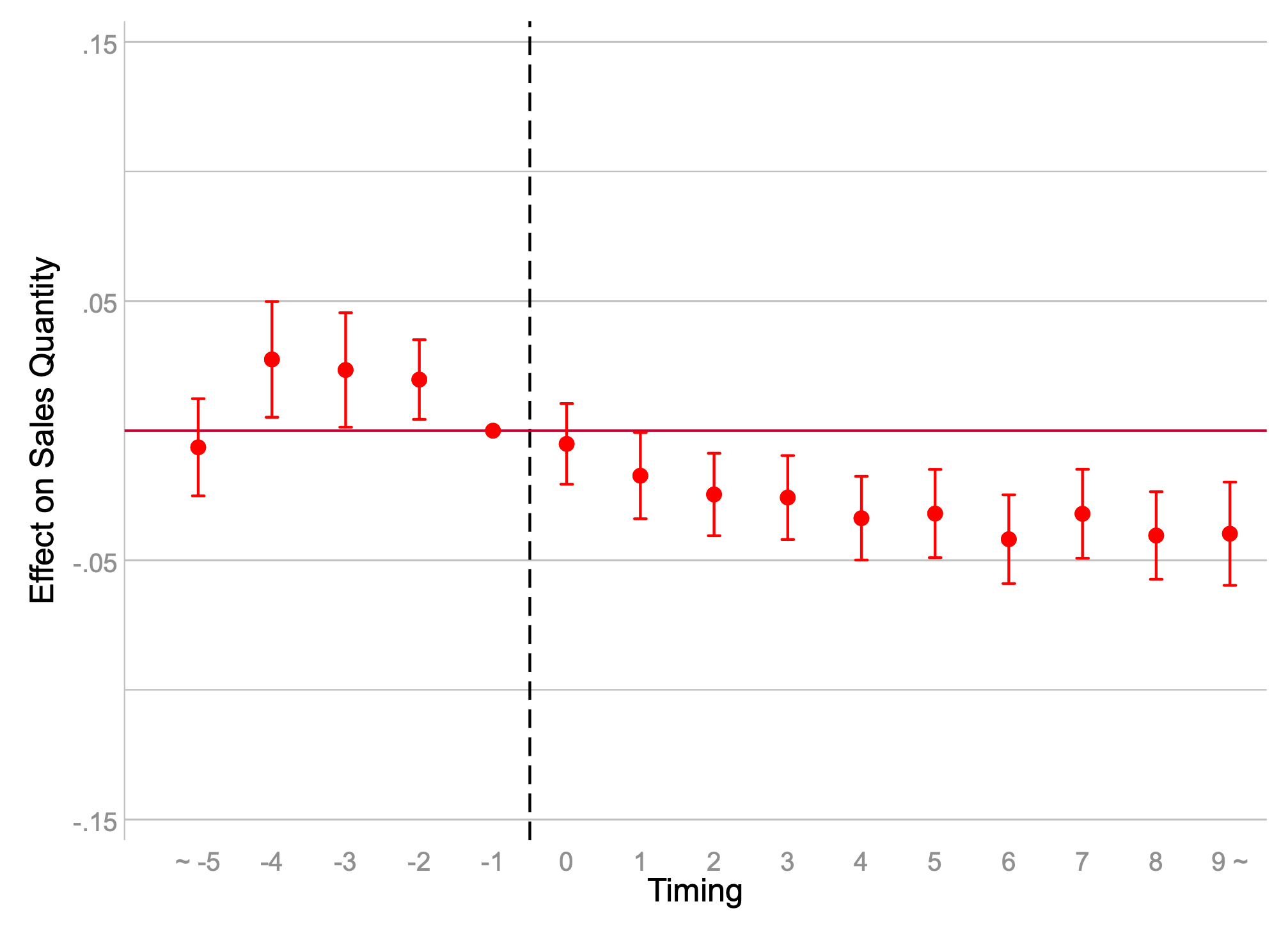}}
    \subfloat[Price]{\includegraphics[width=0.5\textwidth]{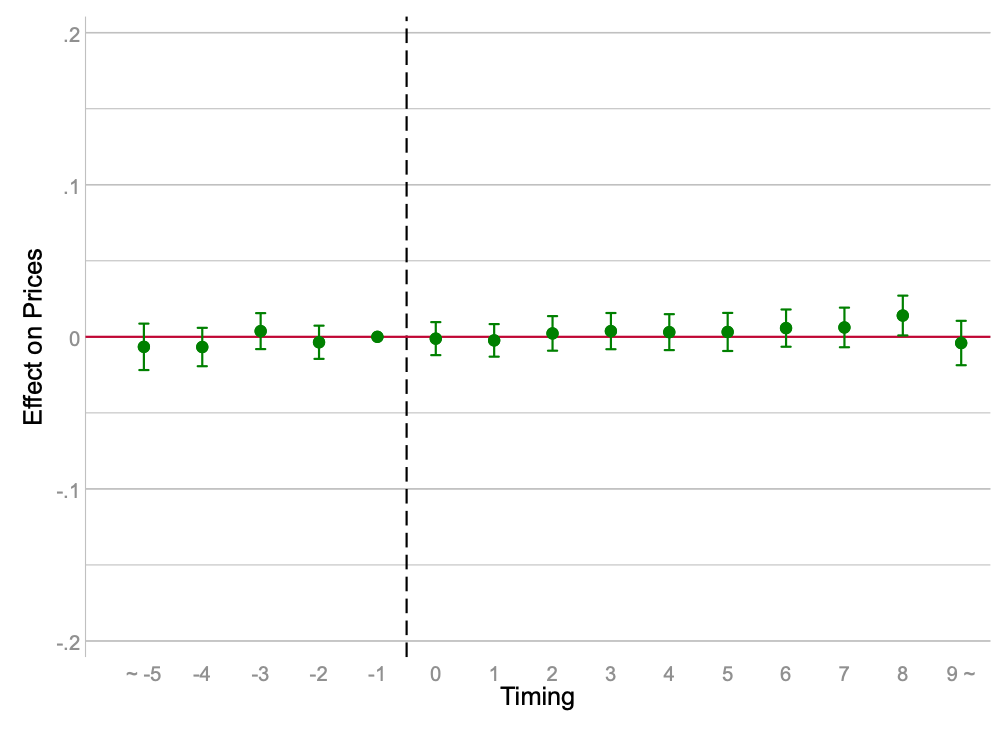}}
    \tabnotes{These figures display the results of the event study regression as described in \eqref{eq:event-4w}. Panels (a), (b) and (c) present regression results for the arsinh of revenue, arsinh of quantity and the log of list prices as the dependent variable, respectively. Solid lines represent the 95\% confidence intervals of the estimates. We use firm-level clustered standard errors.}
    \label{fig:event-ols}
\end{figure}

Figure \ref{fig:event-ols} presents the results. We find that the business stealing effects are significant and enduring for revenue and quantity, suggesting strong substitution between visually similar products after new entries. The substantial loss of revenue due to entry displayed in Panel (a) is mostly driven by a decrease in sales quantity as shown in Panel (b). A treated incumbent loses 5\% decrease in revenues compared to the month before the entry of the visually similar product. However, price is not responsive as expected. All the pre-trend coefficients are statistically insignificant, supporting the validity of the parallel trend assumption.

Due to the nature of online marketplace, products rarely exit from the market, while entry constantly occurs (Figure \ref{fig:num-entry}(a)). This makes the treatment staggered. It is known that, with staggered adoption, ordinary least squares (OLS) estimates may not be a convex combination of treatment effects on the treated in different timing \citep{de2020two,goodman2021difference}. To circumvent this problem, we impute the unobserved potential outcome value under the linear fixed effect model and then takes a simple average of all treated observations to calculate the causal effect \citep{borusyak2021revisiting}. The results from this alternative analysis (Figure \ref{fig:event-impute} in the Appendix) are qualitatively similar to those in Figure \ref{fig:event-ols}, suggesting that our estimates are robust to staggered timing of the treatment.

Results from several different specifications of model \eqref{eq:event-4w} also show that the findings are robust.\footnote{We conduct robustness check for 10\% randomly sampled observations due to the computational reason.} First, we run the event study with additional control variables, including the age of the product measured by months after entry into the marketplace, the log of glyphs, and interaction dummies between time and image cluster (Figure \ref{fig:event-ols-add}). Second, we use an alternative definition of the treatment: a change in one of the \emph{four} visually closest competitors due to a new entry (Figure \ref{fig:event-ols-1of4}). The estimation results in both exercises are qualitatively similar to the previous findings.

Overall, the empirical analysis of spatial competition reveals that competition is mainly local, with substantial business stealing occurring among nearby products. This suggests that a copyright policy that provides ``local protection'' in the characteristics space could significantly influence how the market functions. In subsequent analyses, we develop demand and supply models and apply them to evaluate the competitive and welfare effects of copyright policy.

\section{Models for Supply and Demand}\label{sec:structural_model}

Guided by the empirical findings in the previous section, we now build empirical models for supply and demand in the font market. On the supply side, our model aims to capture the entry decision-making process, especially in terms of the unstructured visual characteristics. The main model primitive to recover is the fixed costs of developing product design. On the demand side, the main objective is to characterize consumers' preferences over the visual characteristics, recovering substitution patterns among products. Our primary goal in building and estimating these models is to conduct counterfactual analyses to understand the role of similarity-based copyright policy and potential shifts in market fundamentals driven by technological advancements, such as the introduction of generative AI.

Throughout the section, the subscript $i$, $j$, $c$, and $t$ denote consumer, font product, country of the marketplace, and time, respectively. We define the market as a combination of country and time. 

\subsection{Discrete Choice Consumer Model} \label{sec:discrete-choice}

We introduce a discrete choice model to describe consumer behaviors. In order to capture heterogeneous preferences on the visual attributes of fonts, we consider the random coefficient logit formulation of
the indirect utility in the spirit of \cite{berry1995automobile}. The characteristics captured by the embeddings enter the model. The indirect utility is specified as
\begin{align} \label{eq:discrete-utility}
    U_{ijct} &:= \Bar{\beta}^{p} p_{jct} + \Bar{\beta}^{str}x_{j}^{str} + x_{j}^{emb\prime}\beta_{i}^{img} + \xi_{jct} + \epsilon_{ig(j)ct} + (1 - \rho)\Bar{\epsilon}_{ijct}
\end{align}and $U_{i0ct} := \epsilon_{i0ct}$, where $j=0$ denotes the outside option that includes free open source fonts, $x_{j}^{str}$ is the vector of structured characteristics, including glyph counts and a constant, $x_{j}^{emb}$ is the vector of visual embeddings, and $p_{jt}$ is the sales price. We assume that $\beta_{i}^{img}$ is a random coefficient that follows the normal distribution $N(\Bar{\beta}, \Sigma)$ where $\Sigma$ is a diagonal matrix. To ensure tractability of the random coefficient model, we choose $x_{j}^{emb}$ to have a relatively low dimension of six.\footnote{This choice of dimension is made based on the scree plot (Figure \ref{fig:scree-plot}). This choice of $x_{j}^{emb}$ explain roughly 99\% of the total variation of the embeddings with 128 dimension. Moreover, due to the linear transformation of PCA, it approximately preserves the distance (i.e., visual similarity) when $x_{j}^{emb}$'s capture most of the variation. Alternatively we can use the \emph{partial least squares}, a supervised alternative to the PCA \citep{hastie2009elements}; see Appendix \ref{sec:PLS}.} In addition, we define nests by using the official product categories and tags that consumers use to browse products on MyFonts.com in order to account for the effects of the website's search system on consumer choices,\footnote{The official categories of fonts are coarse product categories defined by the industry; they are Serif, San Serif, Slab Serif, Script, Display, and Handwritten.} with $g(j)$ denoting the nest that product $j$ is in. Finally, $\epsilon_{ijct} := \epsilon_{ig(j)ct} + (1 - \rho)\Bar{\epsilon}_{ijct}$ is an i.i.d. shock, following the type I extreme value distribution, which is similar to the demand model of \citet{brenkers2006liberalizing}. As $\rho \rightarrow 1$, substitutions would happen mostly within nests. The market share is defined as
\begin{equation*}
    s_{jct} := \int \frac{\exp \left[V_{ijct}/(1 - \rho)\right]}{\exp\left[I_{ig(j)ct}/(1 - \rho)\right]} \cdot \frac{\exp I_{ig(j)ct}}{1 + \sum_{h \in G}\exp I_{ihct}}d\Phi(\beta_{i}^{img}),
\end{equation*} where $G$ is the set of the nests, $V_{ijct} := \Bar{\beta}^{p} p_{jct} + \Bar{\beta}^{str}x_{j}^{str} + x_{j}^{emb'}\beta_{i}^{img} + \xi_{jct}$, and $I_{ig(j)ct} := (1-\rho)\log \sum_{j' \in J_{g(j)ct}} \exp \left[ V_{ij'ct}/(1 -\rho)\right]$ is the inclusive value within the nest $g(j)$.

\subsection{Model for Entry and Product Positioning} \label{sec:entry}

For the supply side, we consider a multi-stage model in which each firm makes an entry decision in the first stage, followed by a product 
 positioning decision subject to a copyright policy in the second stage, and a pricing decision in the third stage. This model serves as an empirical counterpart to the theory of spatial location choice \citep{hotelling1929stability,salop1979monopolistic}, but with some key differences. First, we do not make any assumptions on the topology of spatial competition, such as linear or circular shapes. Instead, we model spatial competition in a characteristics space constructed from the neural network embeddings. Second, we do not assume symmetry among firms and their equilibrium outcomes. Instead, we aim to estimate model primitives that reflect firm heterogeneity. 

In each period $t$, a firm $f$ makes a decision on whether to introduce product $k$ into the marketplace. We assume that each firm can introduce at most one product each period. Let $A_{ft,k}$ denote the entry decision ($A_{ft,k}=1$ if $k$ enters and $0$ otherwise). Let $J_{ft}$ be the portfolio of products offered by the firm $f$ available at time $t$, which excludes product $k$ that the firm currently considers launching or not.

The total profit $\Pi_{ft,k}$ of firm $f$ at time $t$ by launching product $k$ is specified as
\begin{equation} \label{eq:total-profit}
    \Pi_{ft,k} := \sum_{j \in J_{ft}} \pi_{jt} + A_{ft,k}\left(\pi_{kt} - F(\boldsymbol{x}_{t},\nu_{k})\right),
\end{equation} where $\pi_{jt}:=\pi_{jt}(A_{ft,k})$ is the variable profit of product $j$ that is implictly a function of $A_{ft,k}$ (and similarly for $\pi_{kt}$) and $F(\boldsymbol{x}_{t},\nu_{k})$ is the fixed costs of developing $k$. The variable profit of each product is expressed as
\begin{equation} \label{eq:variable-profit}
    \pi_{jt} := \pi_{jt}(A_{ft,k}) := s_{jt}M_{t}(p_{jt} - mc_{jt} ),
\end{equation}
where $p_{jt}$ and $mc_{jt}$ are the price and marginal cost of product $j$ at time $t$, respectively, and $s_{jt}$ and $M_{t}$ are the market share and size at time $t$, respectively. The observed market share is mapped from the collection of characteristics via the demand function which is derived from the aggregated consumer choices in Section \ref{sec:discrete-choice}, incorporating the dimension-reduction restrictions discussed therein. Therefore, $s_{jt}$ is also a function of $A_{ft,k}$ since the entry decision affects all products through demand-side consumer choices. We define the market size $M_{t}$ as the number of active users registered in the marketplace.\footnote{A market size for each country, $M_{ct}$, is calculated by counting the number of active users at country $c$ (including consumers who only purchase free fonts) and multiplying the fraction of desktop sales to it. We define active users as those whose activity falls between their registration date and last login date. The overall market share, $s_{jt}$, at time $t$ is calculated as $s_{jt} := \sum_{c}s_{jct}M_{ct} / M_{t}$.}

In addition, we model the fixed cost of development $F(\boldsymbol{x}_{t},\nu_{k})$ in \eqref{eq:total-profit} as a function of the characteristics $\boldsymbol{x}_{t}$ across all products in the marketplace:
\begin{align} 
     F(\boldsymbol{x}_{t},\nu_{k})&:= \nu_{k0} + \sum_{\ell} \left[ (\eta_{0\ell} + \nu_{k\ell})x_{k\ell}^{emb} + \sum_{j\neq k} \left( \eta_{1\ell}d^{\ell}_{jk} + \eta_{2\ell}(d^{\ell}_{jk})^{2}  + \eta_{3\ell}(d^{\ell}_{jk})^{3} \right) \right], \label{eq:fixed-cost}
\end{align} 
where $d^{\ell}_{jk} := \|x^{emb}_{k\ell} - x^{emb}_{j\ell}\|_{2}$ for each dimension $\ell$ of the embedding is the distance of incumbent product $j$ from product $k$ and $\nu_{k} := (\nu_{k0},\nu_{k1},...,\nu_{k6})'$ are i.i.d. random shocks. We specify $F(\boldsymbol{x}_{t},\nu_{k})$ to be a nonlinear function of the characteristics of product $k$ and the distances to its competitors.\footnote{This specification is akin to the distance-based demand model as in \citet{pinkse2002spatial} and \citet{magnolfi2022triplet}, but our problem is fundamentally different from theirs as we focus on supply-side behaviors.} This specification is designed to capture a reduced fixed cost associated with the presence of visually similar products---arising from mimicking advantages---while maintaining parsimony. However, the fixed cost need not decrease monotonically. One possible reason is that subtle differentiation from other competitors can be more costly, as it requires more sophisticated product design strategies. Consequently, the overall shape of the cost function is an empirical question.

In each period $t$, each firm makes a sequence of decisions along the following timeline: in the first stage, the firm makes an entry decision after the cost shock $\nu_{kt}$ is realized (and before the demand shock is realized). Upon entry, the firm chooses the optimal location of product $k$ subject to similarity constraints imposed by a copyright policy. Lastly, the unobserved demand shock $(\xi_{kt})$ is realized and the firm conducts pricing.

We specify the model in a backward fashion. In the final stage, a firm solves the pricing problem for given product characteristics and unobserved demand shocks:
\begin{equation*}
    \boldsymbol{p}_{ft}^{*} = \arg\max_{p_{jt} \in \{p_{jt}: j\in J_{ft} \cup \{k\}\}} \sum_{j \in J_{ft} \cup \{k\}} s_{jt}M_{t}(p_{jt}- mc_{jt}),
\end{equation*} where we suppress the arguments of $s_{jt}$ for simplicity. The standard first-order condition with respect to the price of product $k$ is given by
\begin{align} \label{eq:FOC-pricing}
    \sum_{j \in J_{ft} \cup \{k\}} \frac{\partial s_{jt}}{\partial p_{kt}}M_{t}(p_{jt}-mc_{jt}) + s_{kt}M_{t} = 0.
\end{align} 

By solving the pricing equation \eqref{eq:FOC-pricing}, one can obtain the optimal pricing function $p_{kt}^{*}(\boldsymbol{p}_{-k,t},\boldsymbol{x}_{t},\boldsymbol{\xi}_{t})$. The optimal price is a nonlinear function of prices and observed and unobserved characteristics of all (possibly neighboring) products in the market through demand, which implies that the pricing equation effectively captures competition in the marketplace.\footnote{We assume the optimal pricing equation gives a single pricing rule.} We can recover marginal costs through the simultaneous equations system of the first order condition \eqref{eq:FOC-pricing}.\footnote{Although distributing fonts per se incurs minimal costs, marginal costs include commission fees paid to the platform. We infer these marginal costs from the pricing model rather than specifying them directly, since the wholesale price (or commission) between a firm and the platform is unobservable.}

In the second stage, firm $f$ decides the positioning of product $k$ given the optimal price $p_{kt}^{*}$. The demand shock $\xi_{kt}$ is not realized yet, hence the firm chooses the location of $k$ to maximize the expected profit as
\begin{align} \label{eq:loc-choice}
    x^{emb,*}_{k} &= \arg \max_{x^{emb}_{k} \in \mathbb{S}^{d}} \mathbb{E}_{\xi_{kt}}\left[\sum_{j \in J_{ft} \cup \{k\}}\pi_{jt}\right] - F(\boldsymbol{x}_{t},\nu_{k}) \\
    &\text{s.t. } \lVert x^{emb}_{k} - x^{emb}_{j'}\rVert_{2} \geq \underbar{d} \text{  for all  } j' \in J_{-ft}, \nonumber
\end{align} where $x^{emb}_{k}$ is the embedding vector of product $k$, $J_{-ft} := J_{t} \setminus \{J_{ft}\cup \{k\}\}$ is the set of products sold by $f$'s competitors at time $t$, and $\underbar{d}$ is the similarity constraint imposed by the copyright policy, forming a local protective boundary of radius $\underbar{d}$ for incumbents. The specification of the similarity constraint is consistent with the modeling of local competition in the reduced-form analysis in Section \ref{sec:spatial}, where the embedding distance between two products is used. The optimization problem in \eqref{eq:loc-choice} is similar to that in \cite{fan2013ownership}, yet with key distinctions. First, the specification of similarity constraint using neural network embeddings is unique to our study, which enables us to model copyright policies. Second, unlike \cite{fan2013ownership}, we consider the maximization of expected net profit. Third, we model fixed costs as dependent on the characteristics of competing products, reflecting the cost-benefit consideration of emulating similar products. The necessary conditions for optimality are written as
\begin{align} \label{eq:second-foc}
    \sum_{j \in J_{ft} \cup \{k\}}&\mathbb{E}_{\xi_{kt}}\left[\frac{\partial \pi_{jt}}{\partial x^{emb}_{k}} + \sum_{j' \in J_{-ft}} \frac{\partial \pi_{jt}}{\partial p_{j'}}\frac{\partial p_{j'}}{\partial x^{emb}_{k}}  \right]  \\
    &+ \sum_{j' \in J_{-ft}}\left[\lambda_{kj'} \left(\frac{\partial \|x^{emb}_{k} - x^{emb}_{j'}\|_{2}}{\partial x^{emb}_{k}} - \underbar{d}\right)\right] = \frac{\partial F(\boldsymbol{x}_{t},\nu_{kt})}{\partial x^{emb}_{k}}, \nonumber
\end{align} where $\lambda_{kj'}$ is a Karush–Kuhn–Tucker (KKT) multiplier for the similarity constraint imposed on $k$ with respect to $j' \in J_{-ft}$.\footnote{The second summation in \eqref{eq:second-foc} omits the term for $j$ by the envelope condition. Also, in \eqref{eq:second-foc}, we assume that the expectation and partial differentiation are interchangeable.} The main product characteristics captured in the embeddings are continuous, which enables differentiation for this optimality conditions.

Finally in the first stage, firm $f$ pays the fixed costs $F(\boldsymbol{x}_{t},\nu_{k})$ if the expected net profit is greater than zero:
\begin{equation} \label{eq:entry-condition}
    \mathbb{E}_{\xi_{kt}}\left[\Pi_{ft,k}(A_{ft,k} =1)\right] - \Pi_{ft,k}(A_{ft,k} =0) \geq 0.
\end{equation}
This follows a revealed profit approach, which has been used by many studies in the entry game literature (e.g., \citealp{bresnahan1991entry,berry1992estimation, berry1999public,seim2006empirical}). The distinctive feature is that we consider a product-level entry instead of a firm-level entry. Also, we do not consider reduced-form parametric specification of the profit function; instead, the profit function is determined by demand- and supply-side primitives.\footnote{These features also appear in \cite{eizenberg2014upstream} and \cite{wollmann2018trucks}.}

In constructing the supply-side model, we assume that firms make sequential decisions of entry and product positioning in each period given other firms' previous decisions and their own existing portfolio positions. This modeling choice is made for the following reasons. It is difficult to argue that, in this marketplace, firms engage in best-response dynamics for high-dimensional product attributes (i.e., product design) by interacting with a large number of opponents. Rather, it may be more plausible to view them as attribute-takers that determine entry and product positioning sequentially. Relatedly, in the subsequent counterfactual analyses, it is extremely difficult to simulate the unlikely firm behavior of simultaneous product positioning.

In making sequential decisions, we assume that firms' forward-looking behaviors are not present, which is consistent with their observed market behaviors. One dynamic action that may be relevant for the copyright policy is entry deterrence: a firm can occupy an area in the product space to prevent the entry of competitors, leveraging copyright protection as a barrier. Nonetheless, we do not reflect this in our model, because it is unlikely that such a prevention motive is prevailing in this marketplace. This can be seen in Figure \ref{fig:scatter-PC1-PC2}, where we observe no apparent areas predominantly possessed by particular firms.\footnote{These arguments are consistent with the views of industry experts we interviewed. For example, Wujin Sim, the former director of Sandol, one of Korea's major font foundries, confirms that the forward-looking and preventive motive is minimal in font markets for at least two reasons: (i) the creative drive of designers is typically the main motive and (ii) labor-intensive font production has high fixed costs.}
Moreover, if one were to develop a dynamic model, there would be practical challenges. Since a firm's key strategic choice is differentiating products from those of competitors, the state space of a dynamic model is desired to incorporate the visual characteristics of all products in the market. Implementing this, however, is practically infeasible and extremely computationally burdensome given the large number of products. One may consider reducing the dimensionality of the state space by limiting firms' considerations on competing products, similar to the approaches in \cite{weintraub2008markov} and \cite{benkard2015oblivious}. Unfortunately, such simplification would be undesirable in our context, as it would restrict the visual differentiation behavior that is key to addressing our policy questions. Instead, our approach is to preserve the richness of firms' location choices, while adopting a static model that can be interpreted as reflecting the behavior of a myopic decision-maker.

\subsection{Identification and Estimation} \label{sec:id-est}

For the identification of parameters in the models for demand and product positioning, we use instrumental variables (IVs) and introduce related conditions. The key condition to identify the demand-side parameters is
\begin{equation} \label{eq:ortho-id}
    \mathbb{E}\left[\xi_{jt} | \boldsymbol{x}_{t}, \boldsymbol{z}_{t} \right] = 0,\qquad j=1,...,J,
\end{equation} where $\boldsymbol{z}_{t}$ is the vector of IVs across all products in the market. Valid IVs should generate shifts in prices across markets and be exogenous from the unobserved demand shock $\xi_{jt}$. To this end, we first use the monthly average spot exchange rates from Federal Reserve Economic Data (FRED).\footnote{Exchange rates include Euro, Britain Pound Sterling, Australian Dollar, Canadian Dollar, Swedish Krona, and Swiss Franc to USD.} These variations in exchange rates generate an exogenous shift in prices across countries and over time periods. Second, we also use the characteristics of competitors as IVs, namely, the ``BLP instruments.'' According to the timing assumption of the supply-side model, the product characteristics are chosen exogenously to unobserved demand shocks as similarly in \cite{eizenberg2014upstream}. This also implies that own product characteristics are exogenous to the demand shock, hence can instrument themselves.

We use optimal IVs in the spirit of \citet{amemiya1977maximum} and \citet{chamberlain1987asymptotic}.\footnote{It is documented in the literature that using optimal IVs not only improves asymptotic efficiency but also may substantially increase the finite sample precision of the estimates \citep{reynaert2014improving, conlon2020best}. Also, our practice of using differentiation IVs for approximating optimal IVs is known to enhance performance, especially against weak IV problem, in the literature \citep{gandhi2019measuring}.} We first construct differentiation IVs as in \citet{gandhi2019measuring} and use them to attain estimates for approximating optimal IVs, following \citet{berry1999voluntary}. To be specific, we count the number of local competitors of each product $j$ by own and rival firms respectively as:
\begin{align}
        &z^{\text{Local,Other}}_{jt\ell} = \sum_{j' \in J_{ft} \setminus \{j\}} \mathbbm{1}\left( d^{\ell}_{jj'} < \text{SD}_\ell \right), \qquad z^{\text{Local,Rival}}_{jt\ell} = \sum_{j' \notin J_{ft}} \mathbbm{1}\left( d^{\ell}_{jj'} < \text{SD}_\ell \right),
\end{align} where $d^{\ell}_{jj'} = |x^{emb}_{j'\ell} - x^{emb}_{j\ell}|$ is the absolute difference between the embedding element $\ell$ of products $j$ and $j'$, and $SD_{\ell}$ is one standard deviation of the component $\ell$. We use differentiation IVs to deal with a potential weak IV problem, which may arise from our ``large'' market setting in the sense of \citet{armstrong2016large}. Since there are many products in our marketplaces, using every product to construct ``BLP instruments'' might lead to weak identifying power. On the other hand, differentiation IVs are reported to be robust against this issue as these IVs are constructed based on \textit{local} competitors' product characteristics instead of whole products in the marketplace \citep{gandhi2019measuring}. In addition to theoretical justification, our empirical findings of local competition in Section \ref{sec:spatial} naturally motivate the use of these IVs.

To mitigate potential bias from universal quantity discounts in the marketplace, we focus on desktop license transactions at base quantity levels, which are exempt from such discounts and make up the majority of sales.\footnote{As shown in Table \ref{tab:qunatity-category} in the Appendix, single-user desktop license transactions are not subject to quantity discounts and account for about half of all desktop license sales.} Quantity reflects the number of users, which is likely to be orthogonal to prices and product characteristics. Restricting to this subsample is therefore unlikely to introduce significant bias into the estimation.

Next, we discuss the supply-side estimation. First, we assume the observed product positions and prices are equilibrium positions and prices. In the second stage, we set $\underbar{d}$ in \eqref{eq:loc-choice} to the minimum value of pairwise distances across all products, which can be regarded as the radius of the protective boundary under the current copyright regime. This simplifies the estimation process, as the location choices of firms in the data become the interior solutions of \eqref{eq:loc-choice} under the current copyright policy. This allows us to disregard the KKT multipliers due to the complementary slackness condition.

The main object to identify in the supply-side model is the fixed cost function $F$ in \eqref{eq:fixed-cost}. Note that the variable profit function \eqref{eq:variable-profit} is identified as long as the demand function is identified. Therefore, once we estimate the demand function, we can treat $\xi_{jt}$ as residuals and calculate $\mathbb{E}_{\xi_{kt}}\left[ \partial \pi_{jt}/\partial x_{k}^{emb} \right]$ in \eqref{eq:second-foc}.\footnote{To reduce computational complexity, we additionally assume $\partial p_{j'}/\partial x^{emb}_{k} = 0$ for all $j' \neq k$ in \eqref{eq:second-foc}. This assumption is supported by our previous findings. The descriptive statistics in Section \ref{sec:data} and the empirical results in Section \ref{sec:spatial} suggest that price adjustments of incumbent products are very rare. This implies that cross-product price responses with respect to visual characteristics are negligible in our model. Furthermore, we verified that computations under this assumption yield results very similar to those obtained with full price responses for a small sample. More details on computing the derivatives are discussed in Appendix \ref{sec:est-slope-detail}.} We consider two alternative assumptions for the unobservable $\nu_{k\ell}$. When $\nu_{k\ell}$ is viewed as a measurement error, we can assume $\mathbb{E}\left[ \nu_{k\ell} | \boldsymbol{x}_{t}\right] = 0$. Alternatively, when $\nu_{k\ell}$ is viewed as a structural error, we can instrument $\boldsymbol{x}_{t}$ using BLP-type IVs (i.e., competitors' product characteristics; see Appendix \ref{subsec:supply-side-IV} for details) that satisfy $\mathbb{E}\left[ \nu_{k\ell} | \boldsymbol{z}_{t}\right] = 0$. Each of these stochastic assumptions identifies the ``slope'' of the fixed cost function $F$ on the right hand side of \eqref{eq:second-foc}.

Unlike the slope of $F$, however, we cannot point-identify the intercept term of $F$ because the entry condition \eqref{eq:entry-condition} is characterized as an inequality restriction. Thus, we take an approach to partially identify the intercept term of the fixed cost by relying on the standard revealed profit rationale for firm $f$. Note that \eqref{eq:entry-condition} provides an upper bound on the constant of $F$. This upper bound, together with the zero lower bound, is used as a benchmark to examine various levels of fixed costs in the subsequent counterfactual analyses.

\section{Structural Estimation Results} \label{sec:est-result}

\subsection{Demand-Side Results}\label{subsec:demand_est_result}

\begin{table}[h!]
\centering
\caption{Fixed Coefficients Demand Estimation Results}
{\small
\begin{tabular}{lcccc}
\toprule
Column              & (1) & (2) & (3)  & (4) \\
Model              & \multicolumn{2}{c}{OLS} & IV (2nd Stage)  & IV (1st Stage) \\
Variables & $\ln(s_{j}/s_{0})$ & $\ln(s_{j}/s_{0})$ & $\ln(s_{j}/s_{0})$ & Prices \\
\midrule
Prices & -0.0196 & -0.0207 & -0.1658 & - \\
 & (0.0002) & (0.0002) & (0.0015) & - \\
Glyph Counts & 0.0004 & 0.0003 & 0.0008 & 0.0034 \\
 & (0.0000) & (0.0000) & (0.0000) & (0.0001) \\
Ex Rate & - & - & - & 0.2482 \\
 & - & - & - & (0.0024) \\
Constant & -8.0149 & -6.8756 & -5.5018 & -3.8495 \\
 & (0.0084) & (0.9761) & (0.0282) & (0.2258) \\
\midrule
Observations & 225,658 & 225,658 & 225,658 & 225,658 \\
6-dim Embeddings          & Yes       & No       & Yes       & Yes \\
128-dim Embeddings   & No        & Yes      & No       & No \\
BLP-type IVs Included   & No        & No      & Yes       & Yes \\
$R^{2}$ & 0.0497 & 0.1128 & - & 0.1110 \\
$F$ stat & 1192 & 203.8 & 1768 & 1174 \\
\bottomrule
\end{tabular}
}
\tabnotes{This table shows results from OLS and IV regression models with fixed coefficients. We use exchange rates and BLP-type IVs for IV regression. We omit the coefficient estimates of BLP-type IVs from the table. Robust standard errors are in parentheses. All coefficient estimates are statistically significant at 1\% level. The Cragg-Donald $F$ statistic of IV regression is estimated to be 1266.}
\label{tab:linear-reg-result}
\end{table}

We first report regression results from fixed-coefficient linear models to check the validity and strength of instruments for prices. Table \ref{tab:linear-reg-result} shows the results. Columns (1) and (2) report simple OLS regression results, which respectively control for the 6- and 128-dimensional embeddings. The two regression results are qualitatively similar in terms of prices and glyph counts coefficients estimates, suggesting that 6-dimensional embeddings sufficiently control for the attributes captured in the embeddings. All the coefficient estimates are statistically significant at 1\% level. The sign of the estimates all seem reasonable. As the number of glyphs (i.e., unique characters in a font family) is associated with functionality and the supported number of languages, the estimates are expected to be positive. The price coefficient estimates are all negative, although they would be biased due to the endogeneity between prices and unobserved preference shocks (e.g., quality). Columns (3) and (4) show IV regression results of the second and first stages, respectively. Again, the sign of estimates all seem reasonable. Our instruments appear to effectively address endogeneity. The estimated price coefficient of the IV regression in column (3) becomes more negative than those in columns (1) and (2). Because higher prices may be correlated with higher quality or taste shocks, this shift suggests that the instruments are valid. In addition, the Cragg-Donald $F$ statistic is estimated to be 1266, indicating that the instruments are strong. 

\begin{table}[h!]
\centering
\caption{Random Coefficents Demand Estimation Results}
{\small
\begin{tabular}{lccc}
\toprule
Variables/Parameters & $\Bar{\beta}$ & $\sigma$ & $\rho$ \\
\midrule
Constant & $-7.148$ & - & - \\
         & $(0.035)$ & - & - \\
Prices   & $-0.156$ & - & - \\
         & $(0.001)$ & - & - \\
Glyph Counts    & $0.001$ & - & - \\
         & $(0.000)$ & - & - \\
Embbeding 1      & $5.292$ & $9.500$ & - \\
         & $(0.082)$ & $(0.096)$ & - \\
Embbeding 2      & $-6.328$ & $2.499$ & - \\
         & $(0.109)$ & $(0.458)$ & - \\
Embbeding 3      & $-11.823$ & $7.652$ & - \\
         & $(0.177)$ & $(0.209)$ & - \\
Embbeding 4      & $-11.661$ & $5.582$ & - \\
         & $(0.226)$ & $(0.720)$ & - \\
Embbeding 5      & $2.374$ & $11.567$ & - \\
         & $(0.140)$ & $(0.504)$ & - \\
Embbeding 6      & $10.145$ & $0.113$ & - \\
         & $(0.242)$ & $(0.005)$ & - \\
Category \& Tag & - & - & $0.317$ \\
                 & - & - & $(0.011)$ \\
\bottomrule
\end{tabular}
}
\tabnotes{This result shows estimation results of the random coefficient nested logit model with dimension reduction in \eqref{eq:discrete-utility}. The number of observations and that of markets are 225,658 and 540, respectively. Heteroscedasticity robust standard errors are shown in parentheses. All coefficient estimates are statistically significant at 1\% level.}
\label{tab:demand-result}
\end{table}

Table \ref{tab:demand-result} reports demand estimation results from our main specification, namely the random coefficient nested logit model with the dimension-reduced visual attributes in \eqref{eq:discrete-utility}. For this specification, motivated from the previous estimation results, we include the 6-dimensional embeddings as the product attributes, since using random coefficients on the 128-dimensional embeddings hampers the estimation process.\footnote{We find that estimating the random coefficient model with the 128-dimensional embeddings is not feasible due to memory issues.} Column ``$\Bar{\beta}$,'' ``$\sigma$,'' and ``$\rho$'' show the estimates of mean, random, and nesting coefficients, respectively. All the estimates are statistically significant at 1\% level. The signs of estimated price and glyph count coefficients are considered reasonable; an increase in price tends to decrease the mean utility, while a decrease in the number of glyphs also lowers the mean utility. The estimate for the nesting parameter $(\rho)$ is 0.317, indicating some degree of substitution within the nest. The sign of the mean coefficient estimates ($\Bar{\beta}$) of embbedings is mixed. For instance, the positive and negative coefficient estimates of embbeding 1 and embbeding 2 indicate that consumers on average tend to prefer bolder fonts and display fonts, conditional on the other characteristics. However, the random coefficient estimates exhibit significant heterogeneity in preferences on product shape. Figure \ref{fig:own-price-elas} displays the distribution of the median own-price elasticity estimates across markets, with each median calculated across products. The median elasticities are centered around -2.4. Overall, the demand estimation results appear economically meaningful and reasonable. 

\begin{figure}
    \centering
    \caption{The Distributions of Own Shape Elasticity}
    \subfloat[Embedding 1]{\includegraphics[width=0.5\linewidth]{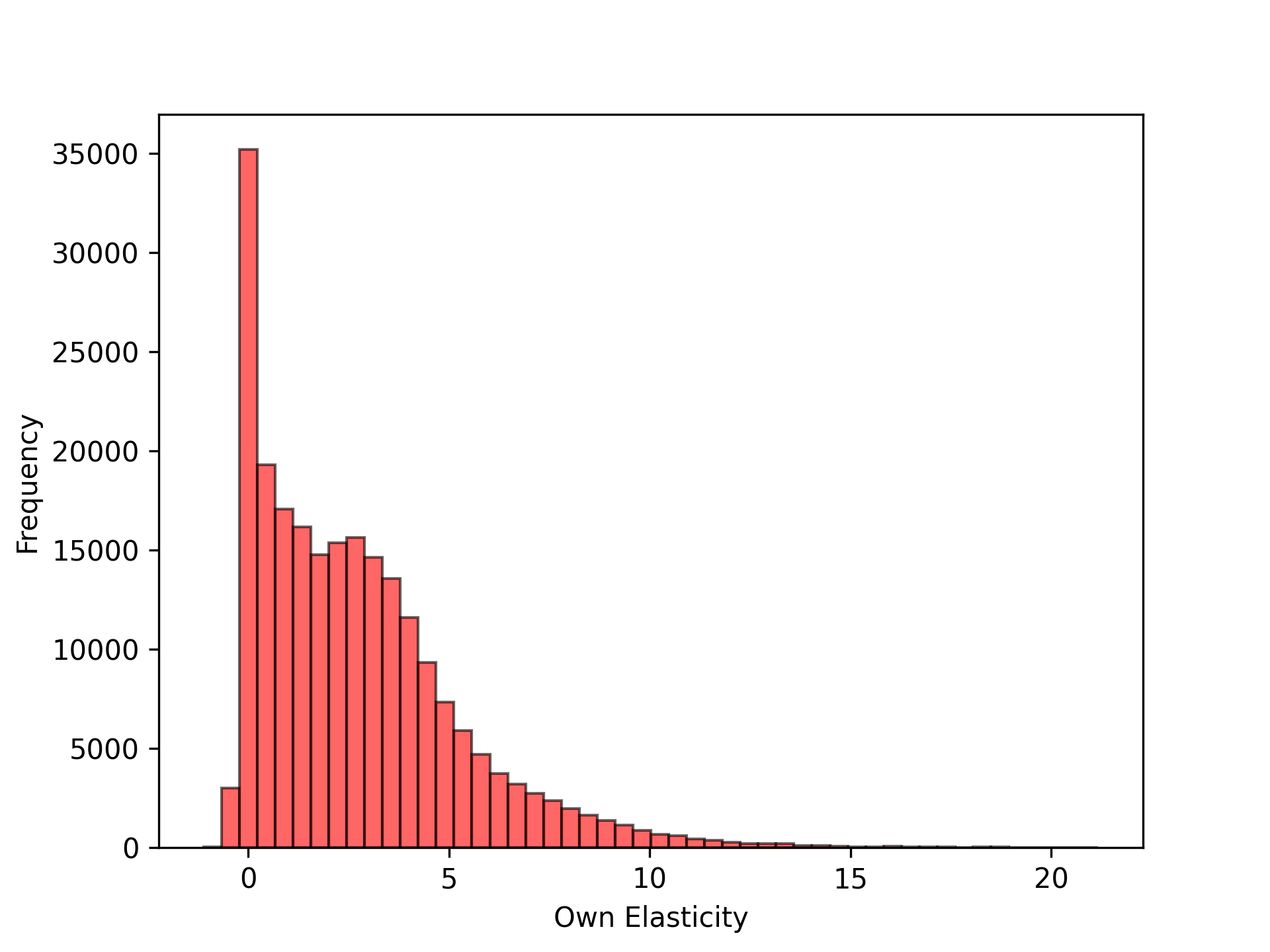}}
    \subfloat[Embedding 2]{\includegraphics[width=0.5\linewidth]{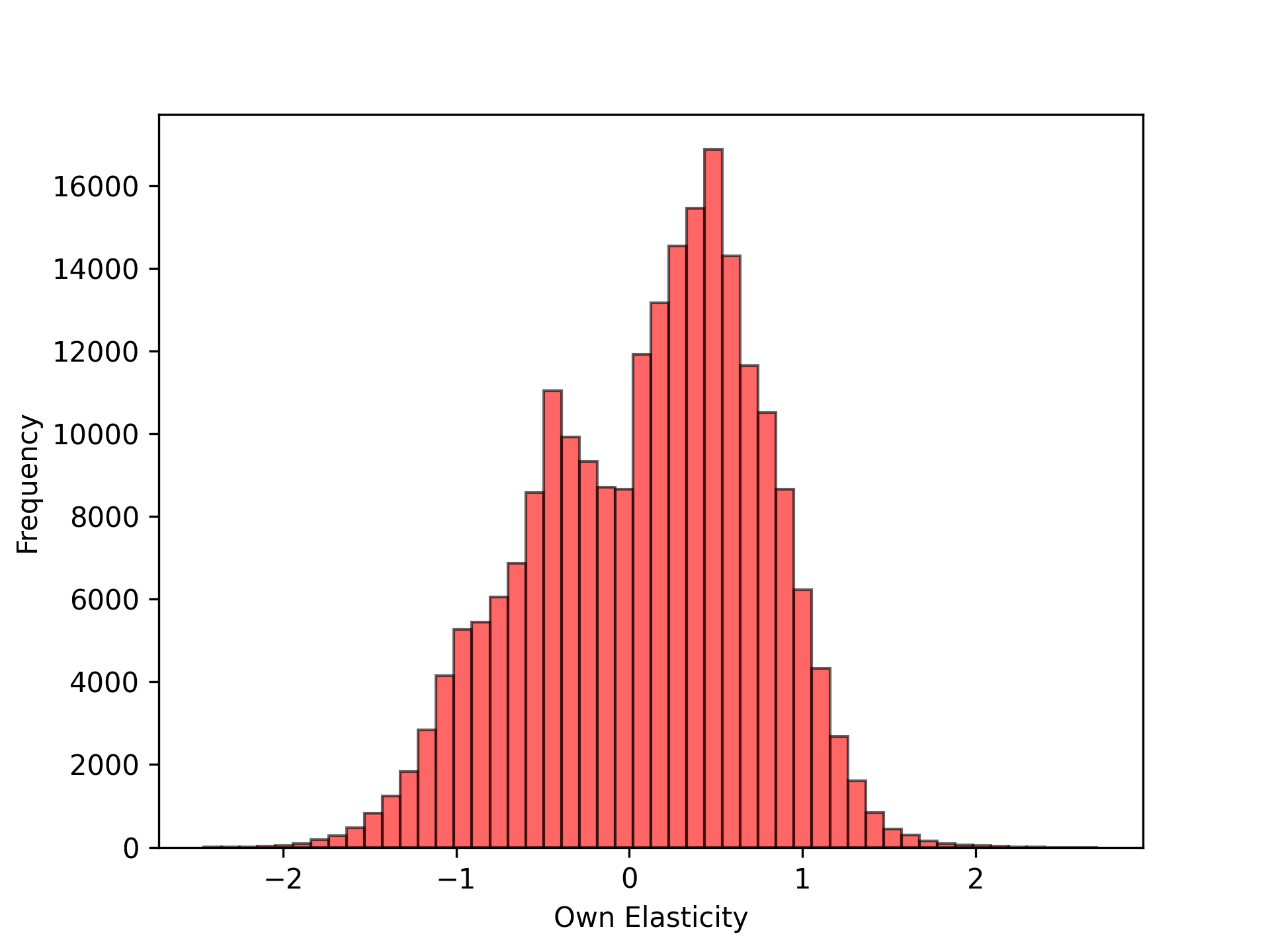}} 
    \tabnotes{This figure shows the distributions of own elasticity with respect to the embbedings. The distribution is plotted across products and markets. Panels (a) and (b) correspond to the distributions with embbedings 1 and 2, respectively. The distributions correspond to embbedings 3 to 6 are shown in Figure \ref{fig:own-shape-elas-pc3-6}.}
    \label{fig:own-shape-elas}
\end{figure}

Figure \ref{fig:own-shape-elas} shows the distributions across products and markets of own elasticities with respect to the embbedings, referred to as \emph{own shape elasticity}; we focus on the first two embeddings; the plots for the rest can be found in Figure \ref{fig:own-shape-elas-pc3-6} in the Appendix. The results indicate that there exists substantial heterogeneity in preferences over product shapes. The own shape elasticities of the first embedding are predominantly positive, suggesting that this component may capture design elements that generally enhance consumer utility.

\begin{figure}[htbp!]
    \centering
    \caption{Price Diversion Ratios and Original Embedding Distances}
    \subfloat[With Embeddings]{\includegraphics[width=0.5\textwidth]{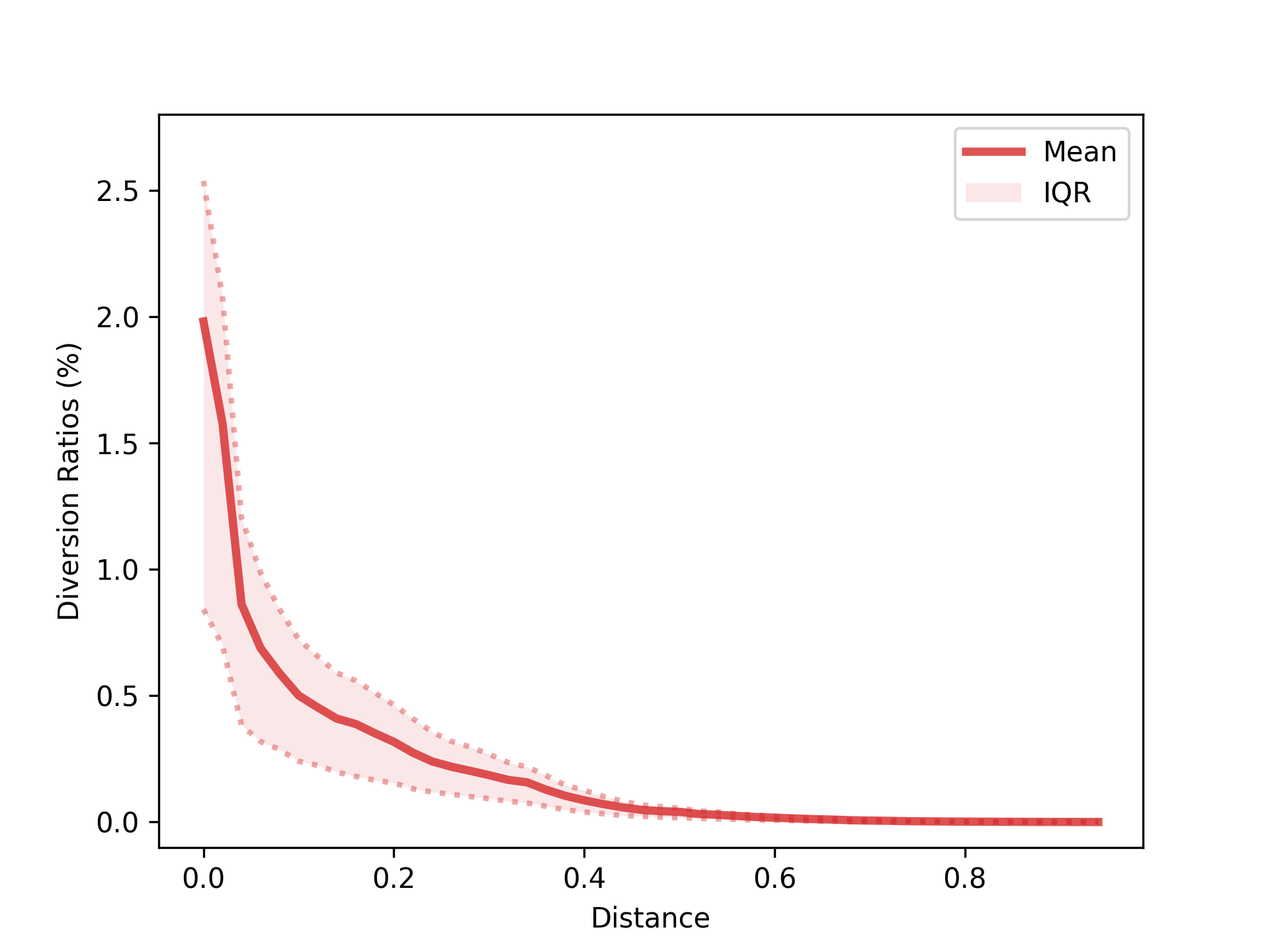}} 
    \subfloat[Without Embeddings]{\includegraphics[width=0.5\textwidth]{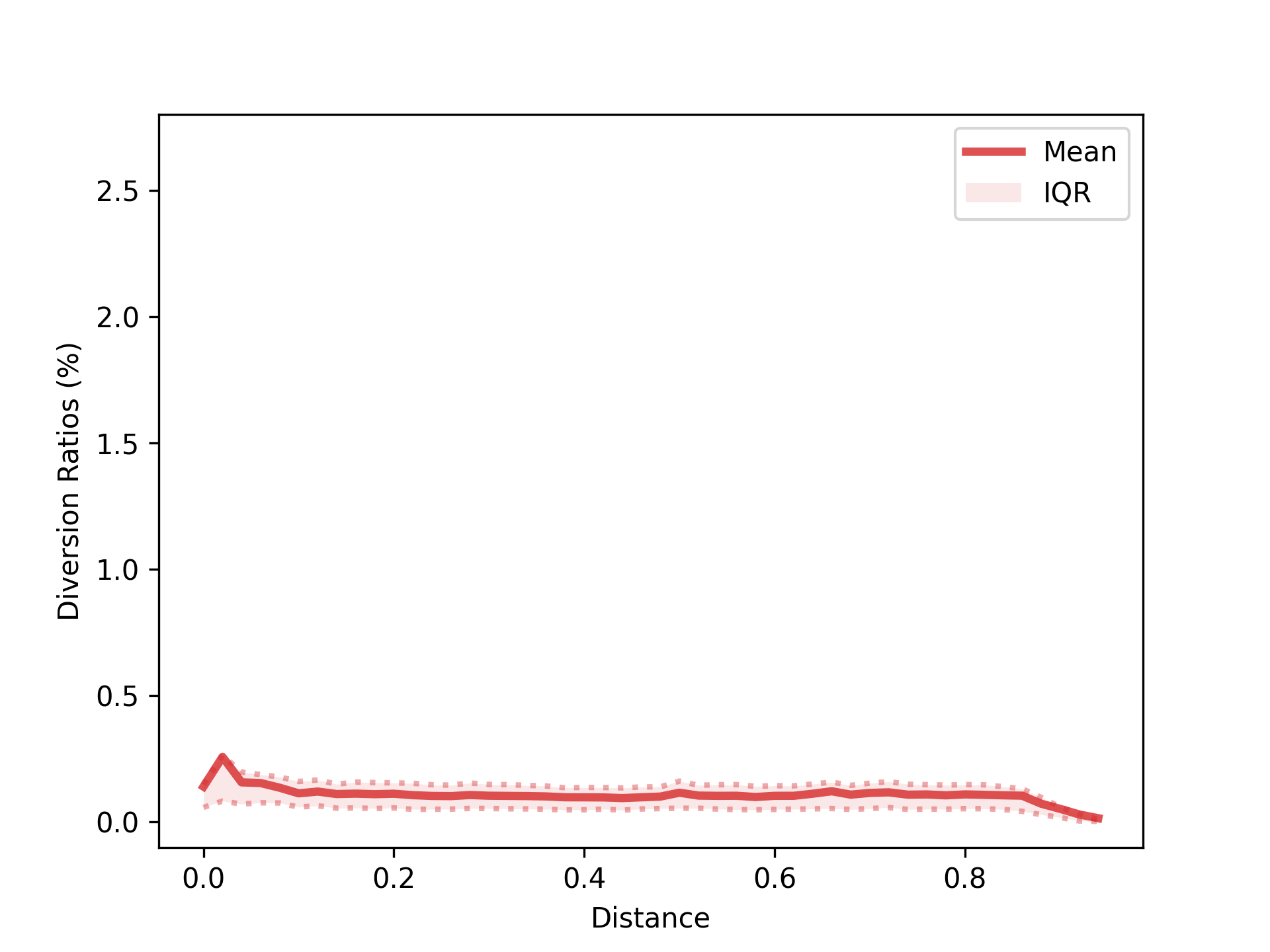}} \\
    \subfloat[Average Number of Products]{\includegraphics[width=0.5\textwidth]{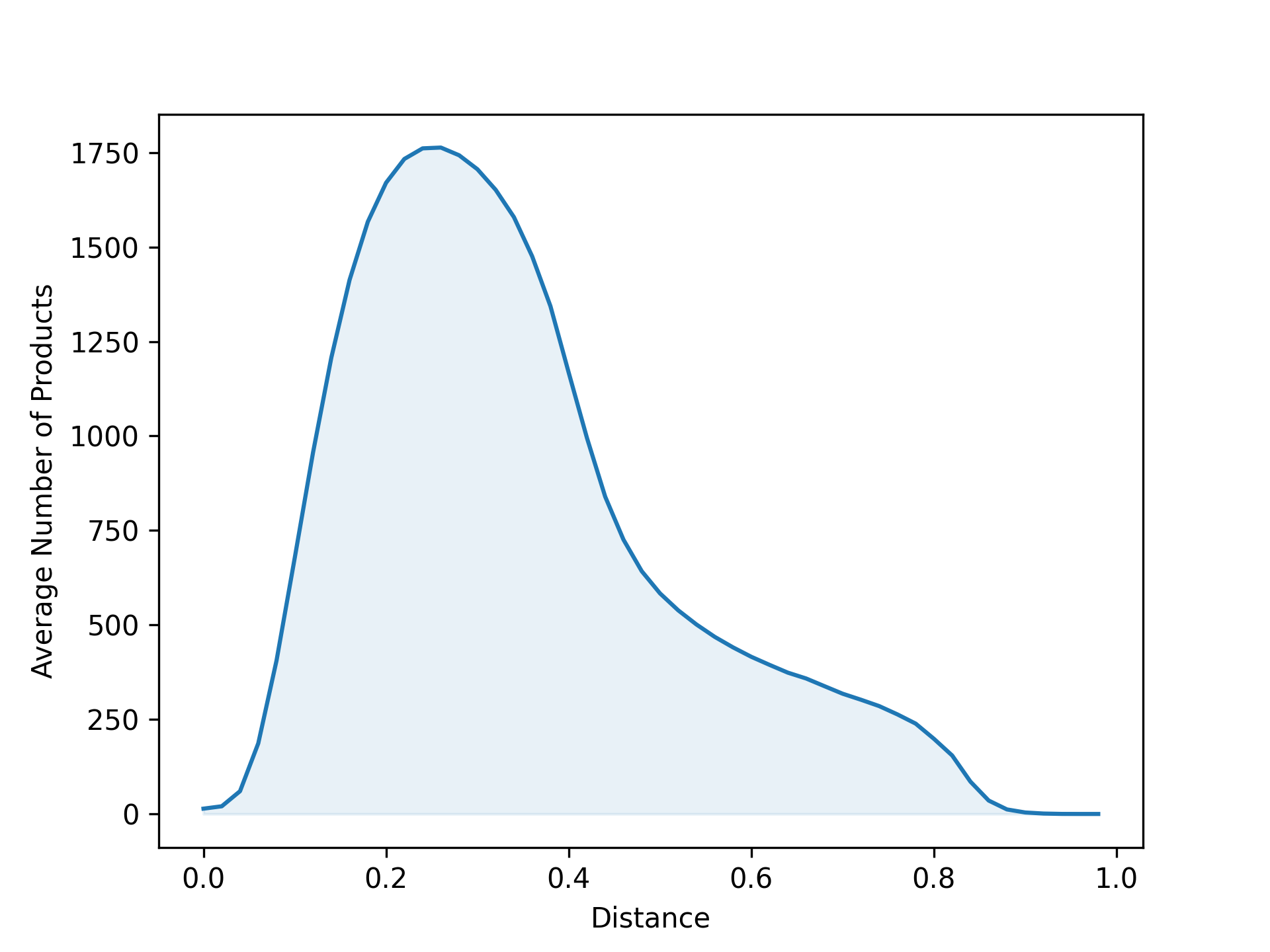}} 
    \tabnotes{Panel (a) shows the average price diversion ratios along the embedding distance $d$ (see equation \eqref{eq:DIV}), based on estimated demand with embeddings $x_{j}^{emb}$ in the utility function; Panel (b) presents the results when $x_{j}^{emb}$ is excluded. As a reference, Panel (c) shows the average number of products along radial areas.} 
    \label{fig:div-dist}
\end{figure}

Using the demand estimates, we examine the patterns of competition in the space of visual characteristics. Specifically, we calculate diversion ratios and display their averages across radial areas of different distances in the embedding space, following the idea of the reduced-form analysis in Section \ref{subsec:num_spatial_competitor}. We define the aggregation, $DIV$, of diversion ratios, $div$, at market $(c,t)$ for a given baseline distance $d$ as:
\begin{equation}\label{eq:DIV}
    DIV_{ct}(div, d) := \frac{1}{|J_{ct}|}\sum_{j \in J_{ct}} \frac{ \sum_{j' \in J_{ct} \setminus \{j\} }div_{jj'}\mathbbm{1}\{x_{j'}^{emb} \in Rad_{j}(d)\} }{\sum_{j' \in J_{ct} \setminus \{j\}} \mathbbm{1}\{x_{j'}^{emb} \in Rad_{j}(d) \} },
\end{equation} where the diversion ratio $div_{jj'}= \frac{\partial s_{j'}}{\partial p_{j}}/\frac{\partial s_{j}}{\partial p_{j}}$ and $Rad_{j}(d) := \{x \in \mathbb{S}^{128} : d \leq \lVert x_{j}^{emb} - x\rVert_{2} < d + 0.02\}$. We set $d = 0, 0.02, ..., 0.98$ and display the mean values and the inter-quantile ranges of $DIV_{ct}(div,d)$ along $d$.

Figure \ref{fig:div-dist}(a) presents the results. The diversion ratio declines sharply as the embedding distance $d$ increases, supporting the aspect of \textit{local} competition in the visual characteristics space documented in the reduced-form analysis. When $d$ is small, as price increases, consumers are more likely to switch to visually similar competing products. Furthermore, recall that we use lower dimensional embeddings for demand estimation, while the radial distances in the reduced-form analysis are calculated using the 128-dimensional embeddings. Figure \ref{fig:div-dist}(a) shows that, despite this additional dimension reduction, the demand estimates remain economically meaningful, suggesting that demand-relevant information is retained after the dimension reduction. 

Incorporating visual characteristics as observables in the demand model is crucial for capturing local competition. To validate this, we recalculate the diversion ratios along radial areas in the embedding space by estimating the demand model \textit{without} the visual attributes, as depicted in Panel (b) of Figure \ref{fig:div-dist}. This specification incorporates only the structured attributes of products with the constructed nests. The diversion ratios estimated without the embbedings are significantly lower than those with the embbedings and do not exhibit the sharp decline observed in Panel (a); instead, they remain flat regardless of increasing distance. This result is not only inconsistent with the empirical evidence of local competition presented in Section \ref{sec:spatial}, but also contradicts industry experts' prevailing understanding of competition.

We further examine various aspects of spatial competition through our demand estimates. First, we find similar patterns as that in Figure \ref{fig:div-dist}(a) for various competition measures such as cross-price elasticity and long-run diversion ratios (Figures \ref{fig:comp-dist} and \ref{fig:comp-dist-nl} in the Appendix). Moreover, we find that the availability of close substitutes causes consumers to substitute less toward outside goods, which include free fonts---the primary competitors of commercial font products (Appendix \ref{subsec:outside}).

\subsection{Supply-Side Results}

In this section, we report the supply-side estimation results, which show that positions in the characteristics space and distances to incumbents are significant determinants of fixed costs. We report the OLS estimation results here; the IV results are qualitatively similar and contained in Appendix \ref{subsec:supply-side-IV}. Table \ref{tab:slope-est} presents the slope estimates of the fixed-cost function. Across all regressions involving different embedding elements, the estimates for the coefficients $\eta_{0\ell}$ on the distance to incumbents are statistically significant at the 1\% level, indicating that location in the characteristics space is an important determinant of fixed costs. While the individual estimates of $\eta_{1\ell}$ to $\eta_{3\ell}$ are not always statistically significant, they are jointly significant at the 1\% level in all regressions except for embedding 4. To assess their joint significance, we conduct a Wald test with the null hypothesis $H_{0}: \eta_{1\ell} = \eta_{2\ell} = \eta_{3\ell} = 0$. The results of the test are reported in the ``$F$-stat'' row. 

\begin{table}[htbp!]
\centering
\caption{Slope Estimation Results}
\resizebox{\textwidth}{!}{%
{\small
\begin{tabular}{lcccccc}
\toprule
& (1) & (2) & (3) & (4) & (5) & (6) \\
Parameters                        & $\partial F/\partial x^{emb}_{1}$                & $\partial F/\partial x^{emb}_{2}$                 & $\partial F/\partial x^{emb}_{3}$                 & $\partial F/\partial x^{emb}_{4}$                 & $\partial F/\partial x^{emb}_{5}$                 & $\partial F/\partial x^{emb}_{6}$                  \\
\midrule
$\eta_{0\ell}$       & 3400.8          & -2916.9          & -6742.9          & -5549.6          & 1699.0           & 5158.2           \\
                        & (218.35)           & (100.01)            & (292.15)            & (185.88)            & (104.96)            & (182.42)            \\
$\eta_{1\ell}$ & 0.15          & 0.12            & 0.41               & -0.04              & 0.10               & -0.19             \\
                        & (0.06)            & (0.05)             & (0.28)             & (0.16)             & (0.15)             & (0.10)             \\
$\eta_{2\ell}$ & 0.41          & -0.34             & -3.61              & 0.23               & 0.43               & 2.59               \\
                        & (0.17)            & (0.29)             & (2.41)             & (1.99)             & (2.23)             & (1.62)             \\
$\eta_{3\ell}$ & -0.22             & 0.57              & 10.85              & -0.28              & -0.98              & -6.65              \\
                        & (0.14)            & (0.48)             & (5.96)             & (6.82)             & (8.92)             & (7.00)             \\
\midrule
$R^{2}$                 & 0.33              & 0.06               & 0.07               & 0.00               & 0.25               & 0.01               \\
$F$-stat            & 271.96         & 33.71           & 40.66           & 0.73               & 177.85          & 4.44            \\
Observations            & \multicolumn{6}{c}{1,630} \\
\bottomrule
\end{tabular}
}
}

\tabnotes{This table shows the estimated slopes of the fixed cost function. Heteroskedasticity robust standard errors are shown in the parentheses. F-statistics of Wald test on $H_{0}: \eta_{1\ell} = \eta_{2\ell} = \eta_{3\ell} = 0$ v.s. $H_{1}:$ the negation of $H_{0}$ are shown in the $F$-stat row. The number of observations is 1,630, which correspond to the number of entrants.}
\label{tab:slope-est}
\end{table}

\begin{figure}[htbp!]
    \centering
    \caption{Fixed Costs Estimation Results}
    \subfloat[Fixed Costs and Distances]{\includegraphics[width=0.5\linewidth]{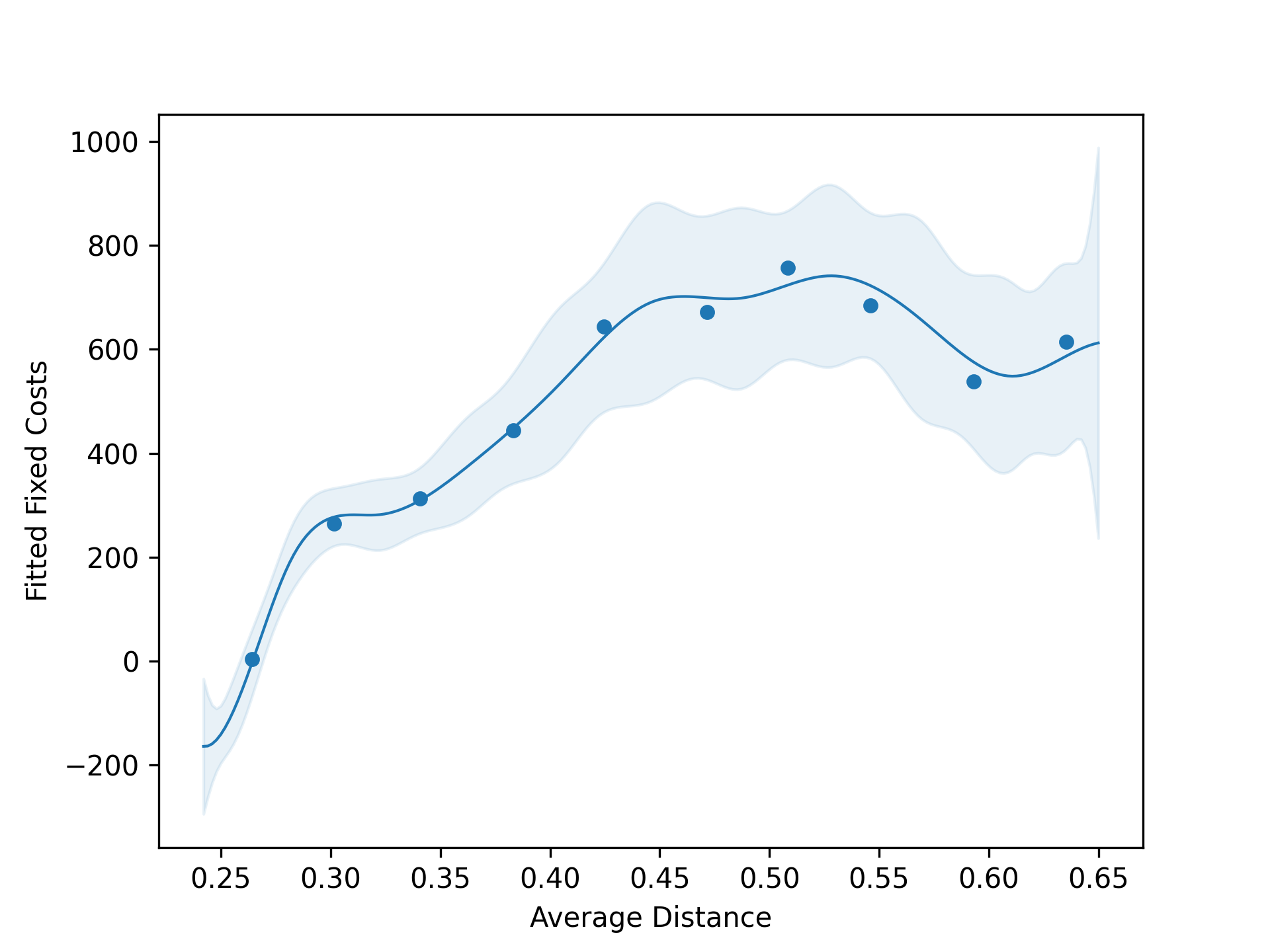}}
    \subfloat[Upper Bound Estimates]{\includegraphics[width=0.5\linewidth]{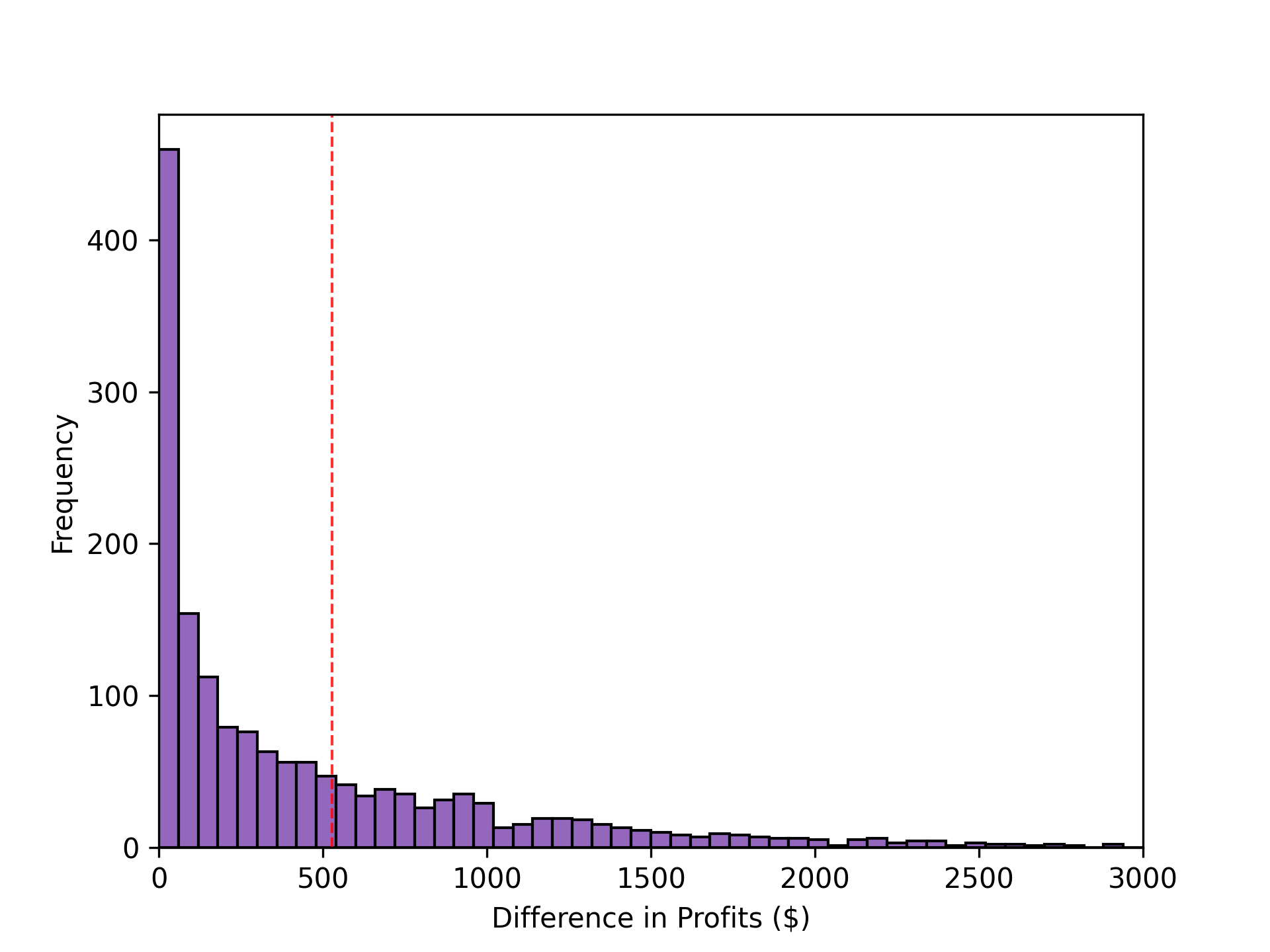}} 
    \tabnotes{Panel (a) presents a binscatter plot of the estimated explained part of fixed costs against average distances. We compute the fitted values by using the estimated slope coefficients which are displayed in Table \ref{tab:slope-est}. The solid line represents a third-order local polynomial fit, with the shaded area indicating the 95\% confidence band. Heteroskedasticity-robust standard errors are applied. The dots represent bin-by-bin averages with the evenly-spaced binning method \citep{cattaneo2024binscatter}. Panel (b) shows the histogram of the estimated differences between the expected and no-entry variable profits indicating the upper bound of the fixed costs, as defined in \eqref{eq:entry-condition}. The red dashed line represents the average value, 523. The standard deviation of the differences is 772. Both figures are based on 1,630 observations, corresponding to the number of entrants.}
    \label{fig:supply-result}
\end{figure}

Next, using the estimation results, we study how proximity to existing products in the characteristics space affects the cost of developing a new product. Specifically, we examine the relationship between the fitted values of fixed costs and the average distances to other competitors. Note that the shape of this relationship is identified from the slope coefficients estimated above. In Figure \ref{fig:supply-result}(a), the explained part of the estimated fixed costs is plotted against the average distance in the embedding space.\footnote{Some fitted values can be lower than zero as we only focus on the explained part of the fixed costs.} The results show an initial rise in fixed costs as the average distance increases, but the trend flattens around an average distance of 0.45. Beyond this point, average distances appear to have only a minor effect on fixed costs. This suggests that having competitors nearby helps entrants reduce their development costs, likely due to positive mimicking externalities gained from producing visually similar products.\footnote{This finding is consistent with industry knowledge. Fonts are vector-based objects and can be easily opened and modified with font-making software, which implies a cost advantage in tweaking existing fonts.} Again, the IV estimates of fixed costs present a similar concave shape (Figure \ref{fig:supply-result-IV}).

Lastly, we present the estimates of the upper bound estimates on the fixed costs. Figure \ref{fig:supply-result}(b) shows the histogram  across entrants of the differences between the expected variable profits upon entry and without entry in the entry condition \eqref{eq:entry-condition}, which exhibits reasonable patterns. The distribution is right-skewed with many values close to zero. This pattern is consistent with the observed product revenue distribution. 

Unlike the demand-side estimation, there is no obvious benchmark without embeddings against which to compare our supply-side estimation results. For the design products we consider, the structured attributes are too coarse to meaningfully define either a model of copyright policy or fixed costs as a function of distance. The analysis of fixed costs relative to distance and the subsequent counterfactual analysis manipulating copyright policy are only possible because of the characteristics space constructed using embeddings.

\section{Counterfactual Policy Analyses} \label{sec:counterfactual}

Using the estimated structural model, we investigate the role of copyright policy in competition and welfare. We conduct two counterfactual analyses: (1) examining the impacts of copyright protection on consumer welfare through infringer responses; (2) exploring the interplay between copyright protection and production costs in determining consumer and producer welfare.

\subsection{Enforcing Stricter Copyright Policies} \label{subsec:protection-study}

As the first counterfactual analysis, we increase the degree of copyright protection to understand its impact on consumer welfare through different infringer responses. Given the similarity constraint in the product positioning equation \eqref{eq:loc-choice}, imposing stricter policy corresponds to increasing the protective boundary radius $\underbar{d}$. Setting a larger $\underbar{d}$ is interpreted as imposing stricter copyright protection in the market. 

We perform two exercises: (i) a baseline simulation that removes entrants within a protective boundary (i.e., infringers) near existing products, and (ii) a relocation simulation that pushes infringers outside the protective boundary, thereby rearranging their products in the characteristic space. In these exercises, we do not account for entrants' counterfactual location choices under stricter copyright protection; endogenous location choices are incorporated in Section \ref{subsec:simul-protection-fclevel}. Nevertheless, these simulations provide valuable insights. The baseline simulation helps illustrate the extent of local monopolistic power that incumbents might exert due to increased protection, and the corresponding reduction in consumer surplus arising from fewer available choices in the marketplace. The relocation simulation allows us to assess whether consumers might benefit from increased diversity in product attributes.

For these analyses, we impose the copyright policy starting from the first period in our dataset (April 2014) and simulate equilibrium prices and market shares for the counterfactual products based on the demand estimates and pricing model. Consumer surplus for each market $ct$ under policy parameter $\underbar{d}$ is calculated as:
\begin{equation}
    CS_{ct}(\underbar{d}) = \frac{1}{N_s}\sum_{i=1}^{N_s} \ln \left(1 + \sum_{j \in J_{ct}(\underbar{d})} \exp V_{ijct}(\underbar{d}) \right)/ (- \Bar{\beta}^{p}), \label{eq:CS}
\end{equation} where $N_s$ is the number of simulated consumers, and $V_{ijct}(\underbar{d}) := \Bar{\beta}^{p} p_{jct}(\underbar{d}) + \Bar{\beta}^{str}x_{j}^{str} + x_{j}^{emb'}\beta_{i}^{img} + \xi_{jct}$ and $J_{ct}(\underbar{d})$ represents the consumers' counterfactual choice set given $\underbar{d}$. According to \eqref{eq:CS}, copyright protection can affect consumer surplus through two channels: (i) increased prices due to greater monopolistic power enter $V_{ijt}$, decreasing consumer utilities; (ii) the reduced choice set is reflected in $J_{ct}(\underbar{d})$.

\begin{figure}[ht!]
    \centering
    \caption{Simulated Consumer Surplus with Copyright Protection}
    \includegraphics[width=4in]{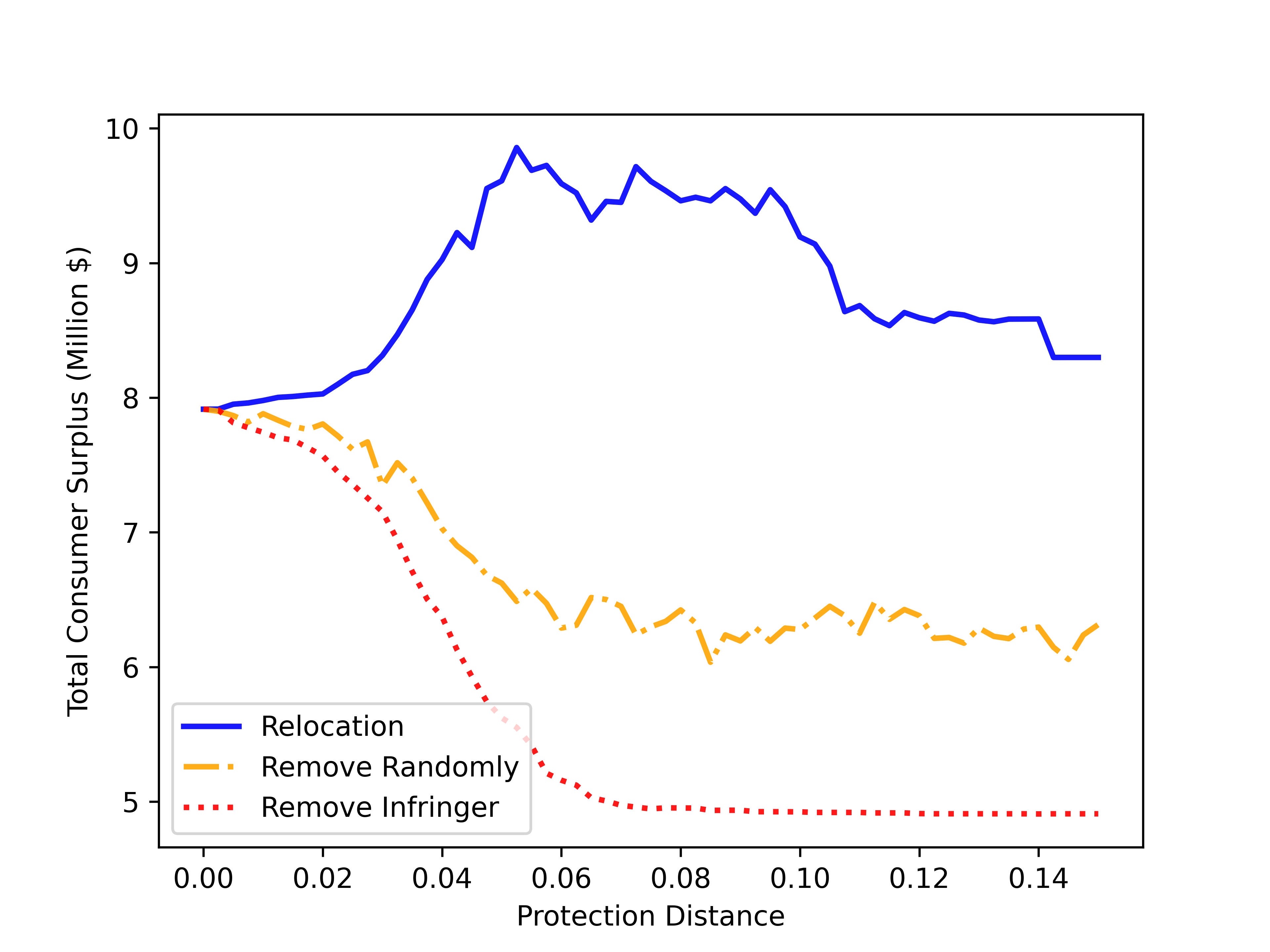}
    \tabnotes{This figure shows changes in consumer surpluses of simulation exercises as protection distance increases. The orange dash-dot and red dot lines show the results of removing random products and infringers, respectively. The blue line presents those of pushing infringers to be outside of protection boundary. As a reference, in our data, the minimum and maximum Euclidean distances among products are 0.0002 and 0.9563, respectively. See Table \ref{tab:dist-examples} for the example of product shapes and corresponding distances.}
    \label{fig:simul-result}
\end{figure}

We find that consumer surplus sharply declines as infringers are removed with increasing protection levels. In Figure \ref{fig:simul-result}, the red dashed line represents the baseline simulation results. For instance, under the strictest protection level ($\underbar{d} = 0.15$), consumer surplus decreases by approximately 39\% compared to the baseline copyright regime ($\underbar{d}=0$). This decline may not be solely driven by the reduced number of products; product attributes can also play a significant role. To isolate this effect, we compare the change in consumer surplus from the baseline exercise to that from a random removal exercise, in which we eliminate a number of randomly selected products equal to the number of infringers removed in the baseline exercise. The random removal simulation reveals a similar pattern, but the reduction in consumer surplus is less prounounced than in the baseline scenario. Since consumer surplus in \eqref{eq:CS} is a monotonic function of the utility provided by each product, this suggests that infringers tend to locate in regions where utility levels are high. In Appendix \ref{subsubsec:local_mono}, we further decompose the decrease in consumer surplus into two components: price increases due to enhanced monopolistic power, and diversity loss resulting from the elimination of new entrants. We find the latter factor significantly dominates. Therefore, to fully understand welfare changes under copyright policy, it is crucial to consider the location choices of designers, as these decisions directly affect the diversity of available products.

In addition, consumers might benefit from the rearrangement of product locations induced by stricter copyright protection. In Figure \ref{fig:simul-result}, the solid blue line illustrates an inverse U-shaped relationship between the level of protection and consumer surplus. Since the number of products remains constant across all copyright regimes in this scenario, this result implies that a moderate level of protection, such as $\underbar{d} = 0.05$, optimally reallocates products toward desirable attributes, leading to an increase in consumer surplus of approximately 24\% compared to the baseline regime. However, consumer surplus declines at higher protection levels, suggesting that consumers do not necessarily prefer products with excessively differentiated attributes.

\subsection{Interplay between Copyright Policies and Fixed Costs} \label{subsec:simul-protection-fclevel}

Next, we investigate the interaction between copyright protection and cost reductions driven by technological advancements in the industry, such as the introduction of generative AI for font design (e.g., \citealp{peong2024typographic}). We conduct simulations examining how various combinations of fixed costs and protection distances ($\underbar{d}$) shapes welfare and market outcomes. These studies take into account not only endogenous consumer decisions but also firms' entry decisions, optimal positioning, and pricing behaviors in response to counterfactual changes in cost structure and policy stringency.

We specifically consider two types of technological changes: the advent of technology that complements human designers (Scenario A) and technology that substitutes them (Scenario B). We regard the former as a more plausible scenario in the current industry. In Scenario A, we conceptualize the technology as an ``assistant'' that does not fundamentally alter the nature of design process but reduces overall costs. To represent this scenario, we consider a cost function that maintains the concave shape of the estimated function (Figure \ref{fig:supply-result}(a)), but whose level is uniformly lower across product characteristics and random shocks:
\begin{equation*}
    F^{assist}(\boldsymbol{x}_{t},\nu_{kt}) = F(\boldsymbol{x}_{t},\nu_{kt}) - C \ \text{ for all } \boldsymbol{x}_{t}, \nu_{kt},
\end{equation*} where $F(\boldsymbol{x}_{t},\nu_{kt})$ is the factual cost function defined in \eqref{eq:fixed-cost} and the cost-reduction is captured by $C > 0$. In this sense, we assume that \emph{mimicking advantage}---the cost reduction associated with designing visually similar products to incumbents---persists with this technology. To incorporate this cost function in simulation, we use the slope estimates of $F(\boldsymbol{x}_{t},\nu_{kt})$ and experiment with various intercept levels including those that correspond to reduced costs.

In Scenario B, we view the technology as a ``substitute'' that almost entirely automates the product design process. Under this scenario, generative AI produces fonts at low costs regardless of their visual similarity to incumbents. Consequently, the new cost structure becomes less dependent on the visual similarity between new and existing products, diminishing the cost advantages of mimicry: for  any order of derivative $\alpha = 0, 1,...$ and the embedding element $\ell$,
\begin{equation*}
    \left | \frac{\partial^{\alpha} F^{subst}(\boldsymbol{x}_{t},\nu_{kt})}{(\partial x^{emb}_{\ell})^{\alpha}} \right |= \left |\frac{\partial^{\alpha} F(\boldsymbol{x}_{t},\nu_{kt})}{(\partial x^{emb}_{\ell})^{\alpha}}\right | - C^{\alpha}_{\ell} \ \text{ for all } \boldsymbol{x}_{t}, \nu_{kt},
\end{equation*} where the cost reduction is captured by $C_{\ell}^{\alpha} > 0$ for all $\alpha,\ell$. We incorporate this cost function by simply setting the slope of $F^{subst}(\boldsymbol{x}_{t},\nu_{kt})$ to zero for all $\boldsymbol{x}_{t}, \nu_{kt}$ and experimenting with various intercept values. Further details on constructing these counterfactual cost functions for both scenarios can be found in Appendix \ref{subsec:details_counterfactual}.

With the counterfactual cost functions specified, we implement the simulations in the following manner. We generate hypothetical embeddings from the existing benchmark and compute the associated costs (using the counterfactual cost functions) and the expected profits (using the demand estimates). Firms are then allowed to optimally choose product locations and make entry decisions. As fully simulating firms' decisions by solving the model is extremely burdensome, we (i) focus on a random sample of firms that sequentially make entry and product-positioning decisions, (ii) utilize an approximation technique to calculate expected profits (\citealp{berry1999voluntary}), and (iii) limit our analysis to markets in April 2014, the initial period of our dataset. Additional details of these simulation exercises can be found in Appendix \ref{subsec:details_counterfactual}.

A policymaker (e.g., a copyright judge) might be interested in either aggregate social welfare or welfare per product. Aggregate producer surplus, $PS_{t}$, at time $t$ is defined as the sum of total profits across firms (i.e., across existing and new products), where each profit reflects the fixed costs associated with introducing new product $k$: $PS_{t}:=\sum_{f,k}\Pi_{ft,k}$ where $\Pi_{ft,k}$ is the total profit defined in \eqref{eq:total-profit}. Aggregate social welfare is then defined as $SW_{t} = PS_{t} + \sum_{c} CS_{ct}$, where $CS_{ct}$ is the aggregate consumer surplus defined in \eqref{eq:CS}.\footnote{In the welfare calculation, we do not include the fixed costs of \textit{incumbents} as we focus on copyright policies targeting entrants given exiting incumbents. Other studies on optimal product variety (e.g., \citealp{mankiw1986free,berry2016optimal}) include fixed costs associated with all products in the calculation within a static setting.} In addition to the aggregate measures, we compute average social welfare per product for two reasons. First, in creative markets characterized by a large number of products---such as the font marketplace---average welfare may be more relevant to policymakers. This is particularly true in legal contexts, where judicial decisions typically focus on infringement cases involving individual products. Second, averaging mitigates potential biases arising from simplifying assumptions made for computational feasibility, such as fixing the number of potential entrants.

\begin{figure}[htbp!]
    \centering
    \caption{Welfare by Varying Fixed Costs and Protection Levels (Scenario A)}
    \subfloat[Aggregate Consumer Surplus]{\includegraphics[width=0.5\linewidth]{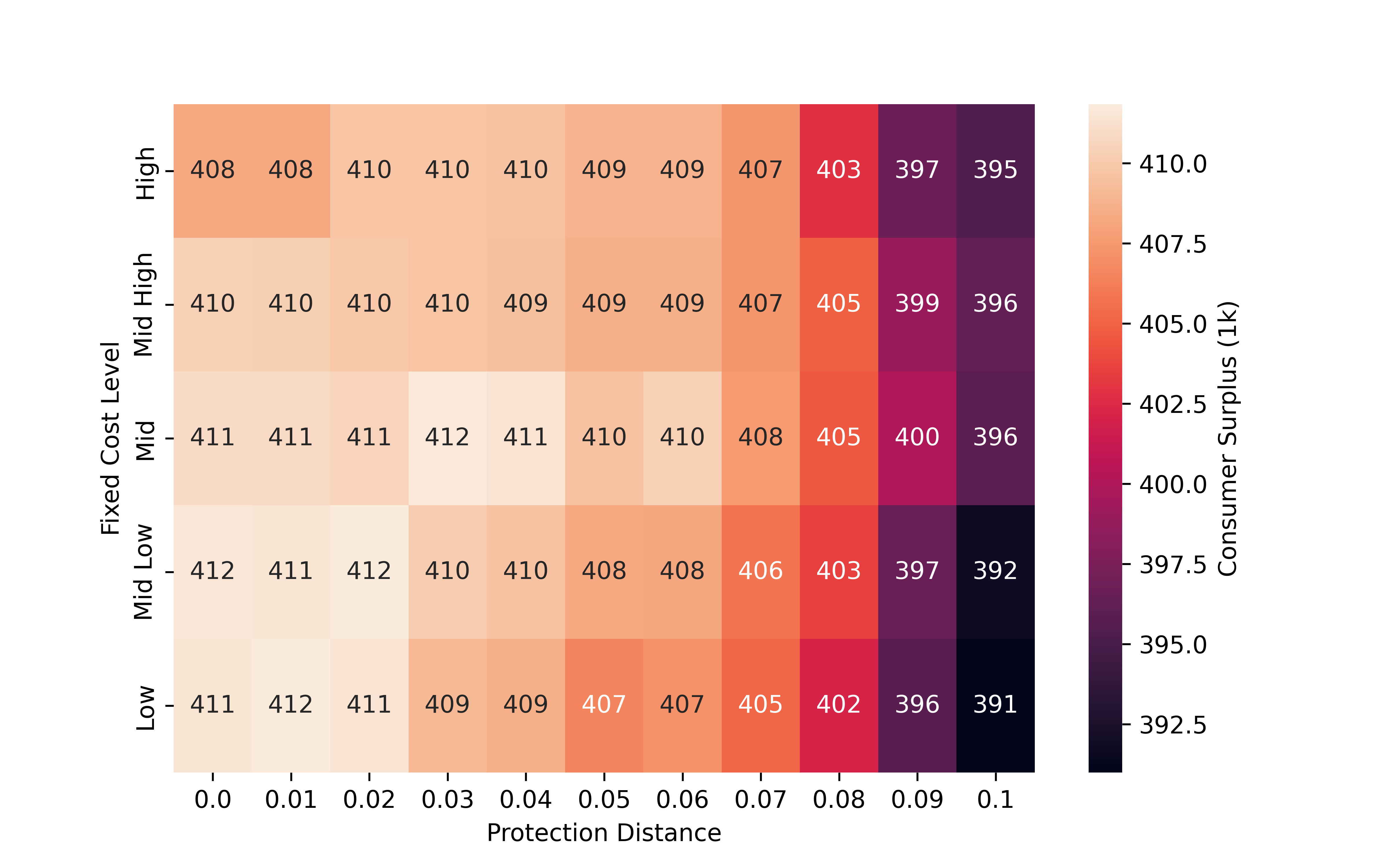}} 
    \subfloat[Average Consumer Surplus]{\includegraphics[width=0.5\linewidth]{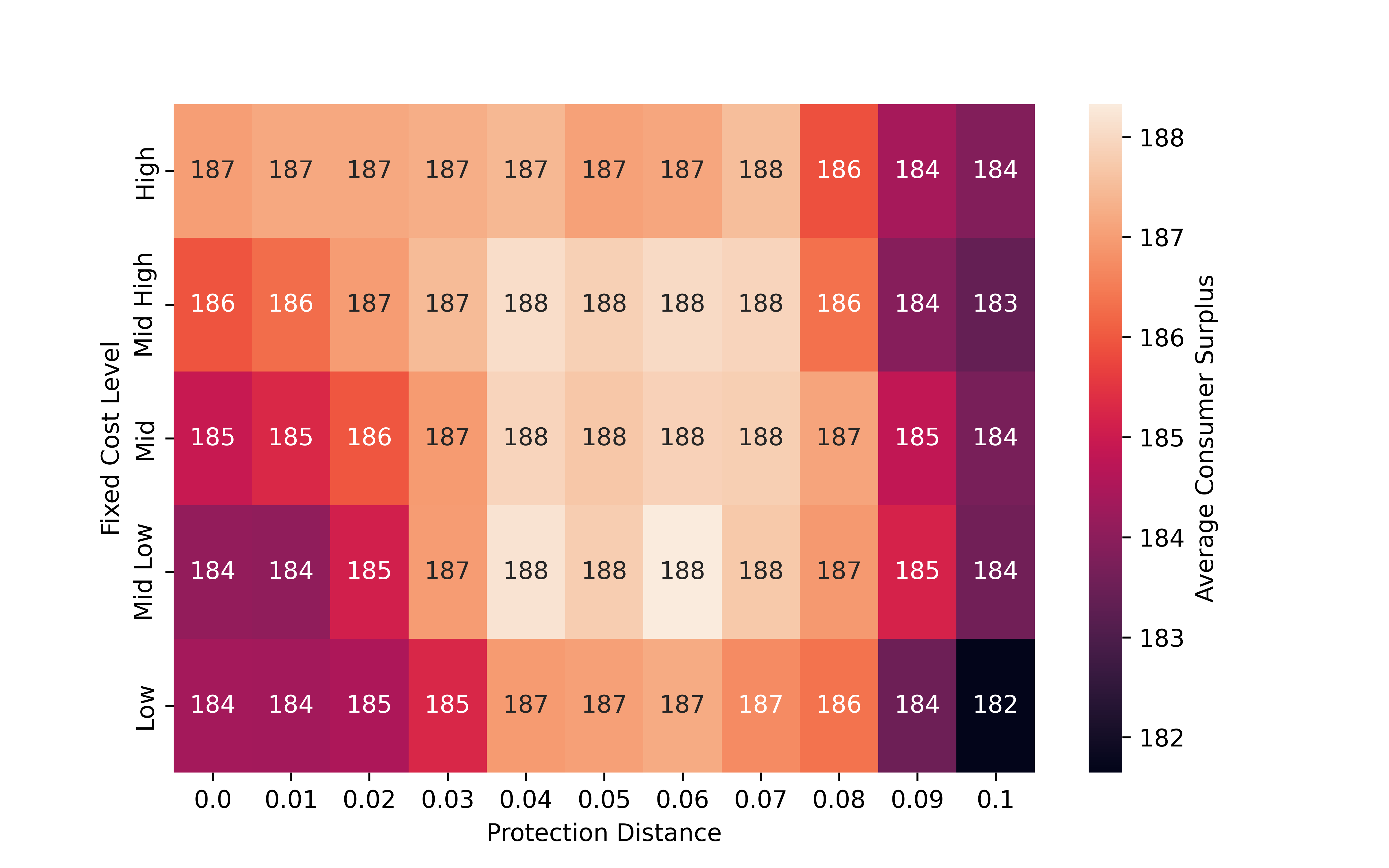}}\\ 
    \subfloat[Aggregate Producer Surplus]{\includegraphics[width=0.5\linewidth]{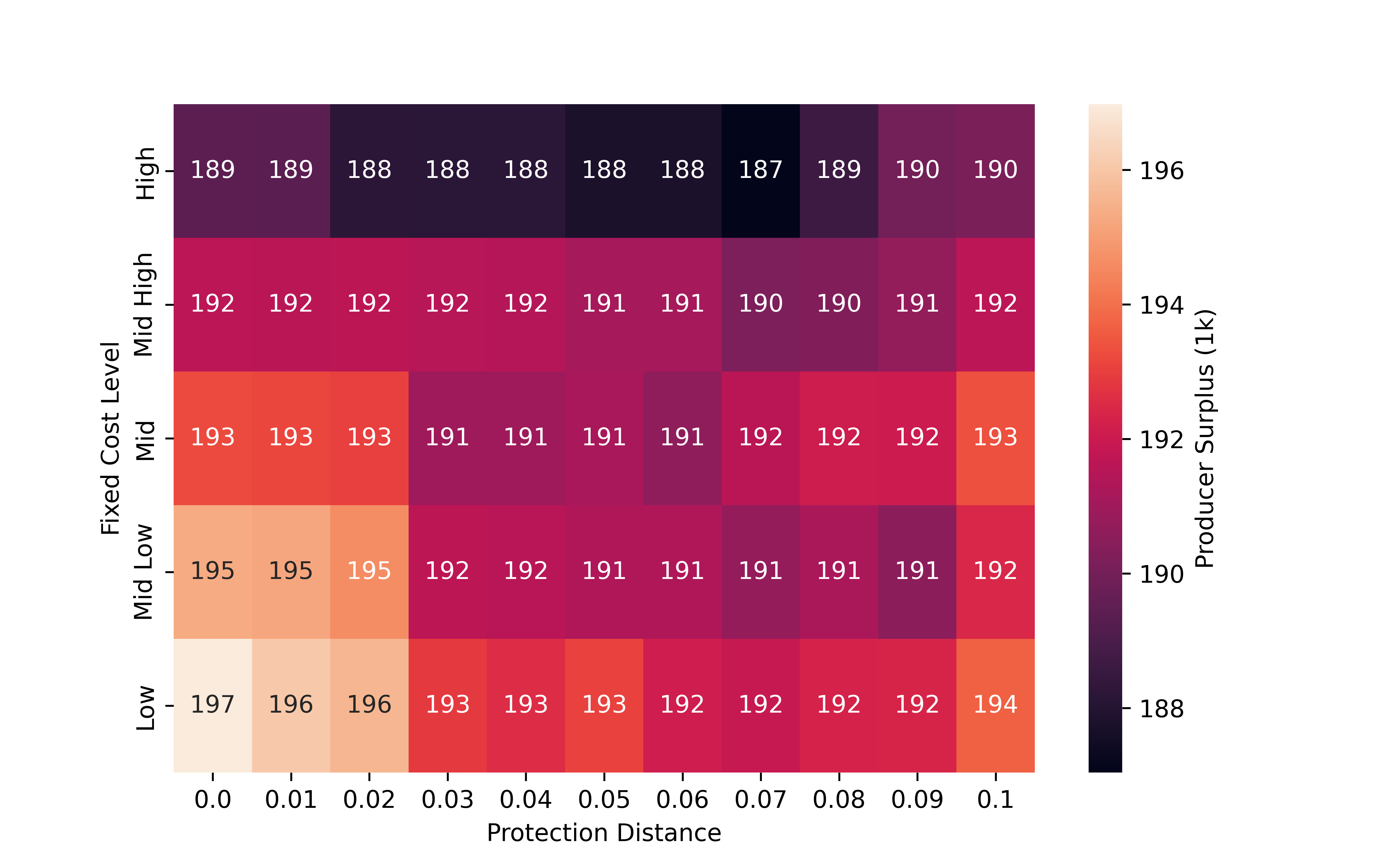}}
    \subfloat[Average Producer Surplus]{\includegraphics[width=0.5\linewidth]{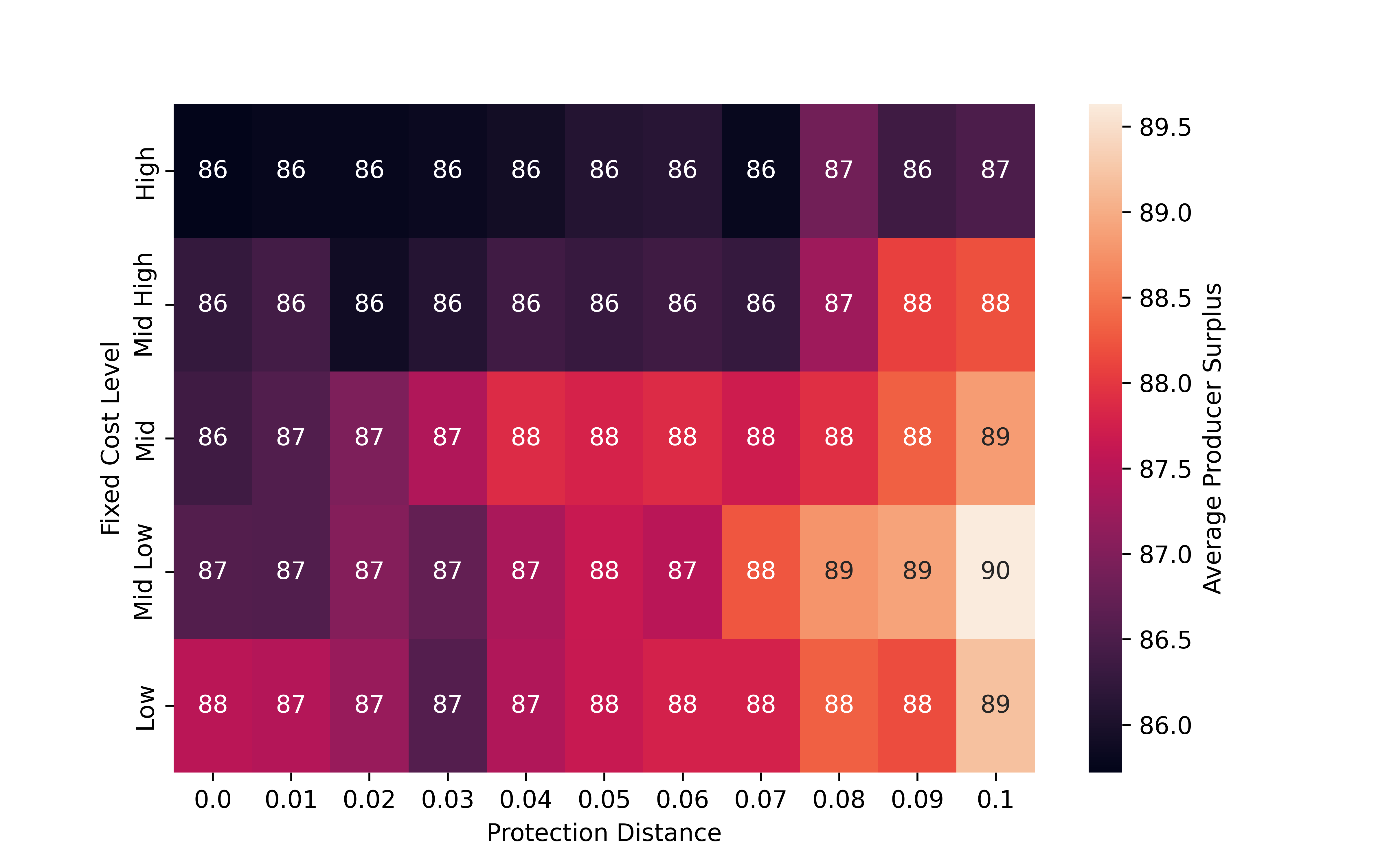}} \\
    \subfloat[Aggregate Social Welfare]{\includegraphics[width=0.5\linewidth]{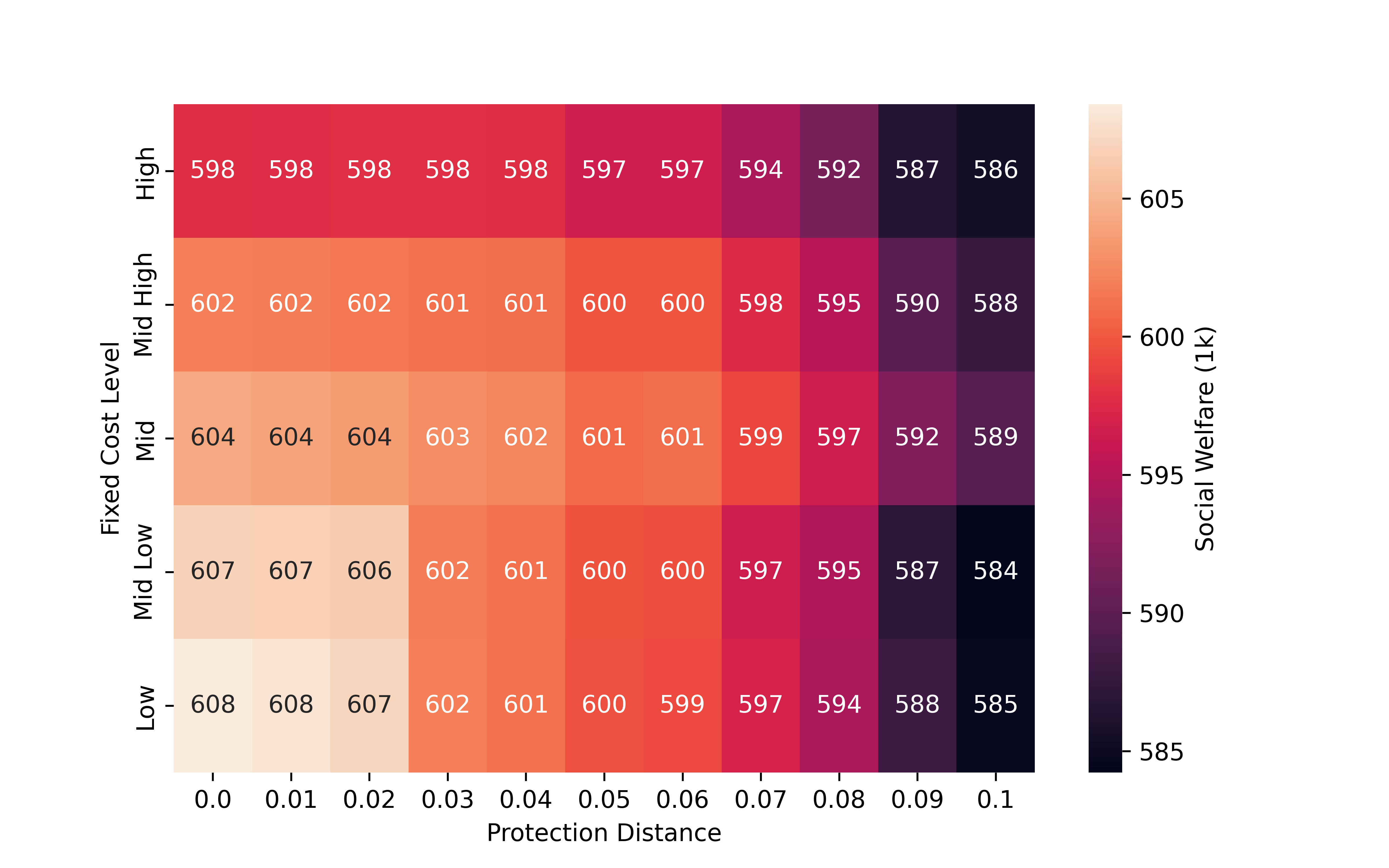}} 
    \subfloat[Average Social Welfare]{\includegraphics[width=0.5\linewidth]{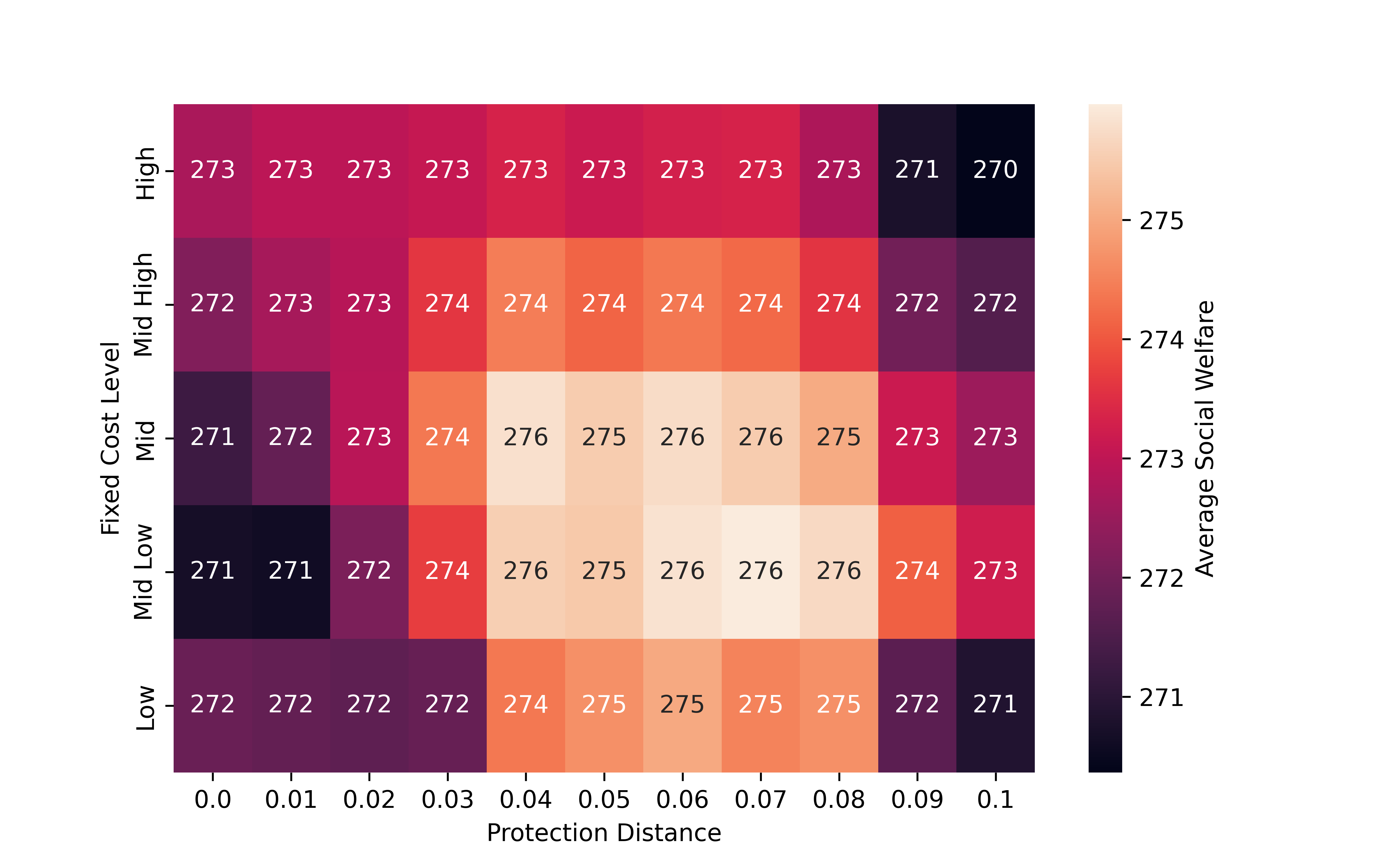}} 

    \tabnotes{This figure shows the counterfactual welfare and market outcomes under various combinations of fixed cost levels and protection distances. Panels (a), (c), (e) show aggregate consumer surplus, producer surplus, social welfare, respectively. Panels (b), (d), (f) display average consumer surplus, producer surplus, social welfare, respectively. A welfare outcome of each simulation is displayed in the corresponding cell, expressed in 1K USD for aggregate measure and USD for average measure. As a reference, in our data, the minimum and maximum Euclidean distances among products are 0.0002 and 0.9563, respectively. See Table \ref{tab:dist-examples} for the example of product shapes and corresponding distances.}
    \label{fig:simul-protection-fclevel-A}
\end{figure}

\begin{figure}
    \centering
    \caption{Entry Points of Counterfactual Simulations (Scenario A)}
    \subfloat[Protection Distance $\underbar{d} = 0$]{\includegraphics[width=0.5\linewidth]{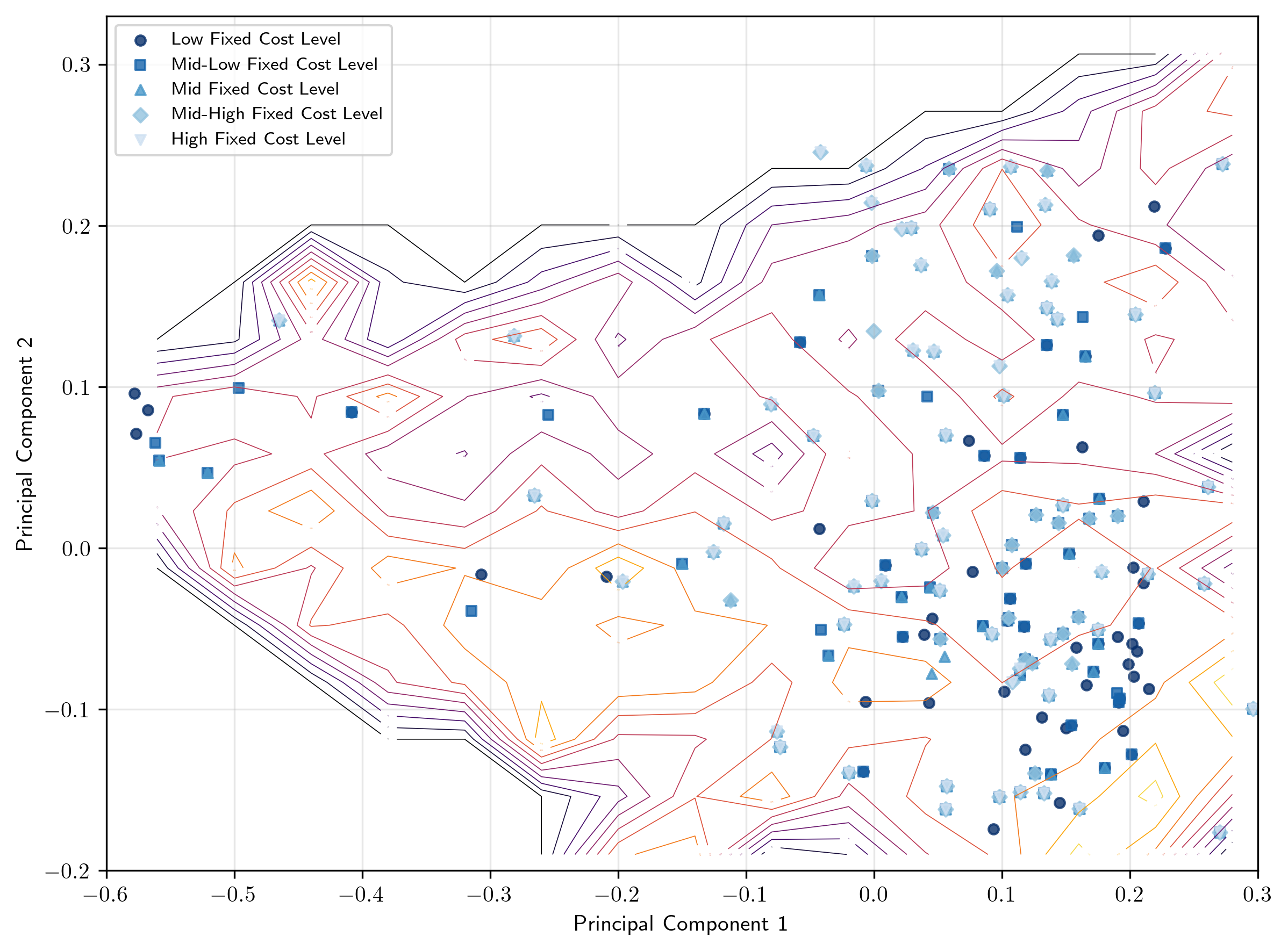}} 
    \subfloat[Protection Distance $\underbar{d} = 0.06$] {\includegraphics[width=0.5\linewidth]{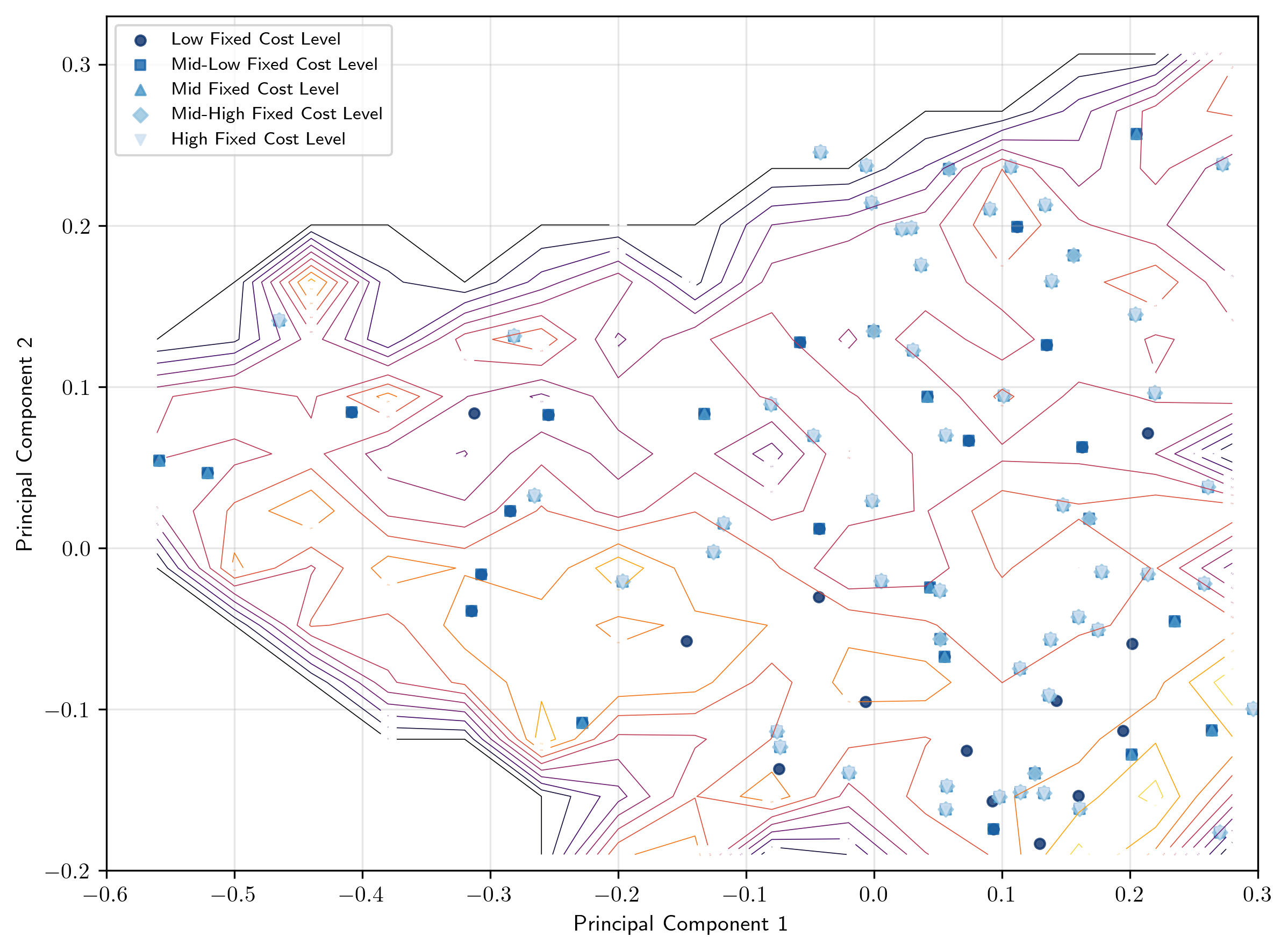}} 
    \tabnotes{Panels (a) and (b) present the entry locations chosen by the firms under protection levels 0 and 0.06, respectively, with a line contour map of median revenues. Note that the incumbents are not shown here. More results are shown in Appendix Figure \ref{fig:more-entry-points-A}}
    \label{fig:entry-points-A}
\end{figure}

Figure \ref{fig:simul-protection-fclevel-A} presents the simulation results for Scenario A, where technology acts as a complement. Across most fixed cost levels, both aggregate and average consumer surplus exhibit an \textit{inverse} U-shaped pattern as the protection level increases. This suggests that moderate copyright protection can incentivize firms to select consumer-favorable locations,\footnote{We show the corresponding entry points in Figures \ref{fig:entry-points-A} and \ref{fig:more-entry-points-A}.} while excessively stringent protection harms consumers. Notably, both aggregate and average surplus decline significantly once protection levels exceed 0.08. This overall pattern is robust and remains consistent with findings from the relocation exercise in Section \ref{subsec:protection-study}, despite the counterfactual shifts of the cost function and the endogenous responses of producers and consumers.

In contrast to consumer surplus, aggregate producer surplus generally exhibits a U-shaped relationship with protection levels across most fixed cost levels. This patterns arises because stricter copyright protection affects incumbents and entrants differently. As protection becomes stricter, incumbents benefit from greater insulation against business stealing, allowing them to secure and retain their original profits. Consequently, incumbents' total profits increase monotonically with the degree of protection. By contrast, entrants' total profits can decrease monotonically, since stricter protection limits potential entry locations. In particular, as copyright becomes more stringent, the loss of mimicry advantages can dominate, even though entrants are encouraged to target higher market-expansion areas---as reflected in consumer surplus---instead of engaging in business stealing. In sum, aggregate producer surplus is determined by the relative magnitudes of incumbents’ profit gains and entrants’ profit losses. Copyright protection can thus be viewed as a redistribution device between incumbents' and entrants' profits. In ranges where aggregate producer surplus diminishes, we find that entrants' losses outweigh incumbents' gains; in ranges where it increases, the reverse holds. By contrast, average producer surplus exhibits a less clear pattern, likely because the average is taken over all firms---both incumbents and entrants---and thus reduces the relative influence of entrants' fixed cost effects on a per-product basis.

In Scenario A, the interaction between copyright protection and production technology plays an important role in determining the optimal protection level. The optimal level varies with fixed costs for both aggregate and average social welfare. The protection level optimizing average welfare ranges between 0.04 to 0.07, underscoring the importance of protective policy. In particular, as technology reduces fixed cost levels, a social planner should implement stricter protection to encourage exploration within the visual characteristics space. This implies that, as far as average welfare is concerned, the stringency of copyright policy observed in the marketplace is suboptimal.\footnote{Recall that the minimum embedding distance observed in the marketplace (i.e., 0.0002) is much smaller than the range of 0.04 to 0.07. In Figure \ref{fig:infringer-dist}, we report the number of products that have at least one close competitors within a given protection distance (i.e., products with infringers) and their shares. For example, 16,101 products have close competitors within a 0.04 radius, accounting for 48\% of all products, while 2,625 products have close competitors within a 0.02 radius, accounting for 8\%.} On the other hand, the optimal protection level for maximizing aggregate social welfare is lower (typically below 0.02), suggesting that the planner should favor looser protection policies to encourage entry, thereby increasing the overall product count and thus aggregate welfare.

\begin{figure}[htbp!]
    \centering
    \caption{Welfare by Varying Fixed Costs and Protection Levels (Scenario B)}
    \subfloat[Aggregate Consumer Surplus]{\includegraphics[width=0.5\linewidth]{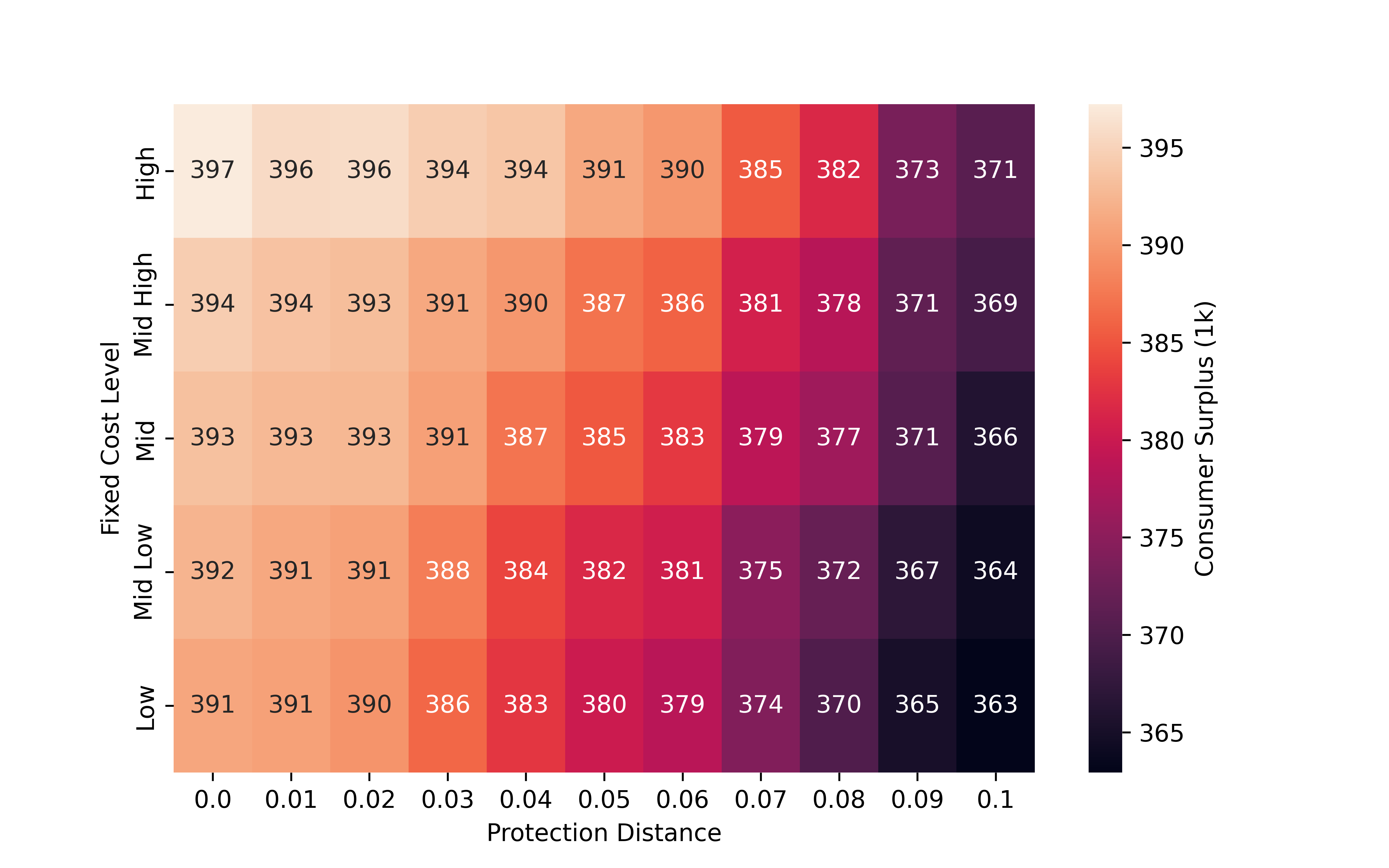}} 
    \subfloat[Average Consumer Surplus]{\includegraphics[width=0.5\linewidth]{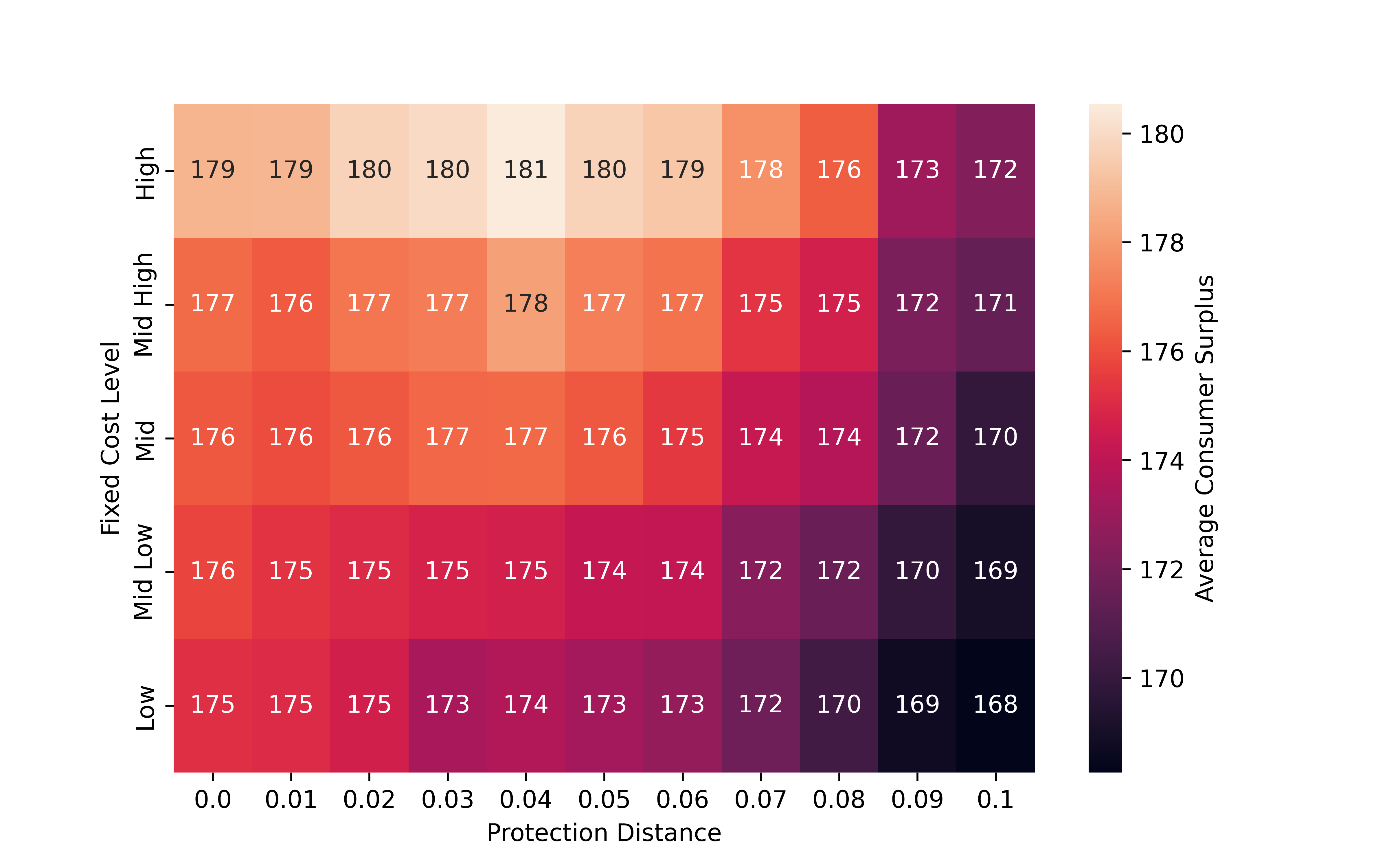}} \\
    \subfloat[Aggregate Producer Surplus]{\includegraphics[width=0.5\linewidth]{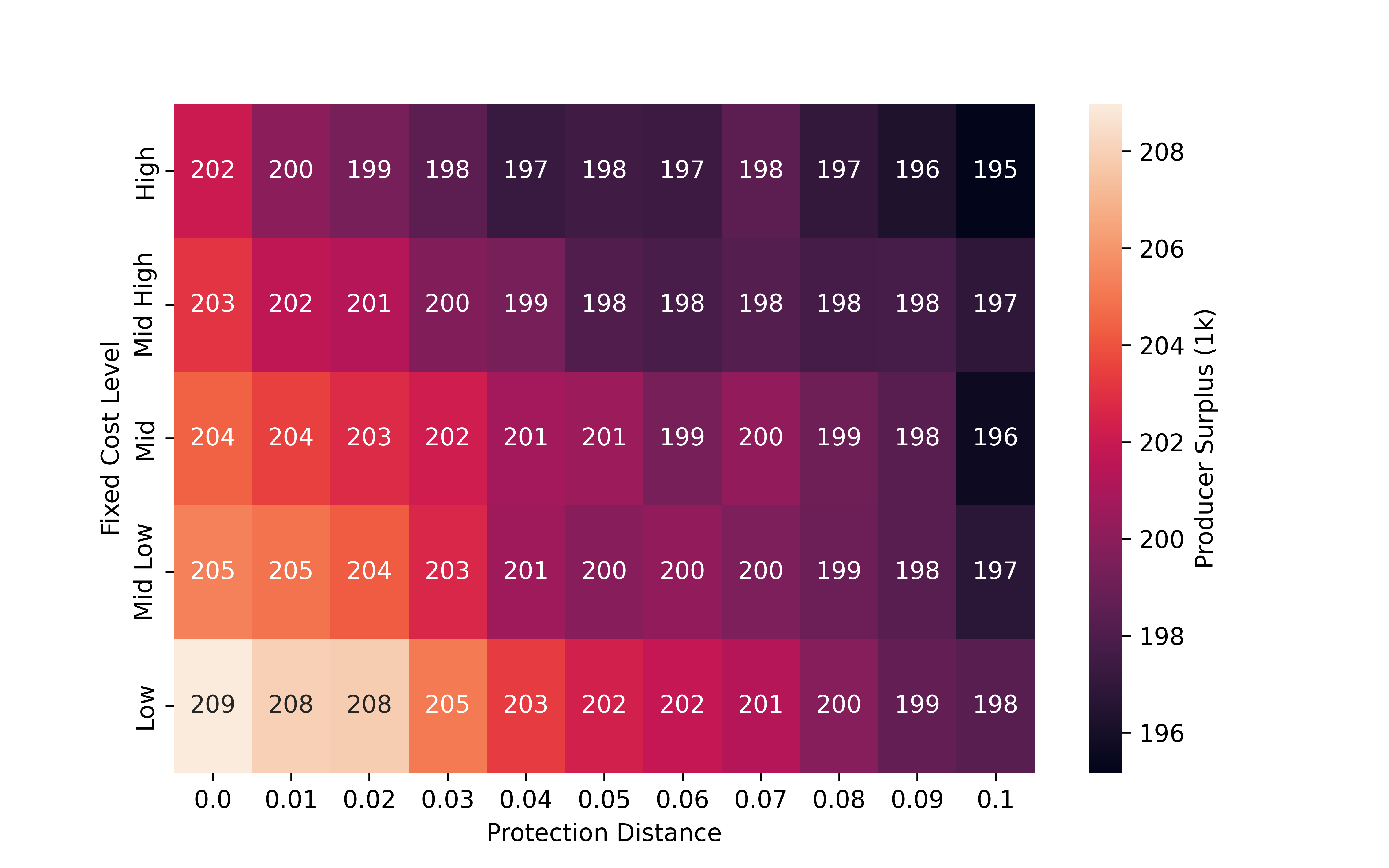}} 
    \subfloat[Average Producer Surplus]{\includegraphics[width=0.5\linewidth]{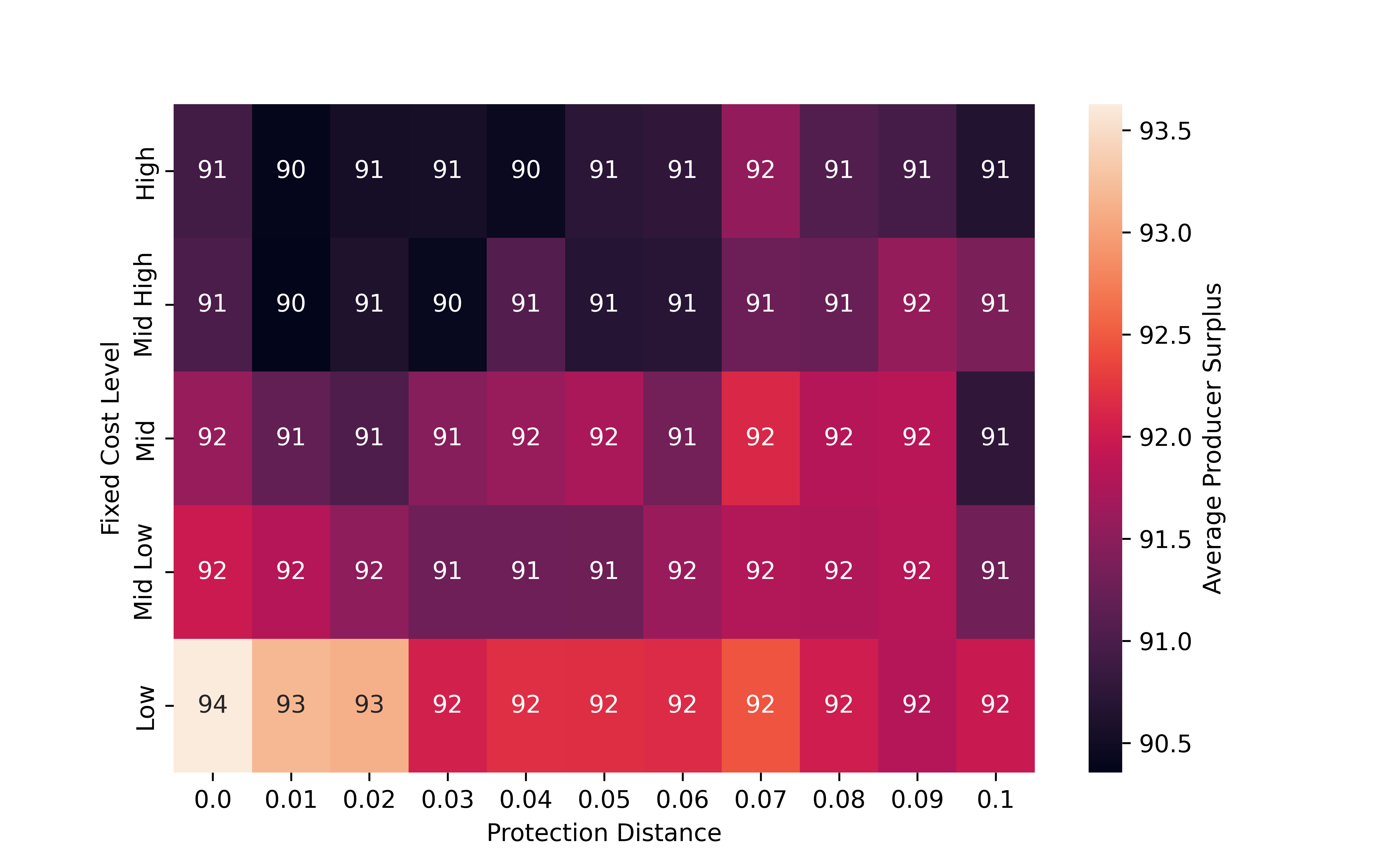}} \\
    \subfloat[Aggregate Social Welfare]{\includegraphics[width=0.5\linewidth]{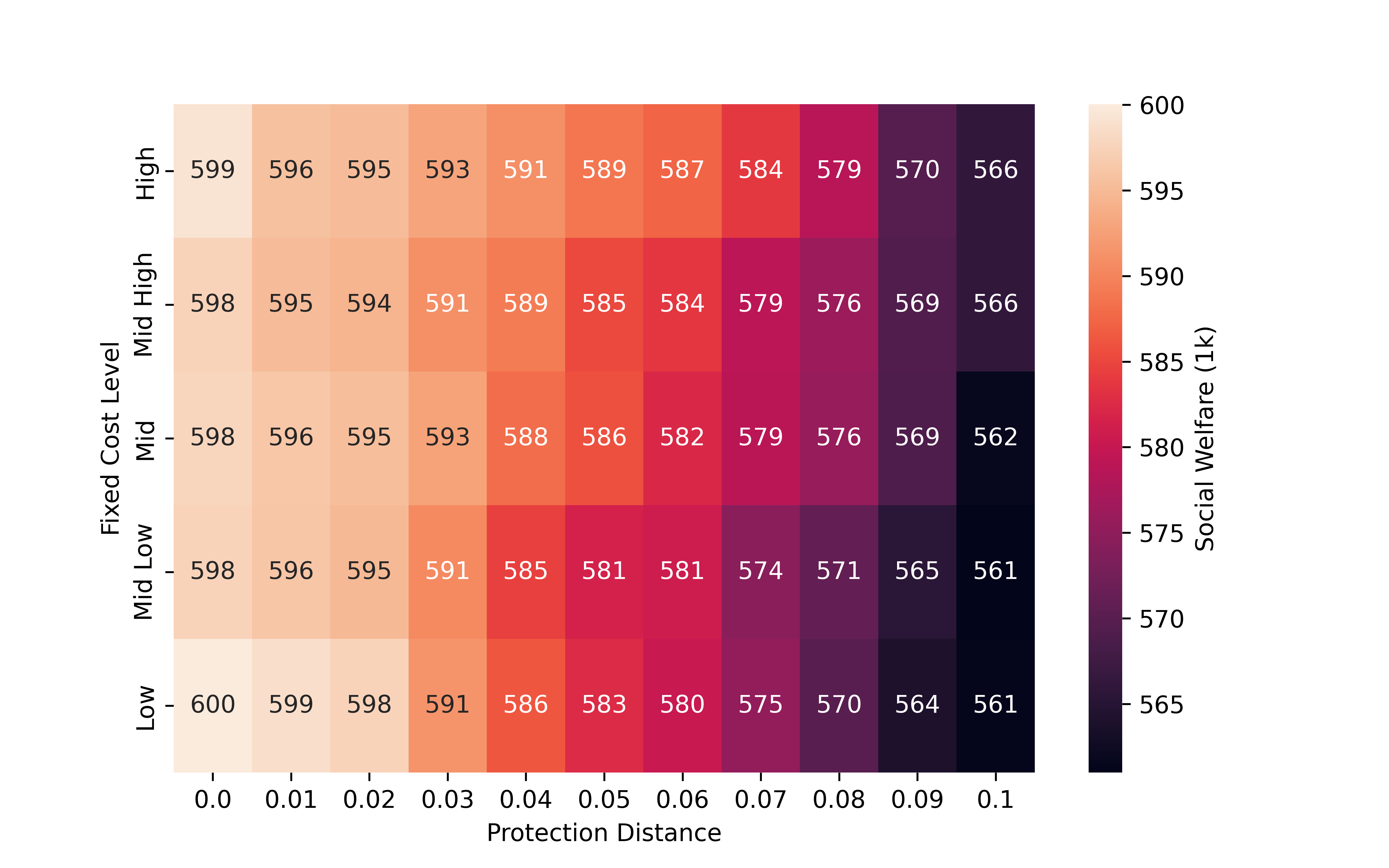}}
    \subfloat[Average Social Welfare]{\includegraphics[width=0.5\linewidth]{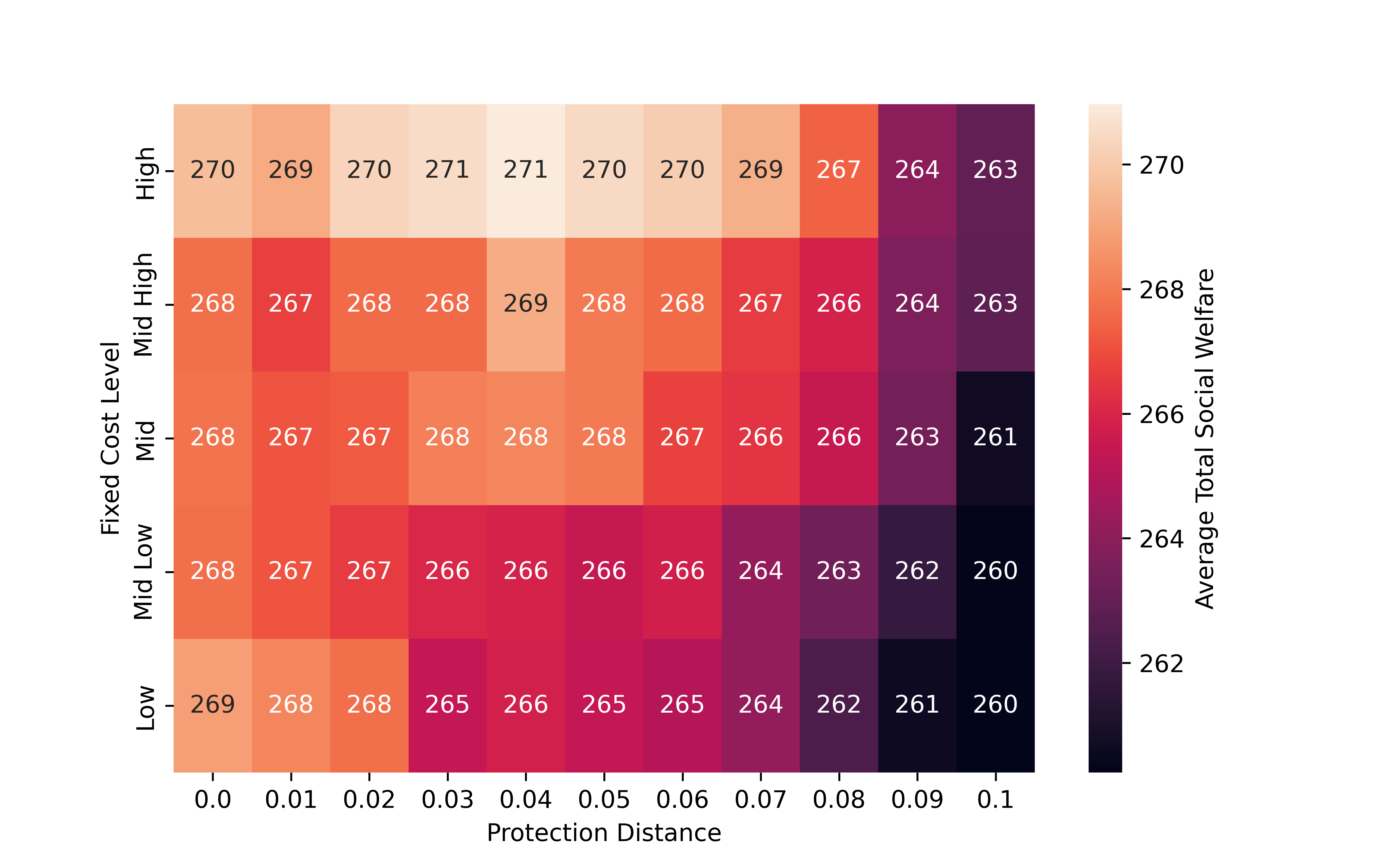}} 
    \tabnotes{This figure shows the counterfactual welfare and market outcomes under various combinations of fixed cost levels and protection distances. Panels (a), (c), (e) show aggregate consumer surplus, producer surplus, social welfare, respectively. Panels (b), (d), (f) display average consumer surplus, producer surplus, social welfare, respectively. A welfare outcome of each simulation is displayed in the corresponding cell, expressed in 1K USD for aggregate measure and USD for average measure. As a reference, in our data, the minimum and maximum Euclidean distances among products are 0.0002 and 0.9563, respectively. See Table \ref{tab:dist-examples} for the example of product shapes and corresponding distances.}
    \label{fig:simul-protection-fclevel-B}
\end{figure}

Technological advancements may not always benefit consumers. Figure \ref{fig:simul-protection-fclevel-B} presents simulation results for Scenario B, where technology acts as a substitute. At any given protection level, consumer surplus decreases with decreasing fixed costs, whereas producer surplus increases. Both aggregate and average measures follow similar trends. This occurs because, under Scenario B, technology results in a flat cost function, eliminating cost advantages related to proximity to existing products. Firms primarily enter low-cost areas, which do not necessarily associated with the highest consumer demand. Specifically, when average fixed costs are generally low, firms prefer these low-cost areas despite modest consumer appeal. Conversely, when fixed costs are higher and fewer low-cost areas exist, firms enter more profitable locations characterized by higher consumer demand. Such patterns are absent in Scenario A, where high-demand areas often coincide with low fixed cost regions due to mimicking advantages.

The optimal copyright policy in Scenario B depends heavily on the planner's objectives and the underlying production technology. When the goal is to maximize \emph{average} social welfare, some level of protection proves beneficial. Figure \ref{fig:simul-protection-fclevel-B} illustrates that the optimal protection level is consistently around 0.04 across all fixed cost levels, except for the lowest-cost scenario, balancing per-product consumer surplus against fixed costs associated with business stealing. However, copyright protection may negatively impact \emph{aggregate} social welfare primarily due to reduced market entry. These results notably differ from Scenario A, which generally favors stricter protection policies both aggregate and average welfare outcomes.

\section{Conclusion} \label{sec:conclusion}

In this paper, we study the role of copyright policy in a creative industry and its interaction with cost-reducing technologies. We combine a modern embedding method for unstructured data with structural economic models to address the policy question. Our focus is on the global font marketplace, which has unique features well-suited to our research purposes. We document localized competition among firms in the characteristic space and the business-stealing effects caused by visually similar entrants. We develop a model of supply and demand that captures firms' entry and positioning behavior within the visual characteristics space, as well as consumers' heterogeneous preferences for visual attributes. Our counterfactual analysis suggests that the stringency of copyright policy could significantly affect welfare through changes in product diversity and potential improvements driven by product relocation. Moreover, it highlights the importance of considering the interplay between copyright protection and technological advancements when determining the optimal level of policy stringency. 

We believe that the counterfactual policy analyses performed using the proposed empirical framework can offer a scientific reference for policymakers in making copyright infringement judgments, which are largely subjective otherwise. Establishing suitable copyright guidelines also have implications for developers of generative AI models, as they have incentives to comply with such policies. On a related note, computer scientists have begun to propose image-generating models that reduces the likelihood of producing outputs that resembles copyrighted training images  \citep{vyas2023provable,wang2024evaluating,ma2024dataset} or produce generic images that are less likely to imitate distinctive features of copyrighted materials \citep{chiba2025tackling}. Interestingly, the notion of similarity they use is based on embedding distances closely related to ours.

The growing availability of unstructured data and machine learning tools is motivating new economic and policy questions. In certain contexts, a structural approach that integrates such data is essential for addressing both positive and normative aspects of an economy and its policies. The empirical models presented in this paper are not confined to our research setting; we believe they are broadly applicable to a wide range of industries where unstructured data can capture important features of products and markets. One important question in using embeddings for economic research is whether the embedding representation captures context-specific economically relevant features (e.g., substitution patterns, local competition), while maintaining general interpretability (e.g., distance, visual similarity). In this paper, we demonstrate how this question can be explored from various angles. It remains valuable to addresses this question more systematically in the current and various other contexts. Overall, we hope this paper shows how structural models can productively engage with unstructured data, opening the door to further important and stimulating research questions.

\bibliographystyle{ecca}
\bibliography{arxiv_091825.bib}

\newpage

\appendix

\renewcommand{\thefigure}{A.\arabic{figure}} 
\renewcommand{\thetable}{A.\arabic{table}}   
\renewcommand{\theequation}{A.\arabic{equation}} 

\setcounter{figure}{0}
\setcounter{table}{0}
\setcounter{equation}{0}

\section{Details on Embeddings in Section \ref{sec:data}}

\subsection{Embedding Construction}\label{subsec:embedding_construction}

To construct embeddings, we train convolutional neural network with triplet loss. Triplet $i$ comprises anchor $x_{i}^{a}$, positive $x_{i}^{p}$,
and negative $x_{i}^{n}$. An anchor is (crops of) a pangram image of a given font family (e.g., Helvetica), positives are (crops of) pangram images of the same or different styles of the same family (e.g., Helvetica Regular, Helvetica Light, Helvetica Bold, Helvetica Italic), and negatives are (crops of) pangram images of different families (e.g., Time New Roman). Then, a triplet-based loss function is defined as
\begin{equation}
L(f;\alpha):=\sum_{i}^{N}[\left\Vert f(x_{i}^{a})-f(x_{i}^{p})\right\Vert _{2}^{2}-\left\Vert f(x_{i}^{a})-f(x_{i}^{n})\right\Vert _{2}^{2}+\alpha]_{+},\label{eq:loss}
\end{equation}
where $f(x)\in\mathbb{S}^{d}$ is the $d$-dimensional embedding of image $x$, $\left\Vert \cdot\right\Vert _{2}$ is the Euclidean norm, and $\alpha$ is a margin. The dimension $d$ is set to be $128$. We minimize this loss function using stochastic gradient descent (SGD). 

We use approximately 20,000 pangrams of fonts to train the neural network. Each pangram is a bitmap with $200\times 1000$ pixels. The training is an iterative process of improving the parameters of the network using small batches of cropped images ($100\times 100$ pixels) to estimate the gradient and then updating the parameters accordingly. As the gradient is evaluated at more batches, the parameters in the network are adjusted. There are 90,000 parameters. Each batch contains 270 cropped images (i.e., 90 triplets). The training of the network is completed when the loss function reaches below a certain threshold (e.g., 0.7). To ensure fast convergence while avoiding bad local minima, we focus on sampling semi-hard triplets, that is, triplets that violate $\left\Vert f(x_{i}^{a})-f(x_{i}^{p})\right\Vert _{2}^{2}+\alpha<\left\Vert f(x_{i}^{a})-f(x_{i}^{n})\right\Vert_{2}^{2}$ with $\alpha=0$. The training takes approximately 24 hours with 4 GPUs (Nvidia 1080-TI).

\subsection{Details of Lasso Estimation} \label{sec:lasso-detail}

In Section \ref{subsec:embedding}, to further interpret the two principal component (PCs), we use Lasso to select tags (i.e., phrases describing font shapes) that explain each PC. To prepare product tags for lasso regression, we first clean the tags by converting them to lowercase, removing non-alphanumeric characters, and eliminating common stopwords such as `and', `or', `the', `a', and `an'. We filter out tags that received no votes from the platform users. We then construct a dummy for each tag. Finally, we merge the cleaned tag data with the principal component data. We use scikit-learn Lasso package for implementation \citep{scikit-learn}. We choose the regularization parameter to be 0.002.

\section{More on Demand Estimation}

\subsection{Diversion to Outside Goods and Embedding Distances}\label{subsec:outside}

\begin{figure}[p]
    \centering
    \caption{Diversion Ratios to Outside Goods and Distance to the Closest}
    \subfloat[Closest Competitors]{    \includegraphics[width=0.5\textwidth]{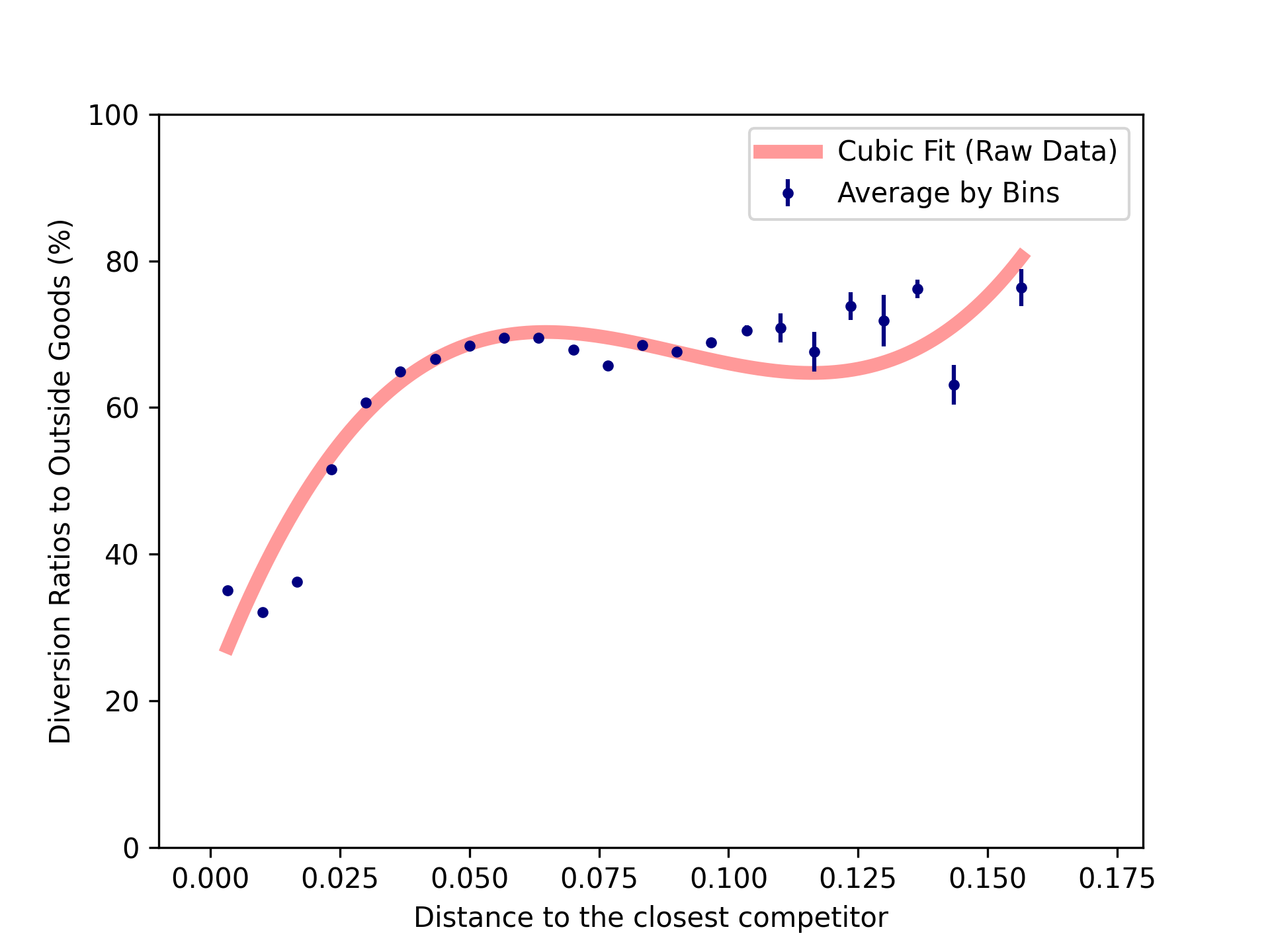}}  \\
    \subfloat[5 Closest Competitors]{\includegraphics[width=0.5\textwidth]{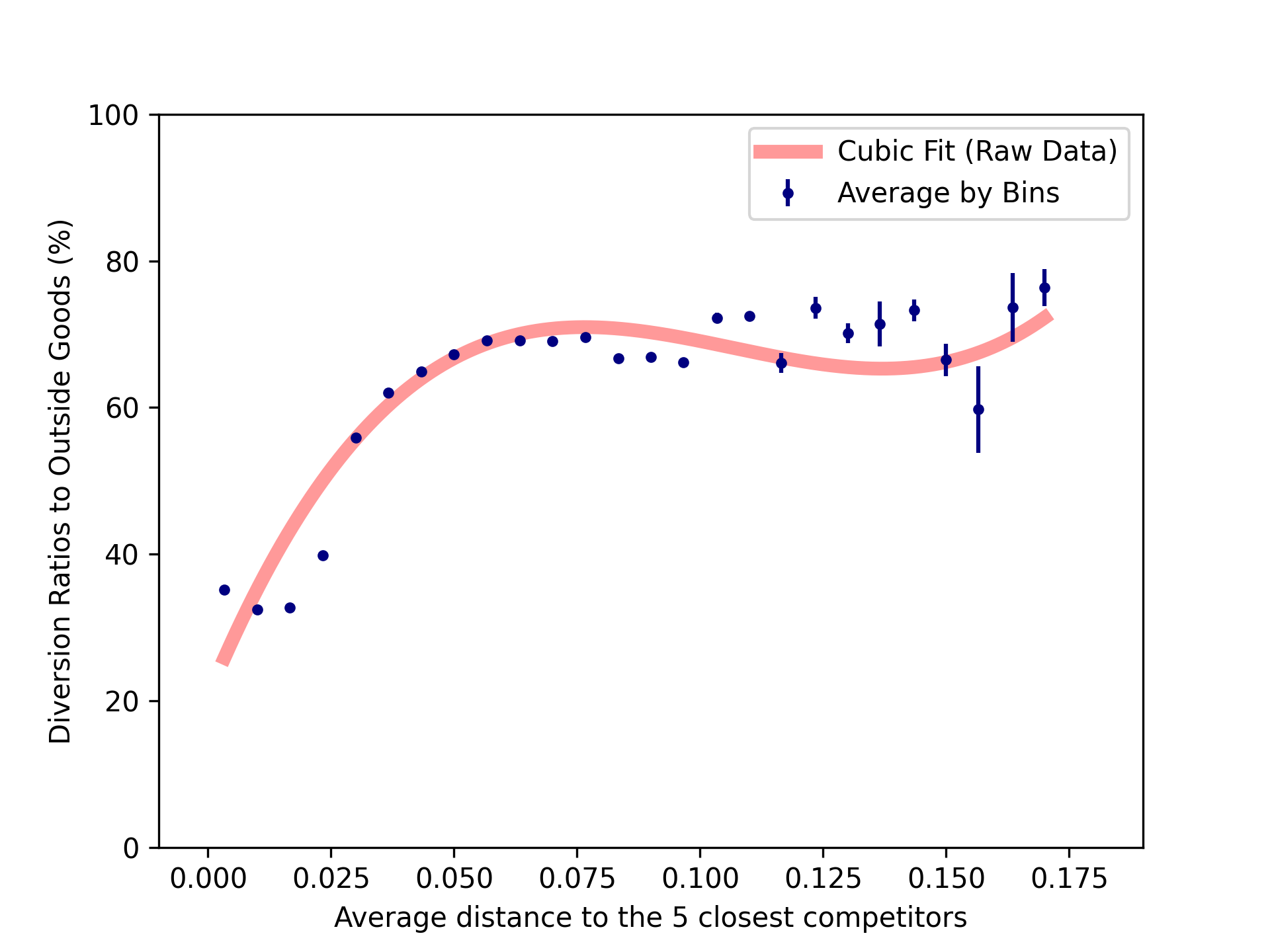}} 
    \subfloat[10 Closest Competitors]{\includegraphics[width=0.5\textwidth]{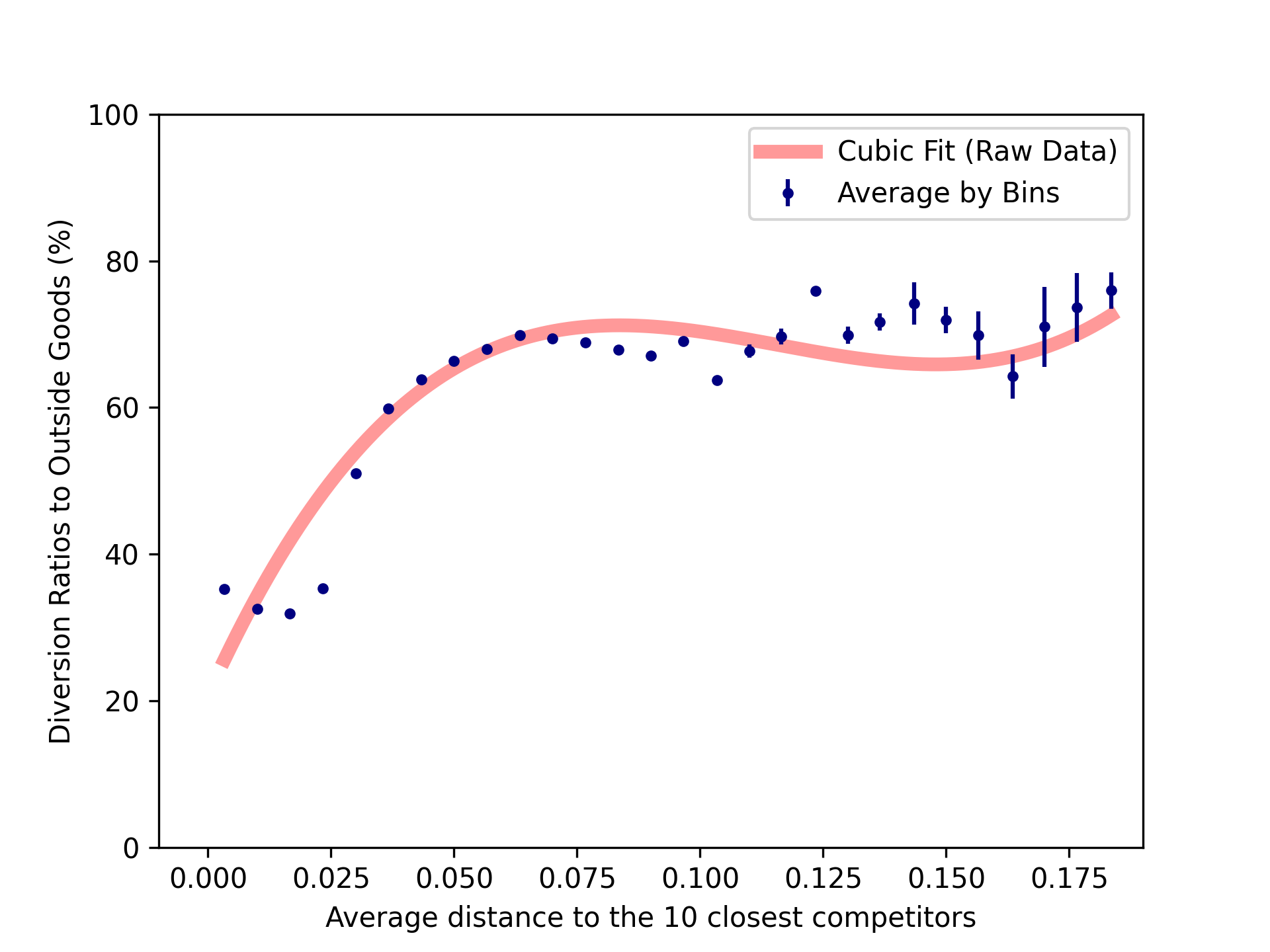}}
    \tabnotes{This figure presents a binscatter plot of diversion ratios to outside goods as a function of distances to the closest competitors or the average distances to the closest competitors. Each dot represents the bin-by-bin average, accompanied by a 95\% confidence interval. The red solid line indicates the third-order polynomial fit of the raw diversion ratios to outside goods data. Panels (a) through (c) display the diversion ratios to outside goods against the distance to the closest competitor, the average distance to the five closest competitors, and the average distance to the ten closest competitors, respectively.}
    \label{fig:div0-avg}
\end{figure}

In addition to the analysis of substitution patterns in relation to the embedding distance (in )Section \ref{subsec:demand_est_result}), we investigate competition between a given product and outside goods, which includes free fonts—the main competitors of commercial font products. To understand how the substitution to outside goods is affected by the availability of substitutes within the market, we calculate diversion ratios to outside goods for each product and examine their relationship with the (average) embedding distance to close competitors within the market. Figure \ref{fig:div0-avg}.(a) displays the binscatter plot of the diversion ratios to outside goods versus the distance to the nearest competitor. The results suggest that the more visually similar a product is within the marketplace, the more likely consumers are to opt for an alternative within the marketplace rather than leaving it entirely. The divergent ratios increase up to around 0.05, and then gradually plateau. Qualitatively similar results are found when examining the binscatter plots of the divergent ratios and the average distances to the 5 and 10 closest competitors (Figures \ref{fig:div0-avg}(b) and (c)).

\newpage 
\section{More on Fixed Cost Estimation} \label{sec:fc-detail}
\subsection{Estimating Upper Bound on Fixed Cost} \label{sec:est-ub-detail}

The entry condition defined in \eqref{eq:entry-condition} requires taking the expectation with respect to $\xi_{kt}$. To address this, we randomly sample $N_{s}$ ($= 30$) demand shocks from the empirical distribution of entrants' shocks at the time of entry. For each sampled $\xi^{s}_{kt}$, we compute the variable profits and approximate the expectation as
\begin{equation}
    \mathbb{E}_{\xi_{kt}}\left[ \sum_{j \in J_{ft} \cup \{k\}} \pi_{jt}(A_{ft,k} = 1)\right] \approx \frac{1}{N_{s}} \sum_{s=1}^{N_{s}}\left[ \sum_{j \in J_{ft} \cup \{k\}} \pi_{jt}(\xi_{k}^{s})\right].
\end{equation}
To compute a variable profit, we simulate prices and shares while fixing all the characteristics, including $\xi^{s}_{kt}$ and the marginal costs recovered from the pricing model. We use the fixed-point iteration method suggested by \cite{morrow2011fixed}, which is known to have a stable convergence property. In fact, we use the observed price as an initial point for the iteration and attain converged prices for every simulation. PyBLP is used for implementation \citep{conlon2020best}. We additionally compute the no-entry variable profit (i.e. $\sum_{j \in J_{ft}}  \pi_{jt}(A_{ft,k} = 0)$) by removing the product and following the same simulation method above.

\subsection{Estimating the Slope of Fixed Cost Function} \label{sec:est-slope-detail}

We use the FOC condition in \eqref{eq:second-foc} to estimate the slope of the fixed cost function. This process involves estimating the expected partial derivative of the profit function with respect to each embedding element. We accomplish this by using numerical differentiation and simulating the average to approximate the expectation. Specifically, we first randomly sample $N_{s} ( = 30)$ vectors $\xi^{s}_{kt} = \{\xi^{s}_{kct}\}_{c=1}^{12}$, compute $\frac{\partial \pi_{jt}(\xi^{s}_{kt})}{\partial x^{emb}_{k\ell}}$ for each $\ell$ by increasing $x^{emb}_{k\ell}$ by a small amount, $h$, and then differentiate while holding all the other characteristics fixed. As a result, the expectation can be approximated as:
\begin{equation} \label{eq:approx-whole}
    \mathbb{E}_{\xi_{kt}}\left[\frac{\partial \pi_{jt}(\xi^{s}_{kt})}{\partial x^{emb}_{k\ell}} \right] \approx \frac{1}{N_{s}}\sum_{s=1}^{N_{s}} \frac{\pi_{jt}(x^{emb}_{k\ell} + h, \xi^{s}_{kt}) - \pi_{jt}(x^{emb}_{k\ell}, \xi^{s}_{kt})}{h} \text{ for each } \ell
\end{equation}
This numerical differentiation, however, is computationally expensive because it requires simulating pricing responses for all products. Instead, we simplify the process by computing the numerical differentiation of the market share function and verifying that this approximation closely matches the results obtained from the full pricing responses. That is, we approximate the expected derivative for given $\xi^{s}_{kt}$ as
\begin{equation} \label{eq:approx-share}
    \frac{1}{N_{s}}\sum_{s=1}^{N_{s}} \sum_{c} (p_{jct} - mc_{jct}) \frac{s_{jct}(x^{emb}_{k\ell} + h, \xi^{s}_{kct}) - s_{jct}(x^{emb}_{k\ell}, \xi^{s}_{kct})}{h}\frac{M_{ct}}{M_{t}}.
\end{equation}
For a small sample, we confirm that \eqref{eq:approx-whole} and \eqref{eq:approx-share} lead to very similar results.

\subsection{IV Estimation}\label{subsec:supply-side-IV}

In estimating the slope of the fixed cost function $F(\boldsymbol{x}_{t},\nu_{k})$, when the unobservable $\nu_{k}$ in the function is viewed as a structural error, we address the issue of simultaneity using IVs. We use BLP-type IVs $Z_{j,\ell}^{cost}$ for each $\ell$. These include the sum of incumbents' embeddings, along with their squared and cubic terms, $\left\{\sum_{j' \in J_{t-1}\setminus\{j\}} x^{emb}_{\ell}, \sum_{j' \in J_{t-1}\setminus\{j\}} (x^{emb}_{\ell})^{2}, \sum_{j' \in J_{t-1}\setminus\{j\}} (x^{emb}_{\ell})^{3}\right\}_{\ell = 1}^{6}$, and the number of local incumbents in the other embedding dimensions, $\left\{\sum_{j'\in J_{t-1}\setminus\{j\}}\mathbbm{1}(d^{\ell'}_{jj'} < \frac{1}{2}SD_{\ell'})\right\}_{\ell' \neq \ell}$.

\begin{figure}[htbp!]
    \centering
    \caption{Fixed Costs Estimation Results with IVs}
    {\includegraphics[width=0.5\linewidth]{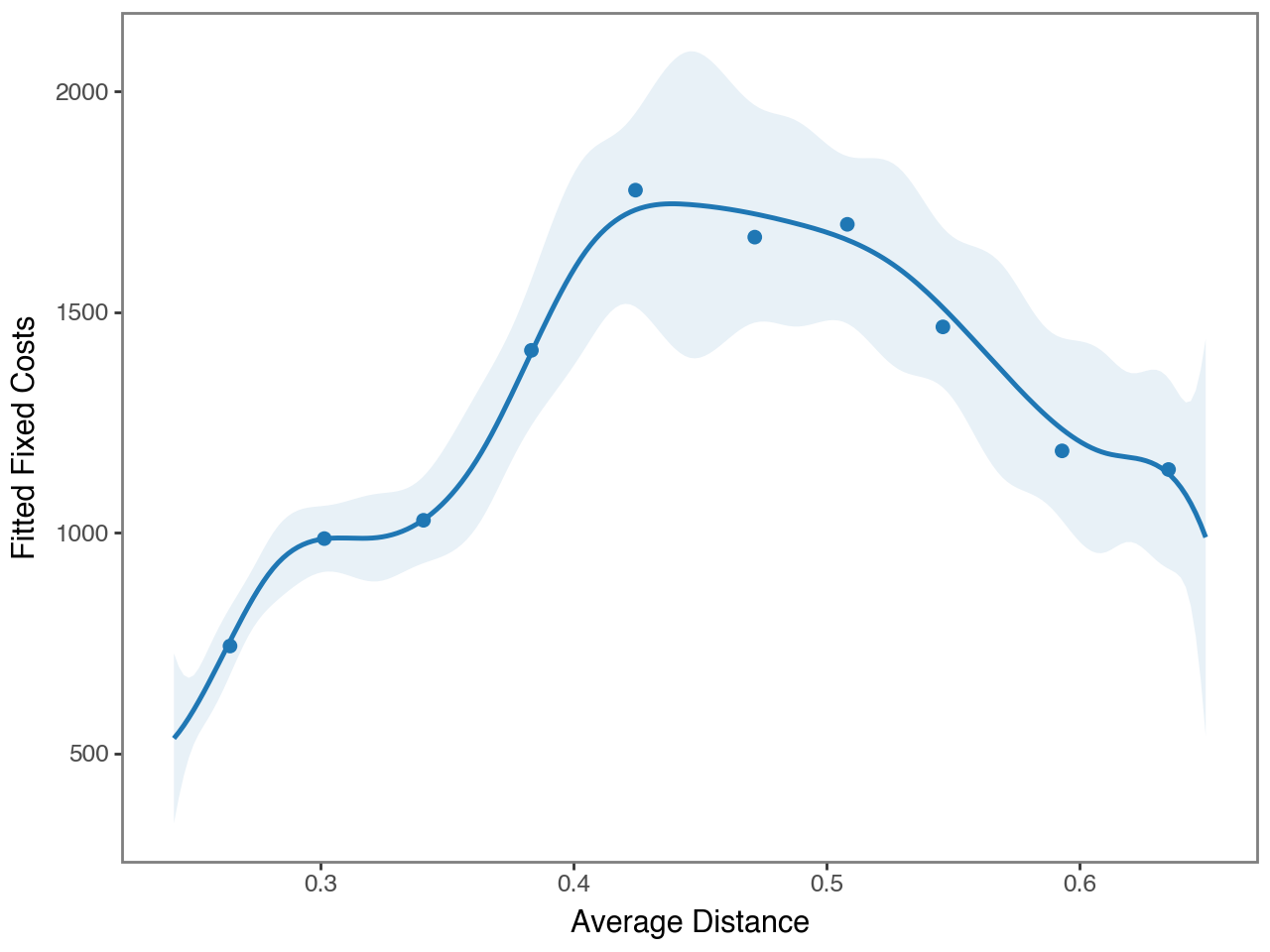}} 
    \tabnotes{The figure presents a binscatter plot of the estimated explained part of fixed costs against average distances. We compute the fitted values by using the estimated slope coefficients which are displayed in Table \ref{tab:cost-IV}. The solid line represents a third-order local polynomial fit, with the shaded area indicating the 95\% confidence band. Heteroskedasticity-robust standard errors are applied. The dots represent bin-by-bin averages with the evenly-spaced binning method \citep{cattaneo2024binscatter}. The figure is based on 1,630 observations, corresponding to the number of entrants.}
    \label{fig:supply-result-IV}
\end{figure}

The IV estimates of fixed costs in Table \ref{tab:cost-IV} are qualitatively similar to the OLS estimates reported in the main text. Figure \ref{fig:supply-result-IV} shows that the estimated fixed cost function is concave, closely resembling the shape of OLS estimates in Figure \ref{fig:supply-result}. This robustness is encouraging, as the concavity of cost function is the main driver of our counterfactual results.

\begin{table}[htbp!]
\centering
\caption{Slope Estimation Results (IV)}
\resizebox{\textwidth}{!}{%
{\small
\begin{tabular}{lcccccc}
\toprule
& (1) & (2) & (3) & (4) & (5) & (6) \\
Parameters & $\partial F/\partial x^{emb}_{1}$ & $\partial F/\partial x^{emb}_{2}$ & $\partial F/\partial x^{emb}_{3}$ & $\partial F/\partial x^{emb}_{4}$ & $\partial F/\partial x^{emb}_{5}$ & $\partial F/\partial x^{emb}_{6}$ \\
\midrule
$\eta_{0\ell}$ 
& 3052.60 & -2383.50 & -6678.60 & -6110.00 & 1562.00 & 4968.60 \\
& (276.65) & (183.67) & (356.82) & (351.16) & (105.99) & (219.39) \\

$\eta_{1\ell}$ 
& 0.16 & 0.04 & 2.90 & -2.65 & 1.39 & -0.09 \\
& (0.25) & (0.14) & (1.12) & (1.13) & (0.67) & (1.17) \\

$\eta_{2\ell}$ 
& 0.85 & -1.38 & -22.47 & 38.54 & -16.44 & -3.17 \\
& (0.78) & (0.95) & (8.90) & (16.49) & (10.28) & (21.79) \\

$\eta_{3\ell}$ 
& -0.83 & 4.53 & 48.77 & -144.49 & 56.18 & 32.95 \\
& (0.61) & (1.89) & (20.13) & (62.56) & (40.62) & (105.27) \\
\midrule
$R^{2}$ 
& 0.31 & -0.08 & -0.01 & -0.24 & 0.11 & -0.06 \\
$F$-stat
& 136.69 & 22.00 & 7.88 & 5.72 & 72.42 & 10.35 \\
Observations & \multicolumn{6}{c}{1,630} \\
\bottomrule
\label{tab:cost-IV}
\end{tabular}
}
}
\tabnotes{This table reports the estimated slopes of the fixed cost function, obtained using two-stage least squares regressions with BLP-type IVs. Heteroskedasticity-robust standard errors are shown in parentheses. The Wald test F-statistics for the null hypothesis $H_{0}: \eta_{1\ell} = \eta_{2\ell} = \eta_{3\ell} = 0$ vs. the alternative $H_{1}$ (the negation of $H_{0}$) are reported in the row labeled F-stat. The number of observations is 1,630, corresponding to the number of entrants. All first-stage regression F-statistics exceed 216.}
\end{table}

\section{Details of Counterfactual Simulations} \label{sec:simul-detail}
\subsection{Relocation Analyses in Section \ref{subsec:protection-study}}

For the relocation exercise, we first generate potential shapes by locally perturbing the embeddings in the actual data. The main reason for local perturbation is that many points in the characteristic space (i.e., the manifold) do not represent actual font shapes, as they lie outside the support of the distribution of font shapes. Local perturbation addresses this issue by ensuring that potential shapes are closer to actual fonts. In addition, given the large number of products in the marketplace, this approach allows us to generate a diverse range of potential shapes. 

Sets of potential locations are created by applying different levels of perturbation, which involve the following steps: First, we randomly sample $\check{N}$ actual font products. Then, different sets of Gaussian noises are generated from 128-dimensional multivariate normal distributions with varying diagonal covariance matrices. To be specific, let $\mathcal{X}$ be the set of actual embeddings. We generate the potential location $\check{x}^{emb} \in \mathbb{R}^{128}$ as:
\begin{equation*}
    \check{x}_{g, level}^{emb} = x^{emb} + e_{g, level},
\end{equation*} where $x^{emb}$ is randomly sampled from $\mathcal{X}$, $e_{g} \sim N(0, \hat{\Sigma}^{e}_{level})$, $\hat{\Sigma}^{e}_{level} = diag(\hat{\sigma}_{1},...,\hat{\sigma}_{128})/c_{level}$, $\hat{\sigma}_{\ell}$ is the standard deviation of $x^{emb}_{\ell}$ for $\ell = 1,...,128$, and $c_{level}$ is the constant governing the degree of perturbation. We set $c_{level}$ to be one of $\{1,2,3,4, 5, 10, 20, 30\}$. As a result, we have in total $8\check{N}$ total potential locations.

\subsection{Local Monopoly and Consumer Surplus}\label{subsubsec:local_mono}

Following the counterfactual analysis in Section \ref{subsec:protection-study}, to understand the degree of local monopolistic power granted by copyright protection, we further decompose the decrease in consumer surplus into two channels: (i) price increases due to enhanced monopolistic power, and (ii) diversity loss resulting from eliminating new entrants. To do this, we first compute the consumer surplus in \eqref{eq:CS} by fixing prices under $\underbar{d} = 0$, while removing entrants according to the protective boundary. We then compare this price-fixed consumer surplus with the price-adjusted one shown in the main text. The price-fixed simulation reflects only the loss from reduced diversity, as it prevents firms from optimizing their prices. Finally, the difference between the price-adjusted and price-fixed consumer surplus for each $\underbar{d}$ captures the loss of consumer surplus due to price increases resulting from the enhanced monopolistic power of incumbent products.

\begin{figure}[h!]
    \centering
    \caption{Decomposition of Consumer Surplus Loss}
    \includegraphics[width=0.7\linewidth]{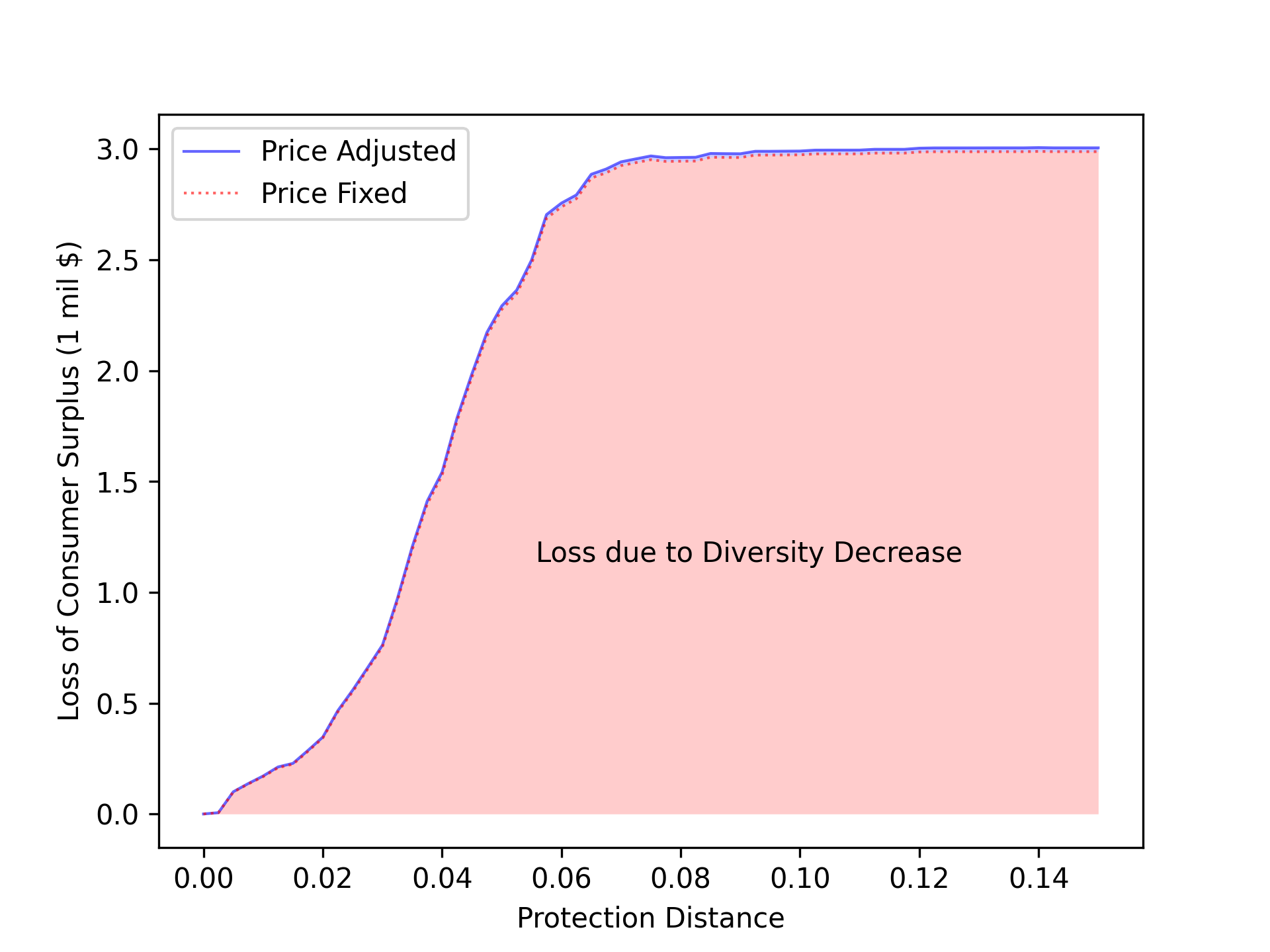}
    \tabnotes{This figure shows the analysis of changes in the loss of consumer surplus from baseline simulation.}
    \label{fig:CS-decomp}
\end{figure}

From this decomposition exercise, we find that most consumer surplus losses can be attributed to reduced diversity. In Figure \ref{fig:CS-decomp}, the losses in consumer surplus, $Loss_{t}^{CS}(\underbar{d}) = CS_{t}(0) - CS_{t}$ for each $\underbar{d}$, are presented. Approximately 99\% of the decrease in consumer surplus is due to the reduction in variety. This implies that pricing is inactive, which is consistent with the empirical findings in Section \ref{sec:spatial}. Here, diversity refers not only to the number of products, but also to the desirability of product attributes, as reflected in the utilities $V_{ijt}$ in \eqref{eq:CS}.

\subsection{Welfare Analyses in Section \ref{subsec:simul-protection-fclevel}}\label{subsec:details_counterfactual}

For the simulation exercises, we need to specify counterfactual fixed cost functions. In Scenario A where technology is an assistant, we compute the fixed cost by summing the fitted cost values from the estimated cost function for a given shape and different levels of $\nu_{k0}$ via $C$, which are assumed to be the same across all firms. In Scenario B where technology is a substitute, there would be no mimicking advantage that is present in Figure \ref{fig:supply-result}(a). Instead, the fixed costs would remain ``flat'' around the designated mean value, with random noise. In this scenario, we set $\eta_{0\ell} = \eta_{1\ell} = \eta_{2\ell} = \eta_{3\ell} = 0 $ for every $\ell \geq 1$ and then sample $\nu_{k0}$ from the distributions with different means.

\begin{figure}
    \centering
    \caption{Potential Product Shapes for Counterfactual Analysis}
    \includegraphics[width=0.8\linewidth]{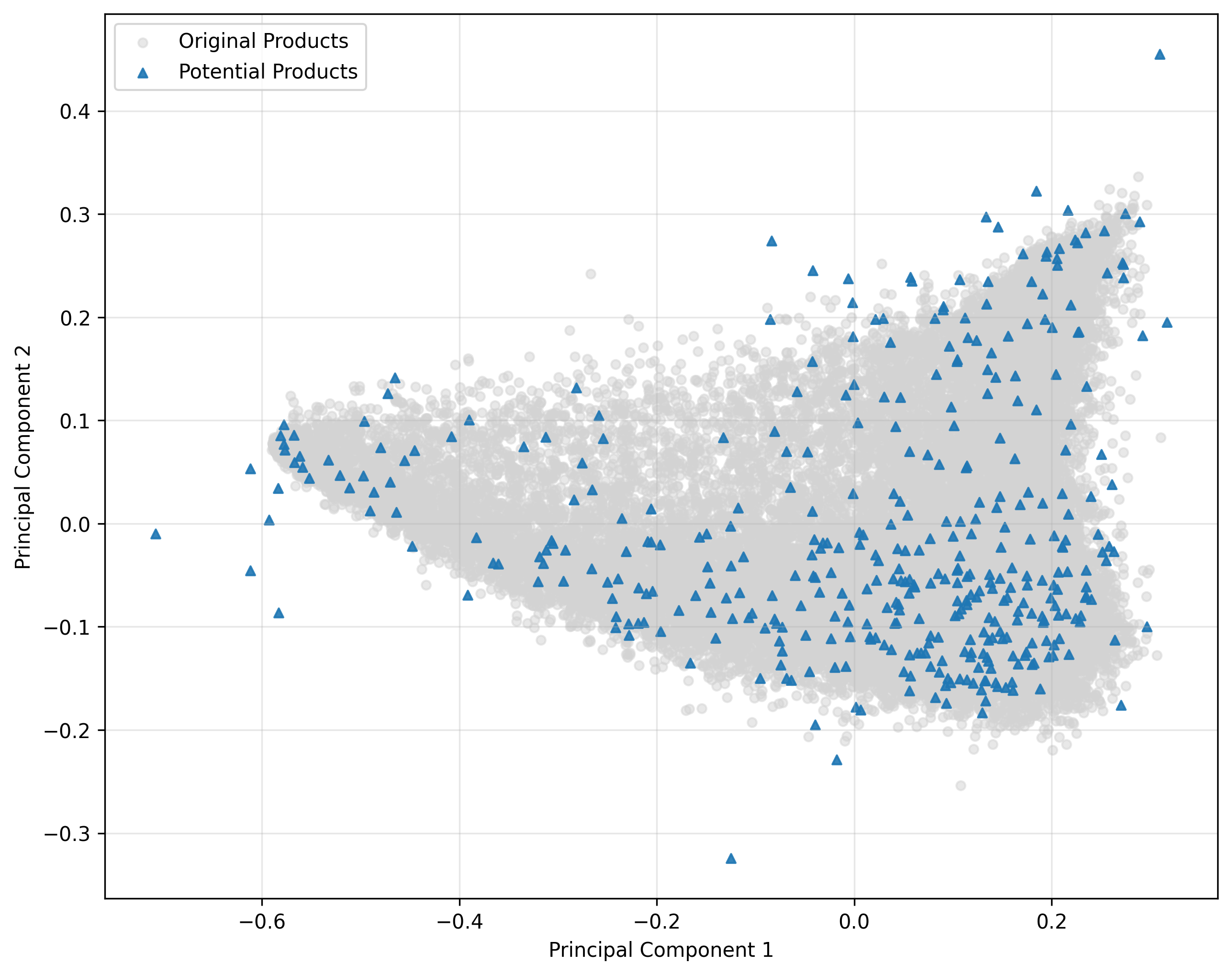}
    \tabnotes{This figure displays the potential entry locations in the counterfactual exercises. The blue triangle dots indicate the potential shapes. We construct the potential shapes by first randomly sampling original embeddings, displayed as gray circle dots, and adding Gaussian noises to them. We explain the further details of potential shapes generation in Appendix Section \ref{sec:simul-detail}.}
    \label{fig:poten-shapes}
\end{figure}

\begin{figure}[p!]
    \centering
    \caption{Histogram of Fixed Costs in the Simulation Exercises (Scenario A)}
    \includegraphics[width=0.8\textwidth]{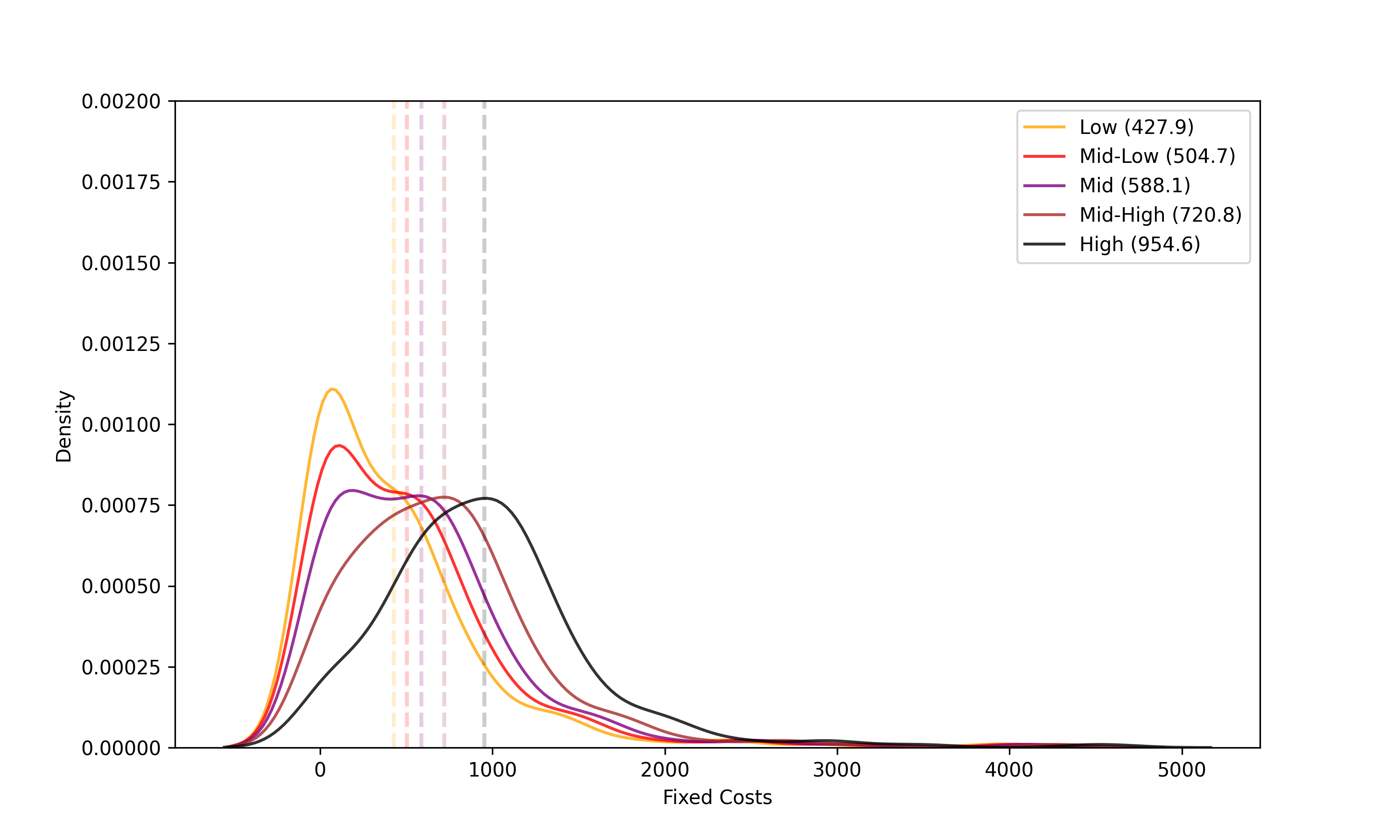}
    \label{fig:hist-fixed-simul-level}
    \tabnotes{This figure shows kernel density estimates of fixed costs $F(\boldsymbol{x}_{t}, \nu_{k})$ (in dollars) by the fixed cost levels specified in Sections \ref{subsec:simul-protection-fclevel} and \ref{subsec:details_counterfactual}. Average values are reported in parentheses. In the counterfactual exercises, the densities are truncated below zero. The distributions for Scenario B are calibrated to resemble those above, even though the fixed costs are generated from a flat $F(\boldsymbol{x}_{t}, \nu_{k})$.}
\end{figure}

With the characterized fixed costs, we implement the simulation exercises in the following way. We first generate 400 potential locations by perturbing randomly sampled embeddings, similar to the relocation analysis in Section \ref{subsec:protection-study}; the generated potential locations are shown in Figure \ref{fig:poten-shapes}.\footnote{We use 8 different standard deviation levels of Gaussian noises by setting $c_{level}$ to be one of $\{1, 2, 3, 4, 5, 10, 20, 30\}.$} We then compute the fixed cost associated with each location. In Scenario A, we define five distinct fixed cost levels: \emph{low}, \emph{mid-low}, \emph{mid}, \emph{mid-high}, and \emph{high}. For each location, we compute the fitted values of cost as a function of the embeddings and add $\nu_{k0}$ based on the assigned fixed cost level. In Scenario B, rather than using fitted values, we randomly draw fixed costs for each location from a distribution calibrated to have a similar mean and variance as in the first scenario. In both cases, we ensure that there are a modest number of entries at high cost levels even under strong protection, and some non-entries at low cost levels under strong protection. Figure \ref{fig:hist-fixed-simul-level} shows the fixed cost distributions. In addition, we calculate the expected profit for each potential location using the demand estimates. For exogenous characteristics and marginal costs of a given firm, we use the corresponding average values across products of that firm.\footnote{We assign the nest of the closest product as the nest for a potential entrant. This is because tags and search menus are based on shapes and functionalities that are reflected in our embeddings.} Finally, we allow firms to search for the optimal location and decide on entry.\footnote{We assume that a new product is introduced to every country.}

Simplifications are necessary in simulating firms' decisions, as computationally solving the full model is extremely burdensome. Simulating a single market requires evaluating each combination of potential entrants, firms, and sampled demand shocks $\xi_{k}$ to calculate the expected profits for all potential products and firms. Consequently, computation time increases rapidly as any of these elements grow. To address this issue, we implement the following. First, we consider a random sample of 100 firms that sequentially decide on entry and product positioning. Second, we impute the expected value of entrants' demand shocks $\xi_{k}$ to approximate the expectation, following a similar approach to \cite{berry1999voluntary}.\footnote{To mitigate concerns about potential approximation errors, we compare simulated expected profits obtained using this imputation method with those based on the empirical distribution, confirming that they yield similar outcomes.} Third, we focus our analysis to markets in April 2014, the first period of our dataset.

\section{Alternative Dimension-Reduction Layer}\label{sec:PLS}

For the structural analysis, we present an alternative way of imposing a dimension-reduction layer for the 128-dimensional neural network embeddings, namely the \emph{parital least squares} (PLS), which can be viewed as a supervised alternative to the PCA \citep{hastie2009elements}. In producing the output (i.e., the 6-dimensional embedding\footnote{Similar to the PCA, we can draw a scree plot (omitted), which suggests to choose the first six components.}), the PLS is supervised using the information from the demand. Specifically, the inputs (i.e., the elements of 128-dimensional embedding) are weighted by the strength of their \emph{univariate} effect on the market share. The idea is to tilt the direction of principal components, which captures the largest variation of the inputs, towards the direction that better explains the market shares of products. Although this adjustment sacrifices the representation of visual attributes as they are, it can be viewed as a fine-tuning of the original neural network that improves the relevance of the representation to consumer preferences. To implement the PLS, we define the dependent variable as the residualized inverse market share purged of the effects from the price and structured attributes using a fixed-coefficient model.

Table \ref{tab:linear-reg-result-PLS} reports the OLS and IV regression results using the 6-dimensional embeddings produced by the PLS. The results largely resemble those in Table \ref{tab:linear-reg-result} where the PCA is used, indicating that the results are robust to the choice between the PCA and PLS as the final dimension-reduction layer.

\begin{figure}[p!]
    \centering
    \caption{Measures of Competition and Embedding Distances (With PLS Embeddings)}
    \subfloat[Prices Diversion Ratios]{\includegraphics[width=0.5\textwidth]{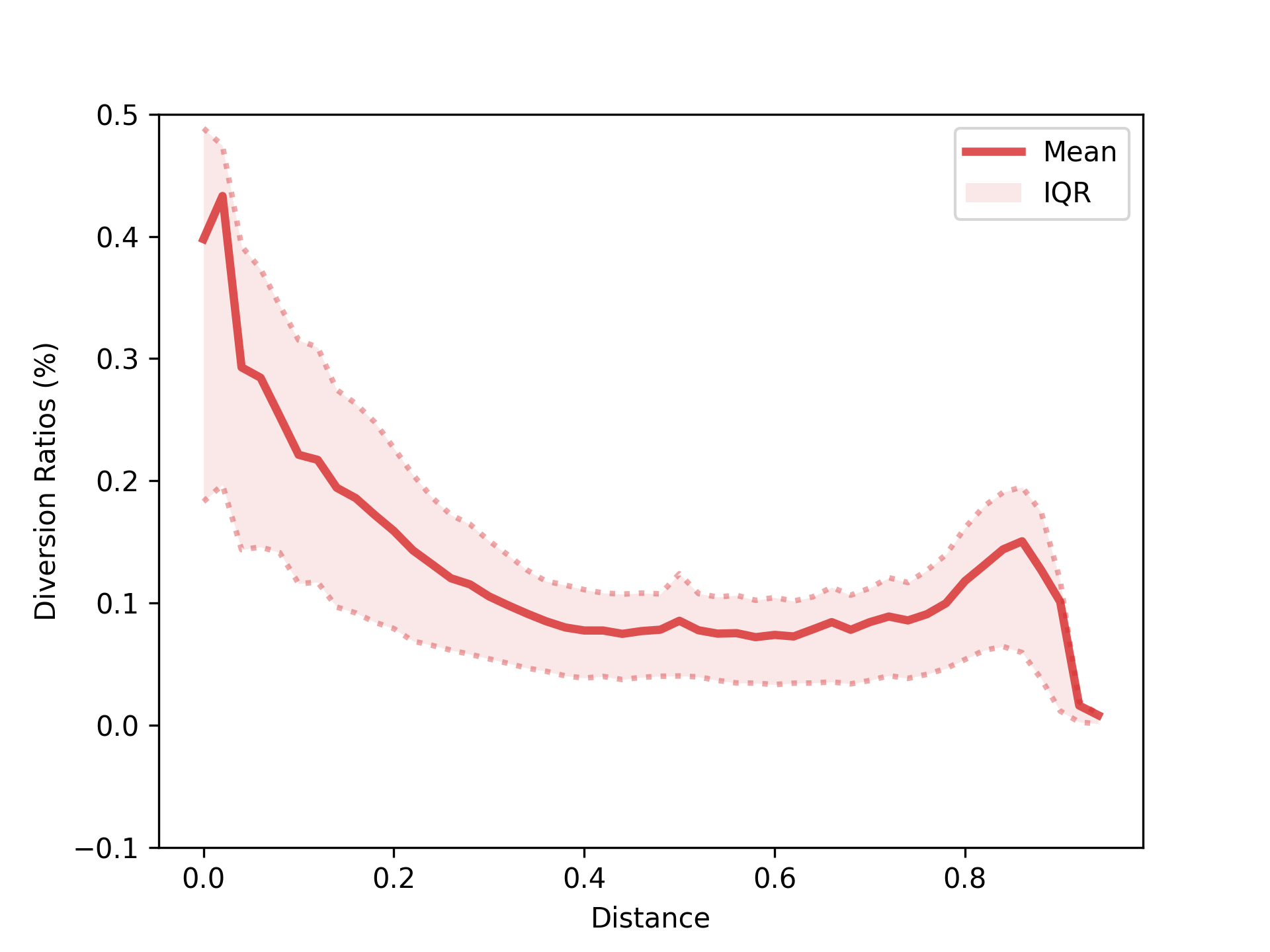}} 
    \subfloat[Long Run Diversion Ratios]{\includegraphics[width=0.5\textwidth]{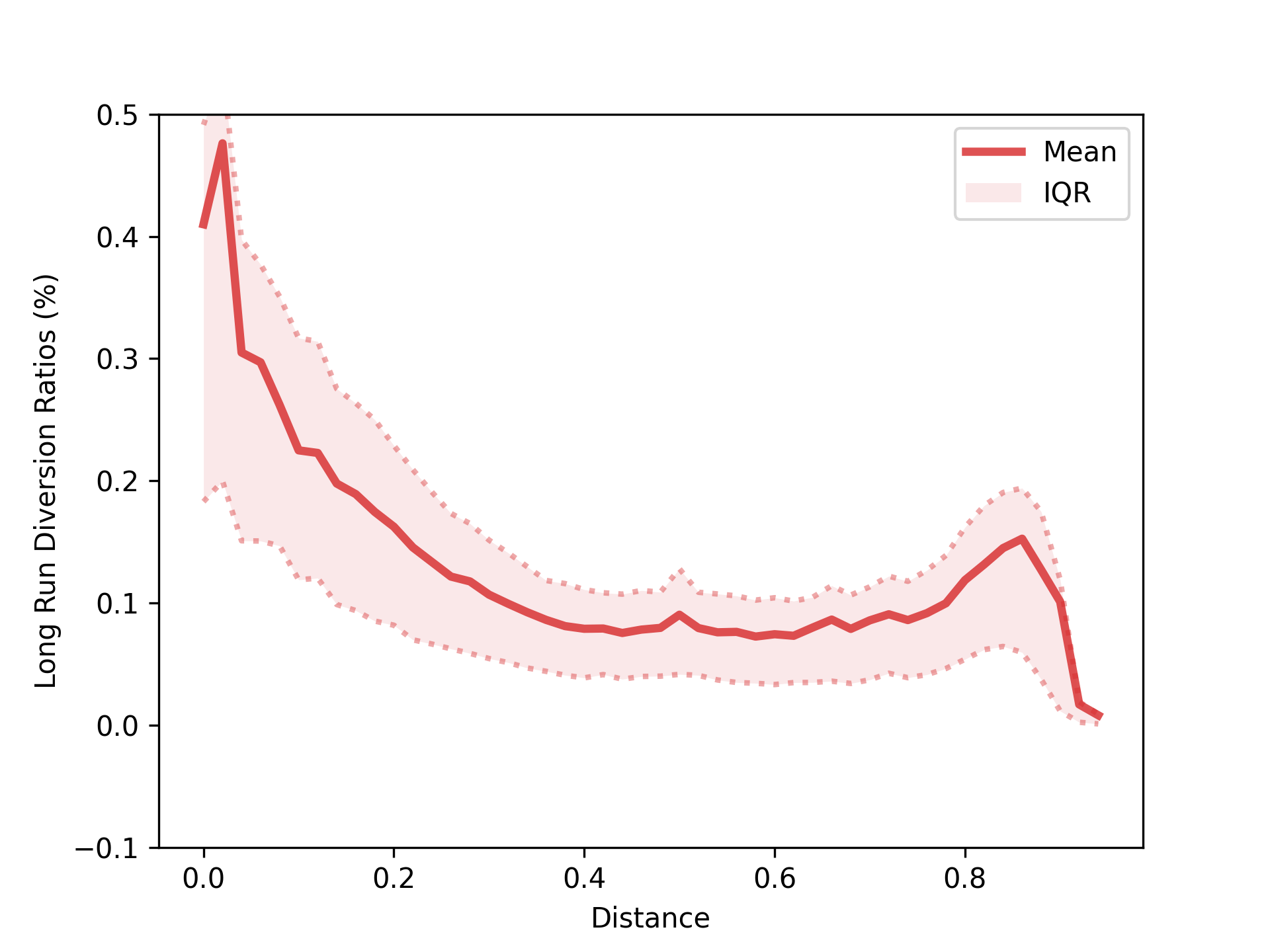}} \\
    \subfloat[Cross Price Elasticity]{\includegraphics[width=0.5\textwidth]{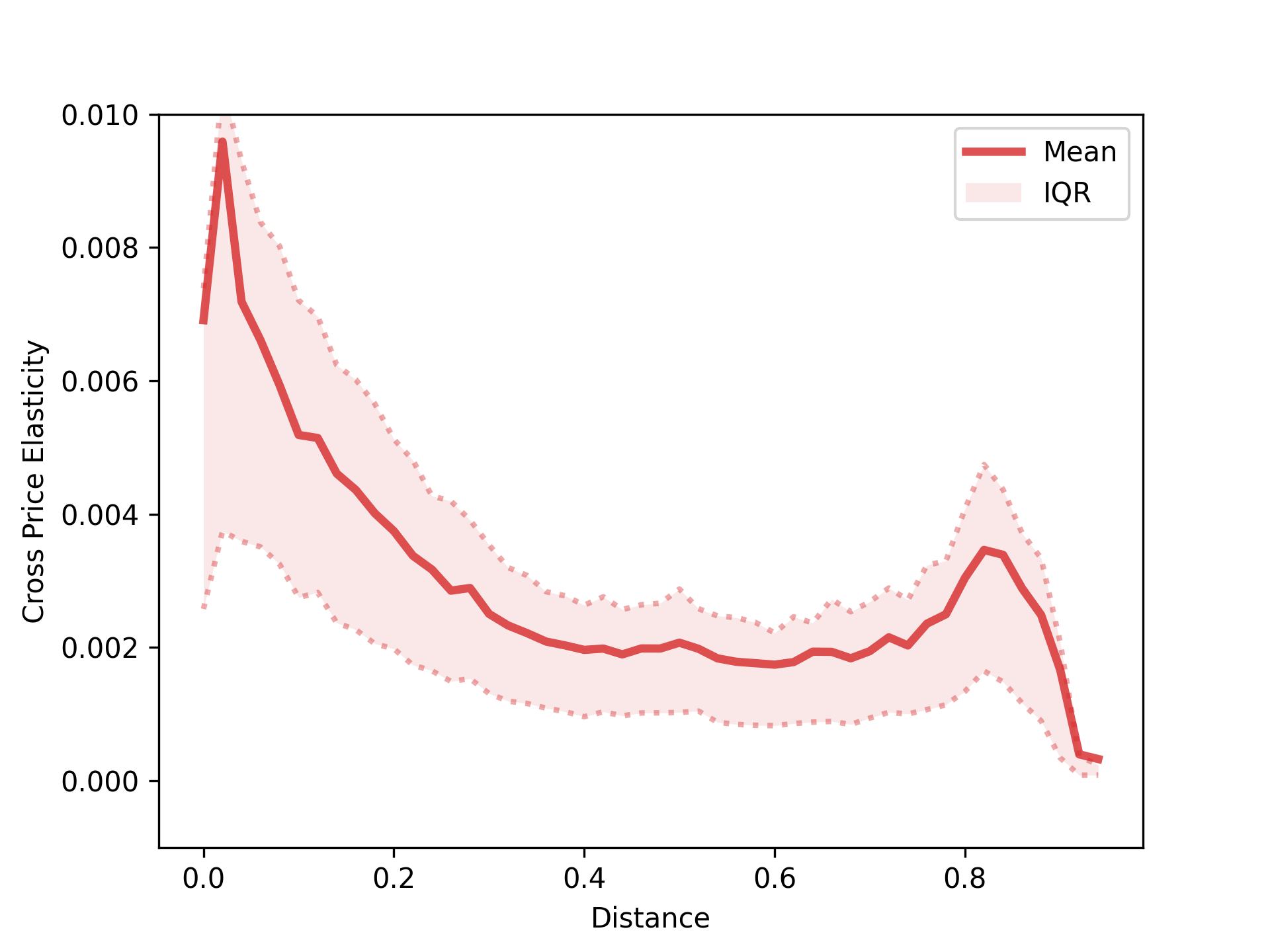}} 
    \subfloat[Average Number of Products]{\includegraphics[width=0.5\textwidth]{image/MeasureCompetition/Avg_Product_Num_240812.png}} 
    \tabnotes{Panels (a) to (c) plot the average price diversion ratios, long-run diversion ratios, and cross-price elasticity along the embedding distance $d$ (see equation \eqref{eq:DIV}), calculated by using demand estimates with random coefficients on the PLS embeddings, respectively. As a reference, Panel (d) shows the average number of products along radial areas.} 
    \label{fig:comp-dist-nl-PLS}
\end{figure}

\begin{figure}[p!]
    \centering
    \caption{Measures of Competition and Embedding Distances}
    \subfloat[Prices Diversion Ratios]{\includegraphics[width=0.5\textwidth]{image/MeasureCompetition/Mkt_Avg_Div_240812.png}} 
    \subfloat[Long Run Diversion Ratios]{\includegraphics[width=0.5\textwidth]{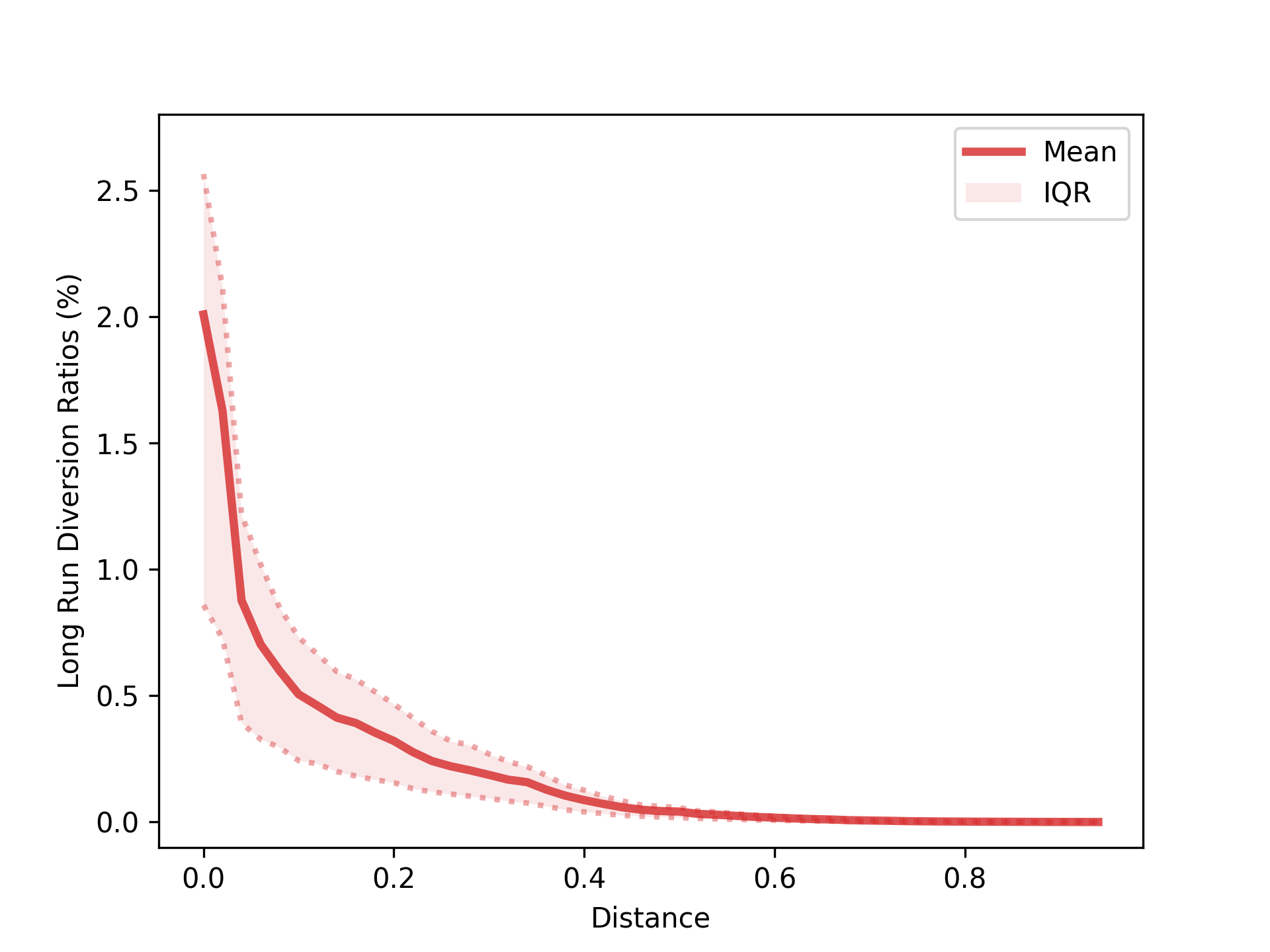}} \\
    \subfloat[Cross Price Elasticity]{\includegraphics[width=0.5\textwidth]{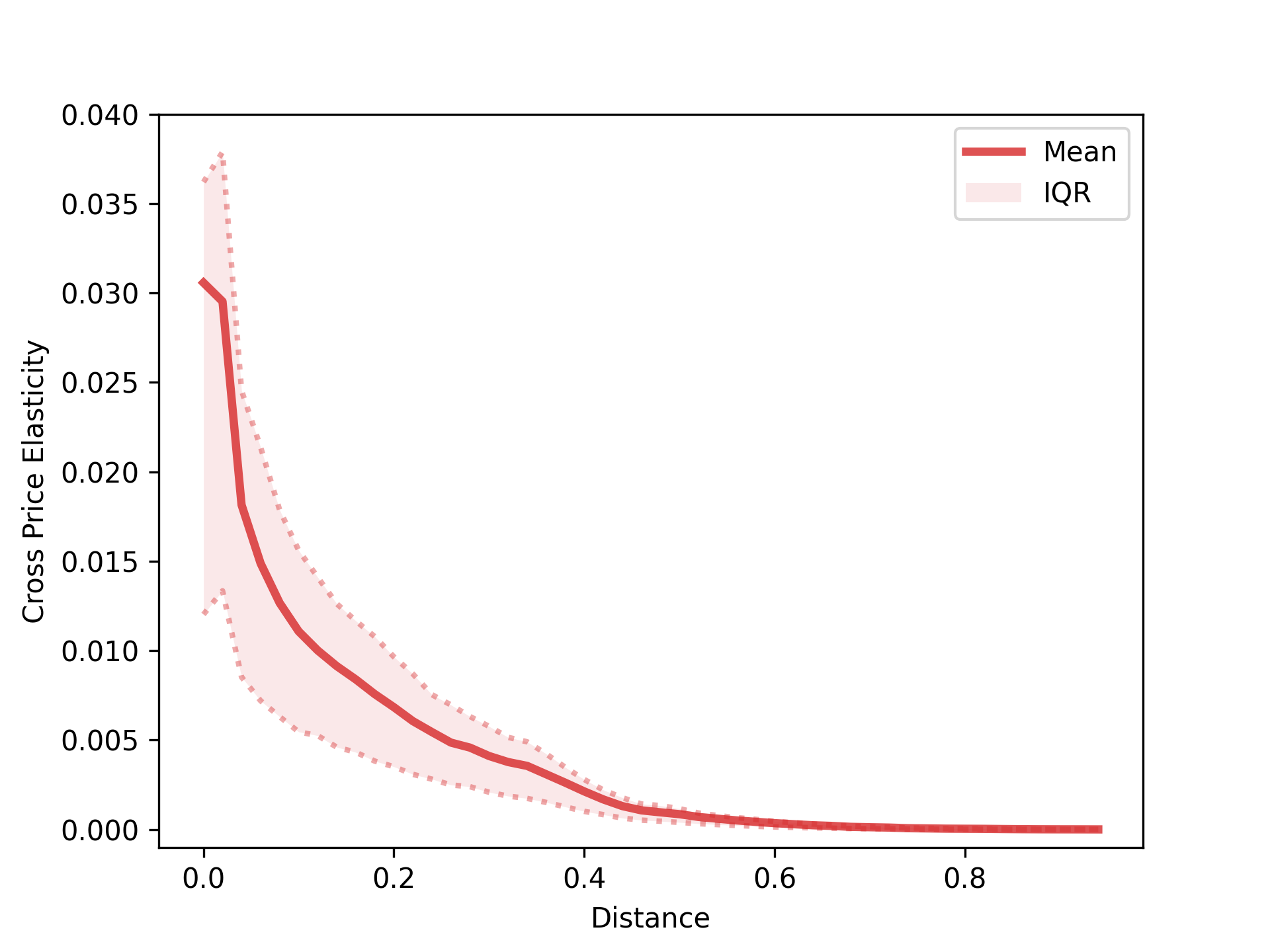}} 
    \subfloat[Average Number of Products]{\includegraphics[width=0.5\textwidth]{image/MeasureCompetition/Avg_Product_Num_240812.png}} 
    \tabnotes{Panels (a) to (c) plot the average price diversion ratios, long-run diversion ratios, and cross-price elasticity along the embedding distance $d$ (see equation \eqref{eq:DIV}), calculated by using demand estimates with random coefficients on the embeddings, respectively. As a reference, Panel (d) shows the average number of products along radial areas.} 
    \label{fig:comp-dist}
\end{figure}

Next, we turn to the main specification in the demand estimation, namely the random-coefficient logit model, and estimate it using the PLS embeddings. Analogous to Figure \ref{fig:comp-dist}, in Figure \ref{fig:comp-dist-nl-PLS} we calculate several competition measures, including diversion ratios, and plot them against the distance in terms of the 128-dimensional embeddings. Again, the overall patterns in Figure \ref{fig:comp-dist-nl-PLS} are similar to those in Figure \ref{fig:comp-dist}, although the measures slightly increase toward the end of the embedding distance. This seems to suggest that, for fonts with substantially different shapes, there is a slight misalignment between distance in the 128-dimensional embeddings and substitutability in the PLS embeddings. Since only a small share of fonts fall into this area, our analysis is not sensitive to this misalignment.

\begin{table}[h!]
\centering
\caption{Fixed Coefficients Demand Estimation Results with PLS}
{\small
\begin{tabular}{lcccc}
\toprule
Column              & (1) & (2)\\
Model              & OLS & IV (2nd Stage) \\
Variables & $\ln(s_{j}/s_{0})$ & $\ln(s_{j}/s_{0})$ \\
\midrule
Prices & -0.0202*** & -0.1571*** \\
 & (0.0002) & (0.0013) \\
Glyph Counts & 0.0004*** & 0.0007*** \\
 & (0.0000) & (0.0000) \\
Constant & -7.9851*** & -5.6200*** \\
 & (0.0083) & (0.0258) \\
\midrule
Observations & 225,658 & 225,658 \\
6-dim Embeddings (PLS)          & Yes       & Yes \\
$R^{2}$ & 0.0686 & - \\
$F$ stat & 1727 & 2080 \\
\bottomrule
\end{tabular}
}
\tabnotes{This table shows results from OLS and IV regression models with fixed coefficients using PLS to transform the 128-dimensional embeddings. Robust standard errors are in parentheses. All coefficient estimates are statistically significant at 1\% level. The Cragg-Donald $F$ test of IV  suggests the IV is strong.}
\label{tab:linear-reg-result-PLS}
\end{table}

\section{Robustness of Event Study Analysis}
\begin{figure}[htbp!]
    \centering
    \caption{Event Study Design ($\beta_{s}$) by Using \cite{borusyak2021revisiting}}
    \subfloat[Revenue]{\includegraphics[width=0.5\textwidth]{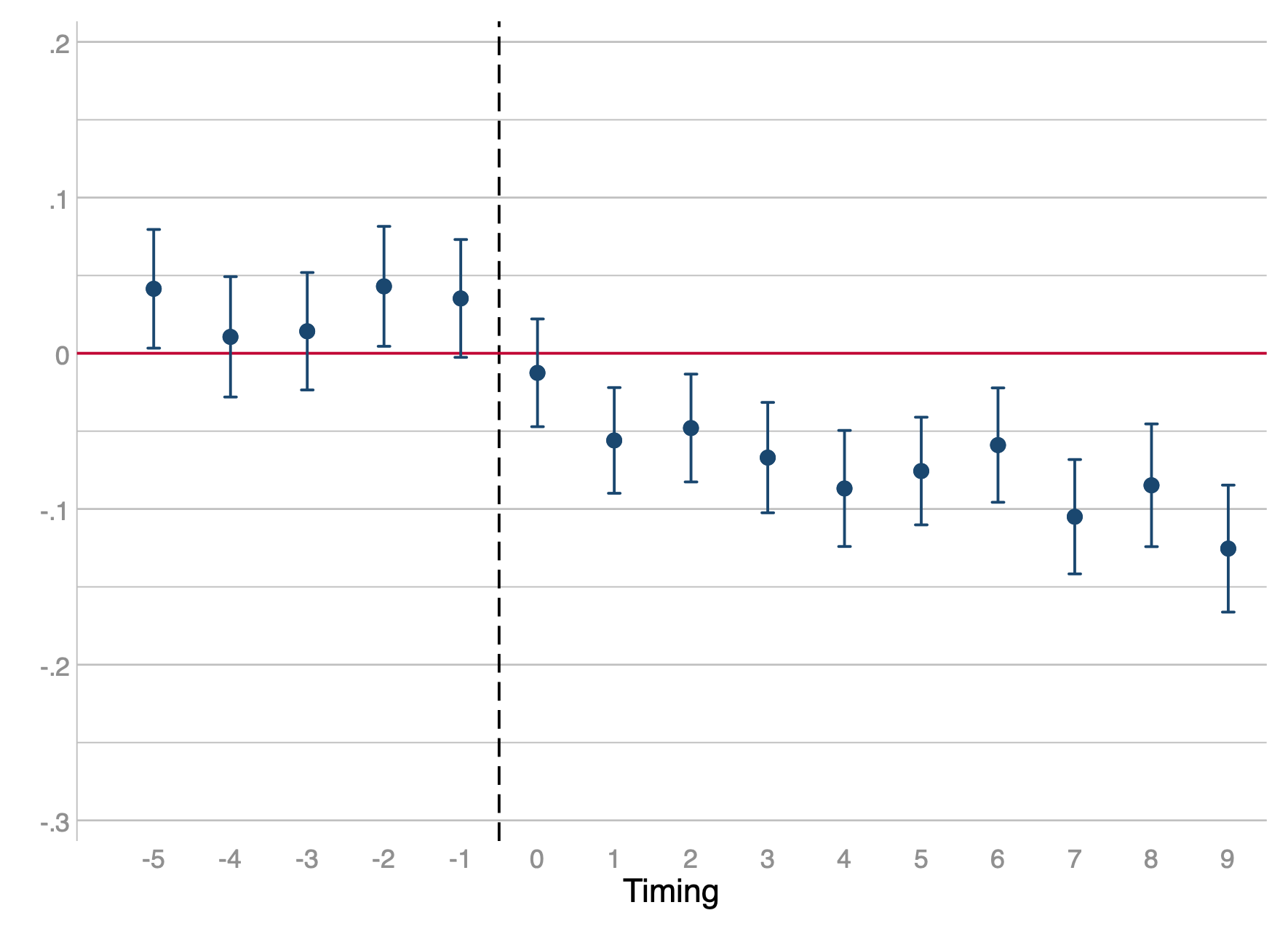}} \\
    \subfloat[Quantity]{\includegraphics[width=0.5\textwidth]{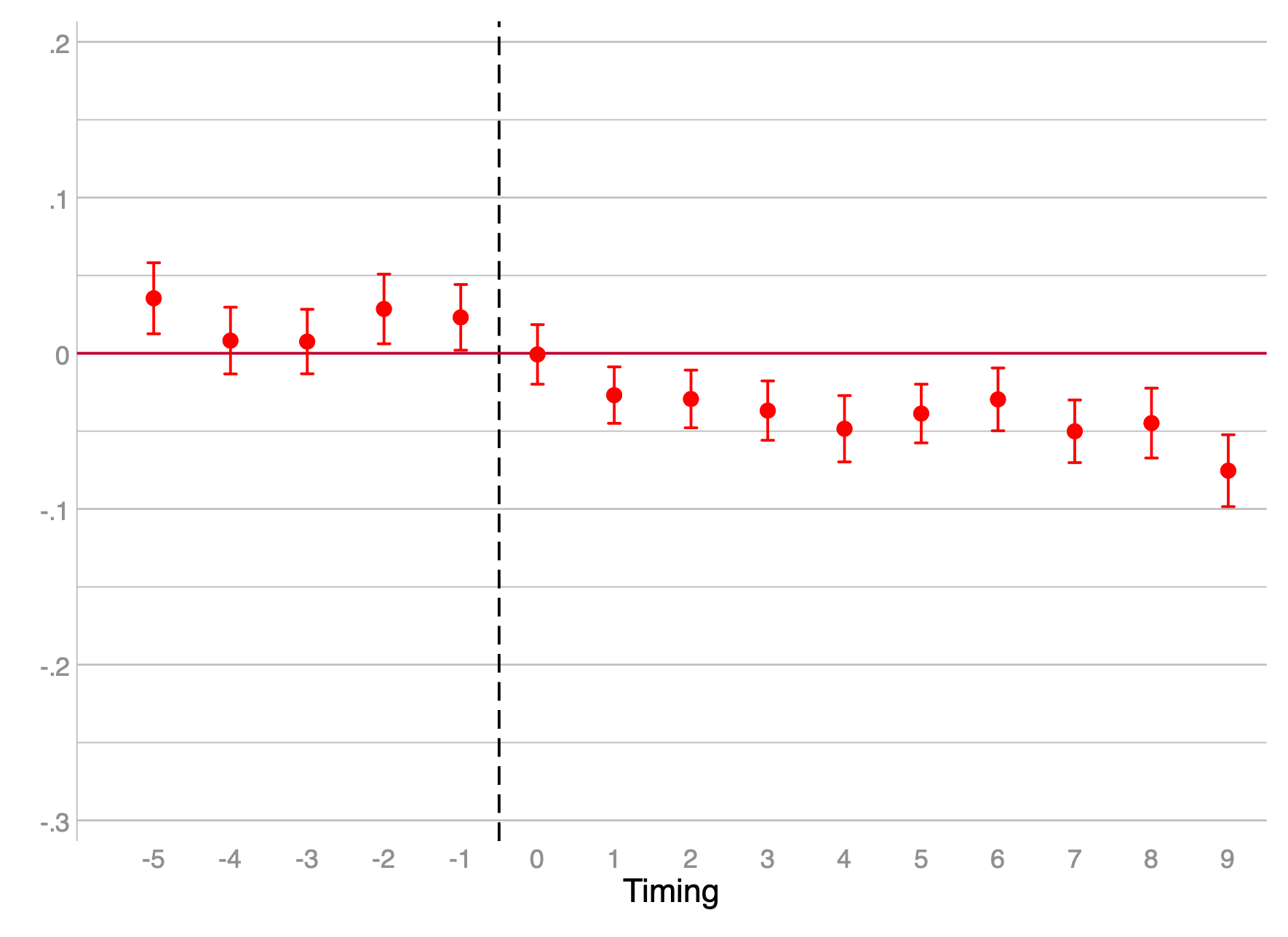}} 
    \subfloat[Price]{\includegraphics[width=0.5\textwidth]{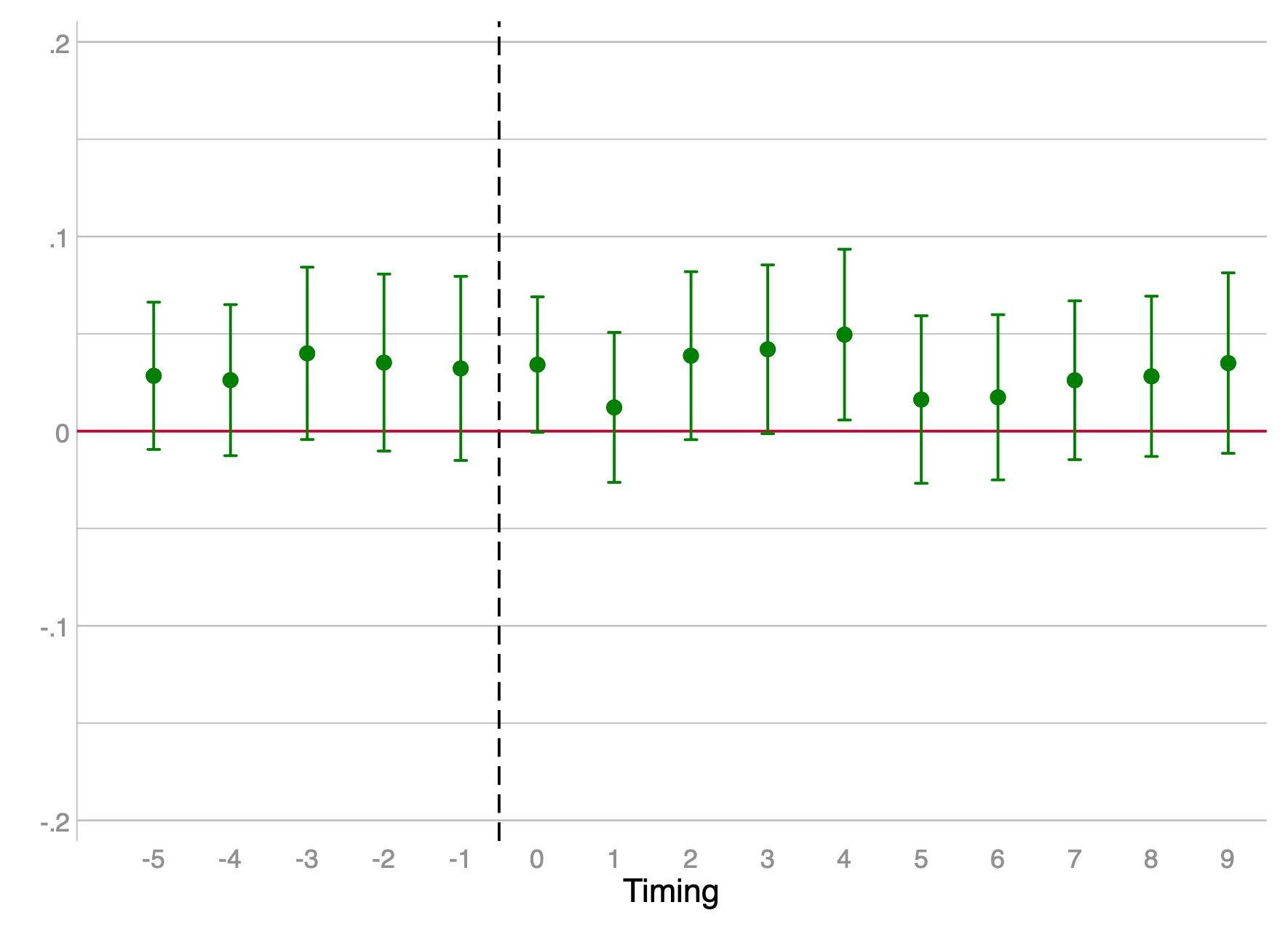}}
    \tabnotes{These figures show results of the event study regression in \eqref{eq:event-4w}, which is implemented via the method by \cite{borusyak2021revisiting}. Panels (a) and (b) contain regression results for arsinh transformation of revenue and quantity as a dependent variable, respectively.  Panel (c) shows the result for log of list prices  as a dependent variable. Solid lines indicate the 95\% confidence intervals of estimates. }
    \label{fig:event-impute}
\end{figure}

\begin{figure}[htbp!]
    \centering
    \caption{Event Study Design ($\beta_{s}$): Additional Control Variables}
    \subfloat[Revenue]{\includegraphics[width=0.5\textwidth]{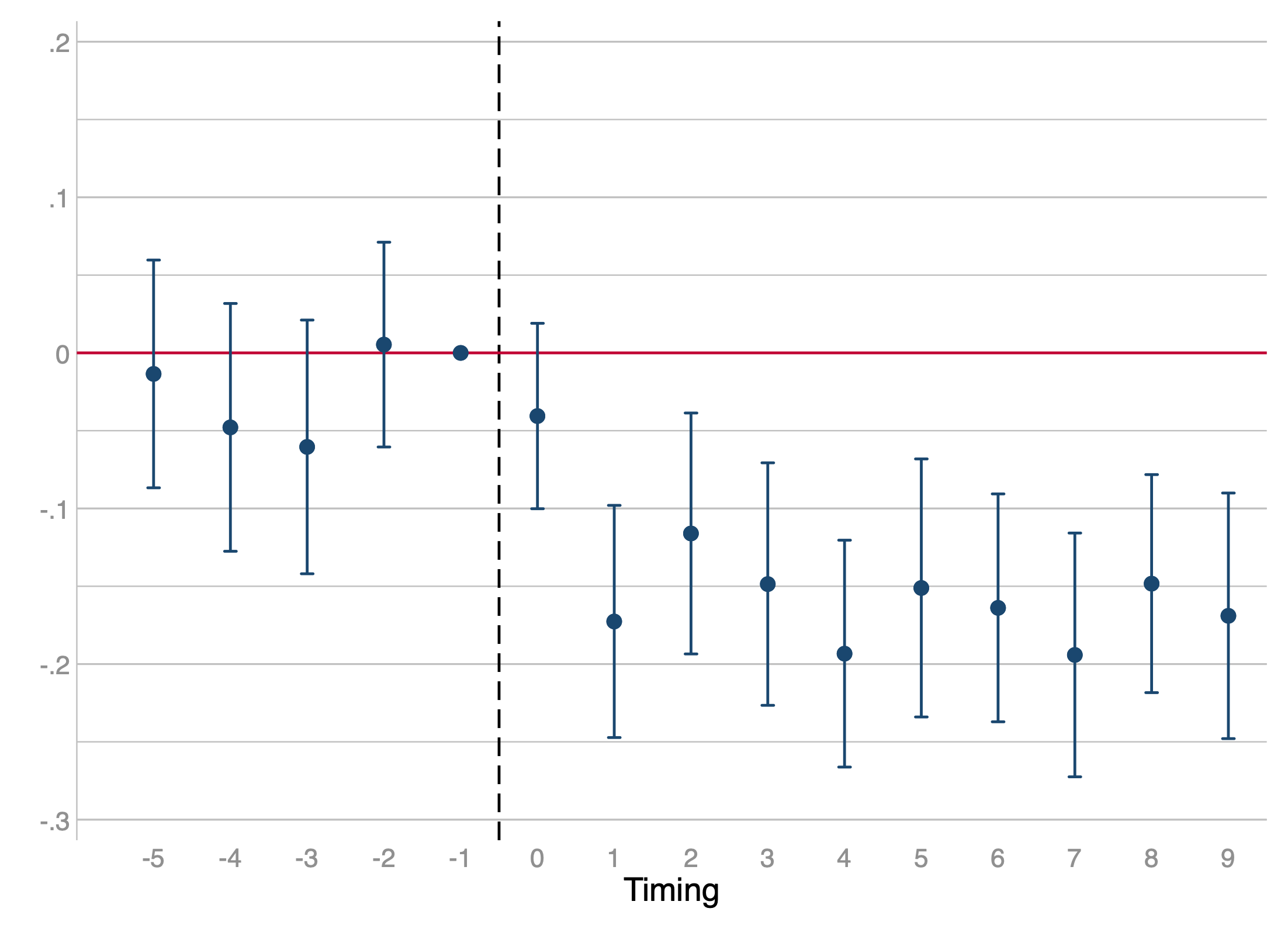}} \\
    \subfloat[Quantity]{\includegraphics[width=0.5\textwidth]{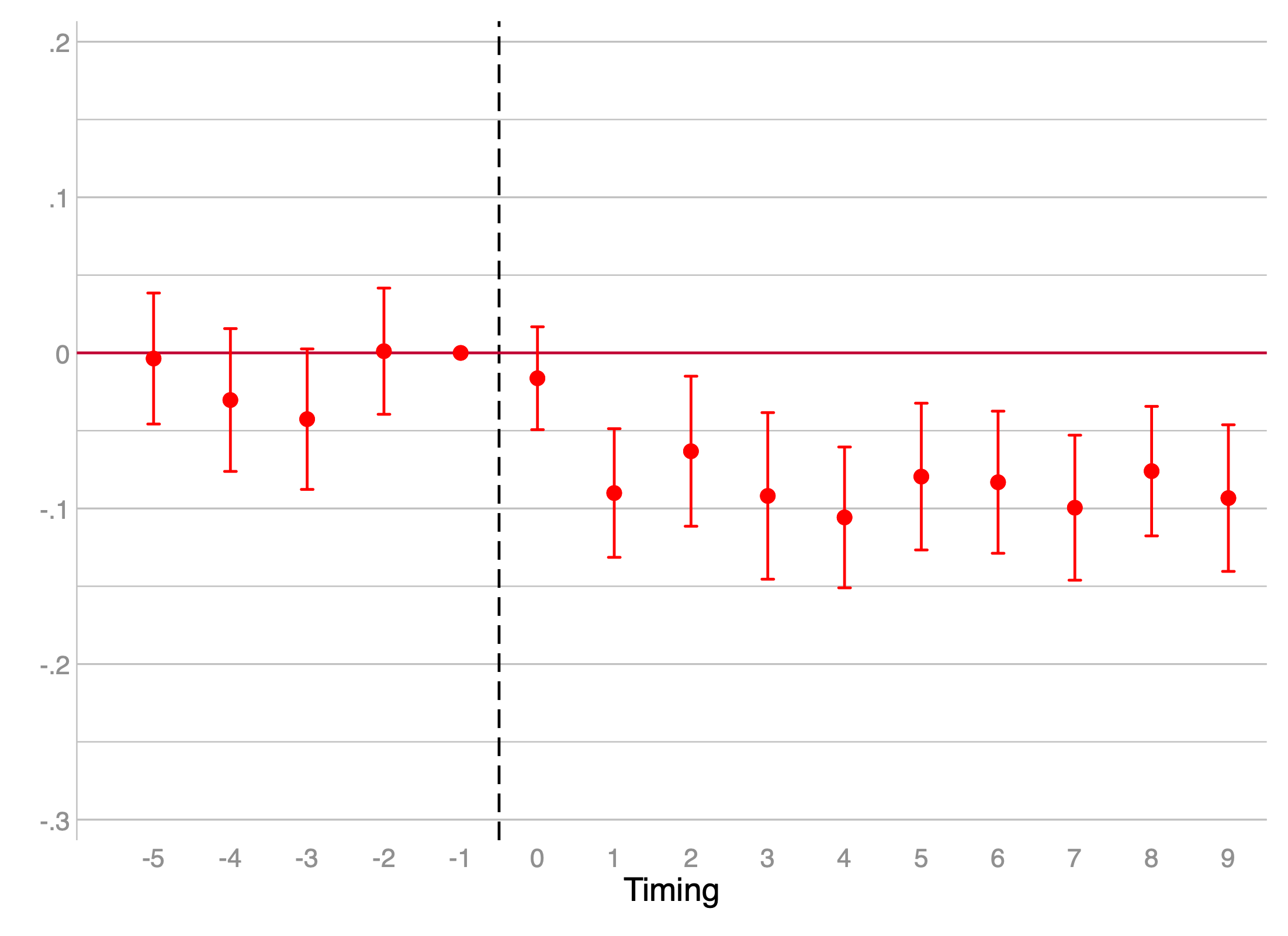}} 
    \subfloat[Price]{\includegraphics[width=0.5\textwidth]{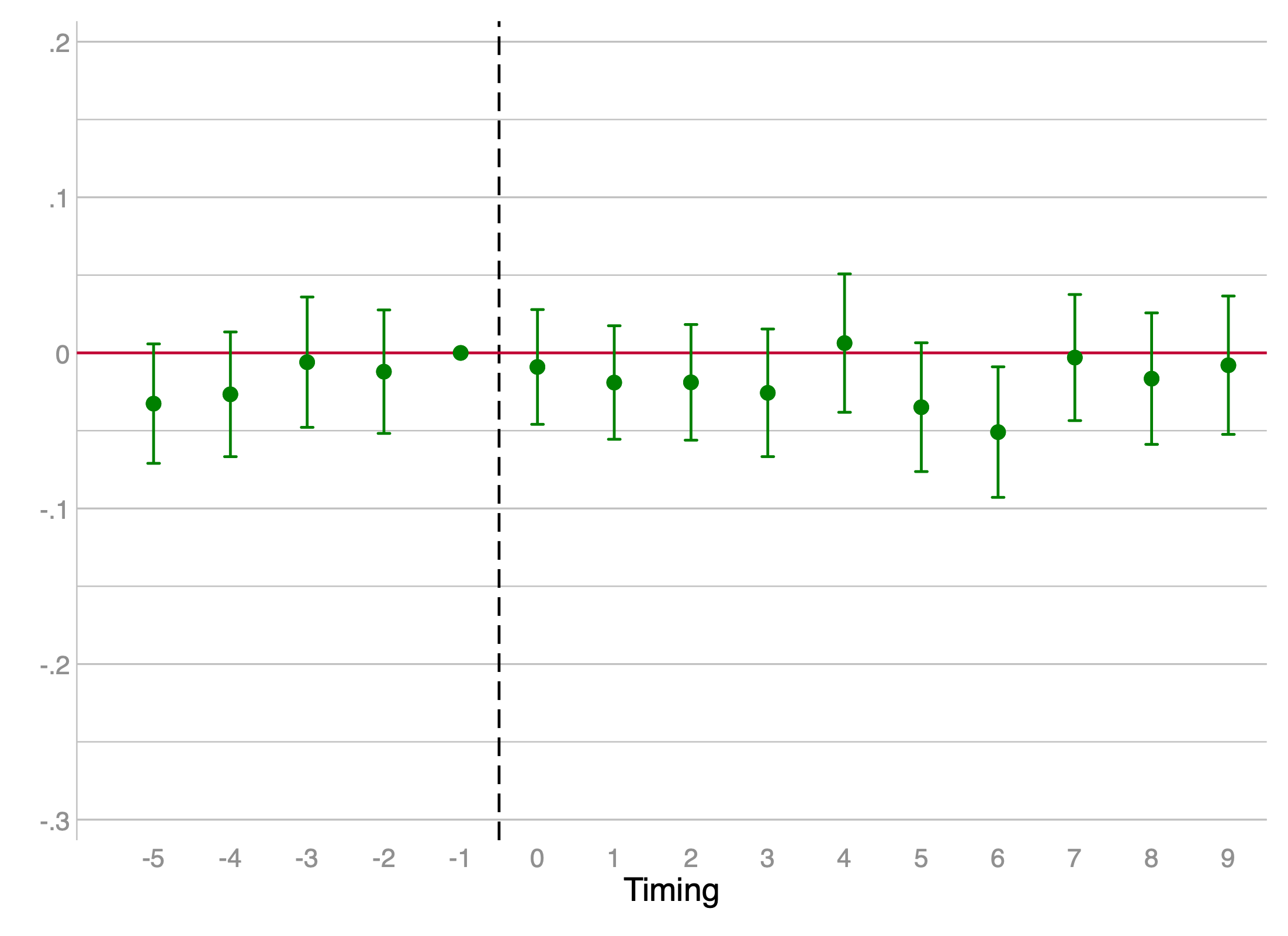}}
    \tabnotes{These figures show results of the event study regression in \eqref{eq:event-4w} with additional control variables; we include 500 image cluster and time interaction dummies, month after the introduction to the marketplace and log of glyphs. Image clusters are attained by $K$ means clustering algorithm. Panels (a) and (b) contain regression results for arsinh transformation of revenue and quantity as a dependent variable, respectively.  Panel (c) shows the result for log of list prices as a dependent variable. Solid lines indicate 95\% confidence intervals of estimates. Firm-level clustered standard errors are used. }
    \label{fig:event-ols-add}
\end{figure}

\begin{figure}[htbp!]
    \centering
    \caption{Event Study Design ($\beta_{s}$): Alternative Treatment Definition}
    \subfloat[Revenue]{\includegraphics[width=0.5\textwidth]{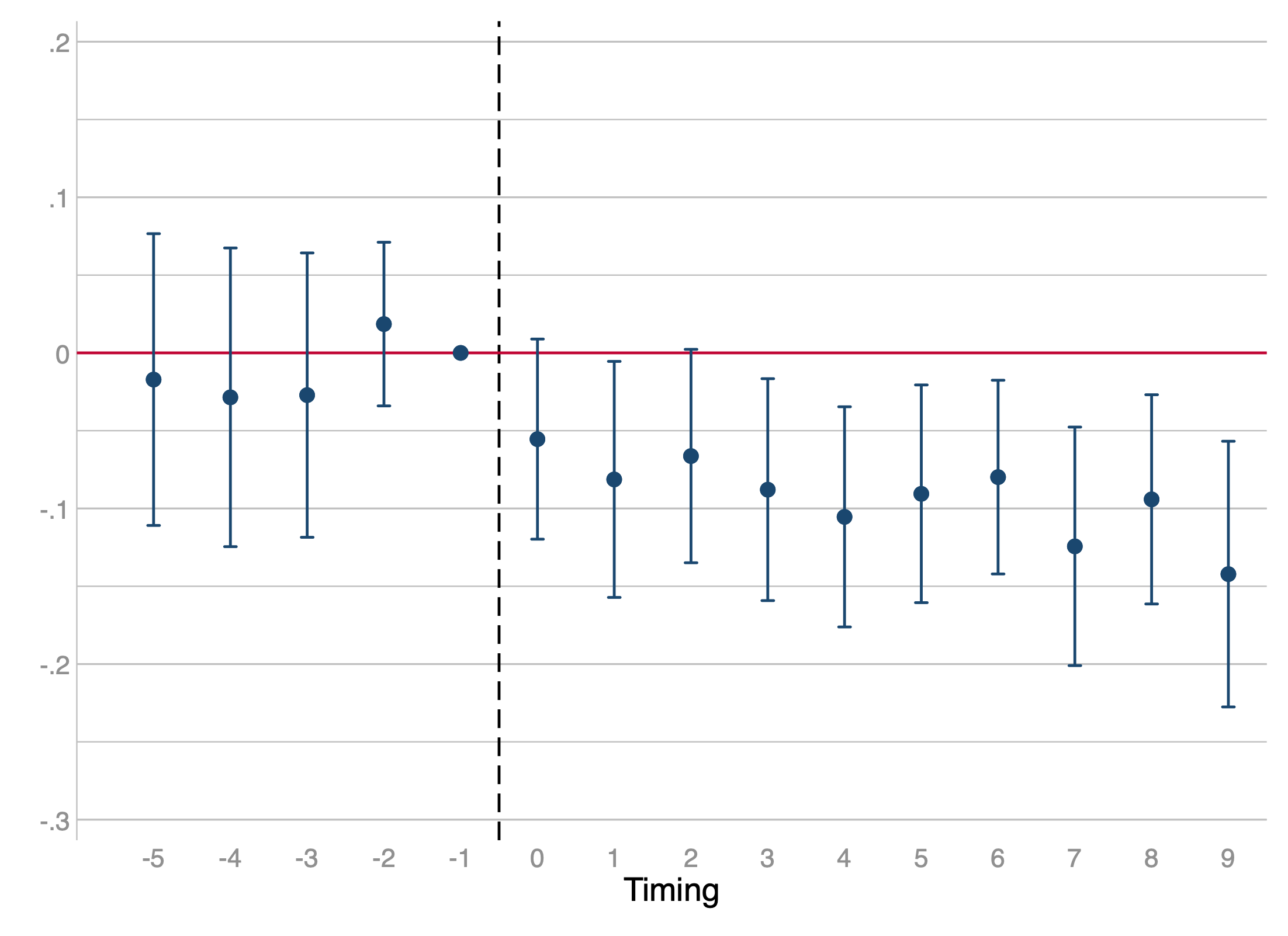}} \\
    \subfloat[Quantity]{\includegraphics[width=0.5\textwidth]{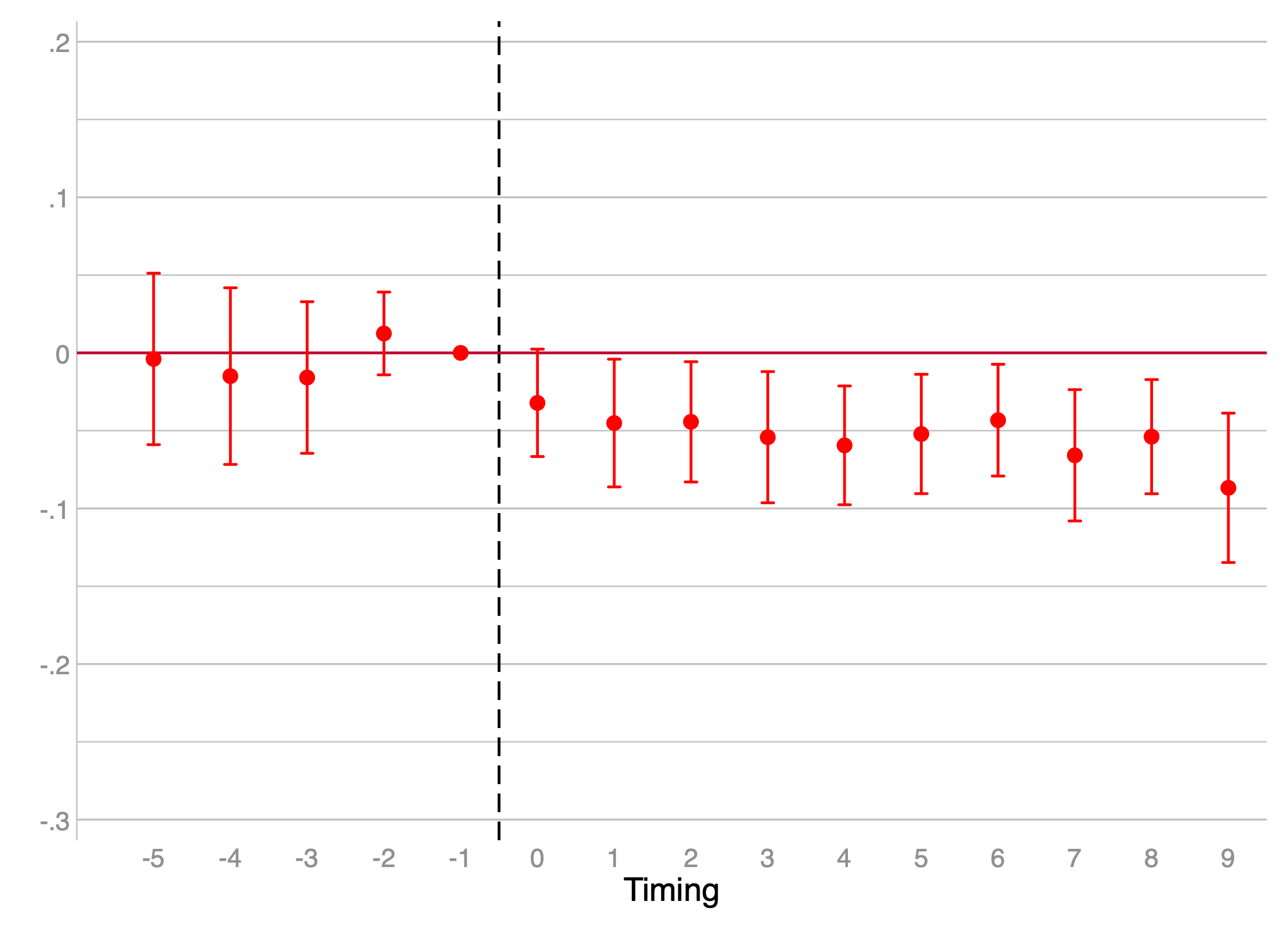}} 
    \subfloat[Price]{\includegraphics[width=0.5\textwidth]{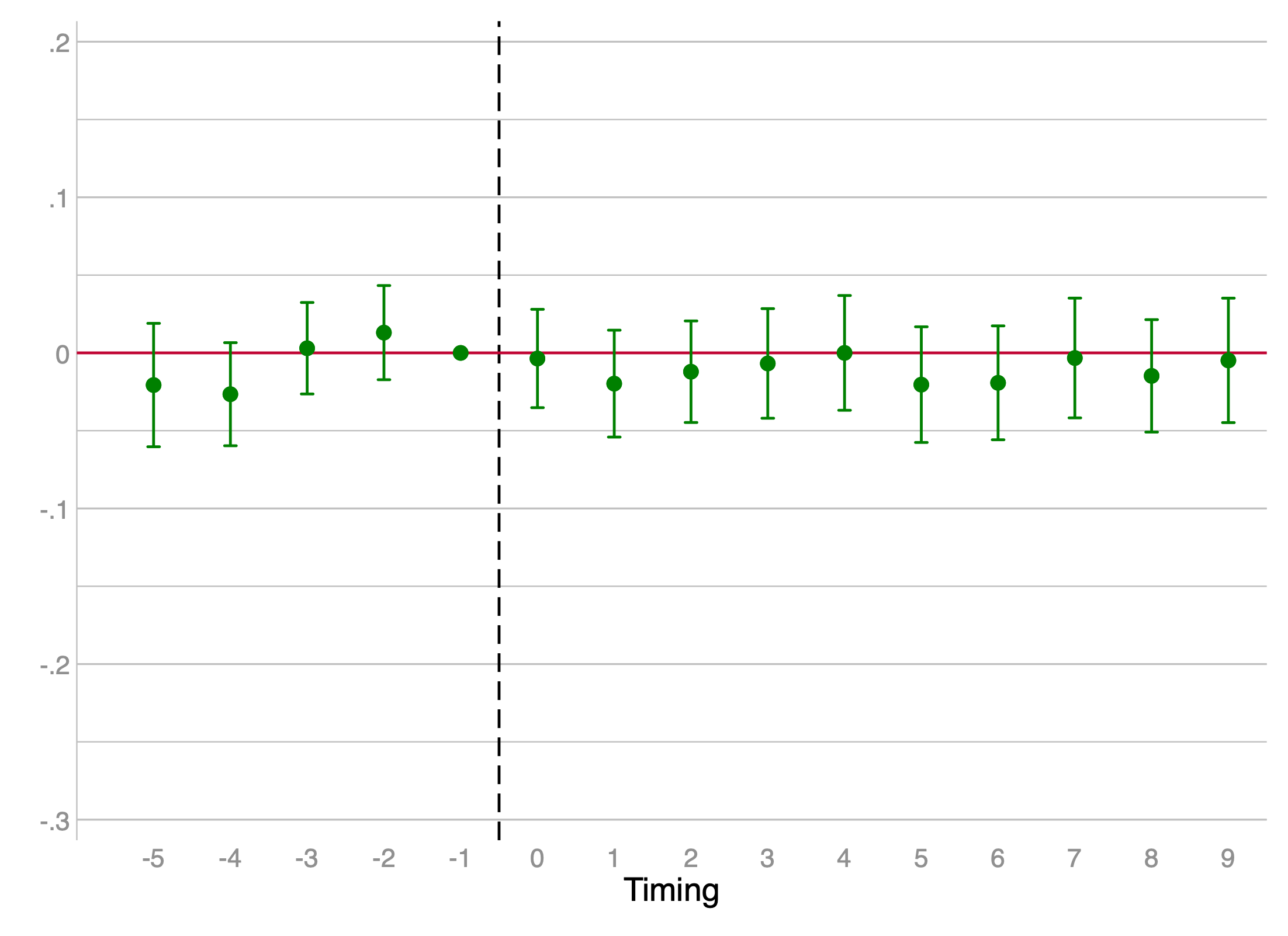}}
    \tabnotes{These figures show results of the event study regression in \eqref{eq:event-4w} with alternative treatment definition; in this figure the treatment is defined to be change in one of four closest competitors due to a new entry. Panels (a) and (b) contain regression results for arsinh transformation of revenue and quantity as a dependent variable, respectively.  Panel (c) shows the results for log of list prices as a dependent variable. Solid lines indicate 95\% confidence intervals of estimates. Firm-level clustered standard errors are used. }
    \label{fig:event-ols-1of4}
\end{figure}

\newpage
\section{Additional Tables and Figures}\label{sec:more_tab_fig}

\begin{figure}[h!]
    \centering
    \caption{Distributions of Log List and Sales Prices}
\subfloat[Every Transaction]{\includegraphics[width = 0.5\textwidth]{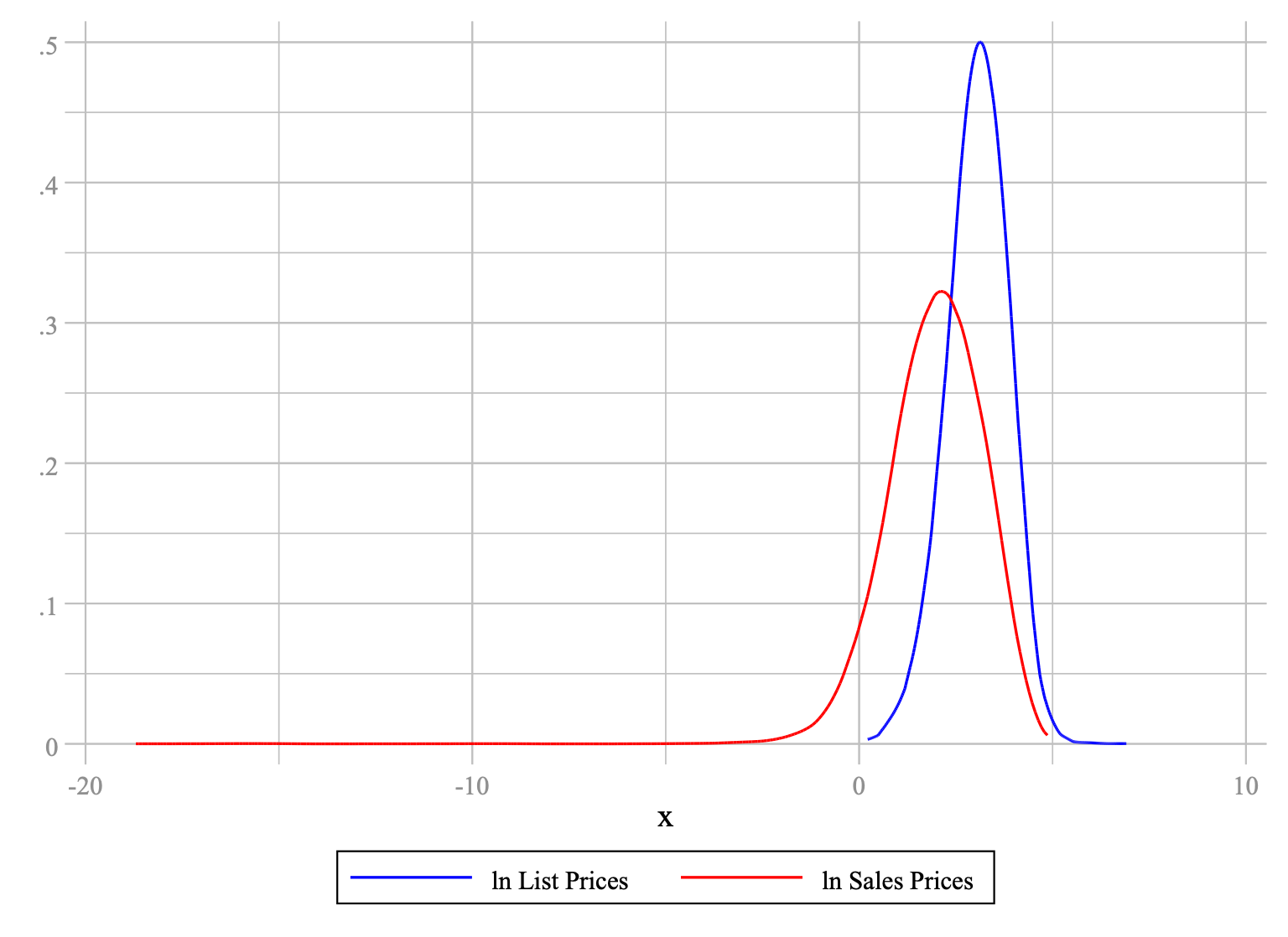}}
\subfloat[Transaction Without Quantity Discount]{\includegraphics[width = 0.5\textwidth]{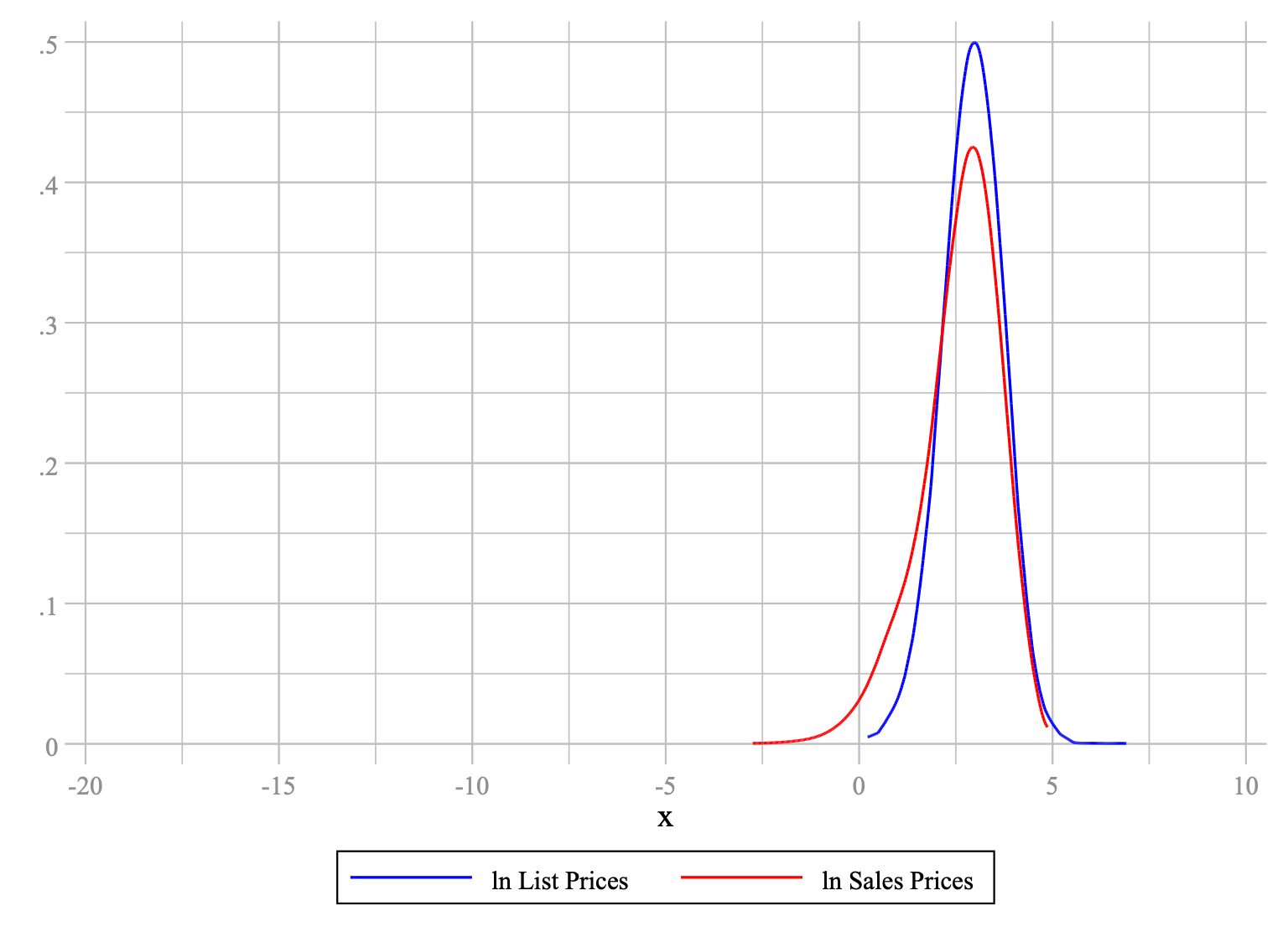}}
\tabnotes{This figure shows the distributions of log list and sales prices. Panels (a) and (b) display the distribution of every transaction and transaction without quantity discount, respectively.}
\label{fig:dist-prices}
\end{figure}

\begin{table}[h!]
\centering
\caption{Descriptive Statistics on Quantity Units in Transactions}
\begin{tabular}{llrrr}
\toprule
License Type & Quantity & Frequency & Percent (\%)
 & Cumulative (\%) \\ \hline
Desktop & 1 user       & 970,882        & 42.06          & 42.06       \\ 
 & 5 users        & 781,630        & 33.86          & 75.91       \\ 
 & Others        &  101,596       &   4.40          & 80.31       \\ \hline
Web & 10k views          & 266,155        & 11.53          & 91.84      \\ 
 & 250k views         & 104,101        & 4.51           & 96.35       \\ 
 & Others        & 84,207        & 3.65           & 100.00      \\ \hline
\multicolumn{2}{c}{Total} & 2,308,571 & 100.00 & 100.00 \\ \bottomrule
\end{tabular}
\caption*{\textit{Notes.} This table shows the number and fraction of transactions in each license type and quantity. Transactions of 12 countries and 2 license types (Desktop and Web) are used.} 
\label{tab:qunatity-category}
\end{table}

\begin{table}[p!]
\centering
\caption{One-Way ANOVA Results (Factor: Product Dummy)}
\label{tab:anova_results}
\begin{tabular}{llrrrrr}
\toprule
Variables & Source & SS & DF & MS & $F$ Stats & P-value \\
\midrule
List Prices & Factor & $2.9 \times 10^8$ & 2,575 & $1.1 \times 10^5$ & $2.3 \times 10^4$ & \textless{} 0.0001 \\
& Residual & $2.2 \times 10^6$ & 466,276 & 4.8 & & \\
& Total & $2.9 \times 10^8$ & 468,851 & 611.7 & & \\
\midrule
\multicolumn{2}{c}{Observations} & \multicolumn{5}{c}{468,852} \\
\multicolumn{2}{c}{$R^{2}$} & \multicolumn{5}{c}{0.9922} \\
\midrule
Revenue & Factor & $2.9 \times 10^9$ & 2,659 & $1.1 \times 10^6$ & 28.24 & \textless{} 0.0001 \\
& Residual & $1.9 \times 10^{10}$ & 484,552 & $3.9 \times 10^4$ & & \\
& Total & $2.2 \times 10^{10}$ & 487,211 & $4.5 \times 10^4$ & & \\
\midrule
\multicolumn{2}{c}{Observations} & \multicolumn{5}{c}{487,212} \\
\multicolumn{2}{c}{$R^{2}$} & \multicolumn{5}{c}{0.1342} \\
\midrule
Quantity & Factor & $4.5 \times 10^7$ & 2,659 & $1.7 \times 10^4$ & 1.46 & \textless{} 0.0001 \\
& Residual & $5.6 \times 10^9$ & 484,552 & $1.1 \times 10^4$ & & \\
& Total & $5.6 \times 10^9$ & 487,211 & $1.2 \times 10^4$ & & \\
\midrule
\multicolumn{2}{c}{Observations} & \multicolumn{5}{c}{487,212} \\
\multicolumn{2}{c}{$R^{2}$} & \multicolumn{5}{c}{0.0079} \\
\midrule
Sales Prices & Model & $3.6 \times 10^7$ & 2,659 & $1.4 \times 10^4$ & 341.69 & \textless{} 0.0001 \\
& Residual & $1.9 \times 10^7$ & 484,552 & 39.9 & & \\
& Total & $5.6 \times 10^7$ & 487,211 & 114 & & \\
\midrule
\multicolumn{2}{c}{Observations} & \multicolumn{5}{c}{487,212} \\
\multicolumn{2}{c}{$R^{2}$} & \multicolumn{5}{c}{0.6522} \\
\bottomrule
\end{tabular}
\tabnotes{This table presents one-way ANOVA results of list price, revenue, quantity and sales price variables. SS stands for sum of square and DF stands for the degree of freedom. MS means model sum (=SS/DF). $F$ Stats and P-value columns contain $F$ statistics and its p-values, respectively. Due to computational constraints, we conduct the ANOVA on a randomly sampled 10\% of observations.}
\end{table}

\begin{table}[p!]
\centering
\caption{Descriptive Statistics of Number of Spatial Competitors}
\begin{adjustbox}{scale = 0.9}
\begin{tabular}{l|rrrrrrrrr}
\toprule
Variables & $R^{0,0.1}$ & $R^{0.1,0.2}$ & $R^{0.2,0.3}$ & $R^{0.3,0.4}$ & $R^{0.4,0.5}$ & $R^{0.5,0.6}$ & $R^{0.6,0.7}$ & $R^{0.7,0.8}$ & $R^{0.8,0.9}$ \\ \midrule
Mean & 700.1 & 5,423.1 & 7,846.9 & 6,930.2 & 3,862.7 & 2,346.5 & 1,774.6 & 1,349.3 & 47.0 \\
S.D. & 545.6 & 2,678.2 & 3,459.4 & 2,671.1 & 2,254.2 & 2,031.5 & 2,178.3 & 2,323.2 & 1,169.9 \\
Min & 0 & 0 & 20 & 1,176 & 1,109 & 87 & 0 & 0 & 0 \\
Max & 2,852 & 13,561 & 21,187 & 22,756 & 20,010 & 12,534 & 12,179 & 12,280 & 8,759 \\
\bottomrule
\end{tabular}
\end{adjustbox}
\tabnotes{This table displays the descriptive statistics for the number of spatial competitors in the visual characteristics space. The number of observations is 4,462,308.}
\label{tab:B_stats}
\end{table}

\begin{figure}[p!]
    \centering
    \caption{Scatter Plots using PCA and UMAP}
    \subfloat[Principal Component Analysis]{\includegraphics[width=0.8\linewidth]{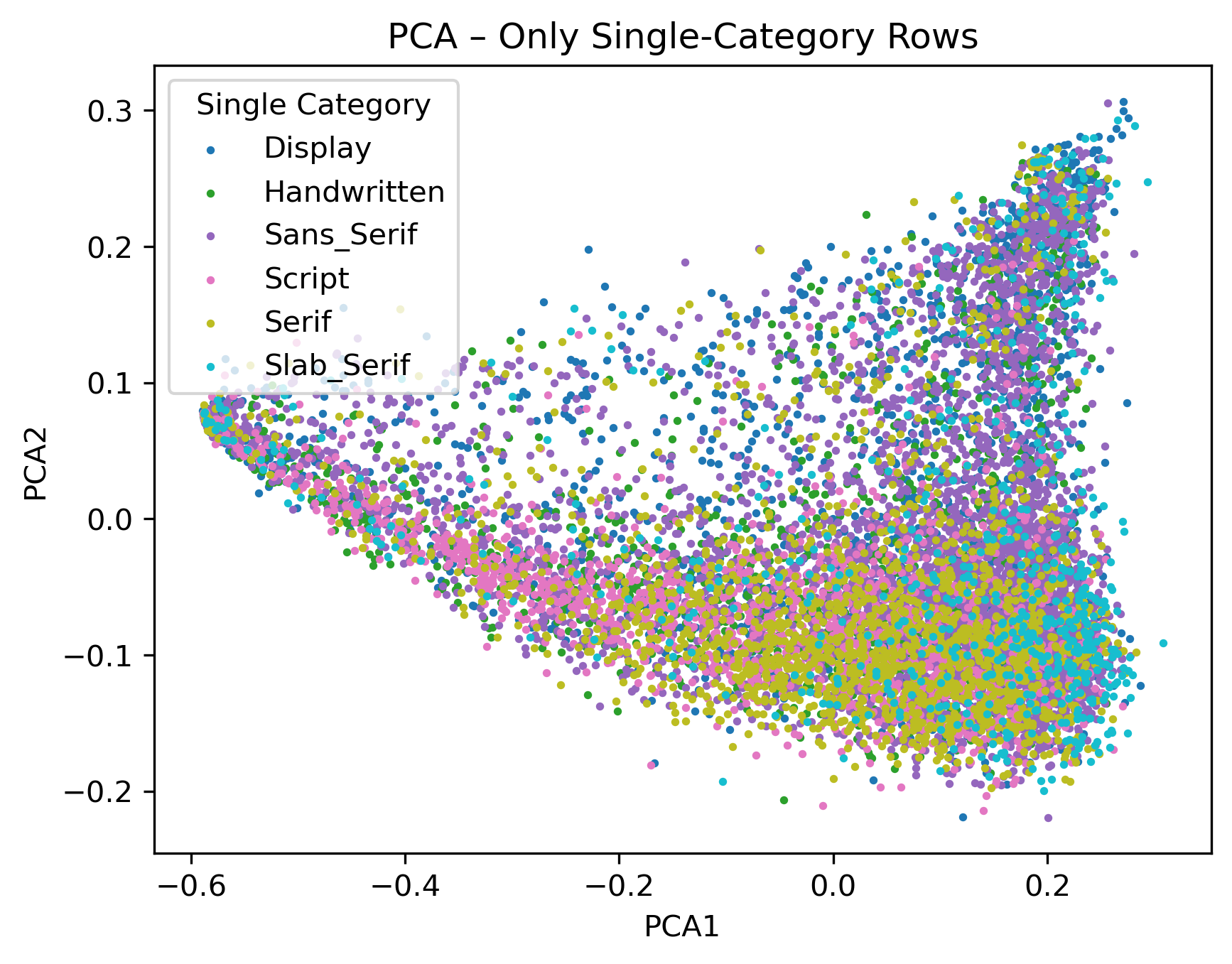}} \\
    \subfloat[Uniform Manifold Approximation and Projection]{\includegraphics[width=0.8\linewidth]{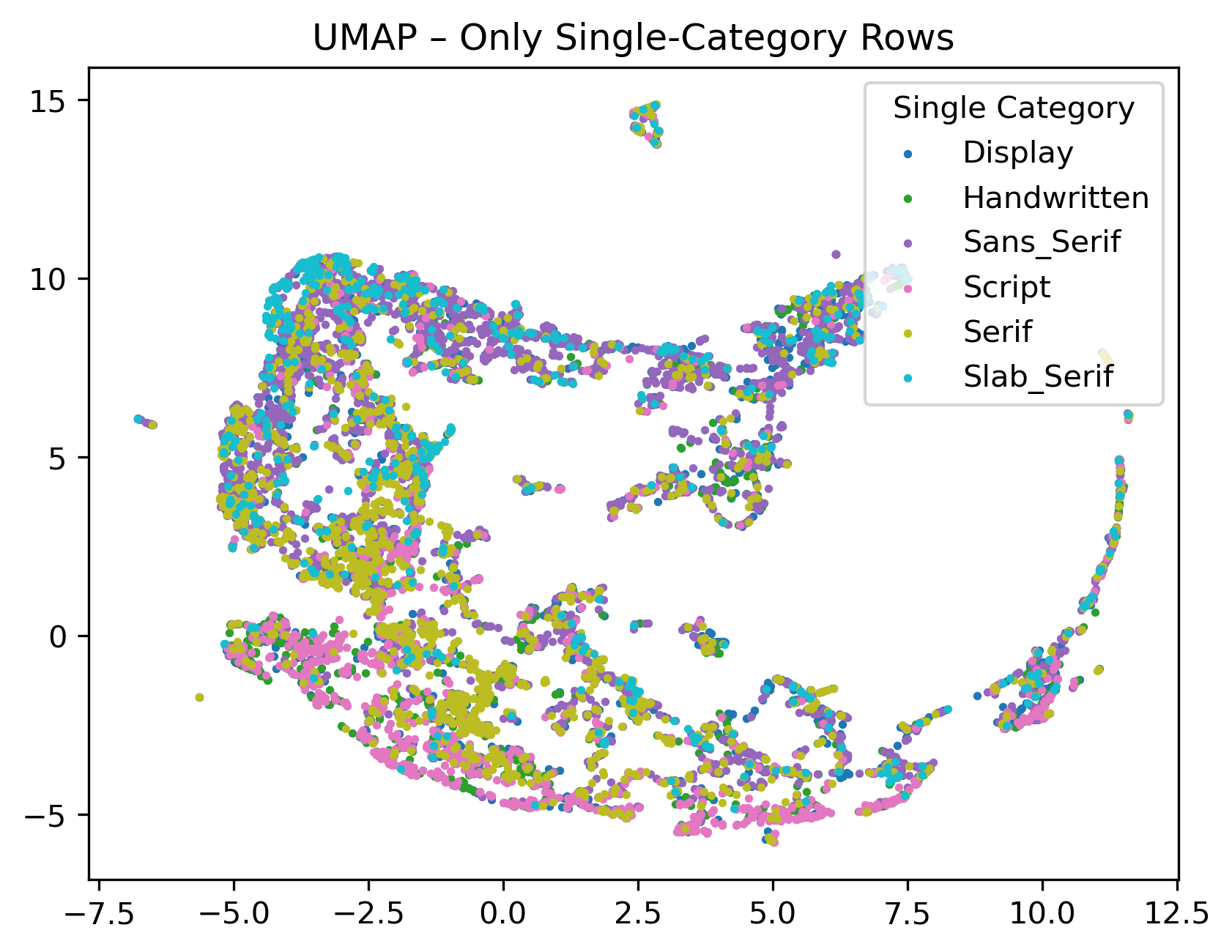}}
    \tabnotes{The colors of the dots represent different industry categories, inferred from product tags. These categories are well separated by PCA and UMAP, with UMAP achieving this more effectively. This is because UMAP has a stronger ability to cluster and preserve global structure \citep{mcinnes2018umap}. However, for structural estimation, we use PCA for dimensionality reduction, as its linearity makes it more interpretable than UMAP. In addition, UMAP is primarily a visualization tool and only produces two-dimensional vectors.}
    \label{fig:scatter-PCA-UMAP}
\end{figure}

\begin{figure}[p!]
    \centering
    \caption{Spatial Regression Results ($\gamma_{r}$) of Prices}\includegraphics[width=0.75\textwidth]{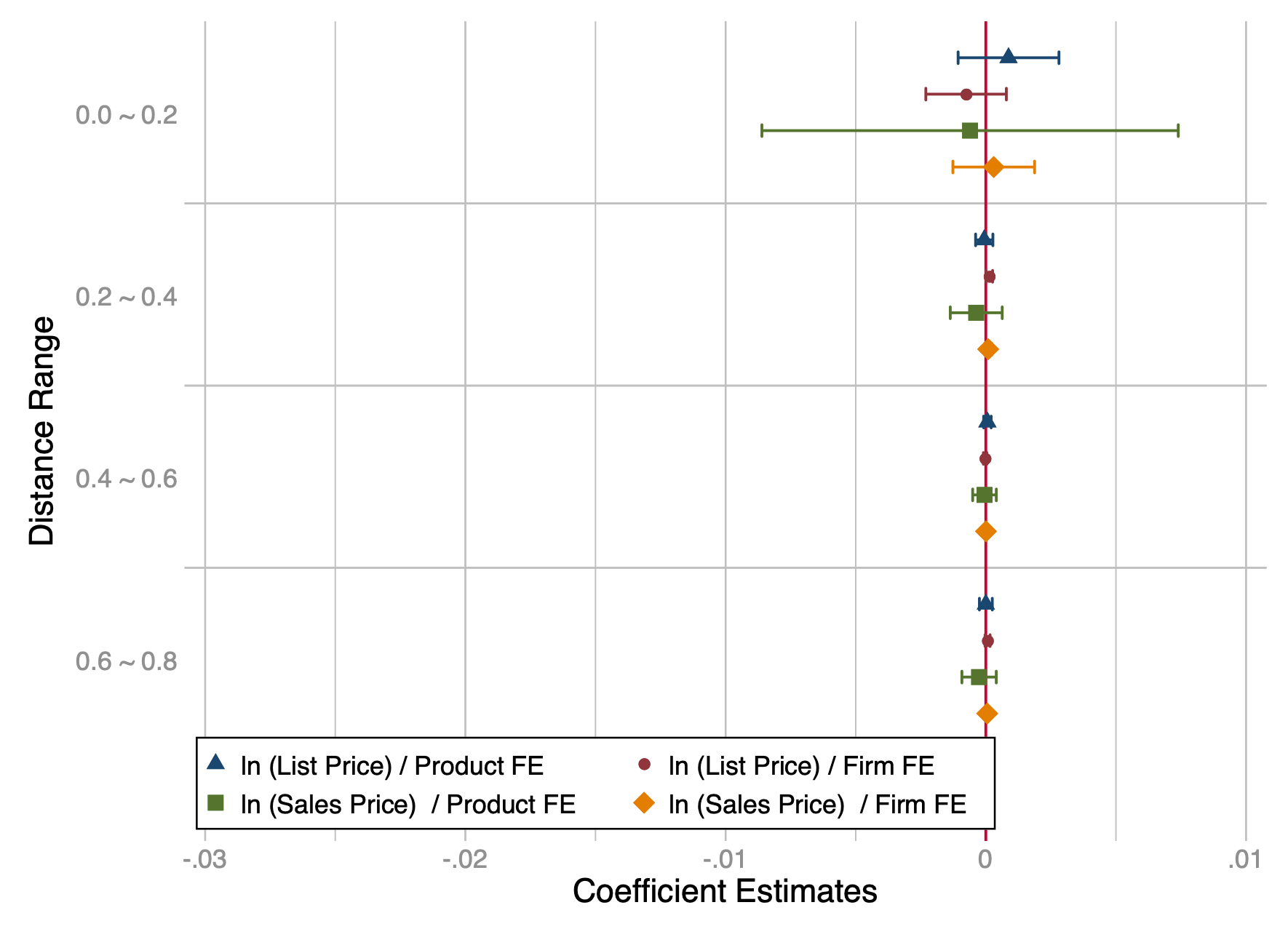}
    \label{fig:prices-TWFE}
    \tabnotes{This figure present coefficient estimates of regression equation \eqref{eq:spatial-reg} for different price measures and fixed effect specifications. The triangle and circle dots indicate results from the log of list prices with product and firm fixed effects, respectively. Similarly, the square and diamond dots indicate results from sales prices. The solid lines show 95\% confidence intervals. Standard errors are clustered at the product level.}
\end{figure}

\begin{figure}[p]
    \centering 
    \caption{Scree Plots}
    \subfloat[Variance Ratio]{\includegraphics[width=0.55\linewidth]{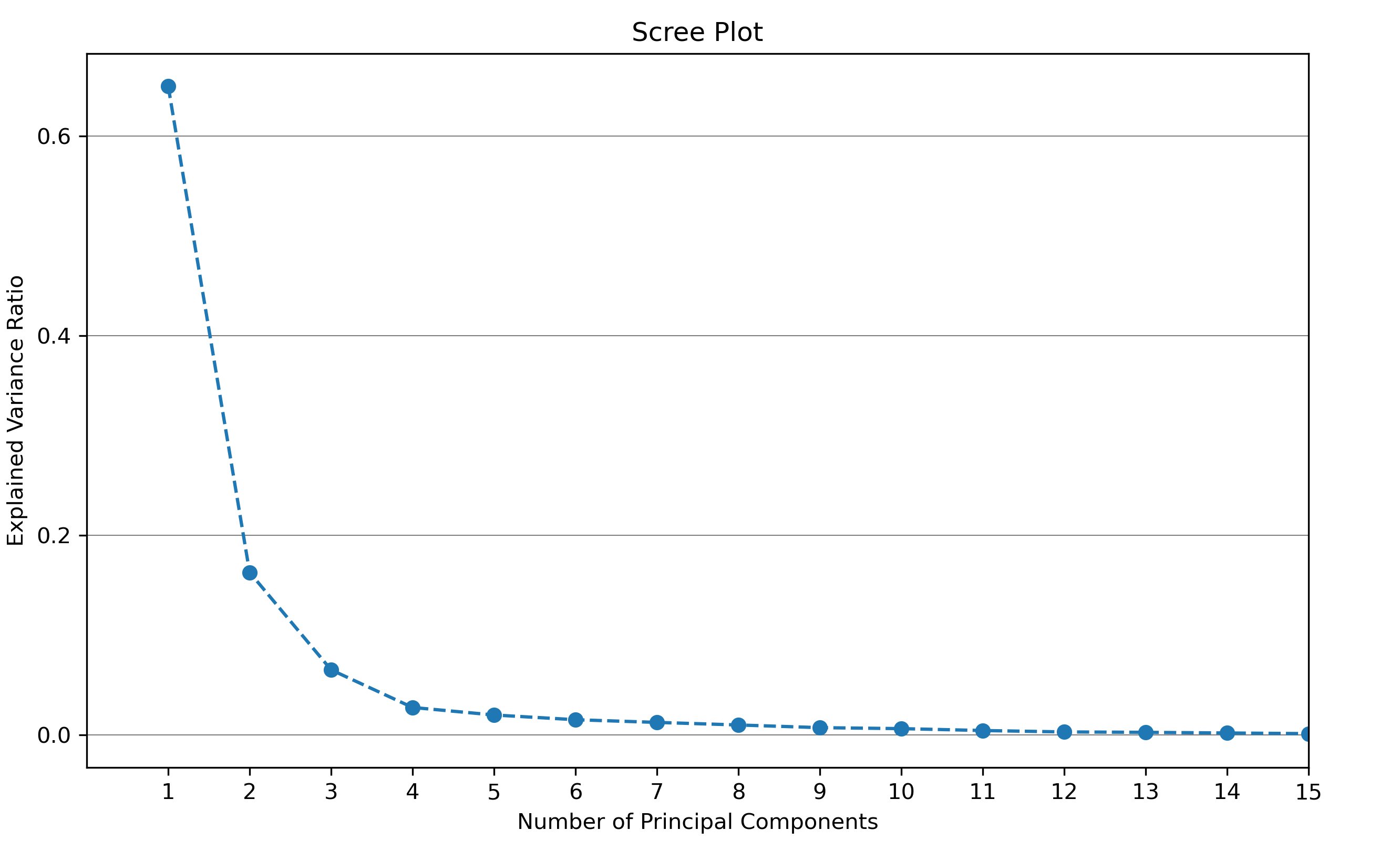}}
    \subfloat[Cumulative Variance Ratio]{\includegraphics[width = 0.55\textwidth]{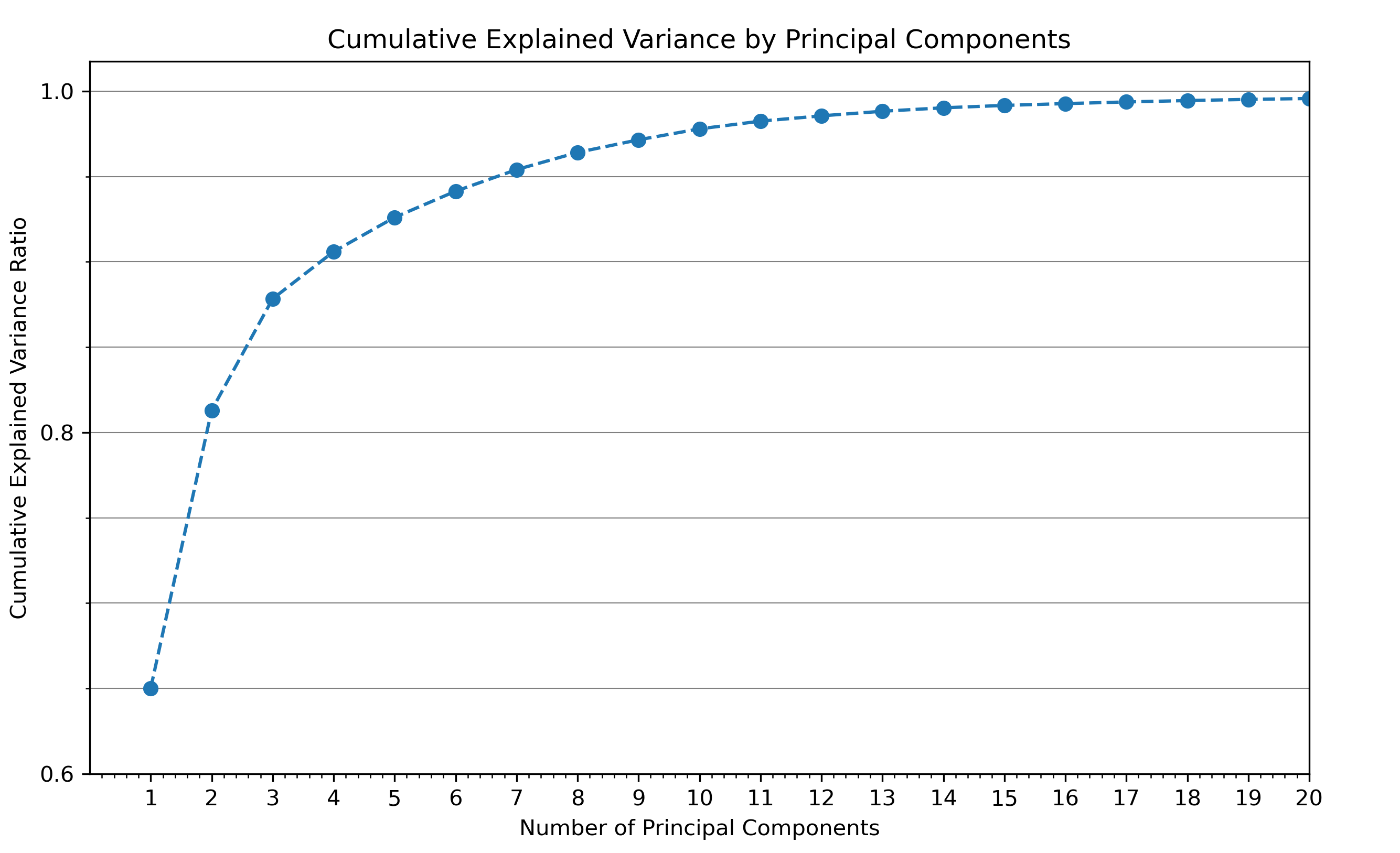}}
    \label{fig:scree-plot}
    \tabnotes{These figures show percentage of variance explained by each principal component. Panel (a) shows the explained variance ratio of each principal component. Panel (b) presents the cumulative explained variance ratio along the number of principal components.}
\end{figure}

\begin{figure}[htbp!]
    \centering
    \caption{The Distribution of Median Own Price Elasticity}
    \includegraphics[width=0.5\textwidth]{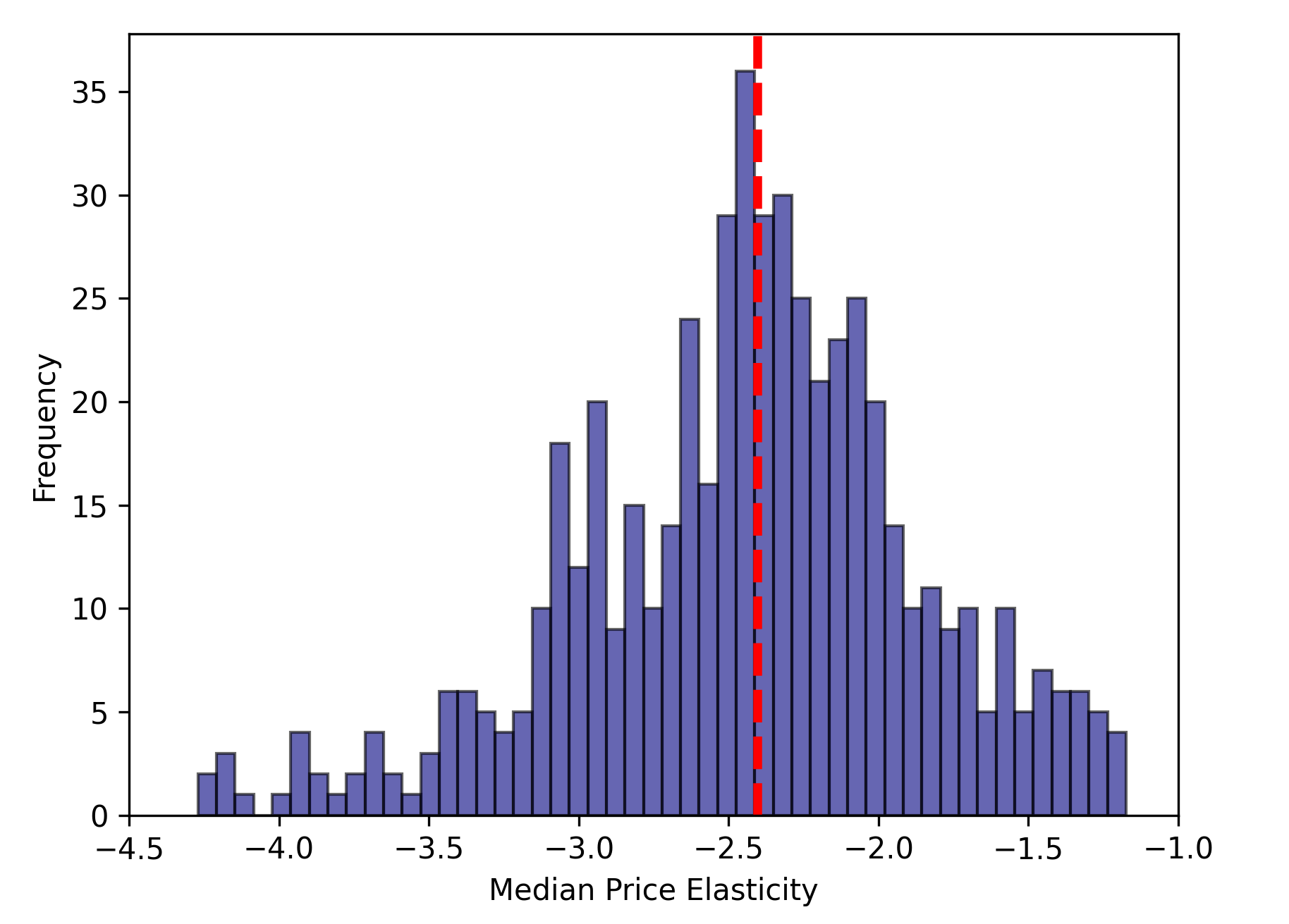}
    \tabnotes{This figure shows median own-price elasticities across markets. A market is a country-month combination. The red dashed vertical line indicates the median value of the entire market.}
    \label{fig:own-price-elas}
\end{figure}
\begin{figure}[htbp!]
    \centering
    \caption{Distribution of Own Shape Elasticities (Embeddings 3 to 6)}
    \subfloat[Embedding 3]{\includegraphics[width=0.5\linewidth]{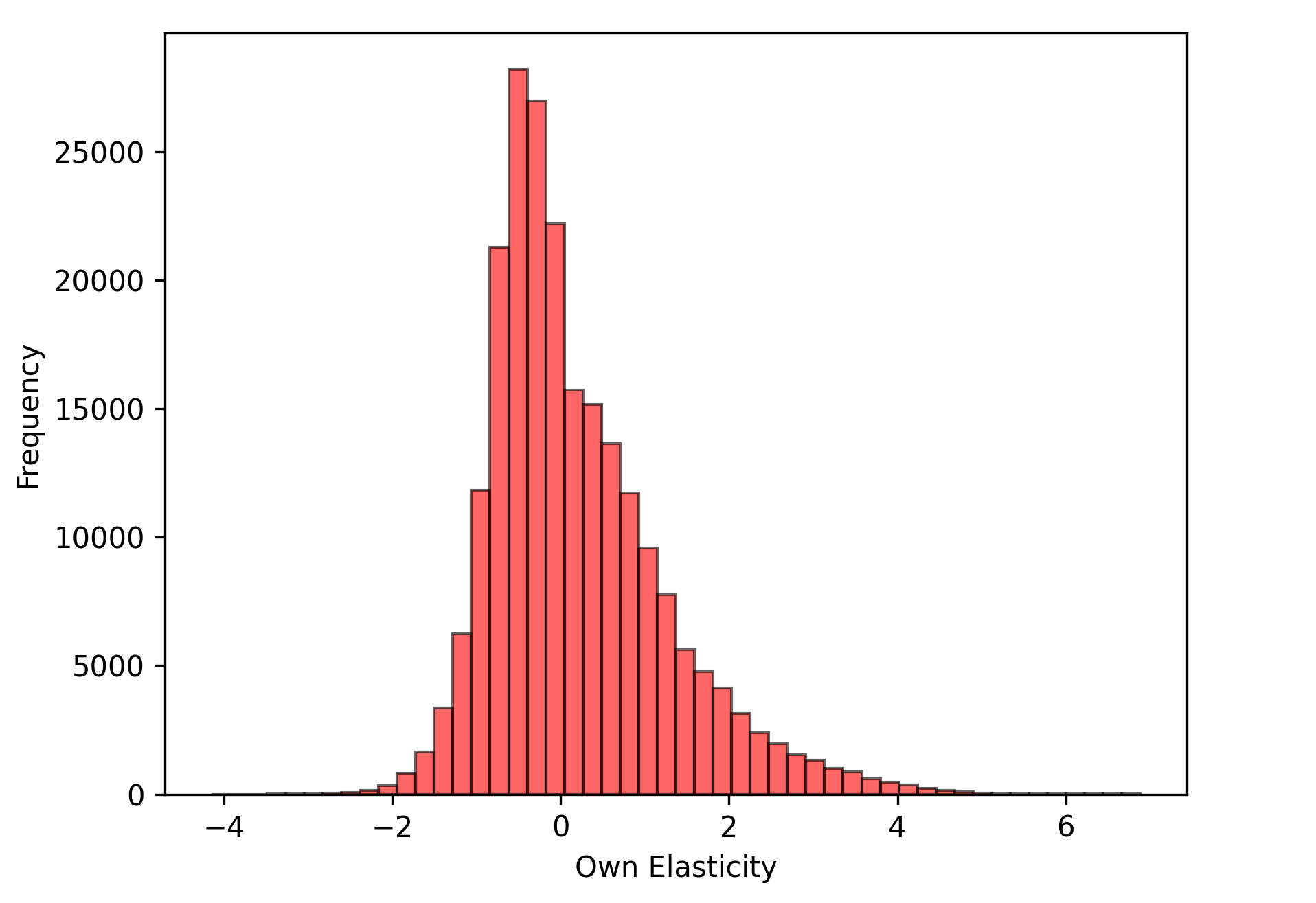}}
    \subfloat[Embedding 4]{\includegraphics[width=0.5\linewidth]{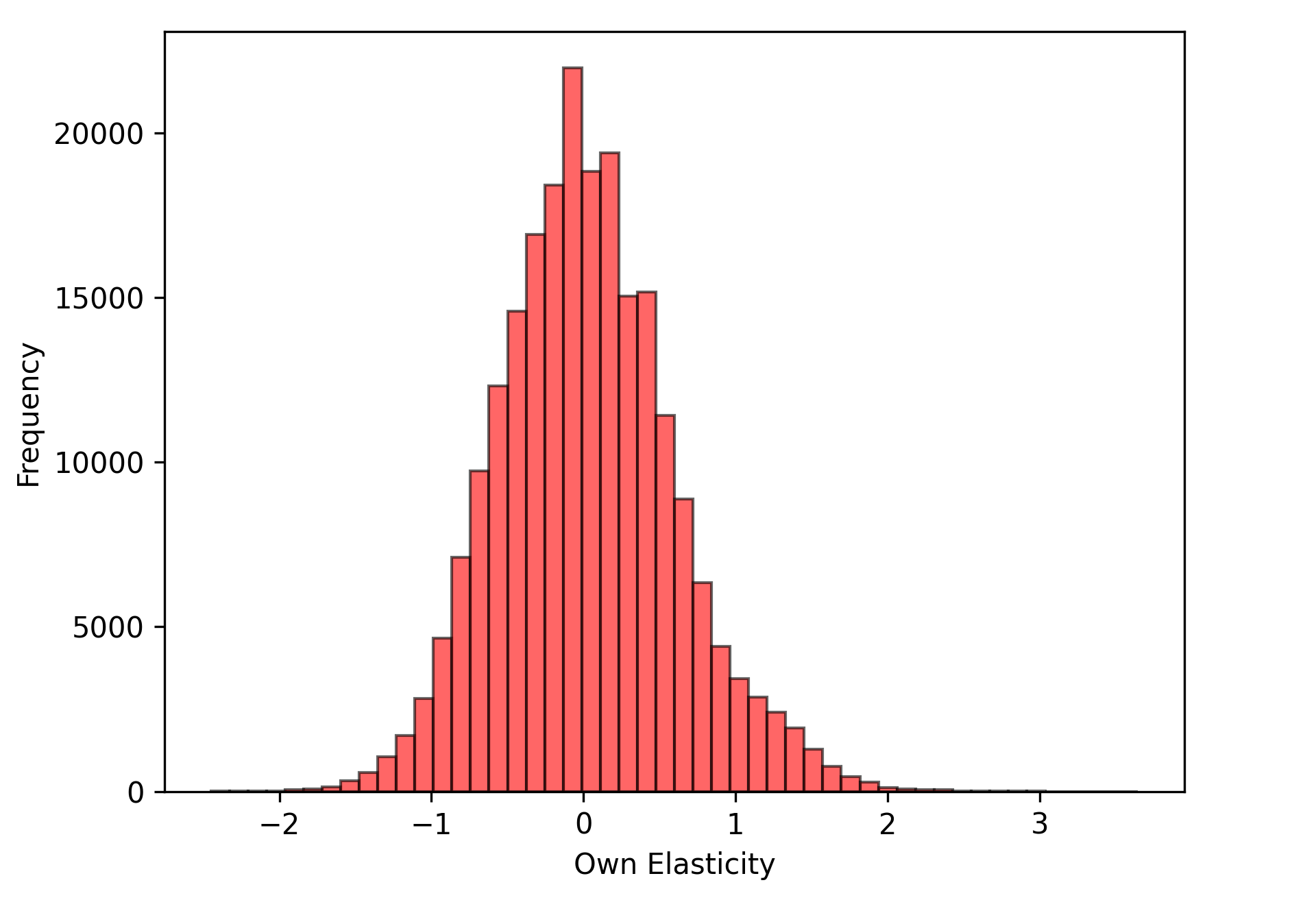}} \\
    \subfloat[Embedding 5]{\includegraphics[width=0.5\linewidth]{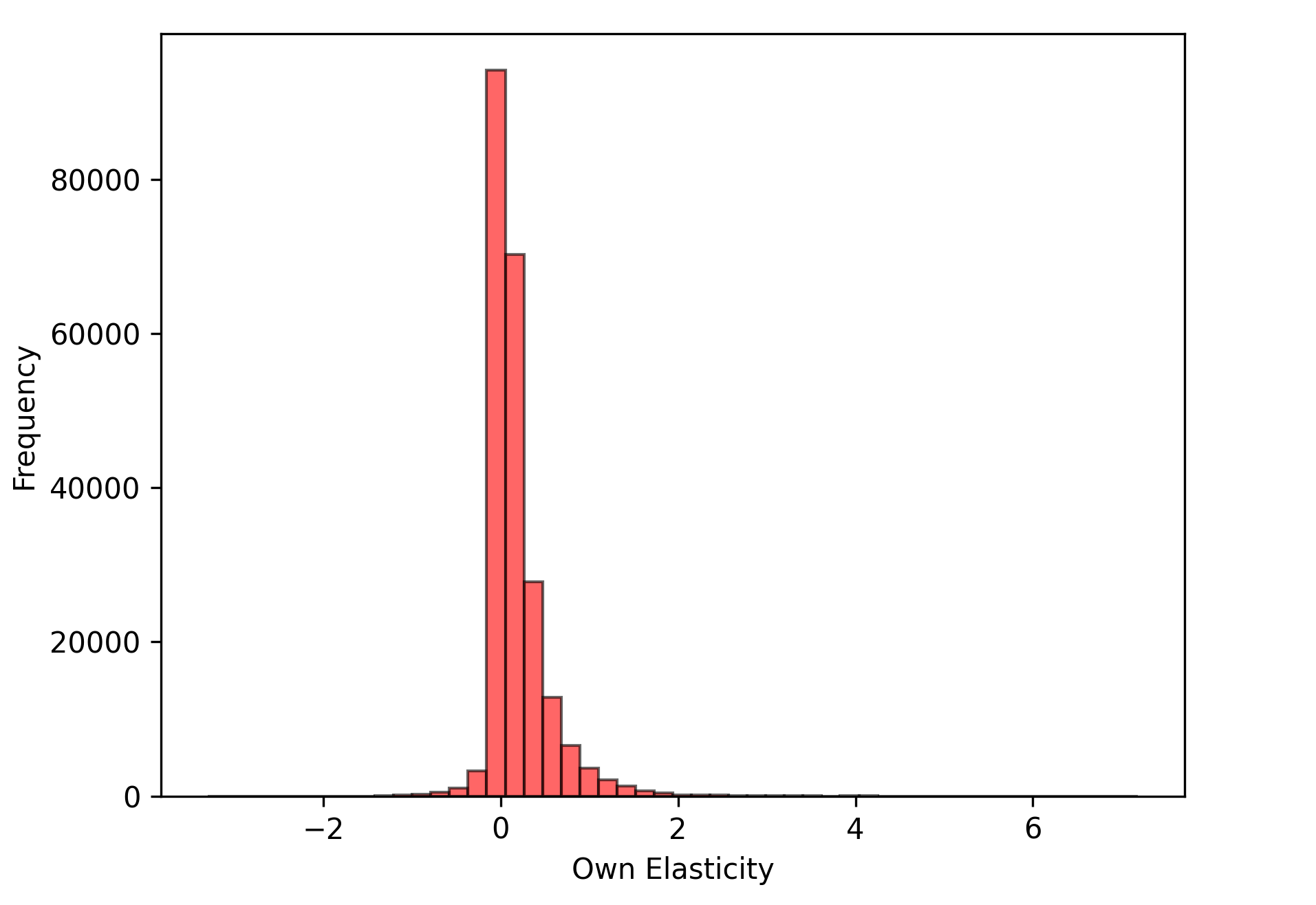}}
    \subfloat[Embedding 6]{\includegraphics[width=0.5\linewidth]{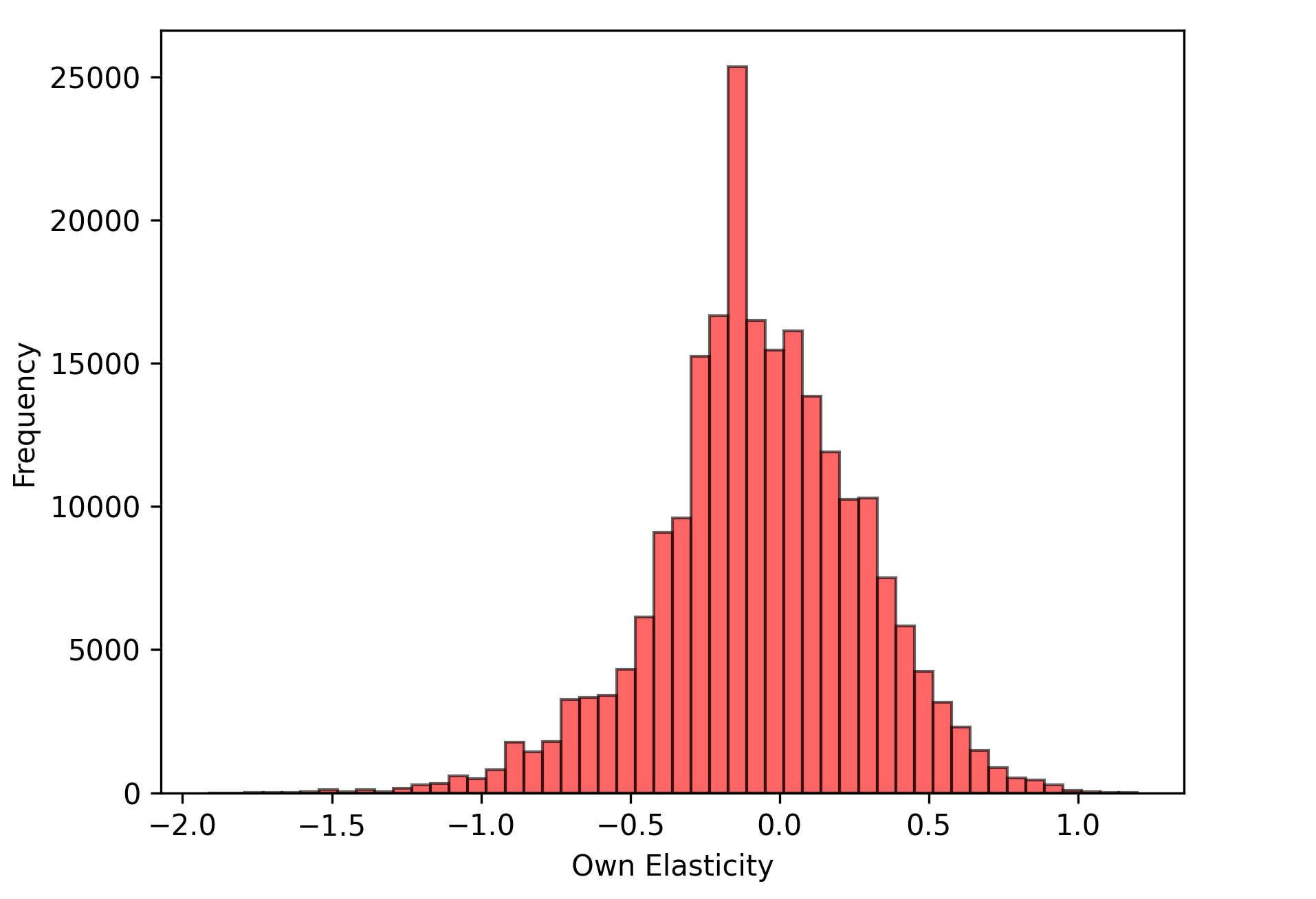}}    
    \tabnotes{This figure shows the distributions of own elasticity with respect to embedding elements. The distribution is plotted across products and markets. Panels (a) and (d) correspond to the distributions with embeddings 3 to 6, respectively. The distributions correspond to embeddings 1 and 2 are shown in Figure \ref{fig:own-shape-elas}.}
    \label{fig:own-shape-elas-pc3-6}
\end{figure}

\begin{figure}[p!]
    \centering
    \caption{Measures of Competition and Embedding Distances (Without Using Visual Characteristics)}
    \subfloat[Prices Diversion Ratios]{\includegraphics[width=0.5\textwidth]{image/MeasureCompetition/Mkt_Avg_Div_nl_240812.png}} 
    \subfloat[Long Run Diversion Ratios]{\includegraphics[width=0.5\textwidth]{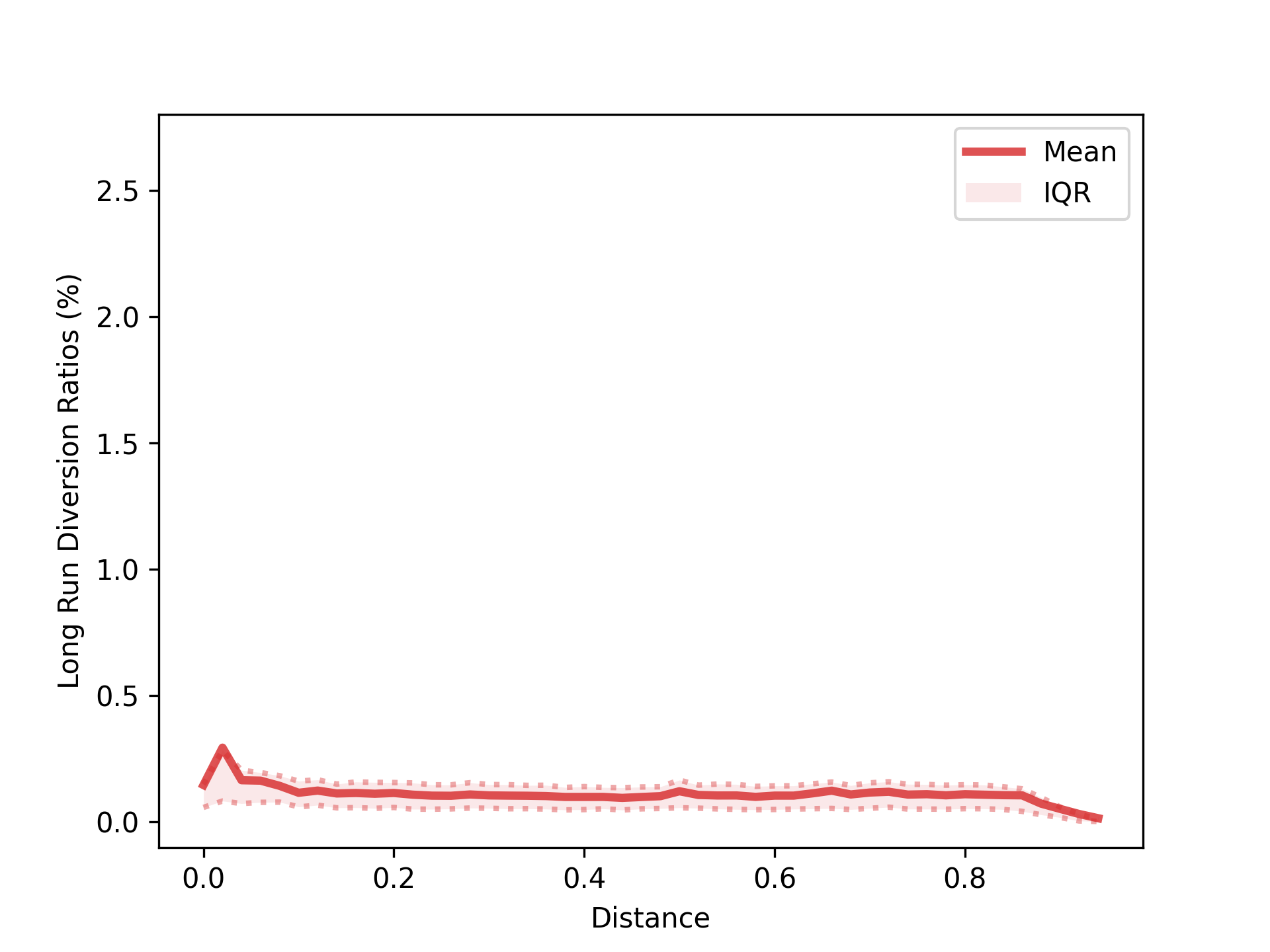}} \\
    \subfloat[Cross Price Elasticity]{\includegraphics[width=0.5\textwidth]{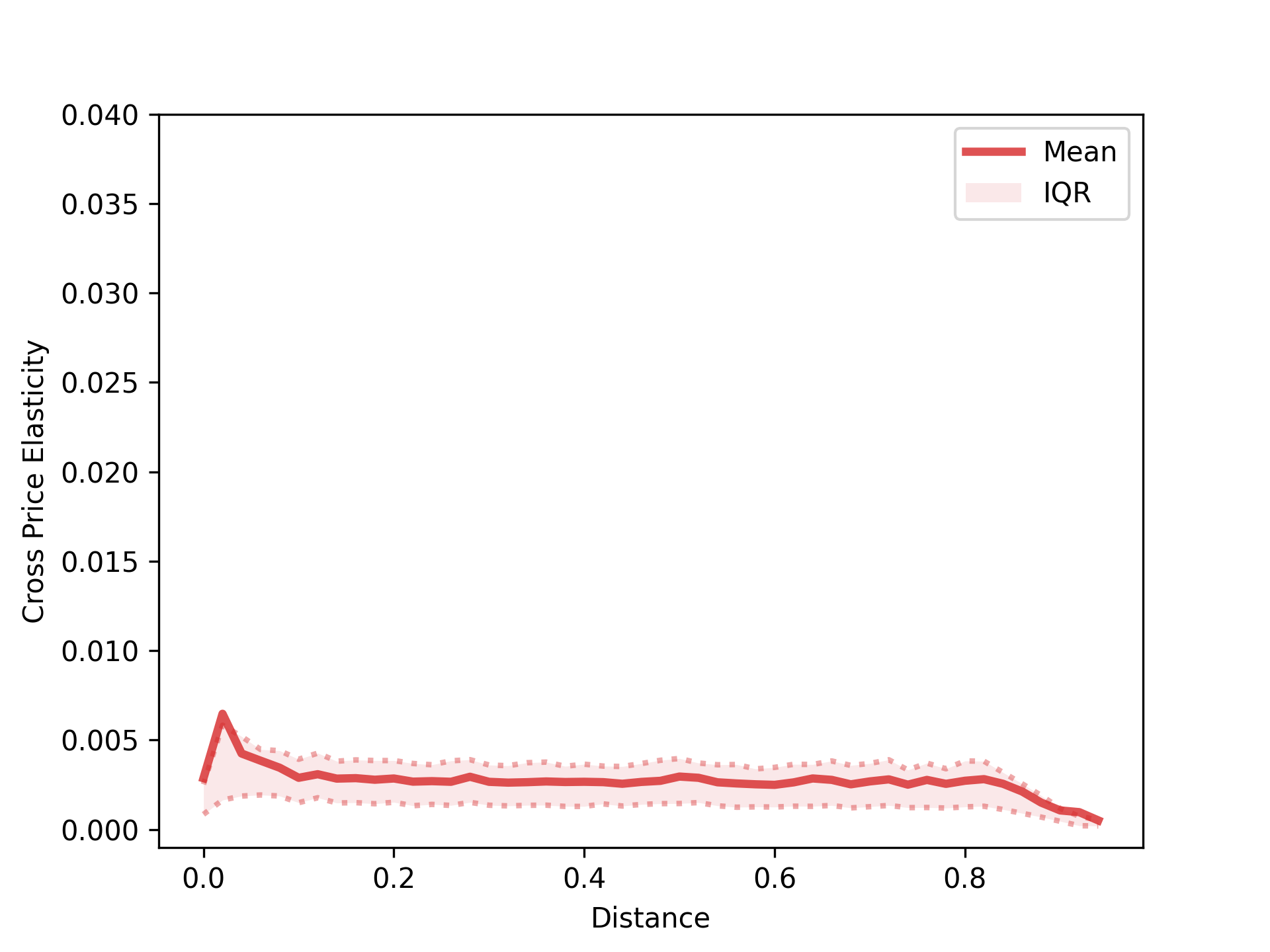}} 
    \subfloat[Average Number of Products]{\includegraphics[width=0.5\textwidth]{image/MeasureCompetition/Avg_Product_Num_240812.png}} 
    \tabnotes{Panels (a) to (c) plot the average price diversion ratios, long-run diversion ratios, and cross-price elasticity along the embedding distance $d$ (see equation \eqref{eq:DIV}), calculated by using demand estimates without random coefficients on the embeddings, respectively. As a reference, Panel (d) shows the average number of products along radial areas.} 
    \label{fig:comp-dist-nl}
\end{figure}

\begin{figure}[htbp!]
    \centering
    \caption{Entry Points of Counterfactual Simulations (Scenario A)}
    \subfloat[Protection Distance $\underbar{d} = 0$]{\includegraphics[width=0.5\linewidth]{image/Counterfactual_Entrypoints/cut_outliers_entry_pattern_d_0.00.png}}
    \subfloat[Protection Distance $\underbar{d} = 0.02$] {\includegraphics[width=0.5\linewidth]{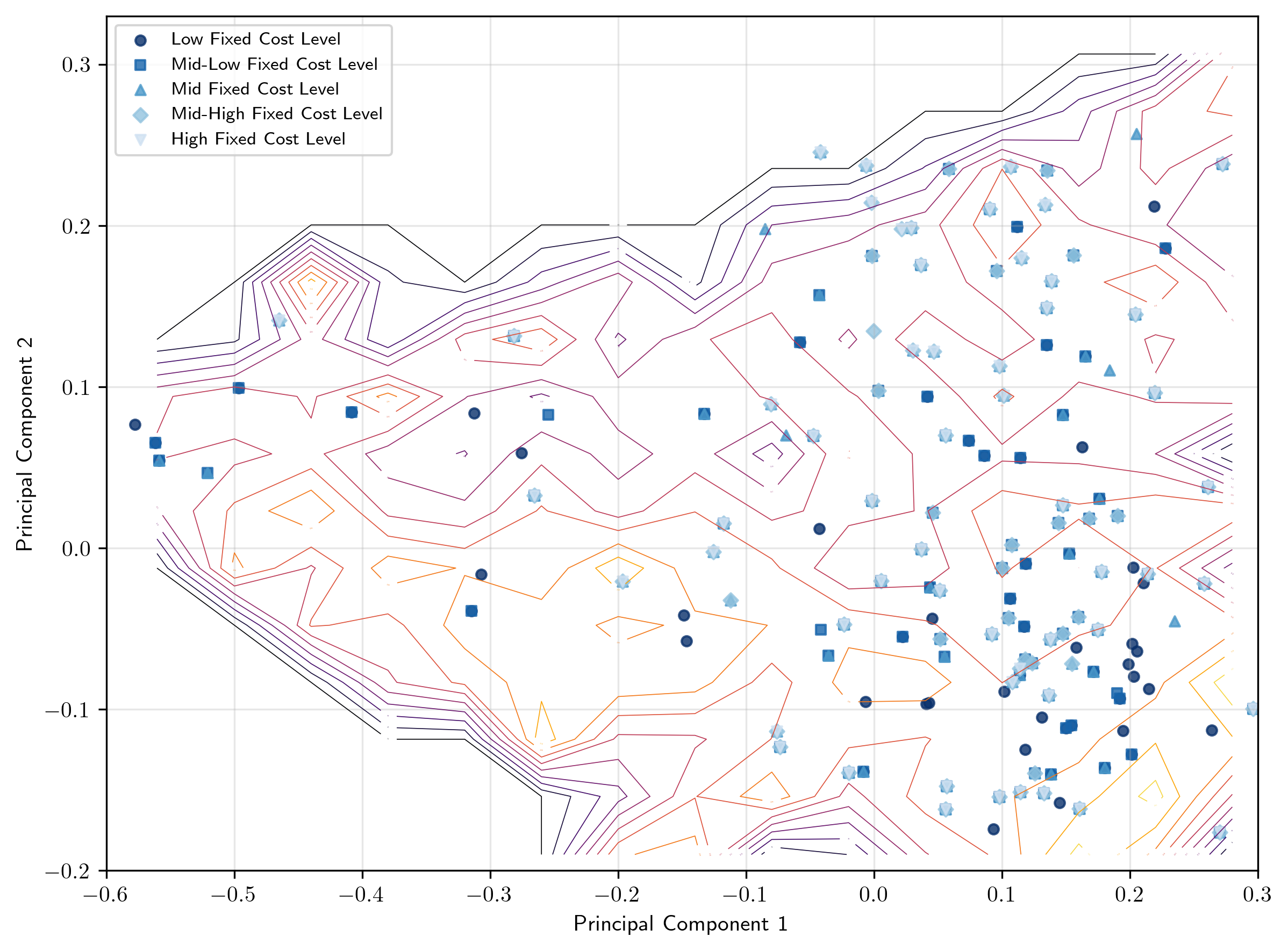}} \\
    \subfloat[Protection Distance $\underbar{d} = 0.04$]{\includegraphics[width=0.5\linewidth]{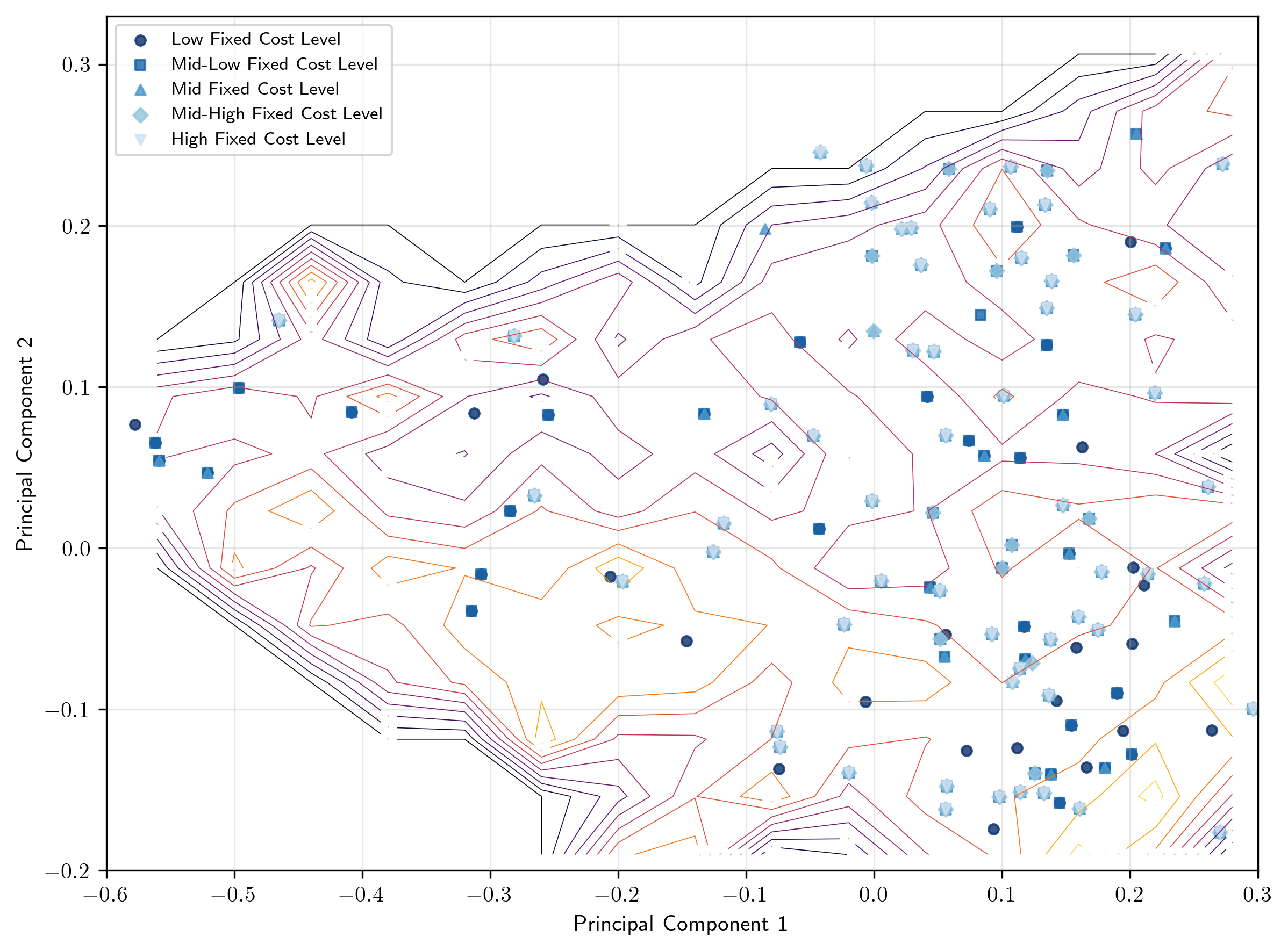}}
    \subfloat[Protection Distance $\underbar{d} = 0.06$] {\includegraphics[width=0.5\linewidth]{image/Counterfactual_Entrypoints/cut_outliers_entry_pattern_d_0.06.png}} \\
    \subfloat[Protection Distance $\underbar{d} = 0.08$]{\includegraphics[width=0.5\linewidth]{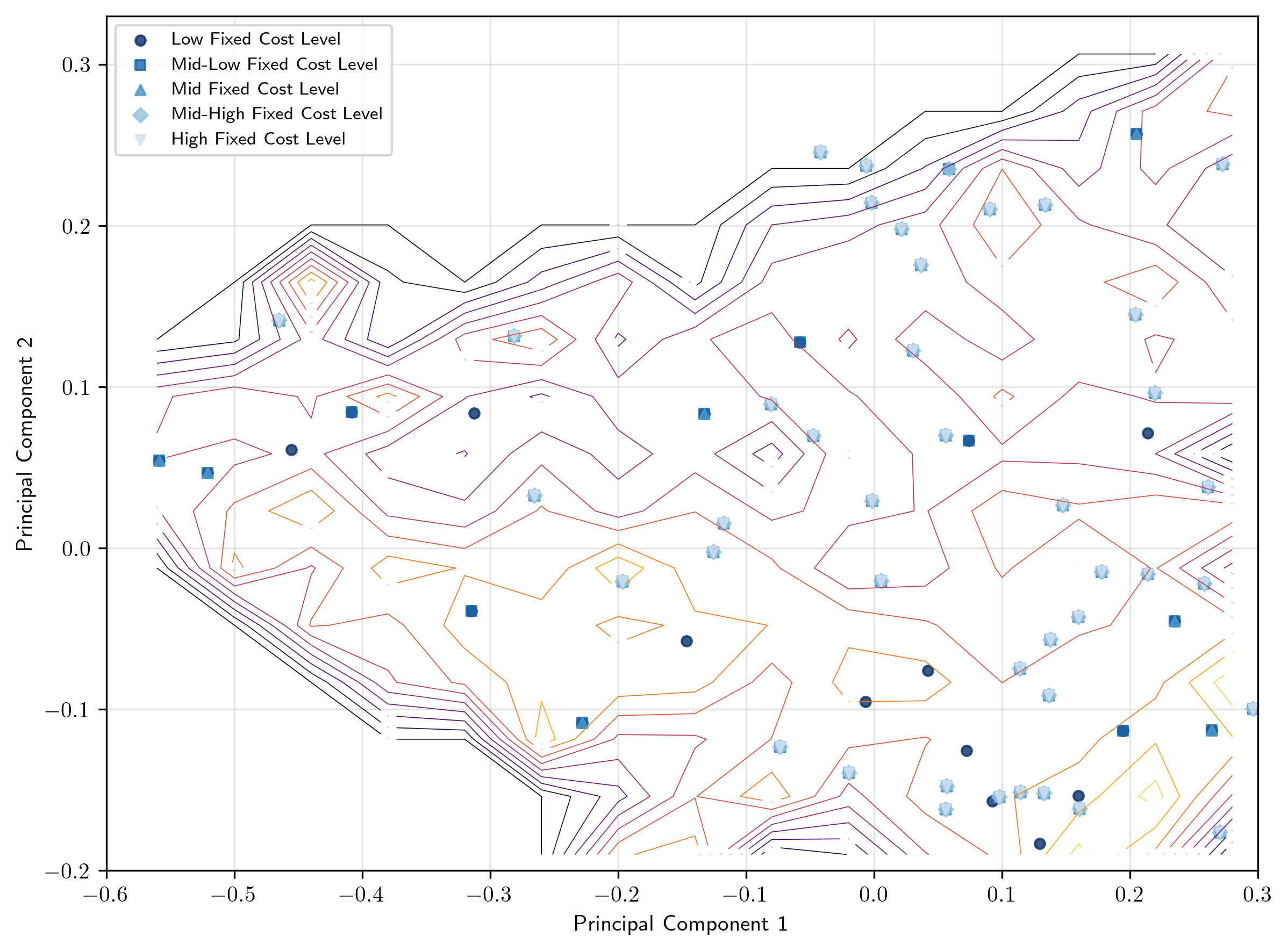}}
    \subfloat[Protection Distance $\underbar{d} = 0.10$]{\includegraphics[width=0.5\linewidth]{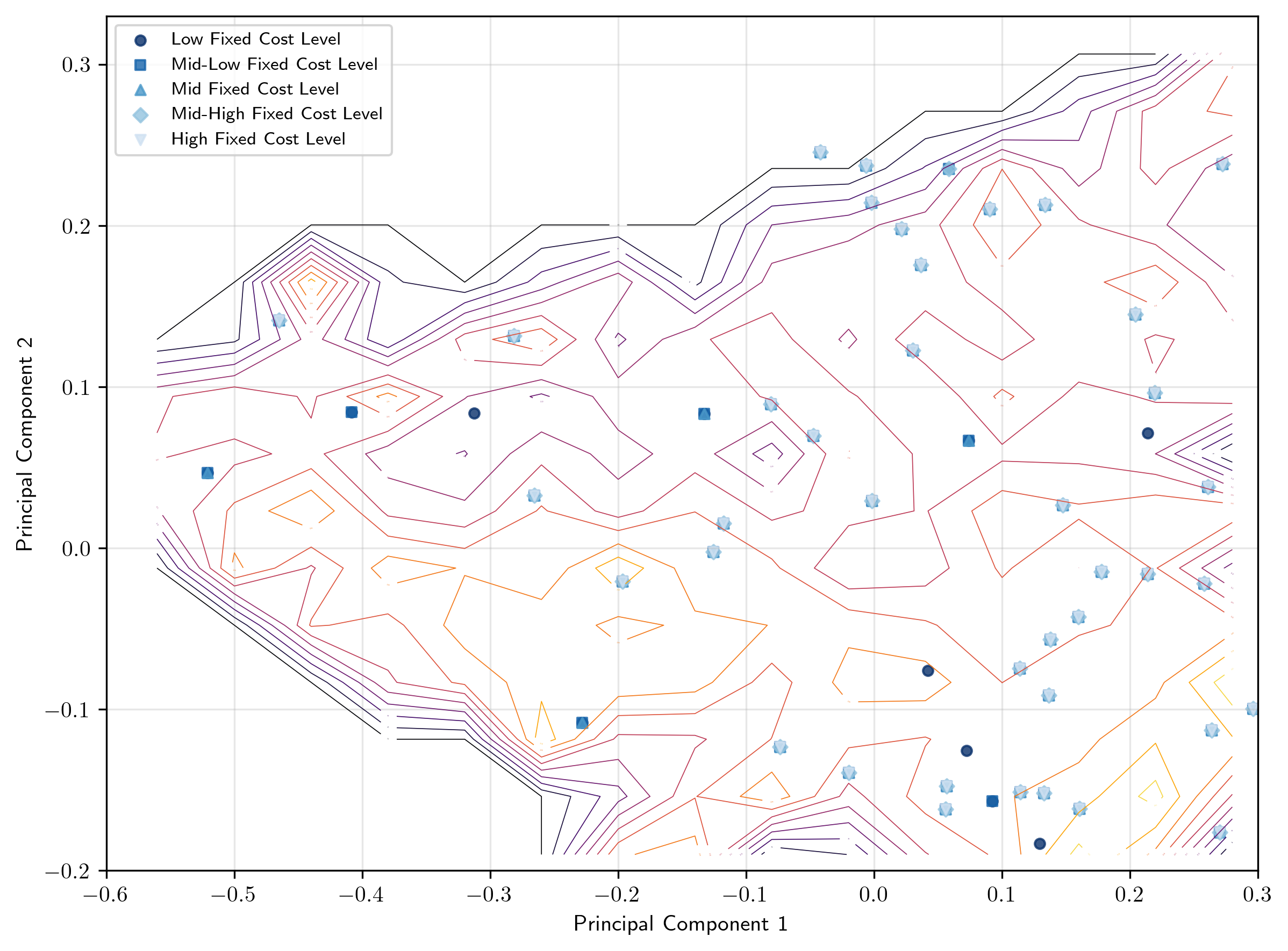}}
    \tabnotes{This figure presents a line contour map of median revenues and the entry points from the counterfactual simulations under scenario A. Panels (a) through (f) display the entry locations chosen by the firms under protection levels $\underline{d} = 0$ to $0.1$, respectively. See Figure \ref{fig:dist-rev-pca} for an alternative depiction of the revenue distribution.}
    \label{fig:more-entry-points-A}
\end{figure}

\begin{figure}
    \centering
    \caption{Spatial Distributions of Revenues}
    \includegraphics[width=0.8\linewidth]{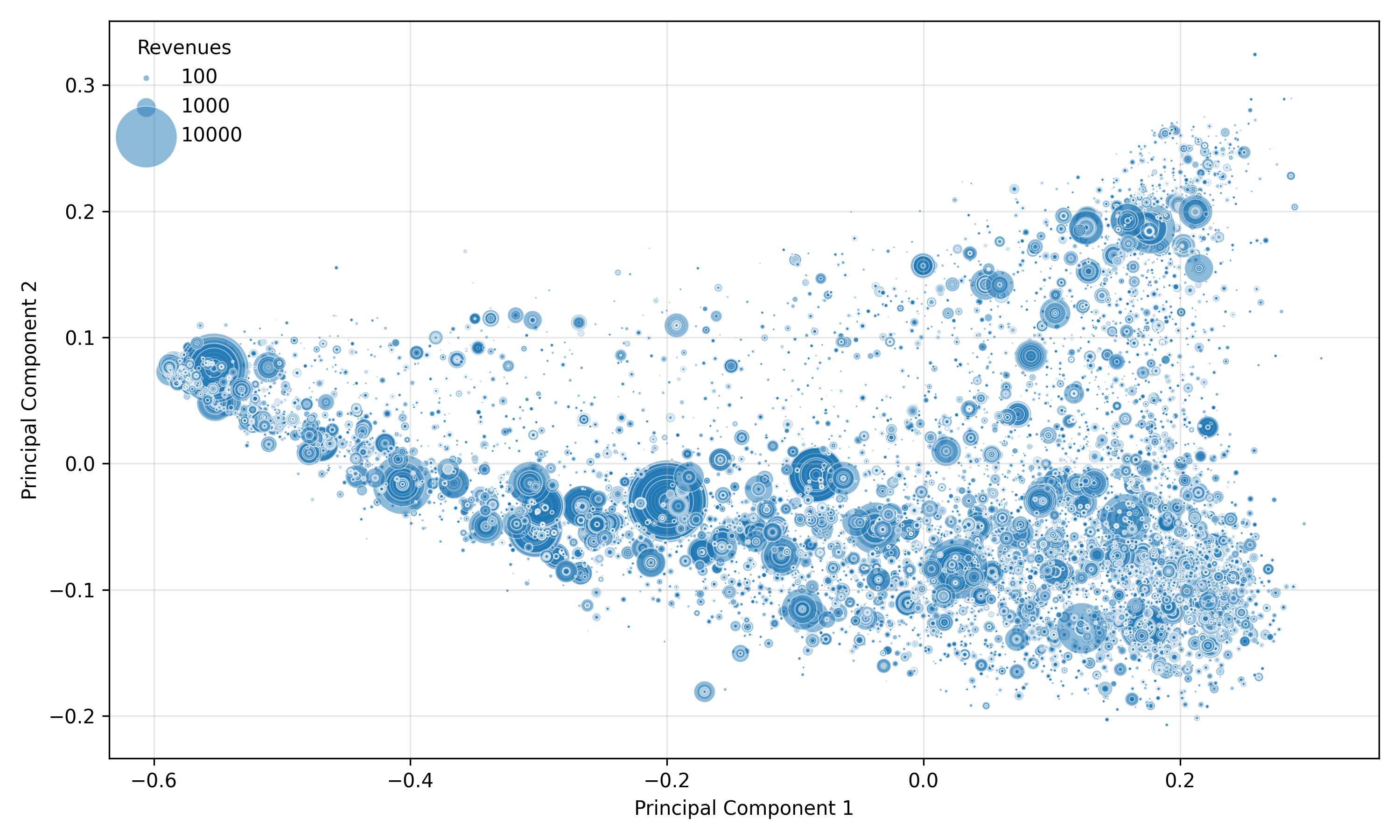} 
    \tabnotes{This figure shows the spatial distribution of revenues in the principal component space.}
    \label{fig:dist-rev-pca}
\end{figure}

\begin{figure}[htbp!]
    \centering
    \caption{Additional Simulation Results by Varying Fixed Costs and Protection Levels (Scenario A)}

    \subfloat[Number of Entrants]{\includegraphics[width=0.5\linewidth]{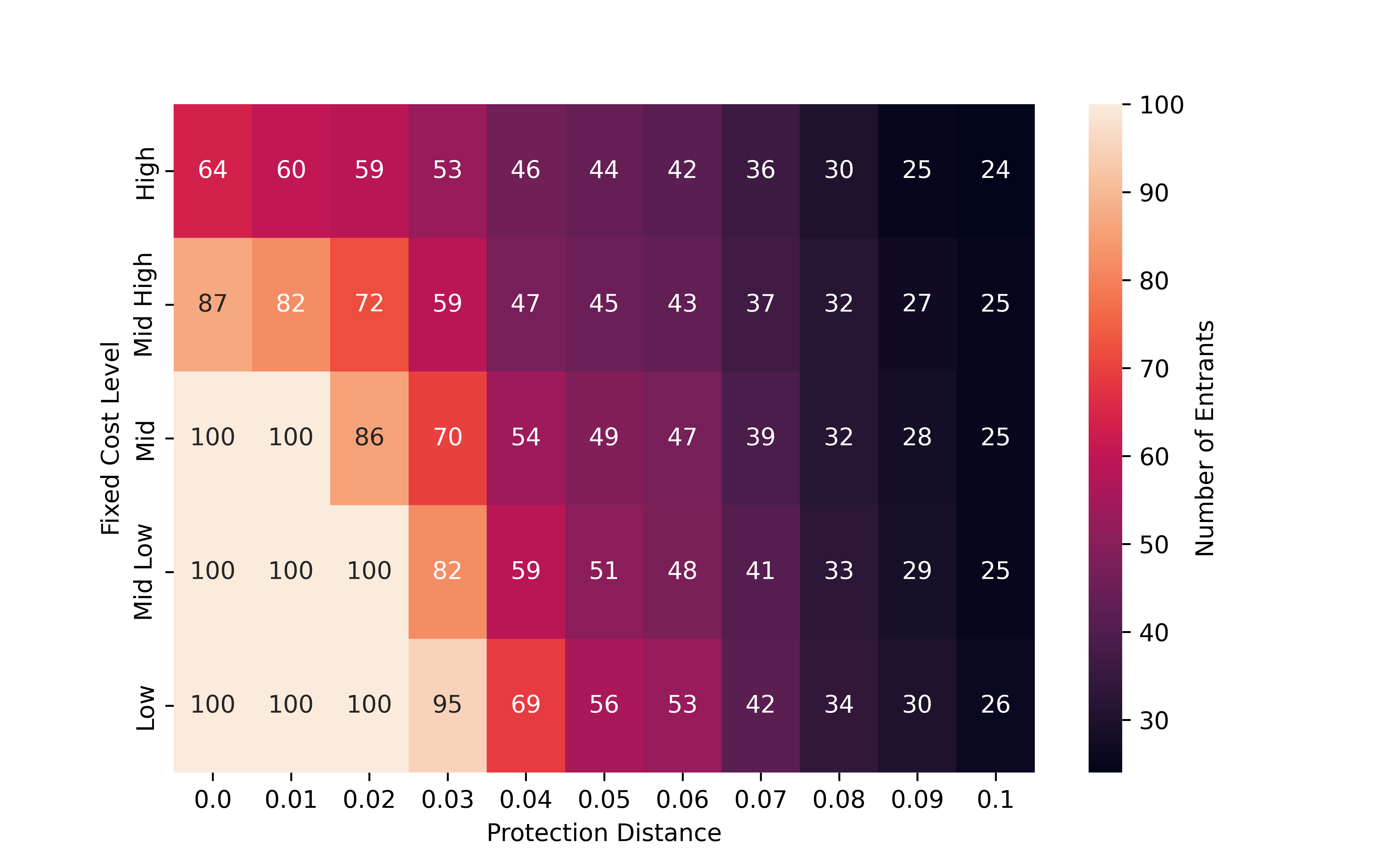}}
    \subfloat[Fixed Costs]{\includegraphics[width=0.5\linewidth]{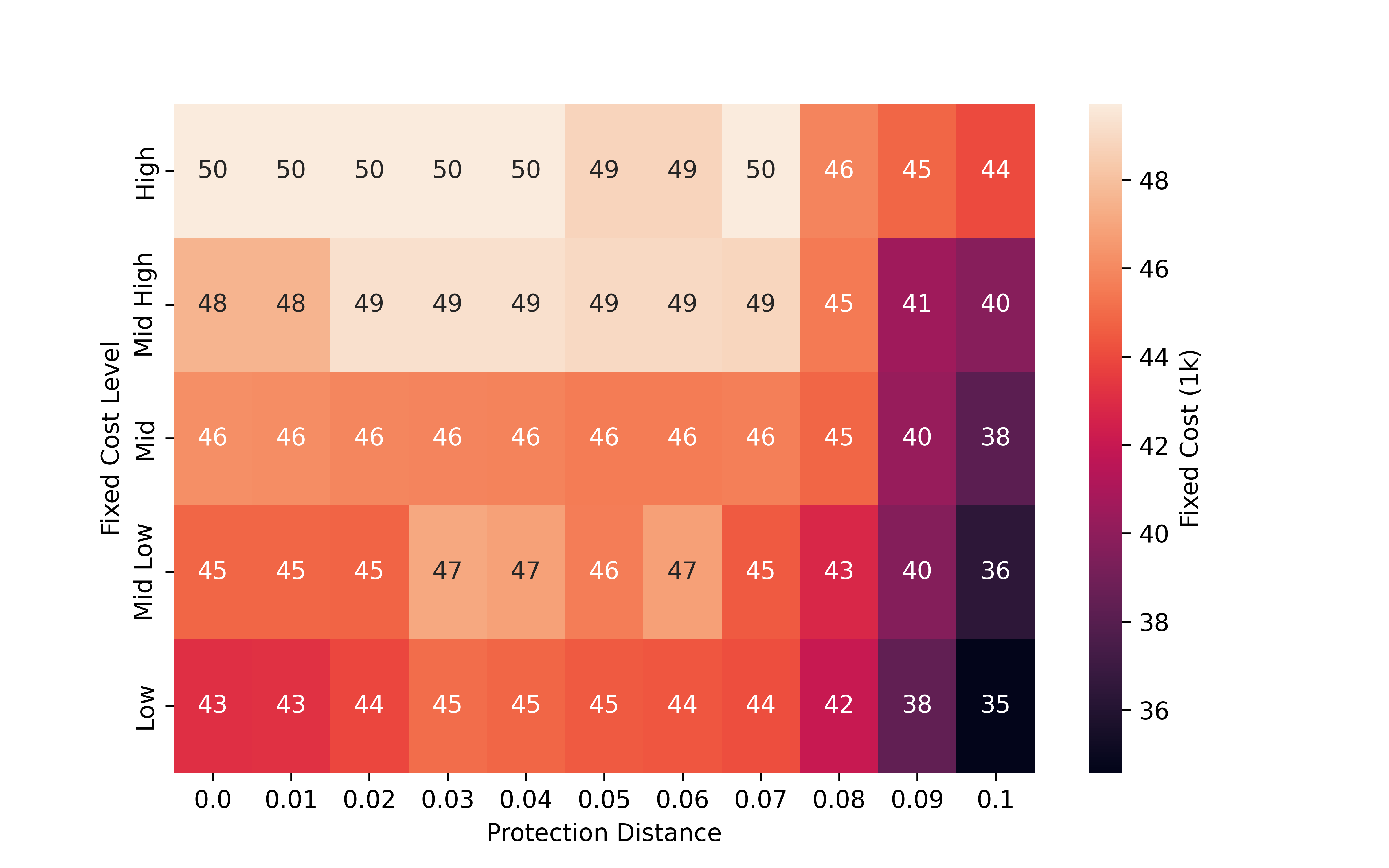}}
    \tabnotes{This figure shows the counterfactual number of entrants and total fixed cost expenditures.}
    \label{fig:simul-protection-fclevel-scenarioA-add}
\end{figure}

\begin{figure}[htbp!]
    \centering
    \caption{Additional Simulation Results by Varying Fixed Costs and Protection Levels (Scenario B)}
    \subfloat[Number of Entrants]{\includegraphics[width=0.5\linewidth]{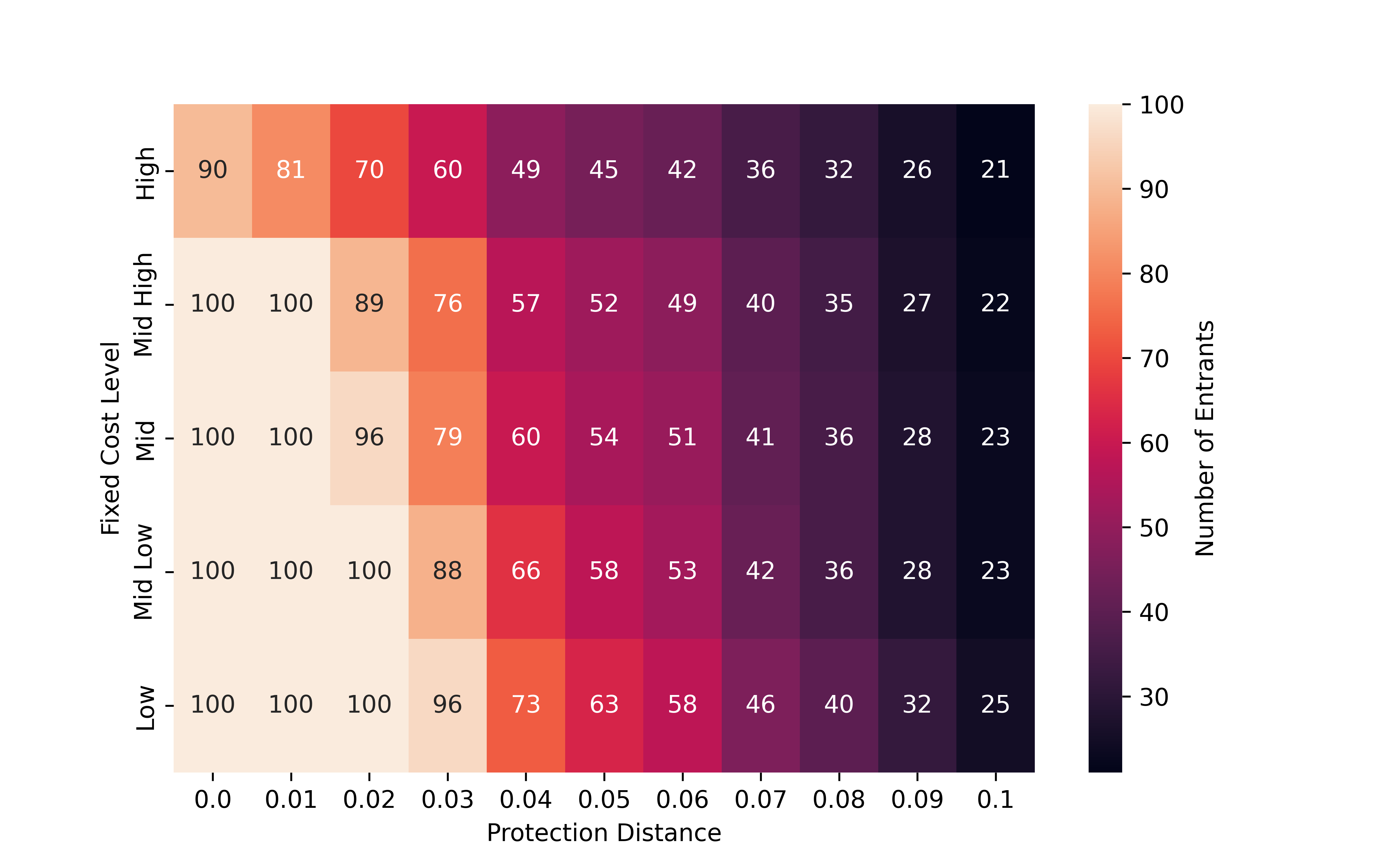}}
    \subfloat[Fixed Costs]{\includegraphics[width=0.5\linewidth]{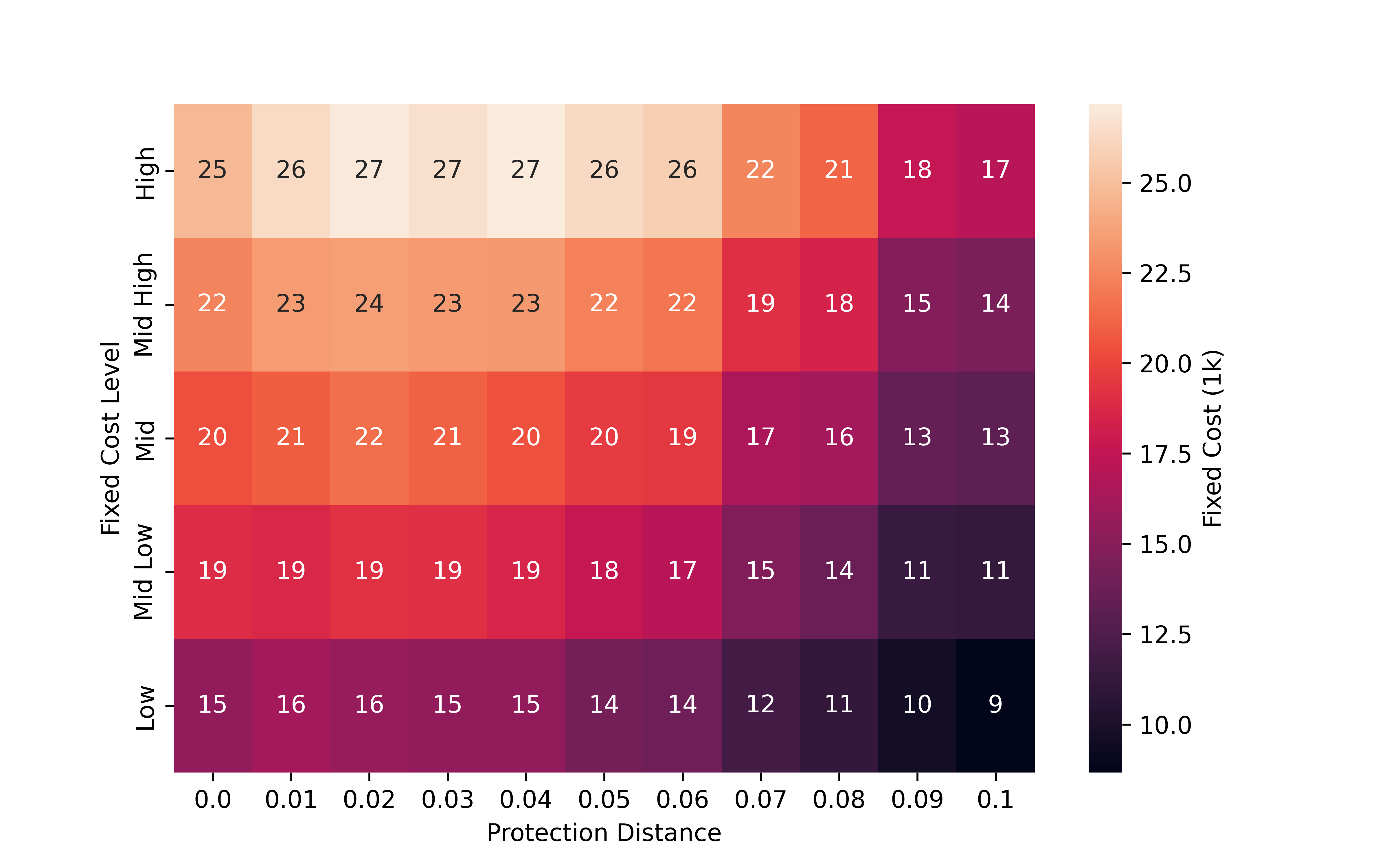}}
    \tabnotes{This figure shows the counterfactual number of entrants and total fixed cost expenditures.}
    \label{fig:simul-protection-fclevel-scenarioB-add}
\end{figure}

\begin{figure}[htbp!]
    \centering
    \caption{Products with Nearby Infringers Given Protection Distance}
    \subfloat[Number of Products]{\includegraphics[width=0.4\linewidth]{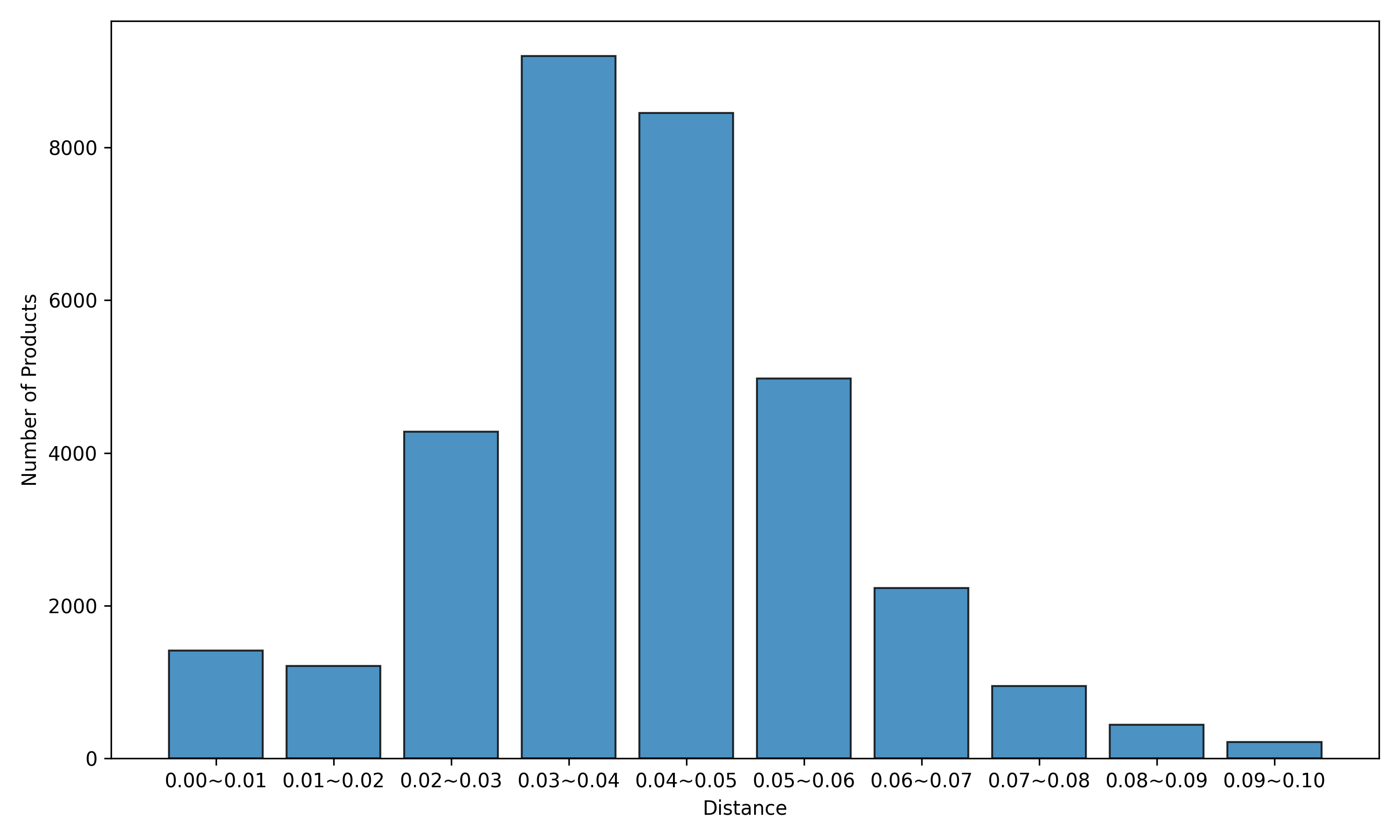}}~~~~~~~~~
    \subfloat[Cumulative Share of Products]{\includegraphics[width=0.4\linewidth]{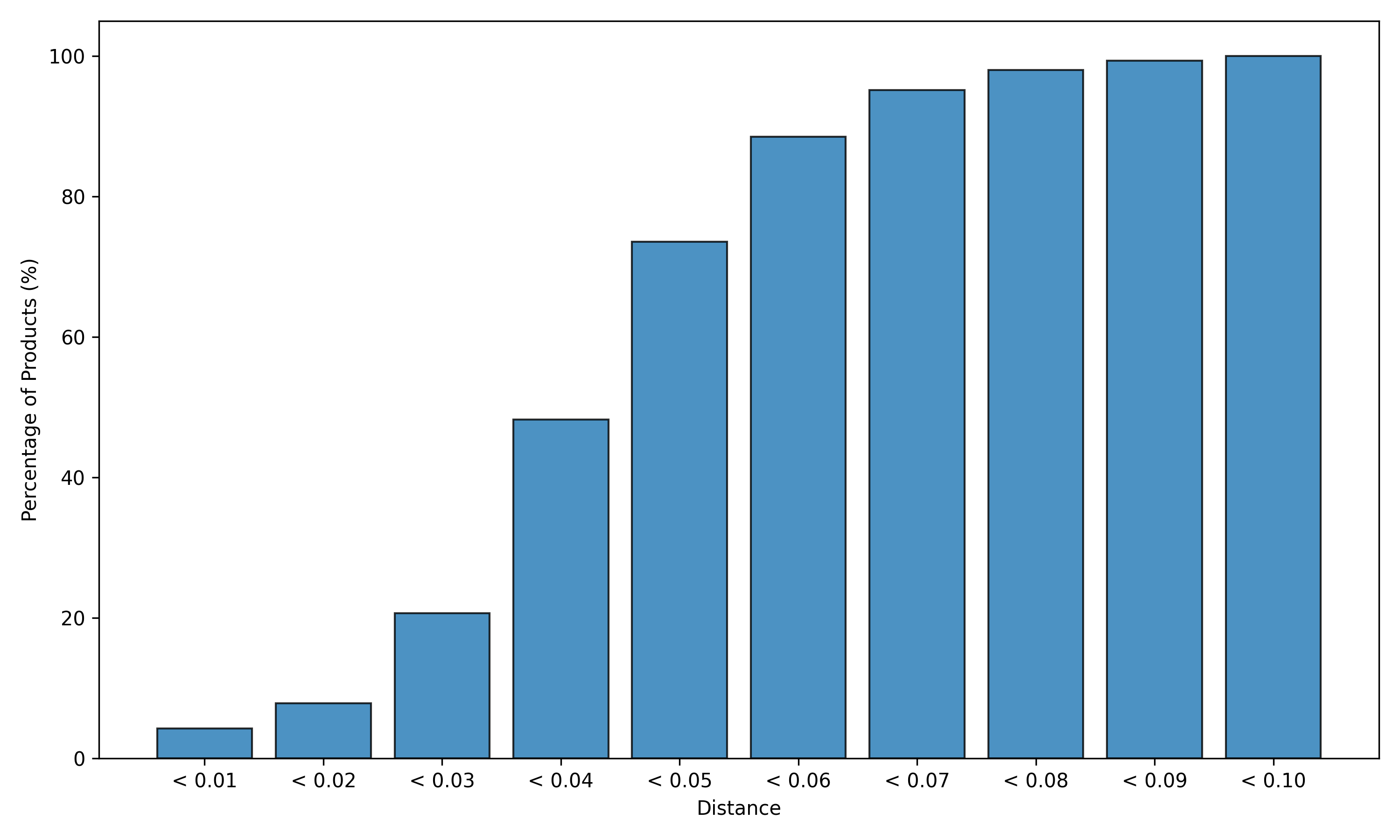}}
    \tabnotes{Panel (a) reports the number of products that have at least one close competitor within a given protection distance (i.e., products with infringers) and Panel (b) depicts the cumulative share of such products.}
    \label{fig:infringer-dist}
\end{figure}

\end{document}